%% file: root.tex
\documentclass[11pt,a4paper]{article}
\usepackage{isabelle,isabellesym}

\usepackage{mathpartir}
\usepackage{amsmath}
\usepackage{amsthm}
\usepackage{natbib}

\usepackage{amssymb}

\usepackage{pdfsetup}

\isabellestyle{it}

\newtheorem{theorem}{Theorem}

\newtheorem{lemma}{Lemma}

\begin{document}

\title{Monotonic References for Gradual Typing}
\author{Jeremy G. Siek and Michael Vitousek}
\maketitle




\input{session}

\bibliographystyle{abbrvnat}
\bibliography{root}

\end{document}

%% file: session.tex
\input{LaTeXsugar.tex}

\input{OptionalSugar.tex}

\input{GTLC_MonoRef_Paper.tex}

\input{GTLC_MonoRef_ECDOnANF.tex}


%% file: LaTeXsugar.tex
%
\begin{isabellebody}%
\def\isabellecontext{LaTeXsugar}%
\isadelimtheory
\isanewline
\endisadelimtheory
\isatagtheory
\endisatagtheory
{\isafoldtheory}%
\isadelimtheory
\endisadelimtheory
\isadelimML
\endisadelimML
\isatagML
\endisatagML
{\isafoldML}%
\isadelimML
\endisadelimML
\isadelimtheory
\endisadelimtheory
\isatagtheory
\endisatagtheory
{\isafoldtheory}%
\isadelimtheory
\endisadelimtheory
\end{isabellebody}%

%% file: OptionalSugar.tex
%
\begin{isabellebody}%
\def\isabellecontext{OptionalSugar}%
\isadelimtheory
\endisadelimtheory
\isatagtheory
\endisatagtheory
{\isafoldtheory}%
\isadelimtheory
\endisadelimtheory
\isadelimML
\endisadelimML
\isatagML
\endisatagML
{\isafoldML}%
\isadelimML
\endisadelimML
\isadelimtheory
\endisadelimtheory
\isatagtheory
\endisatagtheory
{\isafoldtheory}%
\isadelimtheory
\endisadelimtheory
\end{isabellebody}%

%% file: GTLC_MonoRef_Paper.tex
%
\begin{isabellebody}%
\def\isabellecontext{GTLC{\isacharunderscore}MonoRef{\isacharunderscore}Paper}%
\isadelimtheory
\endisadelimtheory
\isatagtheory
\endisatagtheory
{\isafoldtheory}%
\isadelimtheory
\endisadelimtheory
\isamarkupsection{Introduction%
}
\isamarkuptrue%
\begin{isamarkuptext}%
We describe an alternative approach to handling mutable references (aka. pointers)
within a gradually typed language that has different efficiency
characteristics than the prior approach of \cite{Herman:2010aa}. In
particular, we hope to reduce the costs of reading and writing through
references in statically typed regions of code. We would like the
costs to be the same as they would in a statically typed language, that is,
simply the cost of a load or store instruction (for primitive data types).
This reduction in cost is especially important for programmers who
would like to use gradual typing to facilitate transitioning from
a dynamically-typed prototype of an algorithm to a
statically-typed, high-performance implementation.
The programmers we have in mind are scientists and engineers
who currently prototype in Matlab and then manually translate
their algorithms into Fortran.

While our alternative approach succeeds in improving the efficiency
of dereference and updates in statically typed code, it does come
with some limitations. The approach requires all heap-allocated values to be tagged
with their runtime type, which may be an added cost in space (though
many languages require such tags for other reasons).  In addition, our
approach requires all values to be of a uniform size, which is true
for many languages (most functional and object-oriented languages) but
not true for some (the C family of languages). Finally, our approach
is more restrictive than prior ones in that certain usage patterns
are not allowed, triggering runtime exceptions, for which we give
examples later in this introduction.

The source of inefficiency in the prior approach of \cite{Herman:2010aa},
which we refer to as ``guarded references'', is that two kinds of values have
reference type: normal references and guarded references.  A guarded
reference consists of the underlying reference (a memory address) and
two coercions, one to apply when reading and another to apply when
writing. A guarded reference is created during runtime when a normal reference is
casted from one reference type to another.
When a compiler for a language with guarded references
generates code for a dereference or an update, the compiler must emit
code to dispatch on the kind of reference.  Consider the following function
$f$ and two calls to it.
\[
\begin{array}{l}
\mathsf{let}\, f = \lambda  x : \mathsf{Ref}\, \mathsf{Int}.\; !x \; \mathsf{in} \\
\quad  f (\mathsf{ref}_{\mathsf{Int}}\,4); \\
\quad  f (\mathsf{ref}_{\star}\,(\mathsf{true} : \mathsf{Bool} \Rightarrow \star ) : \mathsf{Ref}\, \star \Rightarrow \mathsf{Ref}\, \mathsf{Int})
\end{array}
\]
In the first call to $f$, a normal reference to an integer flows
into the dereference $!x$ whereas in the second call, a guarded reference
flows into the dereference $!x$. The code generated for the dereference in the body
of $f$ needs to be general enough to handle both situations.

Our new approach, called ``monotonic references'', has only one kind
of value at reference type, normal references. When a cast is
applied to a reference, instead of turning the reference into a
guarded reference, we cast the underlying value on the heap, so long
as the cast is to an equally or less dynamic type. Otherwise the cast results in a
runtime error. Thus, the heap is allowed to change monotonically with
respect to the less-dynamic relation. This relation is formally defined
in Section~\ref{sec:types}, but roughly speaking, the fewer occurences
of the dynamic type, the less dynamic a type is. 
\citet{Swamy:2014aa} have independently developed a similar idea, though
in their formulation, the monotonicity is with respect to subtyping
instead of the less dynamic relation, which reflects a difference in goals compared
to this work (security vs. efficiency).

In the above example, a runtime error would be triggered by the cast
from \textsf{Ref} $\star$ to \textsf{Ref Int} because it would attempt
to cast the Boolean \textsf{true} to \textsf{Int} and fail.  On the
other hand, the following example in which an integer is passed into
$f$ in both calls, would terminate without error.
\[
\begin{array}{l}
\mathsf{let}\, f = \lambda  x : \mathsf{Ref}\, \mathsf{Int}.\; !x \; \mathsf{in} \\
\quad  f (\mathsf{ref}_{\mathsf{Int}}\,4); \\
\quad  f (\mathsf{ref}_{\star}\,(4 : \mathsf{Int} \Rightarrow \star ) : \mathsf{Ref}\, \star \Rightarrow \mathsf{Ref}\, \mathsf{Int})
\end{array}
\]
In this example, just prior to the second call to $f$, the cast
from \textsf{Ref} $\star$ to \textsf{Ref Int} would cause the heap
cell containing $4 : \mathsf{Int} \Rightarrow \star$ to be updated with that 
value cast to \textsf{Int}.

In general, monotonic references maintain the invariant that the type of
a value in the heap is at less dynamic than the type of any reference
that points to that value. Thus, if a reference has a fully static type,
such as \textsf{Ref Int}, the corresponding value on the heap must be an actual integer
(and not an injection to $\star$). If a reference does not have a fully static
type, then the corresponding value on the heap might be less dynamic,
and a cast needs to be performed during a read or write to mediate between
the value's type and the reference's type. However, these two situations
can be distinguished during compilation, based on whether the reference 
expression in the dereference or update has a fully static type or not.
Thus, we can generate efficient code for the first case and less efficient
code in the second case.

The behavior of monotonic references is rather different than that of
guarded references. We conjecture that monotonic references are more
picky than guarded references, that is, they trigger runtime errors
strictly more often than guarded references. Here we show two examples
of this phenomenon. In this first example, we cast a reference of type
\textsf{Ref} $\star$ to both \textsf{Ref Int} and \textsf{Ref Bool}.
We dereference at \textsf{Ref Int} then write and subsequently read
at \textsf{Ref Bool}.
\[
\begin{array}{l}
\mathsf{let}\, x = \mathsf{ref}_{\star}\, (4 : \mathsf{Int} \Rightarrow \star ) \,\mathsf{in}\\
\mathsf{let}\, y = x : \mathsf{Ref}\,\star  \Rightarrow \mathsf{Ref}\,\mathsf{Int} \,\mathsf{and}\\
\quad    z = x : \mathsf{Ref}\,\star \Rightarrow \mathsf{Ref}\, \mathsf{Bool} \,\mathsf{in}\\
\quad  !y; z := \mathsf{true}; !z
\end{array}
\]
With guarded references, the above program terminates without error whereas
with monotonic references, it halts with an error during the cast
from \textsf{Ref} $\star$ to \textsf{Ref Bool}. Of course, if we re-order the
sequence of operations so that the write of the Boolean occurs before
the read of the integer, then the guarded references approach triggers
an error on the read from $y$.
\[
\begin{array}{l}
\ldots z := \mathsf{true}; !y; !z
\end{array}
\]

It is worth emphasizing that when a cast on a reference causes a value
on the heap to be changed, the change is rather permanent.
Consider the following example in which a function $f$ takes
a reference of type \textsf{Ref} $\star$ and casts it to \textsf{Ref Int}.
The caller passes in a reference to an injected integer, which works fine,
but then after the call, tries to write an injected Boolean to the reference.
\[
\begin{array}{l}
\mathsf{fun}\, f(y : \mathsf{Ref}\,\star ) = \\
\quad  \mathsf{let}\, z = (y : \mathsf{Ref}\,\star \Rightarrow \mathsf{Ref}\,\mathsf{Int})\,\mathsf{in} \, !z \\
\\
\mathsf{let}\, x = \mathsf{ref}_{\star} \, (4 : \mathsf{Int} \Rightarrow \star )\, \mathsf{in} \\
\quad  f \, x; \\
\quad  x := (\mathsf{true} : \mathsf{Bool} \Rightarrow \star)
\end{array}
\]
With guarded references, the above program terminates without error,
whereas with monotonic references, the write to $x$ triggers
an error.

One interesting challenged in defining the dynamic semantics of
monotonic references is that references may form cycles and we need to
make sure that a cast applied to a reference that is in a cyle does
not cause the program to diverge. Consider the following example in which
we create a pair whose second element is a reference back
to itself.
\[
\begin{array}{l}
\mathsf{let}\,r_1 = \mathsf{ref}_{\mathsf{Int}\times\star}\,(\langle 42,0 : \mathsf{Int} \Rightarrow\star \rangle) \, \mathsf{in} \\
\quad r_1 := \langle 42, r_1 : \mathsf{Ref}(\mathsf{Int}\times \star) \Rightarrow \star \rangle; \\
\quad \mathsf{let}\,r_2 = r_1 : \mathsf{Ref}(\mathsf{Int}\times \star) \Rightarrow \mathsf{Ref}(\mathsf{Int} \times \mathsf{Ref}\,\star) \, \mathsf{in} \\
\quad\quad \mathsf{fst}(!r_2)
\end{array}
\]
Once the pair with the cycle is created, we cast the reference to it from
type $\mathsf{Ref}(\mathsf{Int} \times \star)$ to
$\mathsf{Ref}(\mathsf{Int} \times \mathsf{Ref}\,\star)$.  The correct
result of this program is $42$ but in a naive dynamic semantics
this program would diverge. Our semantics avoids divergence by
checking whether the new type for a heap cell is no less dynamic than
the old type; in such cases the heap cell is left unchanged.

The rest of this paper gives a formal definition of the static and
dynamic semantics for monotonic references and proves type safety. The
formal setting for this work is in an intermediate language that is an
extension of the simply-typed lambda calculus with casts, a dynamic
type, and mutable references. It is straightforward using standard
techniques to compile from a gradually-typed source language to this
intermediate language.%
\end{isamarkuptext}%
\isamarkuptrue%
\isamarkupsection{Types%
}
\isamarkuptrue%
\begin{isamarkuptext}%
\label{sec:types}

For the purposes of studying monotonic references, the types of our language
consist of integers, Booleans, functions, pairs, references, and the dynamic type.
\[
\begin{array}{ll}
\mathrm{types} & A,B,C,D ::= \mathsf{Int} \mid \mathsf{Bool} \mid A \to B \mid A \times B \mid
   \mathsf{Ref}\,A \mid  \star
\end{array}
\]

The less or equally dynamic relation on types is defined by the following equations
for its characteristic function. (This relation is also known as naive subtyping.)
\begin{center}
\begin{tabular}{l@ {~~\isa{{\isacharequal}}~~}l}
\isa{A\ {\isasymsqsubseteq}\ {\isasymstar}} & \isa{True} \\
\isa{\textsf{Int}\ {\isasymsqsubseteq}\ \textsf{Int}} & \isa{True} \\
\isa{\textsf{Bool}\ {\isasymsqsubseteq}\ \textsf{Bool}} & \isa{True} \\
\isa{A\ {\isasymtimes}\ B\ {\isasymsqsubseteq}\ C\ {\isasymtimes}\ D} & \isa{A\ {\isasymsqsubseteq}\ C\ {\isasymand}\ B\ {\isasymsqsubseteq}\ D} \\
\isa{A\ {\isasymrightarrow}\ B\ {\isasymsqsubseteq}\ C\ {\isasymrightarrow}\ D} & \isa{A\ {\isasymsqsubseteq}\ C\ {\isasymand}\ B\ {\isasymsqsubseteq}\ D} \\
\isa{{\isacharparenleft}\textsf{Ref}\ A{\isacharparenright}\ {\isasymsqsubseteq}\ {\isacharparenleft}\textsf{Ref}\ B{\isacharparenright}} & \isa{A\ {\isasymsqsubseteq}\ B} 
\end{tabular}
\end{center}
In all other cases, the less or equally dynamic function returns False.

\begin{lemma}
\isa{A\ {\isasymsqsubseteq}\ A}
\end{lemma}

\begin{lemma}
\isa{{\normalsize{}If\,}\ \mbox{A\ {\isasymsqsubseteq}\ B}\ {\normalsize \,and\,}\ \mbox{B\ {\isasymsqsubseteq}\ C}\ {\normalsize \,then\,}\ A\ {\isasymsqsubseteq}\ C{\isachardot}}
\end{lemma}%
\end{isamarkuptext}%
\isamarkuptrue%
\isadelimproof
\endisadelimproof
\isatagproof
\endisatagproof
{\isafoldproof}%
\isadelimproof
\endisadelimproof
\begin{isamarkuptext}%
The meet function on types is defined below.
(This corresponds to the meet operator of \citet{Siek:2010ya}.) 
Many of the function definitions in this development
use monadic notation in which the combination of
:= and semicolon serve as the notation for the bind operation.
\begin{center}
\begin{tabular}{l@ {~~\isa{{\isacharequal}}~~}l}
\isa{{\isasymstar}\ {\isasymsqinter}\ A} & \isa{\textit{return}\ A} \\
\isa{A\ {\isasymsqinter}\ {\isasymstar}} & \isa{\textit{return}\ A} \\
\isa{\textsf{Int}\ {\isasymsqinter}\ \textsf{Int}} & \isa{\textit{return}\ \textsf{Int}} \\
\isa{\textsf{Bool}\ {\isasymsqinter}\ \textsf{Bool}} & \isa{\textit{return}\ \textsf{Bool}} \\
\isa{A\ {\isasymtimes}\ B\ {\isasymsqinter}\ C\ {\isasymtimes}\ D} & \isa{A{\isacharprime}\ {\isacharcolon}{\isacharequal}\ A\ {\isasymsqinter}\ C{\isacharsemicolon}\ B{\isacharprime}\ {\isacharcolon}{\isacharequal}\ B\ {\isasymsqinter}\ D{\isacharsemicolon}\ \textit{return}\ A{\isacharprime}\ {\isasymtimes}\ B{\isacharprime}} \\
\isa{A\ {\isasymrightarrow}\ B\ {\isasymsqinter}\ C\ {\isasymrightarrow}\ D} & \isa{A{\isacharprime}\ {\isacharcolon}{\isacharequal}\ A\ {\isasymsqinter}\ C{\isacharsemicolon}\ B{\isacharprime}\ {\isacharcolon}{\isacharequal}\ B\ {\isasymsqinter}\ D{\isacharsemicolon}\ \textit{return}\ A{\isacharprime}\ {\isasymrightarrow}\ B{\isacharprime}} \\
\isa{\textsf{Ref}\ A\ {\isasymsqinter}\ \textsf{Ref}\ B} & \isa{A{\isacharprime}\ {\isacharcolon}{\isacharequal}\ A\ {\isasymsqinter}\ B{\isacharsemicolon}\ \textit{return}\ \textsf{Ref}\ A{\isacharprime}} 
\end{tabular}
\end{center}
In all other cases, the meet function returns a cast error.

\begin{lemma}
\isa{{\normalsize{}If\,}\ A\ {\isasymsqinter}\ B\ {\isacharequal}\ C\ {\normalsize \,then\,}\ C\ {\isasymsqsubseteq}\ A\ {\isasymand}\ C\ {\isasymsqsubseteq}\ B{\isachardot}}
\end{lemma}

We say that a type is ``static'' if it does not contain the
dynamic type.
\begin{center}
\begin{tabular}{l@ {~~\isa{{\isacharequal}}~~}l}
\isa{static\ {\isasymstar}} & \isa{False} \\
\isa{static\ \textsf{Int}} & \isa{True} \\
\isa{static\ \textsf{Bool}} & \isa{True} \\
\isa{static\ {\isacharparenleft}A\ {\isasymtimes}\ B{\isacharparenright}} & \isa{static\ A\ {\isasymand}\ static\ B} \\
\isa{static\ {\isacharparenleft}A\ {\isasymrightarrow}\ B{\isacharparenright}} & \isa{static\ A\ {\isasymand}\ static\ B} \\
\isa{static\ {\isacharparenleft}\textsf{Ref}\ A{\isacharparenright}} & \isa{static\ A} 
\end{tabular}
\end{center}

Static types are the least dynamic.
\begin{lemma}[Static is Least Dynamic]
\label{lem:static-most-precise}
\isa{{\normalsize{}If\,}\ \mbox{static\ A}\ {\normalsize \,and\,}\ \mbox{B\ {\isasymsqsubseteq}\ A}\ {\normalsize \,then\,}\ A\ {\isacharequal}\ B{\isachardot}}
\end{lemma}%
\end{isamarkuptext}%
\isamarkuptrue%
\isamarkupsection{Association Lists and Type Environments%
}
\isamarkuptrue%
\begin{isamarkuptext}%
We represent environments, type environments, and heap typings as association lists.
\begin{center}
\begin{tabular}{l@ {~~\isa{{\isacharequal}}~~}l}
\isa{lookup\ x\ {\isacharbrackleft}{\isacharbrackright}} & \isa{stuck} \\
\isa{lookup\ x\ {\isacharparenleft}{\isacharparenleft}y{\isacharcomma}\ v{\isacharparenright}{\isasymcdot}bs{\isacharparenright}} & \isa{\textsf{if}\ x\ {\isacharequal}\ y\ \textsf{then}\ \textit{return}\ v\ \textsf{else}\ lookup\ x\ bs} 
\end{tabular}
\end{center}

The domain of an association list is the set keys.
\begin{center}
\isa{dom\ A\ {\isasymequiv}\ map\ fst\ A}
\end{center}

A heap typing is less or equally dynamic as another heap typing if
each of its components are.

\begin{center}
\isa{{\isasymSigma}{\isacharprime}\ {\isasymsqsubseteq}\ {\isasymSigma}\ {\isasymequiv}\ dom\ {\isasymSigma}\ {\isacharequal}\ dom\ {\isasymSigma}{\isacharprime}\ {\isasymand}\ {\isacharparenleft}{\isasymforall}a\ A{\isachardot}\ lookup\ a\ {\isasymSigma}\ {\isacharequal}\ A\ {\isasymlongrightarrow}\ {\isacharparenleft}{\isasymexists}B{\isachardot}\ lookup\ a\ {\isasymSigma}{\isacharprime}\ {\isacharequal}\ B\ {\isasymand}\ B\ {\isasymsqsubseteq}\ A{\isacharparenright}{\isacharparenright}}
\end{center}

This ordering relation is transitive.
\begin{lemma}[Transitive]
\isa{{\normalsize{}If\,}\ \mbox{{\isasymSigma}{\isacharprime}\ {\isasymsqsubseteq}\ {\isasymSigma}{\isacharprime}{\isacharprime}}\ {\normalsize \,and\,}\ \mbox{{\isasymSigma}\ {\isasymsqsubseteq}\ {\isasymSigma}{\isacharprime}}\ {\normalsize \,then\,}\ {\isasymSigma}\ {\isasymsqsubseteq}\ {\isasymSigma}{\isacharprime}{\isacharprime}{\isachardot}}
\end{lemma}%
\end{isamarkuptext}%
\isamarkuptrue%
\isamarkupsection{Type Sytem%
}
\isamarkuptrue%
\begin{isamarkuptext}%
\begin{figure}[tbp]
Typing rules for expressions
\begin{center}
\isa{\mbox{}\inferrule{\mbox{lookup\ x\ {\isasymGamma}\ {\isacharequal}\ A}}{\mbox{{\isasymGamma}\ {\isasymturnstile}\ x\ {\isacharcolon}\ A}}}\qquad
\isa{{\isasymGamma}\ {\isasymturnstile}\ c\ {\isacharcolon}\ typeof\ c}\\[2ex]
\isa{\mbox{}\inferrule{\mbox{typeof{\isacharunderscore}opr\ f\ {\isacharequal}\ A\ {\isasymrightarrow}\ B}\\\ \mbox{{\isasymGamma}\ {\isasymturnstile}\ e\ {\isacharcolon}\ A}}{\mbox{{\isasymGamma}\ {\isasymturnstile}\ f(e)\ {\isacharcolon}\ B}}}\qquad
\isa{\mbox{}\inferrule{\mbox{{\isasymGamma}\ {\isasymturnstile}\ e{\isadigit{1}}\ {\isacharcolon}\ A}\\\ \mbox{{\isasymGamma}\ {\isasymturnstile}\ e{\isadigit{2}}\ {\isacharcolon}\ B}}{\mbox{{\isasymGamma}\ {\isasymturnstile}\ {\isasymlangle}e{\isadigit{1}}{\isacharcomma}e{\isadigit{2}}{\isasymrangle}\ {\isacharcolon}\ A\ {\isasymtimes}\ B}}}\\[2ex]
\isa{\mbox{}\inferrule{\mbox{{\isacharparenleft}x{\isacharcomma}\ A{\isacharparenright}{\isasymcdot}{\isasymGamma}\ {\isasymturnstile}\ s\ {\isacharcolon}\ B}}{\mbox{{\isasymGamma}\ {\isasymturnstile}\ {\isasymlambda}x{\isacharcolon}A{\isachardot}\ s\ {\isacharcolon}\ A\ {\isasymrightarrow}\ B}}}\qquad
\isa{\mbox{}\inferrule{\mbox{{\isasymGamma}\ {\isasymturnstile}\ e\ {\isacharcolon}\ \textsf{Ref}\ A}\\\ \mbox{static\ A}}{\mbox{{\isasymGamma}\ {\isasymturnstile}\ {\isacharbang}e\ {\isacharcolon}\ A}}}
\end{center}
Typing rules for statements
\begin{center}
\isa{\mbox{}\inferrule{\mbox{{\isasymGamma}\ {\isasymturnstile}\ e\ {\isacharcolon}\ A}\\\ \mbox{{\isacharparenleft}x{\isacharcomma}\ A{\isacharparenright}{\isasymcdot}{\isasymGamma}\ {\isasymturnstile}\ s\ {\isacharcolon}\ B}}{\mbox{{\isasymGamma}\ {\isasymturnstile}\ \textsf{let}\ x{\isacharequal}e\ \textsf{in}\ s\ {\isacharcolon}\ B}}}\qquad
\isa{\mbox{}\inferrule{\mbox{{\isasymGamma}\ {\isasymturnstile}\ e\ {\isacharcolon}\ A}}{\mbox{{\isasymGamma}\ {\isasymturnstile}\ \textsf{return}\ e\ {\isacharcolon}\ A}}}\\[2ex]
\isa{\mbox{}\inferrule{\mbox{{\isasymGamma}\ {\isasymturnstile}\ e\ {\isacharcolon}\ A\ {\isasymrightarrow}\ B}\\\ \mbox{{\isasymGamma}\ {\isasymturnstile}\ e{\isacharprime}\ {\isacharcolon}\ A}\\\ \mbox{{\isacharparenleft}x{\isacharcomma}\ B{\isacharparenright}{\isasymcdot}{\isasymGamma}\ {\isasymturnstile}\ s\ {\isacharcolon}\ C}}{\mbox{{\isasymGamma}\ {\isasymturnstile}\ \textsf{let}\ x{\isacharequal}e(e{\isacharprime})\ \textsf{in}\ s\ {\isacharcolon}\ C}}}\\[2ex]
\isa{\mbox{}\inferrule{\mbox{{\isasymGamma}\ {\isasymturnstile}\ e\ {\isacharcolon}\ A\ {\isasymrightarrow}\ B}\\\ \mbox{{\isasymGamma}\ {\isasymturnstile}\ e{\isacharprime}\ {\isacharcolon}\ A}}{\mbox{{\isasymGamma}\ {\isasymturnstile}\ \textsf{return}\ e(e{\isacharprime})\ {\isacharcolon}\ B}}}\\[2ex]
\isa{\mbox{}\inferrule{\mbox{{\isasymGamma}\ {\isasymturnstile}\ e\ {\isacharcolon}\ A}\\\ \mbox{{\isacharparenleft}x{\isacharcomma}\ \textsf{Ref}\ A{\isacharparenright}{\isasymcdot}{\isasymGamma}\ {\isasymturnstile}\ s\ {\isacharcolon}\ B}}{\mbox{{\isasymGamma}\ {\isasymturnstile}\ \textsf{let}\ x{\isacharequal}\ \textsf{ref}\isactrlbsub A\ \isactrlesub e\ \textsf{in}\ s\ {\isacharcolon}\ B}}}\\[2ex]
\isa{\mbox{}\inferrule{\mbox{{\isasymGamma}\ {\isasymturnstile}\ e\ {\isacharcolon}\ \textsf{Ref}\ A}\\\ \mbox{static\ A}\\\ \mbox{{\isasymGamma}\ {\isasymturnstile}\ e{\isacharprime}\ {\isacharcolon}\ A}\\\ \mbox{{\isasymGamma}\ {\isasymturnstile}\ s\ {\isacharcolon}\ B}}{\mbox{{\isasymGamma}\ {\isasymturnstile}\ e{\isacharcolon}{\isacharequal}e{\isacharprime}{\isacharsemicolon}\ s\ {\isacharcolon}\ B}}}\\[2ex]
\isa{\mbox{}\inferrule{\mbox{{\isasymGamma}\ {\isasymturnstile}\ e\ {\isacharcolon}\ \textsf{Ref}\ A}\\\ \mbox{{\isasymGamma}\ {\isasymturnstile}\ e{\isacharprime}\ {\isacharcolon}\ A}\\\ \mbox{{\isasymGamma}\ {\isasymturnstile}\ s\ {\isacharcolon}\ B}}{\mbox{{\isasymGamma}\ {\isasymturnstile}\ e{\isacharcolon}{\isacharequal}e{\isacharprime}{\isacharat}A{\isacharsemicolon}\ s\ {\isacharcolon}\ B}}}\\[2ex]
\isa{\mbox{}\inferrule{\mbox{{\isasymGamma}\ {\isasymturnstile}\ e\ {\isacharcolon}\ A}\\\ \mbox{{\isacharparenleft}x{\isacharcomma}\ B{\isacharparenright}{\isasymcdot}{\isasymGamma}\ {\isasymturnstile}\ s\ {\isacharcolon}\ C}}{\mbox{{\isasymGamma}\ {\isasymturnstile}\ \textsf{let}\ x{\isacharequal}e{\isacharcolon}A{\isasymRightarrow}B\ \textsf{in}\ s\ {\isacharcolon}\ C}}}\\[2ex]
\isa{\mbox{}\inferrule{\mbox{{\isasymGamma}\ {\isasymturnstile}\ e\ {\isacharcolon}\ \textsf{Ref}\ A}\\\ \mbox{{\isacharparenleft}x{\isacharcomma}\ A{\isacharparenright}{\isasymcdot}{\isasymGamma}\ {\isasymturnstile}\ s\ {\isacharcolon}\ B}}{\mbox{{\isasymGamma}\ {\isasymturnstile}\ \textsf{let}\ x{\isacharequal}{\isacharbang}e{\isacharat}A\ \textsf{in}\ s\ {\isacharcolon}\ B}}}
\end{center}
\caption{Typing rules for expressions and statements}
\label{fig:wt-stmt}
\end{figure}

Figure~\ref{fig:wt-stmt} defines the typing rules for expressions and statements.
The separation of expressions and statements achieves a kind of A-normal form
that we found convenient for the purposes of proving type safety. In particular,
it avoids the need for evaluation contexts. The expressions include only
trivially terminating operations that do not update the heap.

There are two syntactic forms for reference update and dereference,
respectively.  One pair of forms requires the reference type to be
``static'' and the other pair includes a type annotation that records
the type of the reference.  We refer to these later forms as the
``dynamic'' version of reference update and dereference. The dynamic
forms use the type annotation to cast from the runtime type of a value
on the heap to the annotated type of the reference. The dynamic
dereference is a statement and not an expression because it performs a
cast which is a side-effecting operation.

\begin{figure}[tbp]
Typing rules for values
\begin{center}
\isa{\mbox{}\inferrule{\mbox{typeof\ c\ {\isacharequal}\ A}}{\mbox{{\isasymSigma}\ {\isasymturnstile}\ c\ {\isacharcolon}\ A}}}\qquad
\isa{\mbox{}\inferrule{\mbox{{\isasymSigma}\ {\isasymturnstile}\ v\ {\isacharcolon}\ A}\\\ \mbox{{\isasymSigma}\ {\isasymturnstile}\ v{\isacharprime}\ {\isacharcolon}\ B}}{\mbox{{\isasymSigma}\ {\isasymturnstile}\ {\isasymlangle}v{\isacharcomma}v{\isacharprime}{\isasymrangle}\ {\isacharcolon}\ A\ {\isasymtimes}\ B}}}\\[2ex]
\isa{\mbox{}\inferrule{\mbox{{\isasymGamma}{\isacharsemicolon}{\isasymSigma}\ {\isasymturnstile}\ {\isasymrho}}\\\ \mbox{{\isacharparenleft}x{\isacharcomma}\ A{\isacharparenright}{\isasymcdot}{\isasymGamma}\ {\isasymturnstile}\ s\ {\isacharcolon}\ B}}{\mbox{{\isasymSigma}\ {\isasymturnstile}\ {\isasymlangle}{\isasymlambda}x{\isacharcolon}A{\isachardot}s{\isacharcomma}\ {\isasymrho}{\isasymrangle}\ {\isacharcolon}\ A\ {\isasymrightarrow}\ B}}}\qquad
\isa{\mbox{}\inferrule{\mbox{lookup\ a\ {\isasymSigma}\ {\isacharequal}\ A}\\\ \mbox{A\ {\isasymsqsubseteq}\ B}}{\mbox{{\isasymSigma}\ {\isasymturnstile}\ \textsf{ref}\ a\ {\isacharcolon}\ \textsf{Ref}\ B}}}\\[2ex]
\isa{\mbox{}\inferrule{\mbox{{\isasymSigma}\ {\isasymturnstile}\ v\ {\isacharcolon}\ A}}{\mbox{{\isasymSigma}\ {\isasymturnstile}\ {\isacharbrackleft}v{\isacharcolon}A{\isasymRightarrow}{\isasymstar}{\isacharbrackright}\isactrlbsup \isactrlesup \ {\isacharcolon}\ {\isasymstar}}}}
\end{center}
Typing rules for environments
\begin{center}
\isa{{\isacharbrackleft}{\isacharbrackright}{\isacharsemicolon}{\isasymSigma}\ {\isasymturnstile}\ {\isacharbrackleft}{\isacharbrackright}}\qquad
\isa{\mbox{}\inferrule{\mbox{{\isasymSigma}\ {\isasymturnstile}\ v\ {\isacharcolon}\ A}\\\ \mbox{{\isasymGamma}{\isacharsemicolon}{\isasymSigma}\ {\isasymturnstile}\ {\isasymrho}}}{\mbox{{\isacharparenleft}x{\isacharcomma}\ A{\isacharparenright}{\isasymcdot}{\isasymGamma}{\isacharsemicolon}{\isasymSigma}\ {\isasymturnstile}\ {\isacharparenleft}x{\isacharcomma}\ v{\isacharparenright}{\isasymcdot}{\isasymrho}}}}
\end{center}
Typing rules for procedure call stacks
\begin{center}
\isa{{\isasymSigma}\ {\isasymturnstile}\ {\isacharbrackleft}{\isacharbrackright}\ {\isacharcolon}\ A\ {\isasymRightarrow}\ A}\\[2ex]
\isa{\mbox{}\inferrule{\mbox{{\isasymGamma}{\isacharsemicolon}{\isasymSigma}\ {\isasymturnstile}\ {\isasymrho}}\\\ \mbox{{\isacharparenleft}x{\isacharcomma}\ A{\isacharparenright}{\isasymcdot}{\isasymGamma}\ {\isasymturnstile}\ s\ {\isacharcolon}\ B}\\\ \mbox{{\isasymSigma}\ {\isasymturnstile}\ k\ {\isacharcolon}\ B\ {\isasymRightarrow}\ C}}{\mbox{{\isasymSigma}\ {\isasymturnstile}\ {\isacharparenleft}x{\isacharcomma}\ s{\isacharcomma}\ {\isasymrho}{\isacharparenright}{\isasymcdot}k\ {\isacharcolon}\ A\ {\isasymRightarrow}\ C}}}
\end{center}
Typing rules for casted values
\begin{center}
\isa{\mbox{}\inferrule{\mbox{{\isasymSigma}\ {\isasymturnstile}\ v\ {\isacharcolon}\ A}}{\mbox{{\isasymSigma}\ {\isasymturnstile}\ val\ v\ {\isacharcolon}\ A}}}\qquad
\isa{\mbox{}\inferrule{\mbox{{\isasymSigma}\ {\isasymturnstile}\ v\ {\isacharcolon}\ A}\\\ \mbox{B\ {\isasymsqsubseteq}\ A}}{\mbox{{\isasymSigma}\ {\isasymturnstile}\ v{\isacharcolon}A{\isasymRightarrow}B\isactrlbsup \isactrlesup \ {\isacharcolon}\ B}}}
\end{center}
Well-typed heaps
\begin{center}
\isa{{\isasymSigma}\ {\isasymturnstile}\ {\isasymmu}\ {\isacharbar}\ as\ {\isasymequiv}\ {\isacharparenleft}{\isasymforall}a\ A{\isachardot}\ lookup\ a\ {\isasymSigma}\ {\isacharequal}\ A\ {\isasymlongrightarrow}\ {\isacharparenleft}{\isasymexists}cv{\isachardot}\ lookup\ a\ {\isasymmu}\ {\isacharequal}\ {\isacharparenleft}cv{\isacharcomma}\ A{\isacharparenright}\ {\isasymand}\ {\isasymSigma}\ {\isasymturnstile}\ cv\ {\isacharcolon}\ A\ {\isasymand}\ {\isacharparenleft}a\ {\isasymnotin}\ as\ {\isasymlongrightarrow}\ {\isacharparenleft}{\isasymexists}v{\isachardot}\ cv\ {\isacharequal}\ val\ v{\isacharparenright}{\isacharparenright}{\isacharparenright}{\isacharparenright}\ {\isasymand}\ {\isacharparenleft}{\isasymforall}a{\isachardot}\ a\ {\isasymin}\ dom\ {\isasymmu}\ {\isasymlongrightarrow}\ a\ {\isacharless}\ {\isacharbar}{\isasymmu}{\isacharbar}{\isacharparenright}\ {\isasymand}\ as\ {\isasymsubseteq}\ dom\ {\isasymSigma}}
\end{center}
Well-typed states
\begin{center}
\isa{\mbox{}\inferrule{\mbox{{\isasymSigma}\ {\isasymturnstile}\ {\isasymmu}\ {\isacharbar}\ as}\\\ \mbox{{\isasymGamma}{\isacharsemicolon}{\isasymSigma}\ {\isasymturnstile}\ {\isasymrho}}\\\ \mbox{{\isasymGamma}\ {\isasymturnstile}\ s\ {\isacharcolon}\ A}\\\ \mbox{{\isasymSigma}\ {\isasymturnstile}\ k\ {\isacharcolon}\ A\ {\isasymRightarrow}\ B}}{\mbox{{\isasymturnstile}\ {\isacharparenleft}s{\isacharcomma}\ {\isasymrho}{\isacharcomma}\ k{\isacharcomma}\ {\isasymmu}{\isacharcomma}\ as{\isacharparenright}\ {\isacharcolon}\ B}}}
\end{center}

\caption{Typing rules for run-time structures.}
\label{fig:wt-values}
\end{figure}

Figure~\ref{fig:wt-values} defines the typing rules for run-time
structures such as values, environments, stacks, heaps, and states.
Most of these typing rules are straightforward and only a few require
comment. The typing rule for references allows the type of the
reference to be more dynamic than the type in the heap.  Our
heaps are unusual in that they do not only store values but sometimes
also store casted values.  We require the values and casted values in
a well-typed heap to have the types given by the heap typing.  Also,
only those addresses in the active address list may contain casted
values. The rest must contain (uncasted) values.  The typing rule for
casted values requires the target of the cast to be at 
less dynamic than the source, reflecting the invariant that the heap is only
allowed to become less dynamic.

Variable lookup always succeeds in well-typed environments.
\begin{lemma}[Lookup Safety]
\label{lem:lookup-safety}
\isa{{\normalsize{}If\,}\ \mbox{{\isasymGamma}{\isacharsemicolon}{\isasymSigma}\ {\isasymturnstile}\ {\isasymrho}}\ {\normalsize \,and\,}\ \mbox{lookup\ x\ {\isasymGamma}\ {\isacharequal}\ A}\ {\normalsize \,then\,}\ {\isasymexists}v{\isachardot}\ lookup\ x\ {\isasymrho}\ {\isacharequal}\ v\ {\isasymand}\ {\isasymSigma}\ {\isasymturnstile}\ v\ {\isacharcolon}\ A{\isachardot}}
\end{lemma}%
\end{isamarkuptext}%
\isamarkuptrue%
\isadelimproof
\endisadelimproof
\isatagproof
\endisatagproof
{\isafoldproof}%
\isadelimproof
\endisadelimproof
\isadelimproof
\endisadelimproof
\isatagproof
\endisatagproof
{\isafoldproof}%
\isadelimproof
\endisadelimproof
\isadelimproof
\endisadelimproof
\isatagproof
\endisatagproof
{\isafoldproof}%
\isadelimproof
\endisadelimproof
\isadelimproof
\endisadelimproof
\isatagproof
\endisatagproof
{\isafoldproof}%
\isadelimproof
\endisadelimproof
\begin{isamarkuptext}%
We can weaken values and environments with respect to the
typing environment for term variables.

\begin{lemma}[Weaken Values]\ \\
\isa{{\normalsize{}If\,}\ \mbox{{\isasymSigma}\ {\isasymturnstile}\ v\ {\isacharcolon}\ A}\ {\normalsize \,and\,}\ \mbox{a\ {\isasymnotin}\ dom\ {\isasymSigma}}\ {\normalsize \,then\,}\ {\isacharparenleft}a{\isacharcomma}\ B{\isacharparenright}{\isasymcdot}{\isasymSigma}\ {\isasymturnstile}\ v\ {\isacharcolon}\ A{\isachardot}}
\end{lemma}

\begin{lemma}[Weaken Environments]\ \\
\isa{{\normalsize{}If\,}\ \mbox{{\isasymGamma}{\isacharsemicolon}{\isasymSigma}\ {\isasymturnstile}\ {\isasymrho}}\ {\normalsize \,and\,}\ \mbox{a\ {\isasymnotin}\ dom\ {\isasymSigma}}\ {\normalsize \,then\,}\ {\isasymGamma}{\isacharsemicolon}{\isacharparenleft}a{\isacharcomma}\ B{\isacharparenright}{\isasymcdot}{\isasymSigma}\ {\isasymturnstile}\ {\isasymrho}{\isachardot}}
\end{lemma}

We can strengthen values and environments with respect 
to the typing of the heap because the typing rule for 
addresses allows the heap-type to be less dynamic
than the static type of the reference.

\begin{lemma}[Strengthen Values]\ \\
\isa{{\normalsize{}If\,}\ \mbox{{\isasymSigma}\ {\isasymturnstile}\ v\ {\isacharcolon}\ A}\ {\normalsize \,and\,}\ \mbox{{\isasymSigma}{\isacharprime}\ {\isasymsqsubseteq}\ {\isasymSigma}}\ {\normalsize \,then\,}\ {\isasymSigma}{\isacharprime}\ {\isasymturnstile}\ v\ {\isacharcolon}\ A{\isachardot}}
\end{lemma}

\begin{lemma}[Strengthen Environments]\ \\
\isa{{\normalsize{}If\,}\ \mbox{{\isasymGamma}{\isacharsemicolon}{\isasymSigma}\ {\isasymturnstile}\ {\isasymrho}}\ {\normalsize \,and\,}\ \mbox{{\isasymSigma}{\isacharprime}\ {\isasymsqsubseteq}\ {\isasymSigma}}\ {\normalsize \,then\,}\ {\isasymGamma}{\isacharsemicolon}{\isasymSigma}{\isacharprime}\ {\isasymturnstile}\ {\isasymrho}{\isachardot}}
\end{lemma}%
\end{isamarkuptext}%
\isamarkuptrue%
\begin{isamarkuptext}%
We can weaken and strengthen stacks as well.

\begin{lemma}[Weaken Stacks]\ \\
\isa{{\normalsize{}If\,}\ \mbox{{\isasymSigma}\ {\isasymturnstile}\ k\ {\isacharcolon}\ A\ {\isasymRightarrow}\ B}\ {\normalsize \,and\,}\ \mbox{a\ {\isasymnotin}\ dom\ {\isasymSigma}}\ {\normalsize \,then\,}\ {\isacharparenleft}a{\isacharcomma}\ T{\isacharparenright}{\isasymcdot}{\isasymSigma}\ {\isasymturnstile}\ k\ {\isacharcolon}\ A\ {\isasymRightarrow}\ B{\isachardot}}
\end{lemma}

\begin{lemma}[Strengthen Stacks]\ \\
\isa{{\normalsize{}If\,}\ \mbox{{\isasymSigma}\ {\isasymturnstile}\ k\ {\isacharcolon}\ A\ {\isasymRightarrow}\ B}\ {\normalsize \,and\,}\ \mbox{{\isasymSigma}{\isacharprime}\ {\isasymsqsubseteq}\ {\isasymSigma}}\ {\normalsize \,then\,}\ {\isasymSigma}{\isacharprime}\ {\isasymturnstile}\ k\ {\isacharcolon}\ A\ {\isasymRightarrow}\ B{\isachardot}}
\end{lemma}%
\end{isamarkuptext}%
\isamarkuptrue%
\begin{isamarkuptext}%
\begin{lemma}[Strengthen Casted Values]\ \\
\isa{{\normalsize{}If\,}\ \mbox{{\isasymSigma}\ {\isasymturnstile}\ cv\ {\isacharcolon}\ A}\ {\normalsize \,and\,}\ \mbox{{\isasymSigma}{\isacharprime}\ {\isasymsqsubseteq}\ {\isasymSigma}}\ {\normalsize \,then\,}\ {\isasymSigma}{\isacharprime}\ {\isasymturnstile}\ cv\ {\isacharcolon}\ A{\isachardot}}
\end{lemma}

One of the defining aspects of monotonic references is that the
semantics performs strong updates on the heap. However, we only
perform updates that make the types less dynamic. The following lemma
shows that we can perform such updates on well-typed heaps and obtain
well-typed heaps.  We make use of the following auxilliary function in
the statement of the lemma.

\begin{center}
\begin{tabular}{l@ {~~\isa{{\isacharequal}}~~}l}
\isa{cval{\isacharunderscore}ads\ {\isacharparenleft}val\ v{\isacharparenright}\ a\ ads} & \isa{ads\ {\isacharminus}\ {\isacharbraceleft}a{\isacharbraceright}} \\
\isa{cval{\isacharunderscore}ads\ {\isacharparenleft}v{\isacharcolon}A{\isasymRightarrow}B\isactrlbsup \isactrlesup {\isacharparenright}\ a\ ads} & \isa{ads\ {\isasymunion}\ {\isacharbraceleft}a{\isacharbraceright}} 
\end{tabular}
\end{center}

\begin{lemma}[Update Heap]
\isa{{\normalsize{}If\,}\ \mbox{{\isasymSigma}\ {\isasymturnstile}\ {\isasymmu}\ {\isacharbar}\ ads}\ {\normalsize \,and\,}\ \mbox{lookup\ a\ {\isasymSigma}\ {\isacharequal}\ A}\ {\normalsize \,and\,}\ \mbox{B\ {\isasymsqsubseteq}\ A}\ {\normalsize \,and\,}\ \mbox{{\isasymSigma}\ {\isasymturnstile}\ cv\ {\isacharcolon}\ B}\ {\normalsize \,then\,}\ {\isacharparenleft}a{\isacharcomma}\ B{\isacharparenright}{\isasymcdot}{\isasymSigma}\ {\isasymturnstile}\ {\isacharparenleft}a{\isacharcomma}\ cv{\isacharcomma}\ B{\isacharparenright}{\isasymcdot}{\isasymmu}\ {\isacharbar}\ cval{\isacharunderscore}ads\ cv\ a\ ads{\isachardot}}
\end{lemma}
The proof of this lemma relies on the above strengthening lemmas.%
\end{isamarkuptext}%
\isamarkuptrue%
\isamarkupsection{Dynamic Semantics and Type Safety%
}
\isamarkuptrue%
\begin{isamarkuptext}%
The following defines the primitive operators.
\begin{center}
\begin{tabular}{l@ {~~\isa{{\isacharequal}}~~}l}
\isa{{\isasymdelta}\ \textsf{succ}\ {\isacharparenleft}{\isacharparenleft}\ n{\isacharparenright}{\isacharparenright}} & \isa{\textit{return}\ {\isacharparenleft}\ n\ {\isacharplus}\ {\isadigit{1}}{\isacharparenright}} \\
\isa{{\isasymdelta}\ \textsf{prev}\ {\isacharparenleft}{\isacharparenleft}\ n{\isacharparenright}{\isacharparenright}} & \isa{\textit{return}\ {\isacharparenleft}\ n\ {\isacharminus}\ {\isadigit{1}}{\isacharparenright}} \\
\isa{{\isasymdelta}\ \textsf{zero?}\ {\isacharparenleft}{\isacharparenleft}\ n{\isacharparenright}{\isacharparenright}} & \isa{\textit{return}\ {\isacharparenleft}\ {\isacharparenleft}n\ {\isacharequal}\ {\isadigit{0}}{\isacharparenright}{\isacharparenright}} \\
\isa{{\isasymdelta}\ {\isacharparenleft}\textsf{fst}\ A\ B{\isacharparenright}\ {\isacharparenleft}{\isasymlangle}v{\isacharcomma}v{\isacharprime}{\isasymrangle}{\isacharparenright}} & \isa{\textit{return}\ v} \\
\isa{{\isasymdelta}\ {\isacharparenleft}\textsf{snd}\ A\ B{\isacharparenright}\ {\isacharparenleft}{\isasymlangle}v{\isacharcomma}v{\isacharprime}{\isasymrangle}{\isacharparenright}} & \isa{\textit{return}\ v{\isacharprime}} \\
\end{tabular}
\end{center}

\begin{lemma}[Delta Safety]
\label{lem:delta-safety}
\isa{{\normalsize{}If\,}\ \mbox{typeof{\isacharunderscore}opr\ f\ {\isacharequal}\ A\ {\isasymrightarrow}\ B}\ {\normalsize \,and\,}\ \mbox{{\isasymSigma}\ {\isasymturnstile}\ v\ {\isacharcolon}\ A}\ {\normalsize \,then\,}\ {\isasymexists}v{\isacharprime}{\isachardot}\ {\isasymdelta}\ f\ v\ {\isacharequal}\ v{\isacharprime}\ {\isasymand}\ {\isasymSigma}\ {\isasymturnstile}\ v{\isacharprime}\ {\isacharcolon}\ B{\isachardot}}
\end{lemma}%
\end{isamarkuptext}%
\isamarkuptrue%
\isadelimproof
\endisadelimproof
\isatagproof
\endisatagproof
{\isafoldproof}%
\isadelimproof
\endisadelimproof
\isadelimproof
\endisadelimproof
\isatagproof
\endisatagproof
{\isafoldproof}%
\isadelimproof
\endisadelimproof
\isadelimproof
\endisadelimproof
\isatagproof
\endisatagproof
{\isafoldproof}%
\isadelimproof
\endisadelimproof
\isadelimproof
\endisadelimproof
\isatagproof
\endisatagproof
{\isafoldproof}%
\isadelimproof
\endisadelimproof
\begin{isamarkuptext}%
The evaluation function uses the following auxilliary function
to obtain the address from a reference
\begin{center}
\begin{tabular}{l}
\isa{to{\isacharunderscore}addr\ {\isacharparenleft}\textsf{ref}\ a{\isacharparenright}\ {\isacharequal}\ \textit{return}\ a} \\
\isa{to{\isacharunderscore}addr\ v\ {\isacharequal}\ stuck} 
  \qquad if \isa{{\isasymnexists}a{\isachardot}\ v\ {\isacharequal}\ \textsf{ref}\ a}
\end{tabular}
\end{center}
and it uses the below function to extract a value from a potentially-casted
value.
\begin{center}
\begin{tabular}{l}
\isa{to{\isacharunderscore}val\ {\isacharparenleft}val\ v{\isacharparenright}\ {\isacharequal}\ \textit{return}\ v} \\
\isa{to{\isacharunderscore}val\ {\isacharparenleft}v{\isacharcolon}A{\isasymRightarrow}B\isactrlbsup \isactrlesup {\isacharparenright}\ {\isacharequal}\ stuck}
\end{tabular}
\end{center}

The following is the evaluation function for expressions.
The first of the two main accomplishments of monotonic
references is that the below equation for dereference is
standard (with respect to statically typed languages), that is,
it does not need to dispatch based on the kind of reference.
\begin{center}
\begin{tabular}{l@ {~~\isa{{\isacharequal}}~~}l}
\isa{{\isasymlbrakk}x{\isasymrbrakk}{\isasymrho}\ {\isasymmu}} & \isa{lookup\ x\ {\isasymrho}} \\
\isa{{\isasymlbrakk}c{\isasymrbrakk}{\isasymrho}\ {\isasymmu}} & \isa{\textit{return}\ c} \\
\isa{{\isasymlbrakk}f(e){\isasymrbrakk}{\isasymrho}\ {\isasymmu}} & \isa{\textit{bind}\ {\isacharparenleft}{\isasymlbrakk}e{\isasymrbrakk}{\isasymrho}\ {\isasymmu}{\isacharparenright}\ {\isacharparenleft}{\isasymdelta}\ f{\isacharparenright}} \\
\isa{{\isasymlbrakk}{\isasymlambda}x{\isacharcolon}T{\isachardot}\ s{\isasymrbrakk}{\isasymrho}\ {\isasymmu}} & \isa{\textit{return}\ {\isacharparenleft}{\isasymlangle}{\isasymlambda}x{\isacharcolon}T{\isachardot}s{\isacharcomma}\ {\isasymrho}{\isasymrangle}{\isacharparenright}} \\
\isa{{\isasymlbrakk}{\isacharbang}e{\isasymrbrakk}{\isasymrho}\ {\isasymmu}} & \isa{v\ {\isacharcolon}{\isacharequal}\ {\isasymlbrakk}e{\isasymrbrakk}{\isasymrho}\ {\isasymmu}{\isacharsemicolon}\ a\ {\isacharcolon}{\isacharequal}\ to{\isacharunderscore}addr\ v{\isacharsemicolon}\ {\isacharparenleft}cv{\isacharcomma}\ A{\isacharparenright}\ {\isacharcolon}{\isacharequal}\ lookup\ a\ {\isasymmu}{\isacharsemicolon}\ to{\isacharunderscore}val\ cv} 
\end{tabular}
\end{center}

\begin{lemma}[Evaluation Safety]
\isa{{\normalsize{}If\,}\ \mbox{{\isasymGamma}\ {\isasymturnstile}\ e\ {\isacharcolon}\ A}\ {\normalsize \,and\,}\ \mbox{{\isasymGamma}{\isacharsemicolon}{\isasymSigma}\ {\isasymturnstile}\ {\isasymrho}}\ {\normalsize \,and\,}\ \mbox{{\isasymSigma}\ {\isasymturnstile}\ {\isasymmu}\ {\isacharbar}\ {\isasymemptyset}}\ {\normalsize \,then\,}\ {\isasymexists}v{\isachardot}\ {\isasymlbrakk}e{\isasymrbrakk}{\isasymrho}\ {\isasymmu}\ {\isacharequal}\ v\ {\isasymand}\ {\isasymSigma}\ {\isasymturnstile}\ v\ {\isacharcolon}\ A{\isachardot}}
\end{lemma}
The case for variables relies on Lookup Safety (Lemma~\ref{lem:lookup-safety})
and the case for primitive operators relies on Delta Safety 
(Lemma~\ref{lem:delta-safety}).
The case for dereference makes use of the well-typed heap and
that static types are least dynamic (Lemma~\ref{lem:static-most-precise}).
Expressions may only be safely evaluated when the set of active addresses
is empty.%
\end{isamarkuptext}%
\isamarkuptrue%
\begin{isamarkuptext}%
We wrap a cast around a function in the following way. We use
integers for variables (but not De Bruijn notation) which makes
the following somewhat difficult to read.
\begin{center}
\isa{wrap\ v\ A\ B\ C\ D\ {\isasymequiv}\ {\isasymlangle}{\isasymlambda}{\isadigit{0}}{\isacharcolon}C{\isachardot}{\isacharparenleft}\textsf{let}\ {\isadigit{3}}{\isacharequal}{\isadigit{0}}{\isacharcolon}C{\isasymRightarrow}A\ \textsf{in}\ {\isacharparenleft}\textsf{let}\ {\isadigit{2}}{\isacharequal}{\isadigit{1}}({\isadigit{3}})\ \textsf{in}\ {\isacharparenleft}\textsf{let}\ {\isadigit{4}}{\isacharequal}{\isadigit{2}}{\isacharcolon}B{\isasymRightarrow}D\ \textsf{in}\ {\isacharparenleft}\textsf{return}\ {\isadigit{4}}{\isacharparenright}{\isacharparenright}{\isacharparenright}{\isacharparenright}{\isacharcomma}\ {\isacharbrackleft}{\isacharparenleft}{\isadigit{1}}{\isacharcomma}\ v{\isacharparenright}{\isacharbrackright}{\isasymrangle}}
\end{center}%
\end{isamarkuptext}%
\isamarkuptrue%
\begin{isamarkuptext}%
The following auxilliary function creates a casted value that can
be stored in the heap. 
\begin{center}
\begin{tabular}{l@ {~~\isa{{\isacharequal}}~~}l}
\isa{mk{\isacharunderscore}vcast\ {\isacharparenleft}val\ v{\isacharparenright}\ A\ B} & \isa{v{\isacharcolon}A{\isasymRightarrow}B\isactrlbsup \isactrlesup } \\
\isa{mk{\isacharunderscore}vcast\ {\isacharparenleft}v{\isacharcolon}A{\isasymRightarrow}B\isactrlbsup \isactrlesup {\isacharparenright}\ C\ D} & \isa{v{\isacharcolon}A{\isasymRightarrow}D\isactrlbsup \isactrlesup } 
\end{tabular}
\end{center}%
\end{isamarkuptext}%
\isamarkuptrue%
\isadelimproof
\endisadelimproof
\isatagproof
\endisatagproof
{\isafoldproof}%
\isadelimproof
\endisadelimproof
\isadelimproof
\endisadelimproof
\isatagproof
\endisatagproof
{\isafoldproof}%
\isadelimproof
\endisadelimproof
\isadelimproof
\endisadelimproof
\isatagproof
\endisatagproof
{\isafoldproof}%
\isadelimproof
\endisadelimproof
\isadelimproof
\endisadelimproof
\isatagproof
\endisatagproof
{\isafoldproof}%
\isadelimproof
\endisadelimproof
\isadelimproof
\endisadelimproof
\isatagproof
\endisatagproof
{\isafoldproof}%
\isadelimproof
\endisadelimproof
\isadelimproof
\endisadelimproof
\isatagproof
\endisatagproof
{\isafoldproof}%
\isadelimproof
\endisadelimproof
\isadelimproof
\endisadelimproof
\isatagproof
\endisatagproof
{\isafoldproof}%
\isadelimproof
\endisadelimproof
\isadelimproof
\endisadelimproof
\isatagproof
\endisatagproof
{\isafoldproof}%
\isadelimproof
\endisadelimproof
\begin{isamarkuptext}%
The cast function is defined below. We discuss the particulars
of this definition in the following paragraph.

\begin{tabular}{p{\textwidth}}
\isa{cast\ v\ \textsf{Int}\ \textsf{Int}\ {\isasymmu}\ as\ {\isacharequal}\ \textit{return}\ {\isacharparenleft}v{\isacharcomma}\ {\isasymmu}{\isacharcomma}\ as{\isacharparenright}} \\
\isa{cast\ v\ \textsf{Bool}\ \textsf{Bool}\ {\isasymmu}\ as\ {\isacharequal}\ \textit{return}\ {\isacharparenleft}v{\isacharcomma}\ {\isasymmu}{\isacharcomma}\ as{\isacharparenright}} \\
\isa{cast\ v\ {\isasymstar}\ {\isasymstar}\ {\isasymmu}\ as\ {\isacharequal}\ \textit{return}\ {\isacharparenleft}v{\isacharcomma}\ {\isasymmu}{\isacharcomma}\ as{\isacharparenright}} \\
\isa{cast\ v\ {\isacharparenleft}A\ {\isasymrightarrow}\ B{\isacharparenright}\ {\isacharparenleft}C\ {\isasymrightarrow}\ D{\isacharparenright}\ {\isasymmu}\ as\ {\isacharequal}\ \textit{return}\ {\isacharparenleft}wrap\ v\ A\ B\ C\ D{\isacharcomma}\ {\isasymmu}{\isacharcomma}\ as{\isacharparenright}} \\
\isa{cast\ {\isacharparenleft}{\isasymlangle}v{\isadigit{1}}{\isacharcomma}v{\isadigit{2}}{\isasymrangle}{\isacharparenright}\ {\isacharparenleft}A\ {\isasymtimes}\ B{\isacharparenright}\ {\isacharparenleft}C\ {\isasymtimes}\ D{\isacharparenright}\ {\isasymmu}\ as\ {\isacharequal}\ \textit{return}\ {\isacharparenleft}{\isasymlangle}v{\isadigit{1}}{\isacharprime}{\isacharcomma}v{\isadigit{2}}{\isacharprime}{\isasymrangle}{\isacharcomma}\ {\isasymmu}{\isadigit{2}}{\isacharcomma}\ as{\isadigit{2}}{\isacharparenright}} \\
  \qquad if \isa{cast\ v{\isadigit{1}}\ A\ C\ {\isasymmu}\ as\ {\isacharequal}\ \textit{return}\ {\isacharparenleft}v{\isadigit{1}}{\isacharprime}{\isacharcomma}\ {\isasymmu}{\isadigit{1}}{\isacharcomma}\ as{\isadigit{1}}{\isacharparenright}} \\
  \qquad and \isa{cast\ v{\isadigit{2}}\ B\ D\ {\isasymmu}{\isadigit{1}}\ as{\isadigit{1}}\ {\isacharequal}\ \textit{return}\ {\isacharparenleft}v{\isadigit{2}}{\isacharprime}{\isacharcomma}\ {\isasymmu}{\isadigit{2}}{\isacharcomma}\ as{\isadigit{2}}{\isacharparenright}} \\
\isa{cast\ {\isacharparenleft}\textsf{ref}\ a{\isacharparenright}\ {\isacharparenleft}\textsf{Ref}\ A{\isacharparenright}\ {\isacharparenleft}\textsf{Ref}\ B{\isacharparenright}\ {\isasymmu}\ as\ {\isacharequal}\ \textit{return}\ {\isacharparenleft}\textsf{ref}\ a{\isacharcomma}\ {\isasymmu}{\isacharcomma}\ as{\isacharparenright}} \\
  \qquad if \isa{lookup\ a\ {\isasymmu}\ {\isacharequal}\ \textit{return}\ {\isacharparenleft}cv{\isacharcomma}\ C{\isacharparenright}},
    \isa{B\ {\isasymsqinter}\ C\ {\isacharequal}\ \textit{return}\ D},
    \isa{C\ {\isasymsqsubseteq}\ D} \\
\isa{cast\ {\isacharparenleft}\textsf{ref}\ a{\isacharparenright}\ {\isacharparenleft}\textsf{Ref}\ A{\isacharparenright}\ {\isacharparenleft}\textsf{Ref}\ B{\isacharparenright}\ {\isasymmu}\ as\ {\isacharequal}\ \textit{return}\ {\isacharparenleft}\textsf{ref}\ a{\isacharcomma}\ {\isacharparenleft}a{\isacharcomma}\ cv{\isacharprime}{\isacharcomma}\ D{\isacharparenright}{\isasymcdot}{\isasymmu}{\isacharcomma}\ a{\isasymcdot}as{\isacharparenright}} \\
  \qquad if \isa{lookup\ a\ {\isasymmu}\ {\isacharequal}\ \textit{return}\ {\isacharparenleft}cv{\isacharcomma}\ C{\isacharparenright}},
    \isa{B\ {\isasymsqinter}\ C\ {\isacharequal}\ \textit{return}\ D},
    \isa{{\isasymnot}\ C\ {\isasymsqsubseteq}\ D},\\
   \qquad and \isa{cv{\isacharprime}\ {\isacharequal}\ mk{\isacharunderscore}vcast\ cv\ C\ D} \\
\isa{cast\ {\isacharparenleft}\textsf{ref}\ a{\isacharparenright}\ {\isacharparenleft}\textsf{Ref}\ A{\isacharparenright}\ {\isacharparenleft}\textsf{Ref}\ B{\isacharparenright}\ {\isasymmu}\ as\ {\isacharequal}\ cast-error} \\
  \qquad if \isa{lookup\ a\ {\isasymmu}\ {\isacharequal}\ \textit{return}\ {\isacharparenleft}cv{\isacharcomma}\ C{\isacharparenright}},
    \isa{B\ {\isasymsqinter}\ C\ {\isacharequal}\ cast-error} \\
\isa{cast\ {\isacharparenleft}{\isacharbrackleft}v{\isacharcolon}A{\isasymRightarrow}{\isasymstar}{\isacharbrackright}\isactrlbsup \isactrlesup {\isacharparenright}\ {\isasymstar}\ B\ {\isasymmu}\ as\ {\isacharequal}\ cast\ v\ A\ B\ {\isasymmu}\ as} \\
  \qquad if \isa{B\ {\isasymnoteq}\ {\isasymstar}},
    \isa{ground\ A\ {\isacharequal}\ ground\ B}\\
\isa{cast\ {\isacharparenleft}{\isacharbrackleft}v{\isacharcolon}A{\isasymRightarrow}{\isasymstar}{\isacharbrackright}\isactrlbsup \isactrlesup {\isacharparenright}\ {\isasymstar}\ B\ {\isasymmu}\ as\ {\isacharequal}\ cast{\isacharunderscore}error} \\
  \qquad if \isa{B\ {\isasymnoteq}\ {\isasymstar}},
    \isa{ground\ A\ {\isasymnoteq}\ ground\ B} \\
\isa{cast\ {\isacharparenleft}{\isacharbrackleft}v{\isacharcolon}A{\isasymRightarrow}{\isasymstar}{\isacharbrackright}\isactrlbsup \isactrlesup {\isacharparenright}\ {\isasymstar}\ B\ {\isasymmu}\ as\ {\isacharequal}\ cast{\isacharunderscore}error} \\
  \qquad if \isa{B\ {\isasymnoteq}\ {\isasymstar}},
    \isa{ground\ A\ {\isacharequal}\ ground\ B},
    \isa{cast\ v\ A\ B\ {\isasymmu}\ as\ {\isacharequal}\ cast{\isacharunderscore}error} \\
\isa{cast\ v\ A\ {\isasymstar}\ {\isasymmu}\ as\ {\isacharequal}\ \textit{return}\ {\isacharparenleft}{\isacharbrackleft}v{\isacharcolon}A{\isasymRightarrow}{\isasymstar}{\isacharbrackright}\isactrlbsup \isactrlesup {\isacharcomma}\ {\isasymmu}{\isacharcomma}\ as{\isacharparenright}} \\
  \qquad if \isa{A\ {\isasymnoteq}\ {\isasymstar}}
\end{tabular} \\
In the remaining cases, the result is a cast error.

The case for casting references is the most important to this
development and is rather subtle. The main idea is that the value at
address $a$ is cast from its current type $C$ to the meet of its
current type and the target type of the cast.  This cast is
accomplished by storing a so-called ``casted value'' on the heap and
returning address $a$ in the list of active addresses (the addresses
with pending casts to be performed).  One extra wrinkle in the
definition of casting a reference is that there may be cycles in the
heap. To guard against infinite loops and to improve efficiency, we
leave the heap unchanged if the heap type is already less or equally
dynamic than the target type of the cast.

The statement of Cast Safety, given below, is rather complex.  Given a
well-typed value and heap, the result of a cast is either a cast error
or a value whose type is the target type and a new heap and active
address list.  The heap is well-typed in some heap typing that is 
less dynamic than the typing of the original heap.

\begin{lemma}[Cast Safety]
\isa{{\normalsize{}If\,}\ \mbox{{\isasymSigma}\ {\isasymturnstile}\ v\ {\isacharcolon}\ A}\ {\normalsize \,and\,}\ \mbox{{\isasymSigma}\ {\isasymturnstile}\ {\isasymmu}\ {\isacharbar}\ ads{\isadigit{1}}}\ {\normalsize \,then\,}\ {\isacharparenleft}{\isasymexists}v{\isacharprime}\ {\isasymSigma}{\isacharprime}\ {\isasymmu}{\isacharprime}\ ads{\isadigit{2}}{\isachardot}\ cast\ v\ A\ B\ {\isasymmu}\ ads{\isadigit{1}}\ {\isacharequal}\ {\isacharparenleft}v{\isacharprime}{\isacharcomma}\ {\isasymmu}{\isacharprime}{\isacharcomma}\ ads{\isadigit{2}}{\isacharparenright}\ {\isasymand}\ {\isasymSigma}{\isacharprime}\ {\isasymturnstile}\ v{\isacharprime}\ {\isacharcolon}\ B\ {\isasymand}\ {\isasymSigma}{\isacharprime}\ {\isasymturnstile}\ {\isasymmu}{\isacharprime}\ {\isacharbar}\ ads{\isadigit{2}}\ {\isasymand}\ {\isasymSigma}{\isacharprime}\ {\isasymsqsubseteq}\ {\isasymSigma}{\isacharparenright}\ {\isasymor}\ cast\ v\ A\ B\ {\isasymmu}\ ads{\isadigit{1}}\ {\isacharequal}\ cast-error{\isachardot}}
\end{lemma}%
\end{isamarkuptext}%
\isamarkuptrue%
\isadelimproof
\endisadelimproof
\isatagproof
\endisatagproof
{\isafoldproof}%
\isadelimproof
\endisadelimproof
\isadelimproof
\endisadelimproof
\isatagproof
\endisatagproof
{\isafoldproof}%
\isadelimproof
\endisadelimproof
\isadelimproof
\endisadelimproof
\isatagproof
\endisatagproof
{\isafoldproof}%
\isadelimproof
\endisadelimproof
\isadelimproof
\endisadelimproof
\isatagproof
\endisatagproof
{\isafoldproof}%
\isadelimproof
\endisadelimproof
\isadelimproof
\endisadelimproof
\isatagproof
\endisatagproof
{\isafoldproof}%
\isadelimproof
\endisadelimproof
\isadelimproof
\endisadelimproof
\isatagproof
\endisatagproof
{\isafoldproof}%
\isadelimproof
\endisadelimproof
\isadelimproof
\endisadelimproof
\isatagproof
\endisatagproof
{\isafoldproof}%
\isadelimproof
\endisadelimproof
\isadelimproof
\endisadelimproof
\isatagproof
\endisatagproof
{\isafoldproof}%
\isadelimproof
\endisadelimproof
\isadelimproof
\endisadelimproof
\isatagproof
\endisatagproof
{\isafoldproof}%
\isadelimproof
\endisadelimproof
\isadelimproof
\endisadelimproof
\isatagproof
\endisatagproof
{\isafoldproof}%
\isadelimproof
\endisadelimproof
\isadelimproof
\endisadelimproof
\isatagproof
\endisatagproof
{\isafoldproof}%
\isadelimproof
\endisadelimproof
\isadelimproof
\endisadelimproof
\isatagproof
\endisatagproof
{\isafoldproof}%
\isadelimproof
\endisadelimproof
\isadelimproof
\endisadelimproof
\isatagproof
\endisatagproof
{\isafoldproof}%
\isadelimproof
\endisadelimproof
\isadelimproof
\endisadelimproof
\isatagproof
\endisatagproof
{\isafoldproof}%
\isadelimproof
\endisadelimproof
\isadelimproof
\endisadelimproof
\isatagproof
\endisatagproof
{\isafoldproof}%
\isadelimproof
\endisadelimproof
\begin{isamarkuptext}%
The following defines the transitions of the abstract machine.
The transitions are defined in terms of a step function but
we use the following abbreviation.
\begin{center}
\isa{s\ {\isasymlongmapsto}\ s{\isacharprime}\ {\isasymequiv}\ step\ s\ {\isacharequal}\ \textit{return}\ s{\isacharprime}}
\end{center}

The first four transition rules, listed below, process addresses in
the active address list. If the address points to a value, then we can
simply remove that address from the active list. If the address points
to a casted value, then we need to perform the cast. If successful,
the cast produces a new value $v'$, an udpated heap, and a list of
addresses that have become active. If the current heap type for $a$ is
still the same, then we commit the result of this cast, installing
$v'$ at address $a$.  If the heap type has changed, then this cast has
been superceded by some other cast, so we do not commit $v'$.
Finally, the cast was unsuccessful, execution halts with a cast error.
\begin{center}
\isa{\mbox{}\inferrule{\mbox{lookup\ a\ {\isasymmu}\ {\isacharequal}\ \textit{return}\ {\isacharparenleft}cv{\isacharcomma}\ A{\isacharparenright}}\\\ \mbox{cv\ {\isacharequal}\ val\ v}}{\mbox{{\isacharparenleft}s{\isacharcomma}\ {\isasymrho}{\isacharcomma}\ k{\isacharcomma}\ {\isasymmu}{\isacharcomma}\ a{\isasymcdot}as{\isacharparenright}\ {\isasymlongmapsto}\ {\isacharparenleft}s{\isacharcomma}\ {\isasymrho}{\isacharcomma}\ k{\isacharcomma}\ {\isasymmu}{\isacharcomma}\ as{\isacharparenright}}}}\\[2ex]
\isa{\mbox{}\inferrule{\mbox{lookup\ a\ {\isasymmu}\ {\isacharequal}\ \textit{return}\ {\isacharparenleft}cv{\isacharcomma}\ A{\isacharparenright}}\\\ \mbox{cv\ {\isacharequal}\ v{\isacharcolon}B{\isasymRightarrow}C\isactrlbsup \isactrlesup }\\\ \mbox{cast\ v\ B\ C\ {\isasymmu}\ {\isacharparenleft}a{\isasymcdot}as{\isacharparenright}\ {\isacharequal}\ \textit{return}\ {\isacharparenleft}v{\isacharprime}{\isacharcomma}\ {\isasymmu}{\isacharprime}{\isacharcomma}\ as{\isacharprime}{\isacharparenright}}\\\ \mbox{lookup\ a\ {\isasymmu}{\isacharprime}\ {\isacharequal}\ \textit{return}\ {\isacharparenleft}cv{\isacharprime}{\isacharcomma}\ A{\isacharprime}{\isacharparenright}}\\\ \mbox{A\ {\isasymsqsubseteq}\ A{\isacharprime}}}{\mbox{{\isacharparenleft}s{\isacharcomma}\ {\isasymrho}{\isacharcomma}\ k{\isacharcomma}\ {\isasymmu}{\isacharcomma}\ a{\isasymcdot}as{\isacharparenright}\ {\isasymlongmapsto}\ {\isacharparenleft}s{\isacharcomma}\ {\isasymrho}{\isacharcomma}\ k{\isacharcomma}\ {\isacharparenleft}a{\isacharcomma}\ val\ v{\isacharprime}{\isacharcomma}\ A{\isacharparenright}{\isasymcdot}{\isasymmu}{\isacharprime}{\isacharcomma}\ removeAll\ a\ as{\isacharprime}{\isacharparenright}}}}\\[2ex]
\isa{\mbox{}\inferrule{\mbox{lookup\ a\ {\isasymmu}\ {\isacharequal}\ \textit{return}\ {\isacharparenleft}cv{\isacharcomma}\ A{\isacharparenright}}\\\ \mbox{cv\ {\isacharequal}\ v{\isacharcolon}B{\isasymRightarrow}C\isactrlbsup \isactrlesup }\\\ \mbox{cast\ v\ B\ C\ {\isasymmu}\ {\isacharparenleft}a{\isasymcdot}as{\isacharparenright}\ {\isacharequal}\ \textit{return}\ {\isacharparenleft}v{\isacharprime}{\isacharcomma}\ {\isasymmu}{\isacharprime}{\isacharcomma}\ as{\isacharprime}{\isacharparenright}}\\\ \mbox{lookup\ a\ {\isasymmu}{\isacharprime}\ {\isacharequal}\ \textit{return}\ {\isacharparenleft}cv{\isacharprime}{\isacharcomma}\ A{\isacharprime}{\isacharparenright}}\\\ \mbox{{\isasymnot}\ A\ {\isasymsqsubseteq}\ A{\isacharprime}}}{\mbox{{\isacharparenleft}s{\isacharcomma}\ {\isasymrho}{\isacharcomma}\ k{\isacharcomma}\ {\isasymmu}{\isacharcomma}\ a{\isasymcdot}as{\isacharparenright}\ {\isasymlongmapsto}\ {\isacharparenleft}s{\isacharcomma}\ {\isasymrho}{\isacharcomma}\ k{\isacharcomma}\ {\isasymmu}{\isacharprime}{\isacharcomma}\ as{\isacharprime}{\isacharparenright}}}}\\[2ex]
\isa{\mbox{}\inferrule{\mbox{lookup\ a\ {\isasymmu}\ {\isacharequal}\ \textit{return}\ {\isacharparenleft}cv{\isacharcomma}\ A{\isacharparenright}}\\\ \mbox{cv\ {\isacharequal}\ v{\isacharcolon}B{\isasymRightarrow}C\isactrlbsup \isactrlesup }\\\ \mbox{cast\ v\ B\ C\ {\isasymmu}\ {\isacharparenleft}a{\isasymcdot}as{\isacharparenright}\ {\isacharequal}\ cast{\isacharunderscore}error}}{\mbox{step\ {\isacharparenleft}s{\isacharcomma}\ {\isasymrho}{\isacharcomma}\ k{\isacharcomma}\ {\isasymmu}{\isacharcomma}\ a{\isasymcdot}as{\isacharparenright}\ {\isacharequal}\ cast{\isacharunderscore}error}}}
\end{center}

The transitions for allocation and ``static'' reference updates are
mostly standard. Note that each value on the heap is paired with its
type. The fact that static update is completely standard (with respect
to statically-typed languages) is the second of the two main achievements
of monotonic references.
\begin{center}
\isa{\mbox{}\inferrule{\mbox{{\isasymlbrakk}e{\isasymrbrakk}{\isasymrho}\ {\isasymmu}\ {\isacharequal}\ \textit{return}\ v}\\\ \mbox{a\ {\isacharequal}\ {\isacharbar}{\isasymmu}{\isacharbar}}}{\mbox{{\isacharparenleft}\textsf{let}\ x{\isacharequal}\ \textsf{ref}\isactrlbsub A\ \isactrlesub e\ \textsf{in}\ s{\isacharcomma}\ {\isasymrho}{\isacharcomma}\ k{\isacharcomma}\ {\isasymmu}{\isacharcomma}\ {\isacharbrackleft}{\isacharbrackright}{\isacharparenright}\ {\isasymlongmapsto}\ {\isacharparenleft}s{\isacharcomma}\ {\isacharparenleft}x{\isacharcomma}\ \textsf{ref}\ a{\isacharparenright}{\isasymcdot}{\isasymrho}{\isacharcomma}\ k{\isacharcomma}\ {\isacharparenleft}a{\isacharcomma}\ val\ v{\isacharcomma}\ A{\isacharparenright}{\isasymcdot}{\isasymmu}{\isacharcomma}\ {\isacharbrackleft}{\isacharbrackright}{\isacharparenright}}}}\\[2ex]
\isa{\mbox{}\inferrule{\mbox{{\isasymlbrakk}e{\isasymrbrakk}{\isasymrho}\ {\isasymmu}\ {\isacharequal}\ \textit{return}\ \textsf{ref}\ a}\\\ \mbox{{\isasymlbrakk}e{\isacharprime}{\isasymrbrakk}{\isasymrho}\ {\isasymmu}\ {\isacharequal}\ \textit{return}\ v}\\\ \mbox{lookup\ a\ {\isasymmu}\ {\isacharequal}\ \textit{return}\ {\isacharparenleft}v{\isacharprime}{\isacharcomma}\ A{\isacharparenright}}}{\mbox{{\isacharparenleft}e{\isacharcolon}{\isacharequal}e{\isacharprime}{\isacharsemicolon}\ s{\isacharcomma}\ {\isasymrho}{\isacharcomma}\ k{\isacharcomma}\ {\isasymmu}{\isacharcomma}\ {\isacharbrackleft}{\isacharbrackright}{\isacharparenright}\ {\isasymlongmapsto}\ {\isacharparenleft}s{\isacharcomma}\ {\isasymrho}{\isacharcomma}\ k{\isacharcomma}\ {\isacharparenleft}a{\isacharcomma}\ val\ v{\isacharcomma}\ A{\isacharparenright}{\isasymcdot}{\isasymmu}{\isacharcomma}\ {\isacharbrackleft}{\isacharbrackright}{\isacharparenright}}}}
\end{center}

The ``dynamic'' dereference and update transitions
perform casts to mediate between the reference's type
and the type on the heap.
\begin{center}
\isa{\mbox{}\inferrule{\mbox{{\isasymlbrakk}e{\isasymrbrakk}{\isasymrho}\ {\isasymmu}\ {\isacharequal}\ \textit{return}\ \textsf{ref}\ a}\\\ \mbox{lookup\ a\ {\isasymmu}\ {\isacharequal}\ \textit{return}\ {\isacharparenleft}val\ v{\isacharcomma}\ B{\isacharparenright}}\\\ \mbox{cast\ v\ B\ A\ {\isasymmu}\ {\isacharbrackleft}{\isacharbrackright}\ {\isacharequal}\ \textit{return}\ {\isacharparenleft}v{\isacharprime}{\isacharcomma}\ {\isasymmu}{\isacharprime}{\isacharcomma}\ as{\isacharparenright}}}{\mbox{{\isacharparenleft}\textsf{let}\ x{\isacharequal}{\isacharbang}e{\isacharat}A\ \textsf{in}\ s{\isacharcomma}\ {\isasymrho}{\isacharcomma}\ k{\isacharcomma}\ {\isasymmu}{\isacharcomma}\ {\isacharbrackleft}{\isacharbrackright}{\isacharparenright}\ {\isasymlongmapsto}\ {\isacharparenleft}s{\isacharcomma}\ {\isacharparenleft}x{\isacharcomma}\ v{\isacharprime}{\isacharparenright}{\isasymcdot}{\isasymrho}{\isacharcomma}\ k{\isacharcomma}\ {\isasymmu}{\isacharprime}{\isacharcomma}\ as{\isacharparenright}}}}\\[2ex]
\isa{\mbox{}\inferrule{\mbox{{\isasymlbrakk}e{\isasymrbrakk}{\isasymrho}\ {\isasymmu}\ {\isacharequal}\ \textit{return}\ \textsf{ref}\ a}\\\ \mbox{lookup\ a\ {\isasymmu}\ {\isacharequal}\ \textit{return}\ {\isacharparenleft}val\ v{\isacharcomma}\ B{\isacharparenright}}\\\ \mbox{cast\ v\ B\ A\ {\isasymmu}\ {\isacharbrackleft}{\isacharbrackright}\ {\isacharequal}\ cast{\isacharunderscore}error}}{\mbox{step\ {\isacharparenleft}\textsf{let}\ x{\isacharequal}{\isacharbang}e{\isacharat}A\ \textsf{in}\ s{\isacharcomma}\ {\isasymrho}{\isacharcomma}\ k{\isacharcomma}\ {\isasymmu}{\isacharcomma}\ {\isacharbrackleft}{\isacharbrackright}{\isacharparenright}\ {\isacharequal}\ cast{\isacharunderscore}error}}}
\isa{\mbox{}\inferrule{\mbox{{\isasymlbrakk}e{\isasymrbrakk}{\isasymrho}\ {\isasymmu}\ {\isacharequal}\ \textit{return}\ \textsf{ref}\ a}\\\ \mbox{{\isasymlbrakk}e{\isacharprime}{\isasymrbrakk}{\isasymrho}\ {\isasymmu}\ {\isacharequal}\ \textit{return}\ v}\\\ \mbox{lookup\ a\ {\isasymmu}\ {\isacharequal}\ \textit{return}\ {\isacharparenleft}v{\isacharprime}{\isacharcomma}\ B{\isacharparenright}}}{\mbox{{\isacharparenleft}e{\isacharcolon}{\isacharequal}e{\isacharprime}{\isacharat}A{\isacharsemicolon}\ s{\isacharcomma}\ {\isasymrho}{\isacharcomma}\ k{\isacharcomma}\ {\isasymmu}{\isacharcomma}\ {\isacharbrackleft}{\isacharbrackright}{\isacharparenright}\ {\isasymlongmapsto}\ {\isacharparenleft}s{\isacharcomma}\ {\isasymrho}{\isacharcomma}\ k{\isacharcomma}\ {\isacharparenleft}a{\isacharcomma}\ v{\isacharcolon}A{\isasymRightarrow}B\isactrlbsup \isactrlesup {\isacharcomma}\ B{\isacharparenright}{\isasymcdot}{\isasymmu}{\isacharcomma}\ {\isacharbrackleft}a{\isacharbrackright}{\isacharparenright}}}}\\[2ex]
\end{center}

The transition for cast statements is straightforward; all the
hard work is all carried out by the auxilliary cast function.
\begin{center}
\isa{\mbox{}\inferrule{\mbox{{\isasymlbrakk}e{\isasymrbrakk}{\isasymrho}\ {\isasymmu}\ {\isacharequal}\ \textit{return}\ v}\\\ \mbox{cast\ v\ A\ B\ {\isasymmu}\ {\isacharbrackleft}{\isacharbrackright}\ {\isacharequal}\ \textit{return}\ {\isacharparenleft}v{\isacharprime}{\isacharcomma}\ {\isasymmu}{\isacharprime}{\isacharcomma}\ as{\isacharparenright}}}{\mbox{{\isacharparenleft}\textsf{let}\ x{\isacharequal}e{\isacharcolon}A{\isasymRightarrow}B\ \textsf{in}\ s{\isacharcomma}\ {\isasymrho}{\isacharcomma}\ k{\isacharcomma}\ {\isasymmu}{\isacharcomma}\ {\isacharbrackleft}{\isacharbrackright}{\isacharparenright}\ {\isasymlongmapsto}\ {\isacharparenleft}s{\isacharcomma}\ {\isacharparenleft}x{\isacharcomma}\ v{\isacharprime}{\isacharparenright}{\isasymcdot}{\isasymrho}{\isacharcomma}\ k{\isacharcomma}\ {\isasymmu}{\isacharprime}{\isacharcomma}\ as{\isacharparenright}}}}\\[2ex]
\isa{\mbox{}\inferrule{\mbox{{\isasymlbrakk}e{\isasymrbrakk}{\isasymrho}\ {\isasymmu}\ {\isacharequal}\ \textit{return}\ v}\\\ \mbox{cast\ v\ A\ B\ {\isasymmu}\ {\isacharbrackleft}{\isacharbrackright}\ {\isacharequal}\ cast{\isacharunderscore}error}}{\mbox{step\ {\isacharparenleft}\textsf{let}\ x{\isacharequal}e{\isacharcolon}A{\isasymRightarrow}B\ \textsf{in}\ s{\isacharcomma}\ {\isasymrho}{\isacharcomma}\ k{\isacharcomma}\ {\isasymmu}{\isacharcomma}\ {\isacharbrackleft}{\isacharbrackright}{\isacharparenright}\ {\isacharequal}\ cast{\isacharunderscore}error}}}
\end{center}

The transition rules for let and for function call and return
are standard.
\begin{center}
\isa{\mbox{}\inferrule{\mbox{{\isasymlbrakk}e{\isasymrbrakk}{\isasymrho}\ {\isasymmu}\ {\isacharequal}\ \textit{return}\ v}}{\mbox{{\isacharparenleft}\textsf{let}\ x{\isacharequal}e\ \textsf{in}\ s{\isacharcomma}\ {\isasymrho}{\isacharcomma}\ k{\isacharcomma}\ {\isasymmu}{\isacharcomma}\ {\isacharbrackleft}{\isacharbrackright}{\isacharparenright}\ {\isasymlongmapsto}\ {\isacharparenleft}s{\isacharcomma}\ {\isacharparenleft}x{\isacharcomma}\ v{\isacharparenright}{\isasymcdot}{\isasymrho}{\isacharcomma}\ k{\isacharcomma}\ {\isasymmu}{\isacharcomma}\ {\isacharbrackleft}{\isacharbrackright}{\isacharparenright}}}}\\[2ex]
\isa{\mbox{}\inferrule{\mbox{{\isasymlbrakk}e{\isasymrbrakk}{\isasymrho}\ {\isasymmu}\ {\isacharequal}\ \textit{return}\ v}\\\ \mbox{{\isasymlbrakk}e{\isacharprime}{\isasymrbrakk}{\isasymrho}\ {\isasymmu}\ {\isacharequal}\ \textit{return}\ v{\isacharprime}}\\\ \mbox{v\ {\isacharequal}\ {\isasymlangle}{\isasymlambda}y{\isacharcolon}A{\isachardot}s{\isacharprime}{\isacharcomma}\ {\isasymrho}{\isacharprime}{\isasymrangle}}}{\mbox{{\isacharparenleft}\textsf{let}\ x{\isacharequal}e(e{\isacharprime})\ \textsf{in}\ s{\isacharcomma}\ {\isasymrho}{\isacharcomma}\ k{\isacharcomma}\ {\isasymmu}{\isacharcomma}\ {\isacharbrackleft}{\isacharbrackright}{\isacharparenright}\ {\isasymlongmapsto}\ {\isacharparenleft}s{\isacharprime}{\isacharcomma}\ {\isacharparenleft}y{\isacharcomma}\ v{\isacharprime}{\isacharparenright}{\isasymcdot}{\isasymrho}{\isacharprime}{\isacharcomma}\ {\isacharparenleft}x{\isacharcomma}\ s{\isacharcomma}\ {\isasymrho}{\isacharparenright}{\isasymcdot}k{\isacharcomma}\ {\isasymmu}{\isacharcomma}\ {\isacharbrackleft}{\isacharbrackright}{\isacharparenright}}}}\\[2ex]
\isa{\mbox{}\inferrule{\mbox{{\isasymlbrakk}e{\isasymrbrakk}{\isasymrho}\ {\isasymmu}\ {\isacharequal}\ \textit{return}\ v}\\\ \mbox{{\isasymlbrakk}e{\isacharprime}{\isasymrbrakk}{\isasymrho}\ {\isasymmu}\ {\isacharequal}\ \textit{return}\ v{\isacharprime}}\\\ \mbox{v\ {\isacharequal}\ {\isasymlangle}{\isasymlambda}y{\isacharcolon}A{\isachardot}s{\isacharprime}{\isacharcomma}\ {\isasymrho}{\isacharprime}{\isasymrangle}}}{\mbox{{\isacharparenleft}\textsf{return}\ e(e{\isacharprime}){\isacharcomma}\ {\isasymrho}{\isacharcomma}\ k{\isacharcomma}\ {\isasymmu}{\isacharcomma}\ {\isacharbrackleft}{\isacharbrackright}{\isacharparenright}\ {\isasymlongmapsto}\ {\isacharparenleft}s{\isacharprime}{\isacharcomma}\ {\isacharparenleft}y{\isacharcomma}\ v{\isacharprime}{\isacharparenright}{\isasymcdot}{\isasymrho}{\isacharprime}{\isacharcomma}\ k{\isacharcomma}\ {\isasymmu}{\isacharcomma}\ {\isacharbrackleft}{\isacharbrackright}{\isacharparenright}}}}\\[2ex]
\isa{\mbox{}\inferrule{\mbox{{\isasymlbrakk}e{\isasymrbrakk}{\isasymrho}\ {\isasymmu}\ {\isacharequal}\ \textit{return}\ v}}{\mbox{{\isacharparenleft}\textsf{return}\ e{\isacharcomma}\ {\isasymrho}{\isacharcomma}\ {\isacharparenleft}x{\isacharcomma}\ s{\isacharcomma}\ {\isasymrho}{\isacharprime}{\isacharparenright}{\isasymcdot}k{\isacharcomma}\ {\isasymmu}{\isacharcomma}\ {\isacharbrackleft}{\isacharbrackright}{\isacharparenright}\ {\isasymlongmapsto}\ {\isacharparenleft}s{\isacharcomma}\ {\isacharparenleft}x{\isacharcomma}\ v{\isacharparenright}{\isasymcdot}{\isasymrho}{\isacharprime}{\isacharcomma}\ k{\isacharcomma}\ {\isasymmu}{\isacharcomma}\ {\isacharbrackleft}{\isacharbrackright}{\isacharparenright}}}}\\[2ex]
\end{center}
All other states are mapped to stuck.%
\end{isamarkuptext}%
\isamarkuptrue%
\begin{isamarkuptext}%
\begin{lemma}
\isa{{\normalsize{}If\,}\ {\isasymturnstile}\ s\ {\isacharcolon}\ A\ {\normalsize \,then\,}\ final\ s\ {\isasymor}\ {\isacharparenleft}{\isasymexists}s{\isacharprime}{\isachardot}\ step\ s\ {\isacharequal}\ s{\isacharprime}\ {\isasymand}\ {\isasymturnstile}\ s{\isacharprime}\ {\isacharcolon}\ A{\isacharparenright}\ {\isasymor}\ step\ s\ {\isacharequal}\ cast-error{\isachardot}}
\end{lemma}
The proof proceeds by cases on the active address list
(empty or not) and then by cases on the statement component
of the state. The proof is long but straightforward given the
above lemmas and the invariants captured in the definition
of a well-typed state.%
\end{isamarkuptext}%
\isamarkuptrue%
\begin{isamarkuptext}%
The following function maps values to observables.

\begin{center}
\begin{tabular}{l@ {~~\isa{{\isacharequal}}~~}l}
\isa{observe\ {\isacharparenleft}c{\isacharparenright}} & \isa{Con\ c} \\
\isa{observe\ {\isacharparenleft}{\isasymlangle}v{\isacharcomma}v{\isacharprime}{\isasymrangle}{\isacharparenright}} & \isa{OPair\ {\isacharparenleft}observe\ v{\isacharparenright}\ {\isacharparenleft}observe\ v{\isacharprime}{\isacharparenright}} \\
\isa{observe\ {\isacharparenleft}{\isasymlangle}{\isasymlambda}x{\isacharcolon}T{\isachardot}s{\isacharcomma}\ {\isasymrho}{\isasymrangle}{\isacharparenright}} & \isa{Fun} \\
\isa{observe\ {\isacharparenleft}\textsf{ref}\ a{\isacharparenright}} & \isa{Addr} \\
\isa{observe\ {\isacharparenleft}{\isacharbrackleft}v{\isacharcolon}T{\isasymRightarrow}{\isasymstar}{\isacharbrackright}\isactrlbsup \isactrlesup {\isacharparenright}} & \isa{Inj} 
\end{tabular}
\end{center}

Well-typed observables:
\begin{center}
\begin{tabular}{l@ {~~\isa{{\isacharequal}}~~}l}
\isa{{\isasymturnstile}\ OPair\ o{\isacharprime}\ o{\isacharprime}{\isacharprime}\ {\isacharcolon}\ A\ {\isasymtimes}\ B} & \isa{{\isasymturnstile}\ o{\isacharprime}\ {\isacharcolon}\ A\ {\isasymand}\ {\isasymturnstile}\ o{\isacharprime}{\isacharprime}\ {\isacharcolon}\ B} \\
\isa{{\isasymturnstile}\ Fun\ {\isacharcolon}\ A\ {\isasymrightarrow}\ B} & \isa{True} \\
\isa{{\isasymturnstile}\ Con\ c\ {\isacharcolon}\ T} & \isa{typeof\ c\ {\isacharequal}\ T} \\
\isa{{\isasymturnstile}\ OStuck\ {\isacharcolon}\ T} & \isa{False} \\
\isa{{\isasymturnstile}\ OTimeOut\ {\isacharcolon}\ T} & \isa{True} \\
\isa{{\isasymturnstile}\ OCastError\ {\isacharcolon}\ T} & \isa{True} \\
\isa{{\isasymturnstile}\ Addr\ {\isacharcolon}\ \textsf{Ref}\ A} & \isa{True} \\
\isa{{\isasymturnstile}\ Inj\ {\isacharcolon}\ {\isasymstar}} & \isa{True} 
\end{tabular}
\end{center}

\begin{lemma}
\isa{{\normalsize{}If\,}\ {\isasymSigma}\ {\isasymturnstile}\ v\ {\isacharcolon}\ A\ {\normalsize \,then\,}\ {\isasymturnstile}\ observe\ v\ {\isacharcolon}\ A{\isachardot}}
\end{lemma}%
\end{isamarkuptext}%
\isamarkuptrue%
\isadelimproof
\endisadelimproof
\isatagproof
\endisatagproof
{\isafoldproof}%
\isadelimproof
\endisadelimproof
\isadelimproof
\endisadelimproof
\isatagproof
\endisatagproof
{\isafoldproof}%
\isadelimproof
\endisadelimproof
\isadelimproof
\endisadelimproof
\isatagproof
\endisatagproof
{\isafoldproof}%
\isadelimproof
\endisadelimproof
\isadelimproof
\endisadelimproof
\isatagproof
\endisatagproof
{\isafoldproof}%
\isadelimproof
\endisadelimproof
\begin{isamarkuptext}%
A final state is one that has finished executing. It is a return statement
with an empty procedure call stack and empty active address list.
\begin{center}
\isa{final\ {\isacharparenleft}\textsf{return}\ e{\isacharcomma}\ {\isasymrho}{\isacharcomma}\ {\isacharbrackleft}{\isacharbrackright}{\isacharcomma}\ {\isasymmu}{\isacharcomma}\ {\isacharbrackleft}{\isacharbrackright}{\isacharparenright}}
\end{center}

The following steps function iterates the step function. We use a counter
as a technical device to make this function terminate (which Isabelle requires)
even though it otherwise not be guaranteed to terminate.

\begin{center}
\begin{tabular}{l@ {~~\isa{{\isacharequal}}~~}l}
\isa{steps\ {\isadigit{0}}\ s} & \isa{OTimeOut} \\
\isa{steps\ {\isacharparenleft}Suc\ n{\isacharparenright}\ {\isacharparenleft}\textsf{return}\ e{\isacharcomma}\ {\isasymrho}{\isacharcomma}\ {\isacharbrackleft}{\isacharbrackright}{\isacharcomma}\ {\isasymmu}{\isacharcomma}\ {\isacharbrackleft}{\isacharbrackright}{\isacharparenright}} & \isa{observe\ v} 
  \qquad if \isa{{\isasymlbrakk}e{\isasymrbrakk}{\isasymrho}\ {\isasymmu}\ {\isacharequal}\ \textit{return}\ v}\\
\isa{steps\ {\isacharparenleft}Suc\ n{\isacharparenright}\ s} & \isa{steps\ n\ s{\isacharprime}} 
  \qquad if \isa{step\ s\ {\isacharequal}\ s{\isacharprime}} \\
\isa{steps\ {\isacharparenleft}Suc\ n{\isacharparenright}\ s} & \isa{OCastError} 
  \qquad if \isa{step\ s\ {\isacharequal}\ cast{\isacharunderscore}error}
\end{tabular}
\end{center}

This language is type safe because, for arbitrary numbers of steps,
the result is always well typed.

\begin{theorem}[Type Safety]
\isa{{\normalsize{}If\,}\ {\isasymturnstile}\ s\ {\isacharcolon}\ A\ {\normalsize \,then\,}\ {\isasymexists}r{\isachardot}\ steps\ n\ s\ {\isacharequal}\ r\ {\isasymand}\ {\isasymturnstile}\ r\ {\isacharcolon}\ A{\isachardot}}
\end{theorem}%
\end{isamarkuptext}%
\isamarkuptrue%
\isadelimtheory
\endisadelimtheory
\isatagtheory
\endisatagtheory
{\isafoldtheory}%
\isadelimtheory
\endisadelimtheory
\end{isabellebody}%

%% file: GTLC_MonoRef_ECDOnANF.tex
%
\begin{isabellebody}%
\def\isabellecontext{GTLC{\isacharunderscore}MonoRef{\isacharunderscore}ECDOnANF}%
\isadelimtheory
\endisadelimtheory
\isatagtheory
\isacommand{theory}\isamarkupfalse%
\ GTLC{\isacharunderscore}MonoRef{\isacharunderscore}ECDOnANF\isanewline
\isakeyword{imports}\ Main\isanewline
\isakeyword{begin}%
\endisatagtheory
{\isafoldtheory}%
\isadelimtheory
\endisadelimtheory
\isamarkupsubsection{Syntax%
}
\isamarkuptrue%
\isacommand{datatype}\isamarkupfalse%
\ ty\isanewline
\ \ {\isacharequal}\ IntT\isanewline
\ \ {\isacharbar}\ BoolT\isanewline
\ \ {\isacharbar}\ PairT\ ty\ ty\ {\isacharparenleft}\isakeyword{infixr}\ {\isachardoublequoteopen}{\isasymtimes}{\isachardoublequoteclose}\ {\isadigit{2}}{\isadigit{0}}{\isadigit{1}}{\isacharparenright}\isanewline
\ \ {\isacharbar}\ ArrowT\ ty\ ty\ {\isacharparenleft}\isakeyword{infixr}\ {\isachardoublequoteopen}{\isasymrightarrow}{\isachardoublequoteclose}\ {\isadigit{2}}{\isadigit{0}}{\isadigit{0}}{\isacharparenright}\isanewline
\ \ {\isacharbar}\ RefT\ ty\isanewline
\ \ {\isacharbar}\ DynT\isanewline
\isanewline
\isacommand{datatype}\isamarkupfalse%
\ const\isanewline
\ \ {\isacharequal}\ IntC\ int\isanewline
\ \ {\isacharbar}\ BoolC\ bool\isanewline
\isanewline
\isacommand{datatype}\isamarkupfalse%
\ opr\isanewline
\ \ {\isacharequal}\ Succ\isanewline
\ \ {\isacharbar}\ Prev\isanewline
\ \ {\isacharbar}\ IsZero\isanewline
\ \ {\isacharbar}\ Fst\ ty\ ty\isanewline
\ \ {\isacharbar}\ Snd\ ty\ ty\isanewline
\isanewline
\isacommand{type{\isacharunderscore}synonym}\isamarkupfalse%
\ name\ {\isacharequal}\ nat\isanewline
\isanewline
\isacommand{datatype}\isamarkupfalse%
\ stmt\isanewline
\ \ {\isacharequal}\ SLet\ name\ expr\ stmt\isanewline
\ \ {\isacharbar}\ SRet\ expr\isanewline
\ \ {\isacharbar}\ SCall\ name\ expr\ expr\ stmt\isanewline
\ \ {\isacharbar}\ STailCall\ expr\ expr\ \isanewline
\ \ {\isacharbar}\ SAlloc\ name\ ty\ expr\ stmt\isanewline
\ \ {\isacharbar}\ SUpdate\ expr\ expr\ stmt\isanewline
\ \ {\isacharbar}\ SDynUpdate\ expr\ expr\ ty\ stmt\isanewline
\ \ {\isacharbar}\ SCast\ name\ expr\ ty\ ty\ stmt\isanewline
\ \ {\isacharbar}\ SDynDeref\ name\ expr\ ty\ stmt\isanewline
\isanewline
\isakeyword{and}\ expr\ \isanewline
\ \ {\isacharequal}\ Var\ name\isanewline
\ \ {\isacharbar}\ Const\ const\isanewline
\ \ {\isacharbar}\ PrimApp\ opr\ expr\isanewline
\ \ {\isacharbar}\ MkPair\ expr\ expr\isanewline
\ \ {\isacharbar}\ Lam\ nat\ ty\ stmt\isanewline
\ \ {\isacharbar}\ Deref\ expr\isanewline
\isanewline
\isacommand{datatype}\isamarkupfalse%
\ val\isanewline
\ \ {\isacharequal}\ VConst\ const\isanewline
\ \ {\isacharbar}\ VPair\ val\ val\isanewline
\ \ {\isacharbar}\ Closure\ nat\ ty\ stmt\ {\isachardoublequoteopen}{\isacharparenleft}nat\ {\isasymtimes}\ val{\isacharparenright}\ list{\isachardoublequoteclose}\isanewline
\ \ {\isacharbar}\ VRef\ nat\isanewline
\ \ {\isacharbar}\ Inject\ val\ ty\isanewline
\isanewline
\isacommand{datatype}\isamarkupfalse%
\ casted{\isacharunderscore}val\isanewline
\ \ {\isacharequal}\ Val\ val\isanewline
\ \ {\isacharbar}\ VCast\ val\ ty\ ty%
\isamarkupsubsection{Result Monad%
}
\isamarkuptrue%
\isacommand{datatype}\isamarkupfalse%
\ {\isacharprime}a\ result\ {\isacharequal}\ Result\ {\isacharprime}a\ {\isacharbar}\ Stuck\ {\isacharbar}\ TimeOut\ {\isacharbar}\ CastError\isanewline
\isanewline
\isacommand{definition}\isamarkupfalse%
\isanewline
\ \ result{\isacharunderscore}bind\ {\isacharcolon}{\isacharcolon}\ {\isachardoublequoteopen}{\isacharbrackleft}{\isacharprime}a\ result{\isacharcomma}\ {\isacharprime}a\ {\isacharequal}{\isachargreater}\ {\isacharprime}b\ result{\isacharbrackright}\ {\isacharequal}{\isachargreater}\ {\isacharprime}b\ result{\isachardoublequoteclose}\ \isakeyword{where}\isanewline
\ \ {\isachardoublequoteopen}result{\isacharunderscore}bind\ m\ f\ {\isacharequal}\ {\isacharparenleft}case\ m\ of\ Stuck\ {\isacharequal}{\isachargreater}\ Stuck\isanewline
\ \ \ \ \ \ \ \ \ \ \ \ \ \ \ \ \ \ \ \ \ {\isacharbar}\ CastError\ {\isasymRightarrow}\ CastError\isanewline
\ \ \ \ \ \ \ \ \ \ \ \ \ \ \ \ \ \ \ \ \ {\isacharbar}\ Result\ r\ {\isacharequal}{\isachargreater}\ f\ r{\isacharparenright}{\isachardoublequoteclose}\isanewline
\isacommand{declare}\isamarkupfalse%
\ result{\isacharunderscore}bind{\isacharunderscore}def{\isacharbrackleft}simp{\isacharbrackright}\isanewline
\isanewline
\isacommand{syntax}\isamarkupfalse%
\ {\isachardoublequoteopen}{\isacharunderscore}result{\isacharunderscore}bind{\isachardoublequoteclose}\ {\isacharcolon}{\isacharcolon}\ {\isachardoublequoteopen}{\isacharbrackleft}pttrns{\isacharcomma}{\isacharprime}a\ result{\isacharcomma}{\isacharprime}b{\isacharbrackright}\ {\isacharequal}{\isachargreater}\ {\isacharprime}c{\isachardoublequoteclose}\ {\isacharparenleft}{\isachardoublequoteopen}{\isacharparenleft}{\isacharunderscore}\ {\isacharcolon}{\isacharequal}\ {\isacharunderscore}{\isacharsemicolon}{\isacharslash}{\isacharslash}{\isacharunderscore}{\isacharparenright}{\isachardoublequoteclose}\ {\isadigit{0}}{\isacharparenright}\isanewline
\isacommand{translations}\isamarkupfalse%
\ {\isachardoublequoteopen}P\ {\isacharcolon}{\isacharequal}\ E{\isacharsemicolon}\ F{\isachardoublequoteclose}\ {\isacharequal}{\isacharequal}\ {\isachardoublequoteopen}CONST\ result{\isacharunderscore}bind\ E\ {\isacharparenleft}{\isacharpercent}P{\isachardot}\ F{\isacharparenright}{\isachardoublequoteclose}\isanewline
\isanewline
\isacommand{definition}\isamarkupfalse%
\ return\ {\isacharcolon}{\isacharcolon}\ {\isachardoublequoteopen}{\isacharprime}a\ {\isasymRightarrow}\ {\isacharprime}a\ result{\isachardoublequoteclose}\ \isakeyword{where}\isanewline
\ \ {\isachardoublequoteopen}return\ x\ {\isasymequiv}\ Result\ x{\isachardoublequoteclose}\isanewline
\isacommand{declare}\isamarkupfalse%
\ return{\isacharunderscore}def{\isacharbrackleft}simp{\isacharbrackright}\isanewline
\isanewline
\isacommand{definition}\isamarkupfalse%
\ stuck\ {\isacharcolon}{\isacharcolon}\ {\isachardoublequoteopen}{\isacharprime}a\ result{\isachardoublequoteclose}\ \isakeyword{where}\isanewline
\ \ {\isachardoublequoteopen}stuck\ {\isasymequiv}\ Stuck{\isachardoublequoteclose}\isanewline
\isacommand{declare}\isamarkupfalse%
\ stuck{\isacharunderscore}def{\isacharbrackleft}simp{\isacharbrackright}\isanewline
\isanewline
\isacommand{definition}\isamarkupfalse%
\ cast{\isacharunderscore}error\ {\isacharcolon}{\isacharcolon}\ {\isachardoublequoteopen}{\isacharprime}a\ result{\isachardoublequoteclose}\ \isakeyword{where}\isanewline
\ \ {\isachardoublequoteopen}cast{\isacharunderscore}error\ {\isasymequiv}\ CastError{\isachardoublequoteclose}\isanewline
\isacommand{declare}\isamarkupfalse%
\ cast{\isacharunderscore}error{\isacharunderscore}def{\isacharbrackleft}simp{\isacharbrackright}%
\isamarkupsubsection{Operational Semantics%
}
\isamarkuptrue%
\isacommand{type{\isacharunderscore}synonym}\isamarkupfalse%
\ env\ {\isacharequal}\ {\isachardoublequoteopen}{\isacharparenleft}nat\ {\isasymtimes}\ val{\isacharparenright}\ list{\isachardoublequoteclose}\isanewline
\isamarkupcmt{Heap contains type-tagged possibly-casted values!%
}
\isanewline
\isacommand{type{\isacharunderscore}synonym}\isamarkupfalse%
\ heap\ {\isacharequal}\ {\isachardoublequoteopen}{\isacharparenleft}nat\ {\isasymtimes}\ {\isacharparenleft}casted{\isacharunderscore}val\ {\isasymtimes}\ ty{\isacharparenright}{\isacharparenright}\ list{\isachardoublequoteclose}\isanewline
\isacommand{type{\isacharunderscore}synonym}\isamarkupfalse%
\ stack\ {\isacharequal}\ {\isachardoublequoteopen}{\isacharparenleft}name\ {\isasymtimes}\ stmt\ {\isasymtimes}\ env{\isacharparenright}\ list{\isachardoublequoteclose}\isanewline
\isacommand{type{\isacharunderscore}synonym}\isamarkupfalse%
\ state\ {\isacharequal}\ {\isachardoublequoteopen}stmt\ {\isasymtimes}\ env\ {\isasymtimes}\ stack\ {\isasymtimes}\ heap\ {\isasymtimes}\ nat\ list{\isachardoublequoteclose}\isanewline
\isanewline
\isacommand{fun}\isamarkupfalse%
\ lookup\ {\isacharcolon}{\isacharcolon}\ {\isachardoublequoteopen}{\isacharprime}a\ {\isasymRightarrow}\ {\isacharparenleft}{\isacharprime}a\ {\isasymtimes}\ {\isacharprime}b{\isacharparenright}\ list\ {\isasymRightarrow}\ {\isacharprime}b\ result{\isachardoublequoteclose}\ \isakeyword{where}\isanewline
\ \ looknil{\isacharcolon}\ {\isachardoublequoteopen}lookup\ x\ {\isacharbrackleft}{\isacharbrackright}\ {\isacharequal}\ stuck{\isachardoublequoteclose}\ {\isacharbar}\isanewline
\ \ lookcons{\isacharcolon}\ {\isachardoublequoteopen}lookup\ x\ {\isacharparenleft}{\isacharparenleft}y{\isacharcomma}v{\isacharparenright}{\isacharhash}bs{\isacharparenright}\ {\isacharequal}\ {\isacharparenleft}if\ x\ {\isacharequal}\ y\ then\ return\ v\ else\ lookup\ x\ bs{\isacharparenright}{\isachardoublequoteclose}\isanewline
\isanewline
\isacommand{fun}\isamarkupfalse%
\ delta\ {\isacharcolon}{\isacharcolon}\ {\isachardoublequoteopen}opr\ {\isasymRightarrow}\ val\ {\isasymRightarrow}\ val\ result{\isachardoublequoteclose}\ \isakeyword{where}\isanewline
\ \ deltas{\isacharcolon}\ {\isachardoublequoteopen}delta\ Succ\ {\isacharparenleft}VConst\ {\isacharparenleft}IntC\ n{\isacharparenright}{\isacharparenright}\ {\isacharequal}\ return\ {\isacharparenleft}VConst\ {\isacharparenleft}IntC\ {\isacharparenleft}n\ {\isacharplus}\ {\isadigit{1}}{\isacharparenright}{\isacharparenright}{\isacharparenright}{\isachardoublequoteclose}\ {\isacharbar}\isanewline
\ \ deltap{\isacharcolon}\ {\isachardoublequoteopen}delta\ Prev\ {\isacharparenleft}VConst\ {\isacharparenleft}IntC\ n{\isacharparenright}{\isacharparenright}\ {\isacharequal}\ return\ {\isacharparenleft}VConst\ {\isacharparenleft}IntC\ {\isacharparenleft}n\ {\isacharminus}\ {\isadigit{1}}{\isacharparenright}{\isacharparenright}{\isacharparenright}{\isachardoublequoteclose}\ {\isacharbar}\isanewline
\ \ deltaz{\isacharcolon}\ {\isachardoublequoteopen}delta\ IsZero\ {\isacharparenleft}VConst\ {\isacharparenleft}IntC\ n{\isacharparenright}{\isacharparenright}\ {\isacharequal}\ return\ {\isacharparenleft}VConst\ {\isacharparenleft}BoolC\ {\isacharparenleft}n\ {\isacharequal}\ {\isadigit{0}}{\isacharparenright}{\isacharparenright}{\isacharparenright}{\isachardoublequoteclose}\ {\isacharbar}\isanewline
\ \ deltafst{\isacharcolon}\ {\isachardoublequoteopen}delta\ {\isacharparenleft}Fst\ A\ B{\isacharparenright}\ {\isacharparenleft}VPair\ v\ v{\isacharprime}{\isacharparenright}\ {\isacharequal}\ return\ v{\isachardoublequoteclose}\ {\isacharbar}\isanewline
\ \ deltasnd{\isacharcolon}\ {\isachardoublequoteopen}delta\ {\isacharparenleft}Snd\ A\ B{\isacharparenright}\ {\isacharparenleft}VPair\ v\ v{\isacharprime}{\isacharparenright}\ {\isacharequal}\ return\ v{\isacharprime}{\isachardoublequoteclose}\ {\isacharbar}\isanewline
\ \ deltastuck{\isacharcolon}\ {\isachardoublequoteopen}delta\ f\ v\ {\isacharequal}\ stuck{\isachardoublequoteclose}\isanewline
\isanewline
\isacommand{fun}\isamarkupfalse%
\ ground\ {\isacharcolon}{\isacharcolon}\ {\isachardoublequoteopen}ty\ {\isasymRightarrow}\ ty{\isachardoublequoteclose}\ \isakeyword{where}\isanewline
\ \ gndi{\isacharcolon}\ {\isachardoublequoteopen}ground\ IntT\ {\isacharequal}\ IntT{\isachardoublequoteclose}\ {\isacharbar}\isanewline
\ \ gndb{\isacharcolon}\ {\isachardoublequoteopen}ground\ BoolT\ {\isacharequal}\ BoolT{\isachardoublequoteclose}\ {\isacharbar}\isanewline
\ \ gndd{\isacharcolon}\ {\isachardoublequoteopen}ground\ DynT\ {\isacharequal}\ DynT{\isachardoublequoteclose}\ {\isacharbar}\isanewline
\ \ gndp{\isacharcolon}\ {\isachardoublequoteopen}ground\ {\isacharparenleft}A\ {\isasymtimes}\ B{\isacharparenright}\ {\isacharequal}\ {\isacharparenleft}DynT\ {\isasymtimes}\ DynT{\isacharparenright}{\isachardoublequoteclose}\ {\isacharbar}\isanewline
\ \ gndf{\isacharcolon}\ {\isachardoublequoteopen}ground\ {\isacharparenleft}A\ {\isasymrightarrow}\ B{\isacharparenright}\ {\isacharequal}\ {\isacharparenleft}DynT\ {\isasymrightarrow}\ DynT{\isacharparenright}{\isachardoublequoteclose}\ {\isacharbar}\isanewline
\ \ gndr{\isacharcolon}\ {\isachardoublequoteopen}ground\ {\isacharparenleft}RefT\ A{\isacharparenright}\ {\isacharequal}\ RefT\ DynT{\isachardoublequoteclose}\isanewline
\isanewline
\isacommand{fun}\isamarkupfalse%
\ to{\isacharunderscore}addr\ {\isacharcolon}{\isacharcolon}\ {\isachardoublequoteopen}val\ {\isasymRightarrow}\ nat\ result{\isachardoublequoteclose}\ \isakeyword{where}\isanewline
\ \ {\isachardoublequoteopen}to{\isacharunderscore}addr\ {\isacharparenleft}VRef\ a{\isacharparenright}\ {\isacharequal}\ return\ a{\isachardoublequoteclose}\ {\isacharbar}\isanewline
\ \ {\isachardoublequoteopen}to{\isacharunderscore}addr\ v\ {\isacharequal}\ stuck{\isachardoublequoteclose}\isanewline
\isanewline
\isacommand{fun}\isamarkupfalse%
\ to{\isacharunderscore}val\ {\isacharcolon}{\isacharcolon}\ {\isachardoublequoteopen}casted{\isacharunderscore}val\ {\isasymRightarrow}\ val\ result{\isachardoublequoteclose}\ \isakeyword{where}\isanewline
\ \ {\isachardoublequoteopen}to{\isacharunderscore}val\ {\isacharparenleft}Val\ v{\isacharparenright}\ {\isacharequal}\ return\ v{\isachardoublequoteclose}\ {\isacharbar}\isanewline
\ \ {\isachardoublequoteopen}to{\isacharunderscore}val\ v\ {\isacharequal}\ stuck{\isachardoublequoteclose}\isanewline
\isanewline
\isacommand{fun}\isamarkupfalse%
\ eval\ {\isacharcolon}{\isacharcolon}\ {\isachardoublequoteopen}expr\ {\isasymRightarrow}\ env\ {\isasymRightarrow}\ heap\ {\isasymRightarrow}\ val\ result{\isachardoublequoteclose}\ \isakeyword{where}\isanewline
\ \ evalv{\isacharcolon}\ {\isachardoublequoteopen}eval\ {\isacharparenleft}Var\ x{\isacharparenright}\ {\isasymrho}\ {\isasymmu}\ {\isacharequal}\ lookup\ x\ {\isasymrho}{\isachardoublequoteclose}\ {\isacharbar}\isanewline
\ \ evalc{\isacharcolon}\ {\isachardoublequoteopen}eval\ {\isacharparenleft}Const\ c{\isacharparenright}\ {\isasymrho}\ {\isasymmu}\ {\isacharequal}\ return\ {\isacharparenleft}VConst\ c{\isacharparenright}{\isachardoublequoteclose}\ {\isacharbar}\isanewline
\ \ evalpa{\isacharcolon}\ {\isachardoublequoteopen}eval\ {\isacharparenleft}PrimApp\ f\ e{\isacharparenright}\ {\isasymrho}\ {\isasymmu}\ {\isacharequal}\ {\isacharparenleft}v\ {\isacharcolon}{\isacharequal}\ eval\ e\ {\isasymrho}\ {\isasymmu}{\isacharsemicolon}\ delta\ f\ v{\isacharparenright}{\isachardoublequoteclose}\ {\isacharbar}\isanewline
\ \ evalp{\isacharcolon}\ {\isachardoublequoteopen}eval\ {\isacharparenleft}MkPair\ e{\isadigit{1}}\ e{\isadigit{2}}{\isacharparenright}\ {\isasymrho}\ {\isasymmu}\ {\isacharequal}\ {\isacharparenleft}v{\isadigit{1}}\ {\isacharcolon}{\isacharequal}\ eval\ e{\isadigit{1}}\ {\isasymrho}\ {\isasymmu}{\isacharsemicolon}\ v{\isadigit{2}}\ {\isacharcolon}{\isacharequal}\ eval\ e{\isadigit{2}}\ {\isasymrho}\ {\isasymmu}{\isacharsemicolon}\ \isanewline
\ \ \ \ \ \ return\ {\isacharparenleft}VPair\ v{\isadigit{1}}\ v{\isadigit{2}}{\isacharparenright}{\isacharparenright}{\isachardoublequoteclose}\ {\isacharbar}\isanewline
\ \ evall{\isacharcolon}\ {\isachardoublequoteopen}eval\ {\isacharparenleft}Lam\ x\ T\ s{\isacharparenright}\ {\isasymrho}\ {\isasymmu}\ {\isacharequal}\ return\ {\isacharparenleft}Closure\ x\ T\ s\ {\isasymrho}{\isacharparenright}{\isachardoublequoteclose}\ {\isacharbar}\isanewline
\ \ evald{\isacharcolon}\ {\isachardoublequoteopen}eval\ {\isacharparenleft}Deref\ e{\isacharparenright}\ {\isasymrho}\ {\isasymmu}\ {\isacharequal}\ {\isacharparenleft}v\ {\isacharcolon}{\isacharequal}\ eval\ e\ {\isasymrho}\ {\isasymmu}{\isacharsemicolon}\ a\ {\isacharcolon}{\isacharequal}\ to{\isacharunderscore}addr\ v{\isacharsemicolon}\isanewline
\ \ \ \ \ \ \ \ \ \ \ \ \ \ \ \ \ \ \ \ \ \ \ \ \ \ \ \ \ \ \ \ {\isacharparenleft}cv{\isacharcomma}A{\isacharparenright}\ {\isacharcolon}{\isacharequal}\ lookup\ a\ {\isasymmu}{\isacharsemicolon}\ to{\isacharunderscore}val\ cv{\isacharparenright}{\isachardoublequoteclose}\isanewline
\isanewline
\isacommand{fun}\isamarkupfalse%
\ lesseq{\isacharunderscore}dyn\ {\isacharcolon}{\isacharcolon}\ {\isachardoublequoteopen}ty\ {\isasymRightarrow}\ ty\ {\isasymRightarrow}\ bool{\isachardoublequoteclose}\ {\isacharparenleft}\isakeyword{infix}\ {\isachardoublequoteopen}{\isasymsqsubseteq}{\isachardoublequoteclose}\ {\isadigit{7}}{\isadigit{9}}{\isacharparenright}\ \isakeyword{where}\isanewline
\ \ lsda{\isacharcolon}\ {\isachardoublequoteopen}A\ {\isasymsqsubseteq}\ DynT\ {\isacharequal}\ True{\isachardoublequoteclose}\ {\isacharbar}\isanewline
\ \ lsii{\isacharcolon}\ {\isachardoublequoteopen}IntT\ {\isasymsqsubseteq}\ IntT\ {\isacharequal}\ True{\isachardoublequoteclose}\ {\isacharbar}\isanewline
\ \ lsbb{\isacharcolon}\ {\isachardoublequoteopen}BoolT\ {\isasymsqsubseteq}\ BoolT\ {\isacharequal}\ True{\isachardoublequoteclose}\ {\isacharbar}\isanewline
\ \ lspp{\isacharcolon}\ {\isachardoublequoteopen}{\isacharparenleft}A\ {\isasymtimes}\ B{\isacharparenright}\ {\isasymsqsubseteq}\ {\isacharparenleft}C\ {\isasymtimes}\ D{\isacharparenright}\ {\isacharequal}\ {\isacharparenleft}A\ {\isasymsqsubseteq}\ C\ {\isasymand}\ B\ {\isasymsqsubseteq}\ D{\isacharparenright}{\isachardoublequoteclose}\ {\isacharbar}\isanewline
\ \ lsff{\isacharcolon}\ {\isachardoublequoteopen}{\isacharparenleft}A\ {\isasymrightarrow}\ B{\isacharparenright}\ {\isasymsqsubseteq}\ {\isacharparenleft}C\ {\isasymrightarrow}\ D{\isacharparenright}\ {\isacharequal}\ {\isacharparenleft}A\ {\isasymsqsubseteq}\ C\ {\isasymand}\ B\ {\isasymsqsubseteq}\ D{\isacharparenright}{\isachardoublequoteclose}\ {\isacharbar}\isanewline
\ \ lsrr{\isacharcolon}\ {\isachardoublequoteopen}{\isacharparenleft}RefT\ A{\isacharparenright}\ {\isasymsqsubseteq}\ {\isacharparenleft}RefT\ B{\isacharparenright}\ {\isacharequal}\ {\isacharparenleft}A\ {\isasymsqsubseteq}\ B{\isacharparenright}{\isachardoublequoteclose}\ {\isacharbar}\isanewline
\ \ lsnot{\isacharcolon}\ {\isachardoublequoteopen}A\ {\isasymsqsubseteq}\ B\ {\isacharequal}\ False{\isachardoublequoteclose}\isanewline
\isanewline
\isacommand{fun}\isamarkupfalse%
\ meet\ {\isacharcolon}{\isacharcolon}\ {\isachardoublequoteopen}ty\ {\isasymRightarrow}\ ty\ {\isasymRightarrow}\ ty\ result{\isachardoublequoteclose}\ \isakeyword{where}\isanewline
\ \ meetda{\isacharcolon}\ {\isachardoublequoteopen}meet\ DynT\ A\ {\isacharequal}\ return\ A{\isachardoublequoteclose}\ {\isacharbar}\isanewline
\ \ meetad{\isacharcolon}\ {\isachardoublequoteopen}meet\ A\ DynT\ {\isacharequal}\ return\ A{\isachardoublequoteclose}\ {\isacharbar}\isanewline
\ \ meetii{\isacharcolon}\ {\isachardoublequoteopen}meet\ IntT\ IntT\ {\isacharequal}\ return\ IntT{\isachardoublequoteclose}\ {\isacharbar}\isanewline
\ \ meetbb{\isacharcolon}\ {\isachardoublequoteopen}meet\ BoolT\ BoolT\ {\isacharequal}\ return\ BoolT{\isachardoublequoteclose}\ {\isacharbar}\isanewline
\ \ meetpp{\isacharcolon}\ {\isachardoublequoteopen}meet\ {\isacharparenleft}A\ {\isasymtimes}\ B{\isacharparenright}\ {\isacharparenleft}C\ {\isasymtimes}\ D{\isacharparenright}\ {\isacharequal}\ \isanewline
\ \ \ \ \ {\isacharparenleft}A{\isacharprime}\ {\isacharcolon}{\isacharequal}\ meet\ A\ C{\isacharsemicolon}\ B{\isacharprime}\ {\isacharcolon}{\isacharequal}\ meet\ B\ D{\isacharsemicolon}\ return\ {\isacharparenleft}A{\isacharprime}\ {\isasymtimes}\ B{\isacharprime}{\isacharparenright}{\isacharparenright}{\isachardoublequoteclose}\ {\isacharbar}\isanewline
\ \ meetff{\isacharcolon}\ {\isachardoublequoteopen}meet\ {\isacharparenleft}A\ {\isasymrightarrow}\ B{\isacharparenright}\ {\isacharparenleft}C\ {\isasymrightarrow}\ D{\isacharparenright}\ {\isacharequal}\ \isanewline
\ \ \ \ \ {\isacharparenleft}A{\isacharprime}\ {\isacharcolon}{\isacharequal}\ meet\ A\ C{\isacharsemicolon}\ B{\isacharprime}\ {\isacharcolon}{\isacharequal}\ meet\ B\ D{\isacharsemicolon}\ return\ {\isacharparenleft}A{\isacharprime}\ {\isasymrightarrow}\ B{\isacharprime}{\isacharparenright}{\isacharparenright}{\isachardoublequoteclose}\ {\isacharbar}\isanewline
\ \ meetrr{\isacharcolon}\ {\isachardoublequoteopen}meet\ {\isacharparenleft}RefT\ A{\isacharparenright}\ {\isacharparenleft}RefT\ B{\isacharparenright}\ {\isacharequal}\ \isanewline
\ \ \ \ \ {\isacharparenleft}A{\isacharprime}\ {\isacharcolon}{\isacharequal}\ meet\ A\ B{\isacharsemicolon}\ return\ {\isacharparenleft}RefT\ A{\isacharprime}{\isacharparenright}{\isacharparenright}{\isachardoublequoteclose}\ {\isacharbar}\isanewline
\ \ meeterr{\isacharcolon}\ {\isachardoublequoteopen}meet\ A\ B\ {\isacharequal}\ cast{\isacharunderscore}error{\isachardoublequoteclose}\isanewline
\isanewline
\isacommand{fun}\isamarkupfalse%
\ mk{\isacharunderscore}vcast\ {\isacharcolon}{\isacharcolon}\ {\isachardoublequoteopen}casted{\isacharunderscore}val\ {\isasymRightarrow}\ ty\ {\isasymRightarrow}\ ty\ {\isasymRightarrow}\ casted{\isacharunderscore}val{\isachardoublequoteclose}\ \isakeyword{where}\isanewline
\ \ vcastv{\isacharcolon}\ {\isachardoublequoteopen}mk{\isacharunderscore}vcast\ {\isacharparenleft}Val\ v{\isacharparenright}\ A\ B\ {\isacharequal}\ VCast\ v\ A\ B{\isachardoublequoteclose}\ {\isacharbar}\isanewline
\ \ vcastcv{\isacharcolon}\ {\isachardoublequoteopen}mk{\isacharunderscore}vcast\ {\isacharparenleft}VCast\ v\ A\ B{\isacharparenright}\ C\ D\ {\isacharequal}\ VCast\ v\ A\ D{\isachardoublequoteclose}\isanewline
\isanewline
\isacommand{definition}\isamarkupfalse%
\ wrap\ {\isacharcolon}{\isacharcolon}\ {\isachardoublequoteopen}val\ {\isasymRightarrow}\ ty\ {\isasymRightarrow}\ ty\ {\isasymRightarrow}\ ty\ {\isasymRightarrow}\ ty\ {\isasymRightarrow}\ val{\isachardoublequoteclose}\ \isakeyword{where}\isanewline
\ \ {\isachardoublequoteopen}wrap\ v\ A\ B\ C\ D\ {\isasymequiv}\ {\isacharparenleft}Closure\ {\isadigit{0}}\ C\isanewline
\ \ \ \ \ \ \ \ \ \ \ \ \ \ \ \ {\isacharparenleft}SCast\ {\isadigit{3}}\ {\isacharparenleft}Var\ {\isadigit{0}}{\isacharparenright}\ C\ A\ \isanewline
\ \ \ \ \ \ \ \ \ \ \ \ \ \ \ \ {\isacharparenleft}SCall\ {\isadigit{2}}\ {\isacharparenleft}Var\ {\isadigit{1}}{\isacharparenright}\ {\isacharparenleft}Var\ {\isadigit{3}}{\isacharparenright}\isanewline
\ \ \ \ \ \ \ \ \ \ \ \ \ \ \ \ {\isacharparenleft}SCast\ {\isadigit{4}}\ {\isacharparenleft}Var\ {\isadigit{2}}{\isacharparenright}\ B\ D\isanewline
\ \ \ \ \ \ \ \ \ \ \ \ \ \ \ \ \ \ \ {\isacharparenleft}SRet\ {\isacharparenleft}Var\ {\isadigit{4}}{\isacharparenright}{\isacharparenright}{\isacharparenright}{\isacharparenright}{\isacharparenright}\isanewline
\ \ \ \ \ \ \ \ \ \ \ \ \ \ \ \ {\isacharbrackleft}{\isacharparenleft}{\isadigit{1}}{\isacharcomma}v{\isacharparenright}{\isacharbrackright}{\isacharparenright}{\isachardoublequoteclose}\isanewline
\isacommand{declare}\isamarkupfalse%
\ wrap{\isacharunderscore}def{\isacharbrackleft}simp{\isacharbrackright}\isanewline
\isanewline
\isacommand{fun}\isamarkupfalse%
\ cast\ {\isacharcolon}{\isacharcolon}\ {\isachardoublequoteopen}val\ {\isasymRightarrow}\ ty\ {\isasymRightarrow}\ ty\ {\isasymRightarrow}\ heap\ \ {\isasymRightarrow}\ nat\ list\ {\isasymRightarrow}\ {\isacharparenleft}val\ {\isasymtimes}\ heap\ {\isasymtimes}\ {\isacharparenleft}nat\ list{\isacharparenright}{\isacharparenright}\ result{\isachardoublequoteclose}\ \isakeyword{where}\isanewline
\ \ castii{\isacharcolon}\ {\isachardoublequoteopen}cast\ v\ IntT\ IntT\ {\isasymmu}\ as\ {\isacharequal}\ return\ {\isacharparenleft}v{\isacharcomma}{\isasymmu}{\isacharcomma}as{\isacharparenright}{\isachardoublequoteclose}\ {\isacharbar}\isanewline
\ \ castbb{\isacharcolon}\ {\isachardoublequoteopen}cast\ v\ BoolT\ BoolT\ {\isasymmu}\ as\ {\isacharequal}\ return\ {\isacharparenleft}v{\isacharcomma}{\isasymmu}{\isacharcomma}as{\isacharparenright}{\isachardoublequoteclose}\ {\isacharbar}\isanewline
\ \ castdd{\isacharcolon}\ {\isachardoublequoteopen}cast\ v\ DynT\ DynT\ {\isasymmu}\ as\ {\isacharequal}\ return\ {\isacharparenleft}v{\isacharcomma}{\isasymmu}{\isacharcomma}as{\isacharparenright}{\isachardoublequoteclose}\ {\isacharbar}\isanewline
\ \ castff{\isacharcolon}\ {\isachardoublequoteopen}cast\ v\ {\isacharparenleft}A\ {\isasymrightarrow}\ B{\isacharparenright}\ {\isacharparenleft}C\ {\isasymrightarrow}\ D{\isacharparenright}\ {\isasymmu}\ as\ {\isacharequal}\ return\ {\isacharparenleft}wrap\ v\ A\ B\ C\ D{\isacharcomma}\ {\isasymmu}{\isacharcomma}\ as{\isacharparenright}{\isachardoublequoteclose}\ {\isacharbar}\isanewline
\ \ castpp{\isacharcolon}\ {\isachardoublequoteopen}cast\ {\isacharparenleft}VPair\ v{\isadigit{1}}\ v{\isadigit{2}}{\isacharparenright}\ {\isacharparenleft}A\ {\isasymtimes}\ B{\isacharparenright}\ {\isacharparenleft}C\ {\isasymtimes}\ D{\isacharparenright}\ {\isasymmu}\ as\ {\isacharequal}\ \isanewline
\ \ \ \ \ {\isacharparenleft}{\isacharparenleft}v{\isadigit{1}}{\isacharprime}{\isacharcomma}{\isasymmu}{\isadigit{1}}{\isacharcomma}as{\isadigit{1}}{\isacharparenright}\ {\isacharcolon}{\isacharequal}\ cast\ v{\isadigit{1}}\ A\ C\ {\isasymmu}\ as{\isacharsemicolon}\isanewline
\ \ \ \ \ \ {\isacharparenleft}v{\isadigit{2}}{\isacharprime}{\isacharcomma}{\isasymmu}{\isadigit{2}}{\isacharcomma}as{\isadigit{2}}{\isacharparenright}\ {\isacharcolon}{\isacharequal}\ cast\ v{\isadigit{2}}\ B\ D\ {\isasymmu}{\isadigit{1}}\ as{\isadigit{1}}{\isacharsemicolon}\isanewline
\ \ \ \ \ \ return\ {\isacharparenleft}VPair\ v{\isadigit{1}}{\isacharprime}\ v{\isadigit{2}}{\isacharprime}{\isacharcomma}\ {\isasymmu}{\isadigit{2}}{\isacharcomma}\ as{\isadigit{2}}{\isacharparenright}{\isacharparenright}{\isachardoublequoteclose}\ {\isacharbar}\isanewline
\ \ castrr{\isacharcolon}\ {\isachardoublequoteopen}cast\ {\isacharparenleft}VRef\ a{\isacharparenright}\ {\isacharparenleft}RefT\ A{\isacharparenright}\ {\isacharparenleft}RefT\ B{\isacharparenright}\ {\isasymmu}\ as\ {\isacharequal}\isanewline
\ \ \ \ \ {\isacharparenleft}{\isacharparenleft}cv{\isacharcomma}C{\isacharparenright}\ {\isacharcolon}{\isacharequal}\ lookup\ a\ {\isasymmu}{\isacharsemicolon}\isanewline
\ \ \ \ \ \ BC\ {\isacharcolon}{\isacharequal}\ meet\ B\ C{\isacharsemicolon}\isanewline
\ \ \ \ \ \ if\ C\ {\isasymsqsubseteq}\ BC\ then\ return\ {\isacharparenleft}VRef\ a{\isacharcomma}\ {\isasymmu}{\isacharcomma}\ as{\isacharparenright}\isanewline
\ \ \ \ \ \ else\ return\ {\isacharparenleft}VRef\ a{\isacharcomma}\ {\isacharparenleft}a{\isacharcomma}{\isacharparenleft}mk{\isacharunderscore}vcast\ cv\ C\ BC{\isacharcomma}\ BC{\isacharparenright}{\isacharparenright}{\isacharhash}{\isasymmu}{\isacharcomma}\ a{\isacharhash}as{\isacharparenright}{\isacharparenright}{\isachardoublequoteclose}\ {\isacharbar}\isanewline
\ \ castinj{\isacharcolon}\ {\isachardoublequoteopen}cast\ {\isacharparenleft}Inject\ v\ T{\isadigit{1}}{\isacharparenright}\ DynT\ T{\isadigit{2}}\ {\isasymmu}\ as\ {\isacharequal}\isanewline
\ \ \ \ {\isacharparenleft}if\ ground\ T{\isadigit{1}}\ {\isacharequal}\ ground\ T{\isadigit{2}}\ then\ cast\ v\ T{\isadigit{1}}\ T{\isadigit{2}}\ {\isasymmu}\ as\isanewline
\ \ \ \ \ else\ cast{\isacharunderscore}error{\isacharparenright}{\isachardoublequoteclose}\ {\isacharbar}\isanewline
\ \ casttd{\isacharcolon}\ {\isachardoublequoteopen}cast\ v\ T\ DynT\ {\isasymmu}\ as\ {\isacharequal}\ return\ {\isacharparenleft}Inject\ v\ T{\isacharcomma}\ {\isasymmu}{\isacharcomma}\ as{\isacharparenright}{\isachardoublequoteclose}\ {\isacharbar}\isanewline
\ \ casterr{\isacharcolon}\ {\isachardoublequoteopen}cast\ v\ T{\isadigit{1}}\ T{\isadigit{2}}\ {\isasymmu}\ as\ {\isacharequal}\ cast{\isacharunderscore}error{\isachardoublequoteclose}\isanewline
\isanewline
\isacommand{fun}\isamarkupfalse%
\ step\ {\isacharcolon}{\isacharcolon}\ {\isachardoublequoteopen}state\ {\isasymRightarrow}\ state\ result{\isachardoublequoteclose}\ \isakeyword{where}\isanewline
\ \ {\isachardoublequoteopen}step\ {\isacharparenleft}s{\isacharcomma}\ {\isasymrho}{\isacharcomma}\ k{\isacharcomma}\ {\isasymmu}{\isacharcomma}\ a{\isacharhash}ads{\isacharparenright}\ {\isacharequal}\ \isanewline
\ \ \ \ \ \ \ \ {\isacharparenleft}{\isacharparenleft}cv{\isacharcomma}\ A{\isacharparenright}\ {\isacharcolon}{\isacharequal}\ lookup\ a\ {\isasymmu}{\isacharsemicolon}\isanewline
\ \ \ \ \ \ \ \ \ {\isacharparenleft}case\ cv\ of\isanewline
\ \ \ \ \ \ \ \ \ \ \ \ Val\ v\ {\isasymRightarrow}\ return\ {\isacharparenleft}s{\isacharcomma}\ {\isasymrho}{\isacharcomma}\ k{\isacharcomma}\ {\isasymmu}{\isacharcomma}\ ads{\isacharparenright}\isanewline
\ \ \ \ \ \ \ \ \ \ {\isacharbar}\ VCast\ v\ B\ C\ {\isasymRightarrow}\isanewline
\ \ \ \ \ \ \ \ \ \ \ \ \ \ {\isacharparenleft}v{\isacharprime}{\isacharcomma}{\isasymmu}{\isacharprime}{\isacharcomma}ads{\isadigit{2}}{\isacharparenright}\ {\isacharcolon}{\isacharequal}\ cast\ v\ B\ C\ {\isasymmu}\ {\isacharparenleft}a{\isacharhash}ads{\isacharparenright}{\isacharsemicolon}\isanewline
\ \ \ \ \ \ \ \ \ \ \ \ \ \ {\isacharparenleft}cv{\isacharprime}{\isacharcomma}A{\isacharprime}{\isacharparenright}\ {\isacharcolon}{\isacharequal}\ lookup\ a\ {\isasymmu}{\isacharprime}{\isacharsemicolon}\isanewline
\ \ \ \ \ \ \ \ \ \ \ \ \ \ if\ A\ {\isasymsqsubseteq}\ A{\isacharprime}\ then\isanewline
\ \ \ \ \ \ \ \ \ \ \ \ \ \ \ \ \ \ return\ {\isacharparenleft}s{\isacharcomma}\ {\isasymrho}{\isacharcomma}\ k{\isacharcomma}\ {\isacharparenleft}a{\isacharcomma}{\isacharparenleft}Val\ v{\isacharprime}{\isacharcomma}A{\isacharparenright}{\isacharparenright}{\isacharhash}{\isasymmu}{\isacharprime}{\isacharcomma}\ removeAll\ a\ ads{\isadigit{2}}{\isacharparenright}\isanewline
\ \ \ \ \ \ \ \ \ \ \ \ \ \ else\ return\ {\isacharparenleft}s{\isacharcomma}\ {\isasymrho}{\isacharcomma}\ k{\isacharcomma}\ {\isasymmu}{\isacharprime}{\isacharcomma}\ ads{\isadigit{2}}{\isacharparenright}{\isacharparenright}{\isacharparenright}{\isachardoublequoteclose}\ {\isacharbar}\isanewline
\ \ {\isachardoublequoteopen}step\ {\isacharparenleft}SLet\ x\ e\ s{\isacharcomma}\ {\isasymrho}{\isacharcomma}\ k{\isacharcomma}\ {\isasymmu}{\isacharcomma}\ {\isacharbrackleft}{\isacharbrackright}{\isacharparenright}\ {\isacharequal}\isanewline
\ \ \ \ {\isacharparenleft}v\ {\isacharcolon}{\isacharequal}\ eval\ e\ {\isasymrho}\ {\isasymmu}{\isacharsemicolon}\ \isanewline
\ \ \ \ \ return\ {\isacharparenleft}s{\isacharcomma}\ {\isacharparenleft}x{\isacharcomma}v{\isacharparenright}{\isacharhash}{\isasymrho}\ {\isacharcomma}\ k{\isacharcomma}\ {\isasymmu}{\isacharcomma}\ {\isacharbrackleft}{\isacharbrackright}{\isacharparenright}{\isacharparenright}{\isachardoublequoteclose}\ {\isacharbar}\isanewline
\ \ {\isachardoublequoteopen}step\ {\isacharparenleft}SRet\ e{\isacharcomma}\ {\isasymrho}{\isacharcomma}\ {\isacharparenleft}x{\isacharcomma}\ s{\isacharcomma}\ {\isasymrho}{\isacharprime}{\isacharparenright}{\isacharhash}k{\isacharcomma}\ {\isasymmu}{\isacharcomma}\ {\isacharbrackleft}{\isacharbrackright}{\isacharparenright}\ {\isacharequal}\isanewline
\ \ \ \ \ \ {\isacharparenleft}v\ {\isacharcolon}{\isacharequal}\ eval\ e\ {\isasymrho}\ {\isasymmu}{\isacharsemicolon}\isanewline
\ \ \ \ \ \ \ return\ {\isacharparenleft}s{\isacharcomma}\ {\isacharparenleft}x{\isacharcomma}v{\isacharparenright}{\isacharhash}{\isasymrho}{\isacharprime}{\isacharcomma}\ k{\isacharcomma}\ {\isasymmu}{\isacharcomma}\ {\isacharbrackleft}{\isacharbrackright}{\isacharparenright}{\isacharparenright}{\isachardoublequoteclose}\ {\isacharbar}\isanewline
\ \ {\isachardoublequoteopen}step\ {\isacharparenleft}SCall\ x\ e{\isadigit{1}}\ e{\isadigit{2}}\ s{\isacharcomma}\ {\isasymrho}{\isacharcomma}\ k{\isacharcomma}\ {\isasymmu}{\isacharcomma}\ {\isacharbrackleft}{\isacharbrackright}{\isacharparenright}\ {\isacharequal}\isanewline
\ \ \ \ \ {\isacharparenleft}v{\isadigit{1}}\ {\isacharcolon}{\isacharequal}\ eval\ e{\isadigit{1}}\ {\isasymrho}\ {\isasymmu}{\isacharsemicolon}\ v{\isadigit{2}}\ {\isacharcolon}{\isacharequal}\ eval\ e{\isadigit{2}}\ {\isasymrho}\ {\isasymmu}{\isacharsemicolon}\isanewline
\ \ \ \ \ \ case\ v{\isadigit{1}}\ of\isanewline
\ \ \ \ \ \ \ \ Closure\ y\ T\ s{\isacharprime}\ {\isasymrho}{\isacharprime}\ {\isasymRightarrow}\isanewline
\ \ \ \ \ \ \ \ \ \ return\ {\isacharparenleft}s{\isacharprime}{\isacharcomma}\ {\isacharparenleft}y{\isacharcomma}v{\isadigit{2}}{\isacharparenright}{\isacharhash}{\isasymrho}{\isacharprime}{\isacharcomma}\ {\isacharparenleft}x{\isacharcomma}\ s{\isacharcomma}{\isasymrho}{\isacharparenright}{\isacharhash}k{\isacharcomma}\ {\isasymmu}{\isacharcomma}\ {\isacharbrackleft}{\isacharbrackright}{\isacharparenright}\isanewline
\ \ \ \ \ \ {\isacharbar}\ {\isacharunderscore}\ {\isasymRightarrow}\ stuck{\isacharparenright}{\isachardoublequoteclose}\ {\isacharbar}\isanewline
\ \ {\isachardoublequoteopen}step\ {\isacharparenleft}STailCall\ e{\isadigit{1}}\ e{\isadigit{2}}{\isacharcomma}\ {\isasymrho}{\isacharcomma}\ k{\isacharcomma}\ {\isasymmu}{\isacharcomma}\ {\isacharbrackleft}{\isacharbrackright}{\isacharparenright}\ {\isacharequal}\isanewline
\ \ \ \ \ {\isacharparenleft}v{\isadigit{1}}\ {\isacharcolon}{\isacharequal}\ eval\ e{\isadigit{1}}\ {\isasymrho}\ {\isasymmu}{\isacharsemicolon}\ v{\isadigit{2}}\ {\isacharcolon}{\isacharequal}\ eval\ e{\isadigit{2}}\ {\isasymrho}\ {\isasymmu}{\isacharsemicolon}\isanewline
\ \ \ \ \ \ case\ v{\isadigit{1}}\ of\isanewline
\ \ \ \ \ \ \ \ Closure\ y\ T\ s{\isacharprime}\ {\isasymrho}{\isacharprime}\ {\isasymRightarrow}\isanewline
\ \ \ \ \ \ \ \ \ \ return\ {\isacharparenleft}s{\isacharprime}{\isacharcomma}\ {\isacharparenleft}y{\isacharcomma}v{\isadigit{2}}{\isacharparenright}{\isacharhash}{\isasymrho}{\isacharprime}{\isacharcomma}\ k{\isacharcomma}\ {\isasymmu}{\isacharcomma}\ {\isacharbrackleft}{\isacharbrackright}{\isacharparenright}\isanewline
\ \ \ \ \ \ {\isacharbar}\ {\isacharunderscore}\ {\isasymRightarrow}\ stuck{\isacharparenright}{\isachardoublequoteclose}\ {\isacharbar}\isanewline
\ \ {\isachardoublequoteopen}step\ {\isacharparenleft}SAlloc\ x\ A\ e\ s{\isacharcomma}\ {\isasymrho}{\isacharcomma}\ k{\isacharcomma}\ {\isasymmu}{\isacharcomma}\ {\isacharbrackleft}{\isacharbrackright}{\isacharparenright}\ {\isacharequal}\isanewline
\ \ \ \ \ {\isacharparenleft}v\ {\isacharcolon}{\isacharequal}\ eval\ e\ {\isasymrho}\ {\isasymmu}{\isacharsemicolon}\isanewline
\ \ \ \ \ \ let\ a\ {\isacharequal}\ length\ {\isasymmu}\ in\isanewline
\ \ \ \ \ \ return\ {\isacharparenleft}s{\isacharcomma}\ {\isacharparenleft}x{\isacharcomma}VRef\ a{\isacharparenright}{\isacharhash}{\isasymrho}{\isacharcomma}\ k{\isacharcomma}\ {\isacharparenleft}a{\isacharcomma}{\isacharparenleft}Val\ v{\isacharcomma}A{\isacharparenright}{\isacharparenright}{\isacharhash}{\isasymmu}{\isacharcomma}\ {\isacharbrackleft}{\isacharbrackright}{\isacharparenright}{\isacharparenright}{\isachardoublequoteclose}\ {\isacharbar}\isanewline
\ \ {\isachardoublequoteopen}step\ {\isacharparenleft}SUpdate\ e{\isadigit{1}}\ e{\isadigit{2}}\ s{\isacharcomma}\ {\isasymrho}{\isacharcomma}\ k{\isacharcomma}\ {\isasymmu}{\isacharcomma}\ {\isacharbrackleft}{\isacharbrackright}{\isacharparenright}\ {\isacharequal}\isanewline
\ \ \ \ \ \ {\isacharparenleft}v{\isadigit{1}}\ {\isacharcolon}{\isacharequal}\ eval\ e{\isadigit{1}}\ {\isasymrho}\ {\isasymmu}{\isacharsemicolon}\ v{\isadigit{2}}\ {\isacharcolon}{\isacharequal}\ eval\ e{\isadigit{2}}\ {\isasymrho}\ {\isasymmu}{\isacharsemicolon}\ a\ {\isacharcolon}{\isacharequal}\ to{\isacharunderscore}addr\ v{\isadigit{1}}{\isacharsemicolon}\isanewline
\ \ \ \ \ \ \ {\isacharparenleft}cv{\isacharcomma}A{\isacharparenright}\ {\isacharcolon}{\isacharequal}\ lookup\ a\ {\isasymmu}{\isacharsemicolon}\isanewline
\ \ \ \ \ \ \ return\ {\isacharparenleft}s{\isacharcomma}\ {\isasymrho}{\isacharcomma}\ k{\isacharcomma}\ {\isacharparenleft}a{\isacharcomma}{\isacharparenleft}Val\ v{\isadigit{2}}{\isacharcomma}A{\isacharparenright}{\isacharparenright}{\isacharhash}{\isasymmu}{\isacharcomma}\ {\isacharbrackleft}{\isacharbrackright}{\isacharparenright}{\isacharparenright}{\isachardoublequoteclose}\ {\isacharbar}\isanewline
\ \ {\isachardoublequoteopen}step\ {\isacharparenleft}SDynUpdate\ e{\isadigit{1}}\ e{\isadigit{2}}\ A\ s{\isacharcomma}\ {\isasymrho}{\isacharcomma}\ k{\isacharcomma}\ {\isasymmu}{\isacharcomma}\ {\isacharbrackleft}{\isacharbrackright}{\isacharparenright}\ {\isacharequal}\isanewline
\ \ \ \ \ \ {\isacharparenleft}v{\isadigit{1}}\ {\isacharcolon}{\isacharequal}\ eval\ e{\isadigit{1}}\ {\isasymrho}\ {\isasymmu}{\isacharsemicolon}\ v{\isadigit{2}}\ {\isacharcolon}{\isacharequal}\ eval\ e{\isadigit{2}}\ {\isasymrho}\ {\isasymmu}{\isacharsemicolon}\ a\ {\isacharcolon}{\isacharequal}\ to{\isacharunderscore}addr\ v{\isadigit{1}}{\isacharsemicolon}\isanewline
\ \ \ \ \ \ \ {\isacharparenleft}cv{\isacharcomma}B{\isacharparenright}\ {\isacharcolon}{\isacharequal}\ lookup\ a\ {\isasymmu}{\isacharsemicolon}\isanewline
\ \ \ \ \ \ \ return\ {\isacharparenleft}s{\isacharcomma}\ {\isasymrho}{\isacharcomma}\ k{\isacharcomma}\ {\isacharparenleft}a{\isacharcomma}\ {\isacharparenleft}VCast\ v{\isadigit{2}}\ A\ B{\isacharcomma}B{\isacharparenright}{\isacharparenright}{\isacharhash}{\isasymmu}{\isacharcomma}\ {\isacharbrackleft}a{\isacharbrackright}{\isacharparenright}{\isacharparenright}{\isachardoublequoteclose}\ {\isacharbar}\isanewline
\ \ {\isachardoublequoteopen}step\ {\isacharparenleft}SCast\ x\ e\ A\ B\ s{\isacharcomma}\ {\isasymrho}{\isacharcomma}\ k{\isacharcomma}\ {\isasymmu}{\isacharcomma}\ {\isacharbrackleft}{\isacharbrackright}{\isacharparenright}\ {\isacharequal}\isanewline
\ \ \ \ \ \ \ {\isacharparenleft}v\ {\isacharcolon}{\isacharequal}\ eval\ e\ {\isasymrho}\ {\isasymmu}{\isacharsemicolon}\ {\isacharparenleft}v{\isacharprime}{\isacharcomma}{\isasymmu}{\isacharprime}{\isacharcomma}ads{\isacharparenright}\ {\isacharcolon}{\isacharequal}\ cast\ v\ A\ B\ {\isasymmu}\ {\isacharbrackleft}{\isacharbrackright}{\isacharsemicolon}\isanewline
\ \ \ \ \ \ \ \ return\ {\isacharparenleft}s{\isacharcomma}\ {\isacharparenleft}x{\isacharcomma}\ v{\isacharprime}{\isacharparenright}{\isacharhash}{\isasymrho}{\isacharcomma}\ k{\isacharcomma}\ {\isasymmu}{\isacharprime}{\isacharcomma}\ ads{\isacharparenright}{\isacharparenright}{\isachardoublequoteclose}\ {\isacharbar}\isanewline
\ \ {\isachardoublequoteopen}step\ {\isacharparenleft}SDynDeref\ x\ e\ A\ s{\isacharcomma}\ {\isasymrho}{\isacharcomma}\ k{\isacharcomma}\ {\isasymmu}{\isacharcomma}\ {\isacharbrackleft}{\isacharbrackright}{\isacharparenright}\ {\isacharequal}\isanewline
\ \ \ \ \ \ \ \ {\isacharparenleft}v\ {\isacharcolon}{\isacharequal}\ eval\ e\ {\isasymrho}\ {\isasymmu}{\isacharsemicolon}\ a\ {\isacharcolon}{\isacharequal}\ to{\isacharunderscore}addr\ v{\isacharsemicolon}\ {\isacharparenleft}cv{\isacharcomma}B{\isacharparenright}\ {\isacharcolon}{\isacharequal}\ lookup\ a\ {\isasymmu}{\isacharsemicolon}\isanewline
\ \ \ \ \ \ \ \ \ v{\isadigit{1}}\ {\isacharcolon}{\isacharequal}\ to{\isacharunderscore}val\ cv{\isacharsemicolon}\ {\isacharparenleft}v{\isadigit{2}}{\isacharcomma}{\isasymmu}{\isacharprime}{\isacharcomma}ads{\isacharparenright}\ {\isacharcolon}{\isacharequal}\ cast\ v{\isadigit{1}}\ B\ A\ {\isasymmu}\ {\isacharbrackleft}{\isacharbrackright}{\isacharsemicolon}\isanewline
\ \ \ \ \ \ \ \ \ return\ {\isacharparenleft}s{\isacharcomma}\ {\isacharparenleft}x{\isacharcomma}v{\isadigit{2}}{\isacharparenright}{\isacharhash}{\isasymrho}{\isacharcomma}\ k{\isacharcomma}\ {\isasymmu}{\isacharprime}{\isacharcomma}\ ads{\isacharparenright}{\isacharparenright}{\isachardoublequoteclose}\ {\isacharbar}\isanewline
\ \ {\isachardoublequoteopen}step\ s\ {\isacharequal}\ stuck{\isachardoublequoteclose}\isanewline
\isanewline
\isacommand{datatype}\isamarkupfalse%
\ observable\ {\isacharequal}\ OPair\ observable\ observable\ {\isacharbar}\ Fun\ {\isacharbar}\ Con\ const\ {\isacharbar}\ Addr\ {\isacharbar}\ Inj\isanewline
\ \ {\isacharbar}\ OStuck\ {\isacharbar}\ OTimeOut\ {\isacharbar}\ OCastError\isanewline
\isanewline
\isacommand{primrec}\isamarkupfalse%
\ observe\ {\isacharcolon}{\isacharcolon}\ {\isachardoublequoteopen}val\ {\isasymRightarrow}\ observable{\isachardoublequoteclose}\ \isakeyword{where}\isanewline
\ \ obsc{\isacharcolon}\ {\isachardoublequoteopen}observe\ {\isacharparenleft}VConst\ c{\isacharparenright}\ {\isacharequal}\ Con\ c{\isachardoublequoteclose}\ {\isacharbar}\isanewline
\ \ obsp{\isacharcolon}\ {\isachardoublequoteopen}observe\ {\isacharparenleft}VPair\ v\ v{\isacharprime}{\isacharparenright}\ {\isacharequal}\ OPair\ {\isacharparenleft}observe\ v{\isacharparenright}\ {\isacharparenleft}observe\ v{\isacharprime}{\isacharparenright}{\isachardoublequoteclose}\ {\isacharbar}\isanewline
\ \ obsf{\isacharcolon}\ {\isachardoublequoteopen}observe\ {\isacharparenleft}Closure\ x\ T\ s\ {\isasymrho}{\isacharparenright}\ {\isacharequal}\ Fun{\isachardoublequoteclose}\ {\isacharbar}\isanewline
\ \ obsr{\isacharcolon}\ {\isachardoublequoteopen}observe\ {\isacharparenleft}VRef\ a{\isacharparenright}\ {\isacharequal}\ Addr{\isachardoublequoteclose}\ {\isacharbar}\isanewline
\ \ obsinj{\isacharcolon}\ {\isachardoublequoteopen}observe\ {\isacharparenleft}Inject\ v\ T{\isacharparenright}\ {\isacharequal}\ Inj{\isachardoublequoteclose}\isanewline
\isanewline
\isacommand{definition}\isamarkupfalse%
\ final\ {\isacharcolon}{\isacharcolon}\ {\isachardoublequoteopen}state\ {\isasymRightarrow}\ bool{\isachardoublequoteclose}\ \isakeyword{where}\isanewline
\ \ {\isachardoublequoteopen}final\ s\ {\isasymequiv}\ {\isacharparenleft}case\ s\ of\ {\isacharparenleft}SRet\ e{\isacharcomma}\ {\isasymrho}{\isacharcomma}\ {\isacharbrackleft}{\isacharbrackright}{\isacharcomma}\ {\isasymmu}{\isacharcomma}\ {\isacharbrackleft}{\isacharbrackright}{\isacharparenright}\ {\isasymRightarrow}\ True\ {\isacharbar}\ {\isacharunderscore}\ {\isasymRightarrow}\ False{\isacharparenright}{\isachardoublequoteclose}\isanewline
\isacommand{declare}\isamarkupfalse%
\ final{\isacharunderscore}def{\isacharbrackleft}simp{\isacharbrackright}\isanewline
\isanewline
\isacommand{fun}\isamarkupfalse%
\ steps\ {\isacharcolon}{\isacharcolon}\ {\isachardoublequoteopen}nat\ {\isasymRightarrow}\ state\ {\isasymRightarrow}\ observable{\isachardoublequoteclose}\ \isakeyword{where}\isanewline
\ \ stepsz{\isacharcolon}\ {\isachardoublequoteopen}steps\ {\isadigit{0}}\ s\ {\isacharequal}\ OTimeOut{\isachardoublequoteclose}\ {\isacharbar}\isanewline
\ \ stepsret{\isacharcolon}\ {\isachardoublequoteopen}steps\ {\isacharparenleft}Suc\ n{\isacharparenright}\ {\isacharparenleft}SRet\ e{\isacharcomma}\ {\isasymrho}{\isacharcomma}\ {\isacharbrackleft}{\isacharbrackright}{\isacharcomma}\ {\isasymmu}{\isacharcomma}\ {\isacharbrackleft}{\isacharbrackright}{\isacharparenright}\ {\isacharequal}\isanewline
\ \ \ \ \ \ \ \ \ \ \ \ {\isacharparenleft}case\ eval\ e\ {\isasymrho}\ {\isasymmu}\ of\isanewline
\ \ \ \ \ \ \ \ \ \ \ \ \ \ \ Stuck\ {\isasymRightarrow}\ OStuck\isanewline
\ \ \ \ \ \ \ \ \ \ \ \ {\isacharbar}\ CastError\ {\isasymRightarrow}\ OCastError\isanewline
\ \ \ \ \ \ \ \ \ \ \ \ {\isacharbar}\ Result\ v\ {\isasymRightarrow}\ observe\ v{\isacharparenright}{\isachardoublequoteclose}\ {\isacharbar}\isanewline
\ \ stepsrec{\isacharcolon}\ {\isachardoublequoteopen}steps\ {\isacharparenleft}Suc\ n{\isacharparenright}\ s\ {\isacharequal}\isanewline
\ \ \ \ \ \ \ \ \ {\isacharparenleft}case\ step\ s\ of\isanewline
\ \ \ \ \ \ \ \ \ \ \ Stuck\ {\isasymRightarrow}\ OStuck\isanewline
\ \ \ \ \ \ \ \ \ {\isacharbar}\ CastError\ {\isasymRightarrow}\ OCastError\isanewline
\ \ \ \ \ \ \ \ \ {\isacharbar}\ Result\ s{\isacharprime}\ {\isasymRightarrow}\ steps\ n\ s{\isacharprime}{\isacharparenright}{\isachardoublequoteclose}\isanewline
\isanewline
\isacommand{definition}\isamarkupfalse%
\ run\ {\isacharcolon}{\isacharcolon}\ {\isachardoublequoteopen}stmt\ {\isasymRightarrow}\ observable{\isachardoublequoteclose}\ \isakeyword{where}\isanewline
\ \ {\isachardoublequoteopen}run\ s\ {\isasymequiv}\ steps\ {\isadigit{1}}{\isadigit{0}}{\isadigit{0}}{\isadigit{0}}{\isadigit{0}}{\isadigit{0}}{\isadigit{0}}\ {\isacharparenleft}s{\isacharcomma}{\isacharbrackleft}{\isacharbrackright}{\isacharcomma}{\isacharbrackleft}{\isacharbrackright}{\isacharcomma}{\isacharbrackleft}{\isacharbrackright}{\isacharcomma}{\isacharbrackleft}{\isacharbrackright}{\isacharparenright}{\isachardoublequoteclose}%
\isamarkupsubsection{Type System%
}
\isamarkuptrue%
\isacommand{type{\isacharunderscore}synonym}\isamarkupfalse%
\ ty{\isacharunderscore}env\ {\isacharequal}\ {\isachardoublequoteopen}{\isacharparenleft}name\ {\isasymtimes}\ ty{\isacharparenright}\ list{\isachardoublequoteclose}\isanewline
\isanewline
\isacommand{primrec}\isamarkupfalse%
\ typeof\ {\isacharcolon}{\isacharcolon}\ {\isachardoublequoteopen}const\ {\isasymRightarrow}\ ty{\isachardoublequoteclose}\ \isakeyword{where}\isanewline
\ \ {\isachardoublequoteopen}typeof\ {\isacharparenleft}IntC\ n{\isacharparenright}\ {\isacharequal}\ IntT{\isachardoublequoteclose}\ {\isacharbar}\isanewline
\ \ {\isachardoublequoteopen}typeof\ {\isacharparenleft}BoolC\ b{\isacharparenright}\ {\isacharequal}\ BoolT{\isachardoublequoteclose}\isanewline
\isanewline
\isacommand{primrec}\isamarkupfalse%
\ typeof{\isacharunderscore}opr\ {\isacharcolon}{\isacharcolon}\ {\isachardoublequoteopen}opr\ {\isasymRightarrow}\ ty{\isachardoublequoteclose}\ \isakeyword{where}\isanewline
\ \ {\isachardoublequoteopen}typeof{\isacharunderscore}opr\ Succ\ {\isacharequal}\ {\isacharparenleft}IntT\ {\isasymrightarrow}\ IntT{\isacharparenright}{\isachardoublequoteclose}\ {\isacharbar}\isanewline
\ \ {\isachardoublequoteopen}typeof{\isacharunderscore}opr\ Prev\ {\isacharequal}\ {\isacharparenleft}IntT\ {\isasymrightarrow}\ IntT{\isacharparenright}{\isachardoublequoteclose}\ {\isacharbar}\isanewline
\ \ {\isachardoublequoteopen}typeof{\isacharunderscore}opr\ IsZero\ {\isacharequal}\ {\isacharparenleft}IntT\ {\isasymrightarrow}\ BoolT{\isacharparenright}{\isachardoublequoteclose}\ {\isacharbar}\isanewline
\ \ {\isachardoublequoteopen}typeof{\isacharunderscore}opr\ {\isacharparenleft}Fst\ A\ B{\isacharparenright}\ {\isacharequal}\ {\isacharparenleft}{\isacharparenleft}A\ {\isasymtimes}\ B{\isacharparenright}\ {\isasymrightarrow}\ A{\isacharparenright}{\isachardoublequoteclose}\ {\isacharbar}\isanewline
\ \ {\isachardoublequoteopen}typeof{\isacharunderscore}opr\ {\isacharparenleft}Snd\ A\ B{\isacharparenright}\ {\isacharequal}\ {\isacharparenleft}{\isacharparenleft}A\ {\isasymtimes}\ B{\isacharparenright}\ {\isasymrightarrow}\ B{\isacharparenright}{\isachardoublequoteclose}\ \isanewline
\isanewline
\isacommand{primrec}\isamarkupfalse%
\ static\ {\isacharcolon}{\isacharcolon}\ {\isachardoublequoteopen}ty\ {\isasymRightarrow}\ bool{\isachardoublequoteclose}\ \isakeyword{where}\isanewline
\ \ sta{\isacharunderscore}d{\isacharcolon}\ {\isachardoublequoteopen}static\ DynT\ {\isacharequal}\ False{\isachardoublequoteclose}\ {\isacharbar}\isanewline
\ \ sta{\isacharunderscore}i{\isacharcolon}\ {\isachardoublequoteopen}static\ IntT\ {\isacharequal}\ True{\isachardoublequoteclose}\ {\isacharbar}\isanewline
\ \ sta{\isacharunderscore}b{\isacharcolon}\ {\isachardoublequoteopen}static\ BoolT\ {\isacharequal}\ True{\isachardoublequoteclose}\ {\isacharbar}\isanewline
\ \ sta{\isacharunderscore}p{\isacharcolon}\ {\isachardoublequoteopen}static\ {\isacharparenleft}A\ {\isasymtimes}\ B{\isacharparenright}\ {\isacharequal}\ {\isacharparenleft}static\ A\ {\isasymand}\ static\ B{\isacharparenright}{\isachardoublequoteclose}\ {\isacharbar}\isanewline
\ \ sta{\isacharunderscore}f{\isacharcolon}\ {\isachardoublequoteopen}static\ {\isacharparenleft}A\ {\isasymrightarrow}\ B{\isacharparenright}\ {\isacharequal}\ {\isacharparenleft}static\ A\ {\isasymand}\ static\ B{\isacharparenright}{\isachardoublequoteclose}\ {\isacharbar}\isanewline
\ \ sta{\isacharunderscore}r{\isacharcolon}\ {\isachardoublequoteopen}static\ {\isacharparenleft}RefT\ A{\isacharparenright}\ {\isacharequal}\ static\ A{\isachardoublequoteclose}\ \isanewline
\isanewline
\isacommand{inductive}\isamarkupfalse%
\isanewline
\ \ wt{\isacharunderscore}expr\ {\isacharcolon}{\isacharcolon}\ {\isachardoublequoteopen}ty{\isacharunderscore}env\ {\isasymRightarrow}\ expr\ {\isasymRightarrow}\ ty\ {\isasymRightarrow}\ bool{\isachardoublequoteclose}\ {\isacharparenleft}{\isachardoublequoteopen}{\isacharunderscore}\ {\isasymturnstile}\isactrlisub e\ {\isacharunderscore}\ {\isacharcolon}\ {\isacharunderscore}{\isachardoublequoteclose}\ {\isacharbrackleft}{\isadigit{6}}{\isadigit{0}}{\isacharcomma}{\isadigit{6}}{\isadigit{0}}{\isacharcomma}{\isadigit{6}}{\isadigit{0}}{\isacharbrackright}\ {\isadigit{5}}{\isadigit{9}}{\isacharparenright}\ \isanewline
\ \ \isakeyword{and}\ wt{\isacharunderscore}stmt\ {\isacharcolon}{\isacharcolon}\ {\isachardoublequoteopen}ty{\isacharunderscore}env\ {\isasymRightarrow}\ stmt\ {\isasymRightarrow}\ ty\ {\isasymRightarrow}\ bool{\isachardoublequoteclose}\ {\isacharparenleft}{\isachardoublequoteopen}{\isacharunderscore}\ {\isasymturnstile}\isactrlisub s\ {\isacharunderscore}\ {\isacharcolon}\ {\isacharunderscore}{\isachardoublequoteclose}\ {\isacharbrackleft}{\isadigit{6}}{\isadigit{0}}{\isacharcomma}{\isadigit{6}}{\isadigit{0}}{\isacharcomma}{\isadigit{6}}{\isadigit{0}}{\isacharbrackright}\ {\isadigit{5}}{\isadigit{9}}{\isacharparenright}\ \isanewline
\ \ \isakeyword{where}\isanewline
\ \ wt{\isacharunderscore}var{\isacharbrackleft}intro{\isacharbang}{\isacharbrackright}{\isacharcolon}\ {\isachardoublequoteopen}lookup\ x\ {\isasymGamma}\ {\isacharequal}\ Result\ A\ {\isasymLongrightarrow}\ {\isasymGamma}\ {\isasymturnstile}\isactrlisub e\ Var\ x\ {\isacharcolon}\ A{\isachardoublequoteclose}\ {\isacharbar}\isanewline
\ \ wt{\isacharunderscore}const{\isacharbrackleft}intro{\isacharbang}{\isacharbrackright}{\isacharcolon}\ {\isachardoublequoteopen}{\isasymGamma}\ {\isasymturnstile}\isactrlisub e\ Const\ c\ {\isacharcolon}\ typeof\ c{\isachardoublequoteclose}\ {\isacharbar}\isanewline
\ \ wt{\isacharunderscore}primapp{\isacharbrackleft}intro{\isacharbang}{\isacharbrackright}{\isacharcolon}\isanewline
\ \ {\isachardoublequoteopen}{\isasymlbrakk}\ typeof{\isacharunderscore}opr\ f\ {\isacharequal}\ A\ {\isasymrightarrow}\ B{\isacharsemicolon}\ {\isasymGamma}\ {\isasymturnstile}\isactrlisub e\ e\ {\isacharcolon}\ A\ {\isasymrbrakk}\ {\isasymLongrightarrow}\ {\isasymGamma}\ {\isasymturnstile}\isactrlisub e\ PrimApp\ f\ e\ {\isacharcolon}\ B{\isachardoublequoteclose}\ {\isacharbar}\isanewline
\ \ wt{\isacharunderscore}mkpair{\isacharbrackleft}intro{\isacharbang}{\isacharbrackright}{\isacharcolon}\ {\isachardoublequoteopen}{\isasymlbrakk}\ {\isasymGamma}\ {\isasymturnstile}\isactrlisub e\ e{\isadigit{1}}\ {\isacharcolon}\ A{\isacharsemicolon}\ {\isasymGamma}\ {\isasymturnstile}\isactrlisub e\ e{\isadigit{2}}\ {\isacharcolon}\ B\ {\isasymrbrakk}\ {\isasymLongrightarrow}\isanewline
\ \ \ \ \ {\isasymGamma}\ {\isasymturnstile}\isactrlisub e\ MkPair\ e{\isadigit{1}}\ e{\isadigit{2}}\ {\isacharcolon}\ A\ {\isasymtimes}\ B{\isachardoublequoteclose}\ {\isacharbar}\isanewline
\ \ wt{\isacharunderscore}lam{\isacharbrackleft}intro{\isacharbang}{\isacharbrackright}{\isacharcolon}\ {\isachardoublequoteopen}{\isasymlbrakk}\ {\isacharparenleft}x{\isacharcomma}A{\isacharparenright}{\isacharhash}{\isasymGamma}\ {\isasymturnstile}\isactrlisub s\ s\ {\isacharcolon}\ B\ {\isasymrbrakk}\ {\isasymLongrightarrow}\isanewline
\ \ \ \ \ {\isasymGamma}\ {\isasymturnstile}\isactrlisub e\ Lam\ x\ A\ s\ {\isacharcolon}\ {\isacharparenleft}A\ {\isasymrightarrow}\ B{\isacharparenright}{\isachardoublequoteclose}\ {\isacharbar}\isanewline
\ \ wt{\isacharunderscore}deref{\isacharbrackleft}intro{\isacharbang}{\isacharbrackright}{\isacharcolon}\ {\isachardoublequoteopen}{\isasymlbrakk}\ {\isasymGamma}\ {\isasymturnstile}\isactrlisub e\ e\ {\isacharcolon}\ RefT\ A{\isacharsemicolon}\ static\ A\ {\isasymrbrakk}\ {\isasymLongrightarrow}\ {\isasymGamma}\ {\isasymturnstile}\isactrlisub e\ Deref\ e\ {\isacharcolon}\ A{\isachardoublequoteclose}\ {\isacharbar}\isanewline
\isanewline
\ \ wt{\isacharunderscore}let{\isacharbrackleft}intro{\isacharbang}{\isacharbrackright}{\isacharcolon}\ {\isachardoublequoteopen}{\isasymlbrakk}\ {\isasymGamma}\ {\isasymturnstile}\isactrlisub e\ e\ {\isacharcolon}\ A{\isacharsemicolon}\ {\isacharparenleft}x{\isacharcomma}A{\isacharparenright}{\isacharhash}{\isasymGamma}\ {\isasymturnstile}\isactrlisub s\ s\ {\isacharcolon}\ B\ {\isasymrbrakk}\ {\isasymLongrightarrow}\isanewline
\ \ \ \ \ {\isasymGamma}\ {\isasymturnstile}\isactrlisub s\ SLet\ x\ e\ s\ {\isacharcolon}\ B{\isachardoublequoteclose}\ {\isacharbar}\isanewline
\ \ wt{\isacharunderscore}ret{\isacharbrackleft}intro{\isacharbang}{\isacharbrackright}{\isacharcolon}\ {\isachardoublequoteopen}{\isasymlbrakk}\ {\isasymGamma}\ {\isasymturnstile}\isactrlisub e\ e\ {\isacharcolon}\ A\ {\isasymrbrakk}\ {\isasymLongrightarrow}\isanewline
\ \ \ \ \ {\isasymGamma}\ {\isasymturnstile}\isactrlisub s\ SRet\ e\ {\isacharcolon}\ A{\isachardoublequoteclose}\ {\isacharbar}\isanewline
\ \ wt{\isacharunderscore}call{\isacharbrackleft}intro{\isacharbang}{\isacharbrackright}{\isacharcolon}\ {\isachardoublequoteopen}{\isasymlbrakk}\ {\isasymGamma}\ {\isasymturnstile}\isactrlisub e\ e\ {\isacharcolon}\ A\ {\isasymrightarrow}\ B{\isacharsemicolon}\ {\isasymGamma}\ {\isasymturnstile}\isactrlisub e\ e{\isacharprime}\ {\isacharcolon}\ A{\isacharsemicolon}\ {\isacharparenleft}x{\isacharcomma}B{\isacharparenright}{\isacharhash}{\isasymGamma}\ {\isasymturnstile}\isactrlisub s\ s\ {\isacharcolon}\ C\ {\isasymrbrakk}\ {\isasymLongrightarrow}\isanewline
\ \ \ \ \ {\isasymGamma}\ {\isasymturnstile}\isactrlisub s\ SCall\ x\ e\ e{\isacharprime}\ s\ {\isacharcolon}\ C{\isachardoublequoteclose}\ {\isacharbar}\isanewline
\ \ wt{\isacharunderscore}tailcall{\isacharbrackleft}intro{\isacharbang}{\isacharbrackright}{\isacharcolon}\ {\isachardoublequoteopen}{\isasymlbrakk}\ {\isasymGamma}\ {\isasymturnstile}\isactrlisub e\ e\ {\isacharcolon}\ A\ {\isasymrightarrow}\ B{\isacharsemicolon}\ {\isasymGamma}\ {\isasymturnstile}\isactrlisub e\ e{\isacharprime}\ {\isacharcolon}\ A\ {\isasymrbrakk}\ {\isasymLongrightarrow}\isanewline
\ \ \ \ \ {\isasymGamma}\ {\isasymturnstile}\isactrlisub s\ STailCall\ e\ e{\isacharprime}\ {\isacharcolon}\ B{\isachardoublequoteclose}\ {\isacharbar}\isanewline
\ \ wt{\isacharunderscore}alloc{\isacharbrackleft}intro{\isacharbang}{\isacharbrackright}{\isacharcolon}\ {\isachardoublequoteopen}{\isasymlbrakk}\ {\isasymGamma}\ {\isasymturnstile}\isactrlisub e\ e\ {\isacharcolon}\ A{\isacharsemicolon}\ {\isacharparenleft}x{\isacharcomma}RefT\ A{\isacharparenright}{\isacharhash}{\isasymGamma}\ {\isasymturnstile}\isactrlisub s\ s\ {\isacharcolon}\ B\ {\isasymrbrakk}\ {\isasymLongrightarrow}\isanewline
\ \ \ \ \ {\isasymGamma}\ {\isasymturnstile}\isactrlisub s\ SAlloc\ x\ A\ e\ s\ {\isacharcolon}\ B{\isachardoublequoteclose}\ {\isacharbar}\isanewline
\ \ wt{\isacharunderscore}update{\isacharbrackleft}intro{\isacharbang}{\isacharbrackright}{\isacharcolon}\ {\isachardoublequoteopen}{\isasymlbrakk}\ {\isasymGamma}\ {\isasymturnstile}\isactrlisub e\ e\ {\isacharcolon}\ RefT\ A{\isacharsemicolon}\ static\ A{\isacharsemicolon}\ {\isasymGamma}\ {\isasymturnstile}\isactrlisub e\ e{\isacharprime}\ {\isacharcolon}\ A{\isacharsemicolon}\ {\isasymGamma}\ {\isasymturnstile}\isactrlisub s\ s\ {\isacharcolon}\ B\ {\isasymrbrakk}\ {\isasymLongrightarrow}\isanewline
\ \ \ \ \ {\isasymGamma}\ {\isasymturnstile}\isactrlisub s\ SUpdate\ e\ e{\isacharprime}\ s\ {\isacharcolon}\ B{\isachardoublequoteclose}\ {\isacharbar}\isanewline
\ \ wt{\isacharunderscore}dynupdate{\isacharbrackleft}intro{\isacharbang}{\isacharbrackright}{\isacharcolon}\ {\isachardoublequoteopen}{\isasymlbrakk}\ {\isasymGamma}\ {\isasymturnstile}\isactrlisub e\ e\ {\isacharcolon}\ RefT\ A{\isacharsemicolon}\ {\isasymGamma}\ {\isasymturnstile}\isactrlisub e\ e{\isacharprime}\ {\isacharcolon}\ A{\isacharsemicolon}\ {\isasymGamma}\ {\isasymturnstile}\isactrlisub s\ s\ {\isacharcolon}\ B\ {\isasymrbrakk}\ {\isasymLongrightarrow}\isanewline
\ \ \ \ \ {\isasymGamma}\ {\isasymturnstile}\isactrlisub s\ SDynUpdate\ e\ e{\isacharprime}\ A\ s\ {\isacharcolon}\ B{\isachardoublequoteclose}\ {\isacharbar}\isanewline
\ \ wt{\isacharunderscore}cast{\isacharbrackleft}intro{\isacharbang}{\isacharbrackright}{\isacharcolon}\ {\isachardoublequoteopen}{\isasymlbrakk}\ {\isasymGamma}\ {\isasymturnstile}\isactrlisub e\ e\ {\isacharcolon}\ A{\isacharsemicolon}\ {\isacharparenleft}x{\isacharcomma}B{\isacharparenright}{\isacharhash}{\isasymGamma}\ {\isasymturnstile}\isactrlisub s\ s\ {\isacharcolon}\ C\ {\isasymrbrakk}\ {\isasymLongrightarrow}\isanewline
\ \ \ \ \ {\isasymGamma}\ {\isasymturnstile}\isactrlisub s\ SCast\ x\ e\ A\ B\ s\ {\isacharcolon}\ C{\isachardoublequoteclose}\ {\isacharbar}\isanewline
\ \ wt{\isacharunderscore}dynderef{\isacharbrackleft}intro{\isacharbang}{\isacharbrackright}{\isacharcolon}\ {\isachardoublequoteopen}{\isasymlbrakk}\ {\isasymGamma}\ {\isasymturnstile}\isactrlisub e\ e\ {\isacharcolon}\ RefT\ A{\isacharsemicolon}\ {\isacharparenleft}x{\isacharcomma}A{\isacharparenright}{\isacharhash}{\isasymGamma}\ {\isasymturnstile}\isactrlisub s\ s\ {\isacharcolon}\ B\ {\isasymrbrakk}\ {\isasymLongrightarrow}\isanewline
\ \ \ \ \ {\isasymGamma}\ {\isasymturnstile}\isactrlisub s\ SDynDeref\ x\ e\ A\ s\ {\isacharcolon}\ B{\isachardoublequoteclose}\isanewline
\isanewline
\isacommand{inductive}\isamarkupfalse%
\ wt{\isacharunderscore}val\ {\isacharcolon}{\isacharcolon}\ {\isachardoublequoteopen}ty{\isacharunderscore}env\ {\isasymRightarrow}\ val\ {\isasymRightarrow}\ ty\ {\isasymRightarrow}\ bool{\isachardoublequoteclose}\ \ {\isacharparenleft}{\isachardoublequoteopen}{\isacharunderscore}\ {\isasymturnstile}v\ {\isacharunderscore}\ {\isacharcolon}\ {\isacharunderscore}{\isachardoublequoteclose}\ {\isacharbrackleft}{\isadigit{6}}{\isadigit{0}}{\isacharcomma}{\isadigit{6}}{\isadigit{0}}{\isacharcomma}{\isadigit{6}}{\isadigit{0}}{\isacharbrackright}\ {\isadigit{5}}{\isadigit{9}}{\isacharparenright}\ \isanewline
\ \ \isakeyword{and}\ wt{\isacharunderscore}env\ {\isacharcolon}{\isacharcolon}\ {\isachardoublequoteopen}ty{\isacharunderscore}env\ {\isasymRightarrow}\ ty{\isacharunderscore}env\ {\isasymRightarrow}\ env\ {\isasymRightarrow}\ bool{\isachardoublequoteclose}\ {\isacharparenleft}{\isachardoublequoteopen}{\isacharunderscore}{\isacharsemicolon}{\isacharunderscore}\ {\isasymturnstile}\ {\isacharunderscore}{\isachardoublequoteclose}\ {\isacharbrackleft}{\isadigit{6}}{\isadigit{0}}{\isacharcomma}{\isadigit{6}}{\isadigit{0}}{\isacharcomma}{\isadigit{6}}{\isadigit{0}}{\isacharbrackright}\ {\isadigit{5}}{\isadigit{9}}{\isacharparenright}\ \isakeyword{where}\isanewline
\ \ wt{\isacharunderscore}vc{\isacharbrackleft}intro{\isacharbang}{\isacharbrackright}{\isacharcolon}\ {\isachardoublequoteopen}typeof\ c\ {\isacharequal}\ A\ {\isasymLongrightarrow}\ {\isasymSigma}\ {\isasymturnstile}v\ VConst\ c\ {\isacharcolon}\ A{\isachardoublequoteclose}\ {\isacharbar}\isanewline
\ \ wt{\isacharunderscore}pair{\isacharbrackleft}intro{\isacharbang}{\isacharbrackright}{\isacharcolon}\ {\isachardoublequoteopen}{\isasymlbrakk}\ {\isasymSigma}\ {\isasymturnstile}v\ v\ {\isacharcolon}\ A{\isacharsemicolon}\ {\isasymSigma}\ {\isasymturnstile}v\ v{\isacharprime}\ {\isacharcolon}\ B\ {\isasymrbrakk}\ {\isasymLongrightarrow}\ {\isasymSigma}\ {\isasymturnstile}v\ {\isacharparenleft}VPair\ v\ v{\isacharprime}{\isacharparenright}\ {\isacharcolon}\ A\ {\isasymtimes}\ B{\isachardoublequoteclose}\ {\isacharbar}\isanewline
\ \ wt{\isacharunderscore}cl{\isacharbrackleft}intro{\isacharbang}{\isacharbrackright}{\isacharcolon}\ {\isachardoublequoteopen}{\isasymlbrakk}\ {\isasymGamma}{\isacharsemicolon}{\isasymSigma}\ {\isasymturnstile}\ {\isasymrho}{\isacharsemicolon}\ {\isacharparenleft}x{\isacharcomma}A{\isacharparenright}{\isacharhash}{\isasymGamma}\ {\isasymturnstile}\isactrlisub s\ s\ {\isacharcolon}\ B\ {\isasymrbrakk}\ {\isasymLongrightarrow}\ \isanewline
\ \ \ \ \ {\isasymSigma}\ {\isasymturnstile}v\ {\isacharparenleft}Closure\ x\ A\ s\ {\isasymrho}{\isacharparenright}\ {\isacharcolon}\ A\ {\isasymrightarrow}\ B{\isachardoublequoteclose}\ {\isacharbar}\isanewline
\ \ wt{\isacharunderscore}ref{\isacharbrackleft}intro{\isacharbang}{\isacharbrackright}{\isacharcolon}\ {\isachardoublequoteopen}{\isasymlbrakk}\ lookup\ a\ {\isasymSigma}\ {\isacharequal}\ Result\ A{\isacharsemicolon}\ A\ {\isasymsqsubseteq}\ B\ {\isasymrbrakk}\ {\isasymLongrightarrow}\ {\isasymSigma}\ {\isasymturnstile}v\ {\isacharparenleft}VRef\ a{\isacharparenright}\ {\isacharcolon}\ RefT\ B{\isachardoublequoteclose}\ {\isacharbar}\isanewline
\ \ wt{\isacharunderscore}inject{\isacharbrackleft}intro{\isacharbang}{\isacharbrackright}{\isacharcolon}\ {\isachardoublequoteopen}{\isasymSigma}\ {\isasymturnstile}v\ v\ {\isacharcolon}\ A\ {\isasymLongrightarrow}\ {\isasymSigma}\ {\isasymturnstile}v\ {\isacharparenleft}Inject\ v\ A{\isacharparenright}\ {\isacharcolon}\ DynT{\isachardoublequoteclose}\ {\isacharbar}\isanewline
\isanewline
\ \ wt{\isacharunderscore}nil{\isacharbrackleft}intro{\isacharbang}{\isacharbrackright}{\isacharcolon}\ {\isachardoublequoteopen}{\isacharbrackleft}{\isacharbrackright}{\isacharsemicolon}{\isasymSigma}\ {\isasymturnstile}\ {\isacharbrackleft}{\isacharbrackright}{\isachardoublequoteclose}\ {\isacharbar}\isanewline
\ \ wt{\isacharunderscore}cons{\isacharbrackleft}intro{\isacharbang}{\isacharbrackright}{\isacharcolon}\ {\isachardoublequoteopen}{\isasymlbrakk}\ {\isasymSigma}\ {\isasymturnstile}v\ v\ {\isacharcolon}\ A{\isacharsemicolon}\ {\isasymGamma}{\isacharsemicolon}{\isasymSigma}\ {\isasymturnstile}\ {\isasymrho}\ {\isasymrbrakk}\ {\isasymLongrightarrow}\ {\isacharparenleft}x{\isacharcomma}A{\isacharparenright}{\isacharhash}{\isasymGamma}{\isacharsemicolon}{\isasymSigma}\ {\isasymturnstile}\ {\isacharparenleft}x{\isacharcomma}v{\isacharparenright}{\isacharhash}{\isasymrho}{\isachardoublequoteclose}\isanewline
\isanewline
\isacommand{inductive}\isamarkupfalse%
\ wt{\isacharunderscore}stack\ {\isacharcolon}{\isacharcolon}\ {\isachardoublequoteopen}ty{\isacharunderscore}env\ {\isasymRightarrow}\ stack\ {\isasymRightarrow}\ ty\ {\isasymRightarrow}\ ty\ {\isasymRightarrow}\ bool{\isachardoublequoteclose}\ {\isacharparenleft}{\isachardoublequoteopen}{\isacharunderscore}\ {\isasymturnstile}\ {\isacharunderscore}\ {\isacharcolon}\ {\isacharunderscore}\ {\isasymRightarrow}\ {\isacharunderscore}{\isachardoublequoteclose}{\isacharparenright}\ \isakeyword{where}\isanewline
\ \ nil{\isacharunderscore}stack{\isacharbrackleft}intro{\isacharbang}{\isacharbrackright}{\isacharcolon}\ {\isachardoublequoteopen}{\isasymSigma}\ {\isasymturnstile}\ {\isacharbrackleft}{\isacharbrackright}\ {\isacharcolon}\ A\ {\isasymRightarrow}\ A{\isachardoublequoteclose}\ {\isacharbar}\isanewline
\ \ cons{\isacharunderscore}stack{\isacharbrackleft}intro{\isacharbang}{\isacharbrackright}{\isacharcolon}\ {\isachardoublequoteopen}{\isasymlbrakk}\ {\isasymGamma}{\isacharsemicolon}{\isasymSigma}\ {\isasymturnstile}\ {\isasymrho}{\isacharsemicolon}\ {\isacharparenleft}x{\isacharcomma}A{\isacharparenright}{\isacharhash}{\isasymGamma}\ {\isasymturnstile}\isactrlisub s\ s\ {\isacharcolon}\ B{\isacharsemicolon}\ {\isasymSigma}\ {\isasymturnstile}\ k\ {\isacharcolon}\ B\ {\isasymRightarrow}\ C\ {\isasymrbrakk}\ {\isasymLongrightarrow}\isanewline
\ \ \ \ \ \ {\isasymSigma}\ {\isasymturnstile}\ {\isacharparenleft}x{\isacharcomma}\ s{\isacharcomma}{\isasymrho}{\isacharparenright}{\isacharhash}k\ {\isacharcolon}\ A\ {\isasymRightarrow}\ C{\isachardoublequoteclose}\isanewline
\isanewline
\isacommand{definition}\isamarkupfalse%
\ dom\ {\isacharcolon}{\isacharcolon}\ {\isachardoublequoteopen}{\isacharparenleft}{\isacharprime}a\ {\isasymtimes}\ {\isacharprime}b{\isacharparenright}\ list\ {\isasymRightarrow}\ {\isacharprime}a\ set{\isachardoublequoteclose}\ \isakeyword{where}\isanewline
\ \ {\isachardoublequoteopen}dom\ A\ {\isasymequiv}\ set\ {\isacharparenleft}map\ fst\ A{\isacharparenright}{\isachardoublequoteclose}\isanewline
\isanewline
\isacommand{inductive}\isamarkupfalse%
\ wt{\isacharunderscore}casted{\isacharunderscore}val\ {\isacharcolon}{\isacharcolon}\ {\isachardoublequoteopen}ty{\isacharunderscore}env\ {\isasymRightarrow}\ casted{\isacharunderscore}val\ {\isasymRightarrow}\ ty\ {\isasymRightarrow}\ bool{\isachardoublequoteclose}\ {\isacharparenleft}{\isachardoublequoteopen}{\isacharunderscore}\ {\isasymturnstile}cv\ {\isacharunderscore}\ {\isacharcolon}\ {\isacharunderscore}{\isachardoublequoteclose}\ {\isacharbrackleft}{\isadigit{6}}{\isadigit{0}}{\isacharcomma}{\isadigit{6}}{\isadigit{0}}{\isacharcomma}{\isadigit{6}}{\isadigit{0}}{\isacharbrackright}\ {\isadigit{5}}{\isadigit{9}}{\isacharparenright}\ \isakeyword{where}\isanewline
\ \ wt{\isacharunderscore}cv{\isacharunderscore}val{\isacharbrackleft}intro{\isacharbang}{\isacharbrackright}{\isacharcolon}\ {\isachardoublequoteopen}{\isasymSigma}\ {\isasymturnstile}v\ v\ {\isacharcolon}\ A\ {\isasymLongrightarrow}\ {\isasymSigma}\ {\isasymturnstile}cv\ Val\ v\ {\isacharcolon}\ A{\isachardoublequoteclose}\ {\isacharbar}\isanewline
\ \ wt{\isacharunderscore}cv{\isacharbrackleft}intro{\isacharbang}{\isacharbrackright}{\isacharcolon}\ {\isachardoublequoteopen}{\isasymlbrakk}\ {\isasymSigma}\ {\isasymturnstile}v\ v\ {\isacharcolon}\ A{\isacharsemicolon}\ B\ {\isasymsqsubseteq}\ A\ {\isasymrbrakk}\ {\isasymLongrightarrow}\ {\isasymSigma}\ {\isasymturnstile}cv\ VCast\ v\ A\ B\ {\isacharcolon}\ B{\isachardoublequoteclose}\isanewline
\isanewline
\isacommand{definition}\isamarkupfalse%
\ wt{\isacharunderscore}heap\ {\isacharcolon}{\isacharcolon}\ {\isachardoublequoteopen}ty{\isacharunderscore}env\ {\isasymRightarrow}\ heap\ {\isasymRightarrow}\ nat\ set\ {\isasymRightarrow}\ bool{\isachardoublequoteclose}\ \isakeyword{where}\isanewline
\ \ {\isachardoublequoteopen}wt{\isacharunderscore}heap\ {\isasymSigma}\ {\isasymmu}\ as\ {\isasymequiv}\ {\isacharparenleft}{\isasymforall}\ a\ A{\isachardot}\ lookup\ a\ {\isasymSigma}\ {\isacharequal}\ Result\ A\ {\isasymlongrightarrow}\isanewline
\ \ \ \ \ \ \ \ \ \ \ \ \ \ \ \ \ \ \ {\isacharparenleft}{\isasymexists}\ cv{\isachardot}\ lookup\ a\ {\isasymmu}\ {\isacharequal}\ Result\ {\isacharparenleft}cv{\isacharcomma}A{\isacharparenright}\ {\isasymand}\ {\isasymSigma}\ {\isasymturnstile}cv\ cv\ {\isacharcolon}\ A\ {\isasymand}\isanewline
\ \ \ \ \ \ \ \ \ \ \ \ \ \ \ \ \ \ \ \ \ {\isacharparenleft}a\ {\isasymnotin}\ as\ {\isasymlongrightarrow}\ {\isacharparenleft}{\isasymexists}\ v{\isachardot}\ cv\ {\isacharequal}\ Val\ v{\isacharparenright}{\isacharparenright}{\isacharparenright}{\isacharparenright}\isanewline
\ \ \ \ \ {\isasymand}\ {\isacharparenleft}{\isasymforall}\ a{\isachardot}\ a\ {\isasymin}\ dom\ {\isasymmu}\ {\isasymlongrightarrow}\ a\ {\isacharless}\ length\ {\isasymmu}{\isacharparenright}\ {\isasymand}\ {\isacharparenleft}as\ {\isasymsubseteq}\ dom\ {\isasymSigma}{\isacharparenright}{\isachardoublequoteclose}\isanewline
\isanewline
\isacommand{definition}\isamarkupfalse%
\ lesseq{\isacharunderscore}tyenv\ {\isacharcolon}{\isacharcolon}\ {\isachardoublequoteopen}ty{\isacharunderscore}env\ {\isasymRightarrow}\ ty{\isacharunderscore}env\ {\isasymRightarrow}\ bool{\isachardoublequoteclose}\ {\isacharparenleft}\isakeyword{infix}\ {\isachardoublequoteopen}{\isasymsqsubseteq}{\isachardoublequoteclose}\ {\isadigit{8}}{\isadigit{0}}{\isacharparenright}\ \isakeyword{where}\isanewline
\ \ {\isachardoublequoteopen}{\isasymSigma}{\isacharprime}\ {\isasymsqsubseteq}\ {\isasymSigma}\ {\isasymequiv}\ {\isacharparenleft}dom\ {\isasymSigma}\ {\isacharequal}\ dom\ {\isasymSigma}{\isacharprime}{\isacharparenright}\ {\isasymand}\ {\isacharparenleft}{\isasymforall}\ a\ A{\isachardot}\ lookup\ a\ {\isasymSigma}\ {\isacharequal}\ Result\ A\ {\isasymlongrightarrow}\isanewline
\ \ \ \ \ \ \ \ \ \ \ \ \ \ \ \ {\isacharparenleft}{\isasymexists}\ B{\isachardot}\ lookup\ a\ {\isasymSigma}{\isacharprime}\ {\isacharequal}\ Result\ B\ {\isasymand}\ B\ {\isasymsqsubseteq}\ A{\isacharparenright}{\isacharparenright}{\isachardoublequoteclose}\isanewline
\isanewline
\isacommand{inductive}\isamarkupfalse%
\ wt{\isacharunderscore}state\ {\isacharcolon}{\isacharcolon}\ {\isachardoublequoteopen}state\ {\isasymRightarrow}\ ty\ {\isasymRightarrow}\ bool{\isachardoublequoteclose}\ \isakeyword{where}\isanewline
\ \ wts{\isacharunderscore}intro{\isacharbrackleft}intro{\isacharbang}{\isacharbrackright}{\isacharcolon}\ {\isachardoublequoteopen}{\isasymlbrakk}\ wt{\isacharunderscore}heap\ {\isasymSigma}\ {\isasymmu}\ {\isacharparenleft}set\ as{\isacharparenright}{\isacharsemicolon}\ {\isasymGamma}{\isacharsemicolon}{\isasymSigma}\ {\isasymturnstile}\ {\isasymrho}{\isacharsemicolon}\ {\isasymGamma}\ {\isasymturnstile}\isactrlisub s\ s\ {\isacharcolon}\ A{\isacharsemicolon}\ {\isasymSigma}\ {\isasymturnstile}\ k\ {\isacharcolon}\ A\ {\isasymRightarrow}\ B\ {\isasymrbrakk}\ {\isasymLongrightarrow}\isanewline
\ \ \ \ wt{\isacharunderscore}state\ {\isacharparenleft}s{\isacharcomma}\ {\isasymrho}{\isacharcomma}\ k{\isacharcomma}\ {\isasymmu}{\isacharcomma}\ as{\isacharparenright}\ B{\isachardoublequoteclose}\isanewline
\isanewline
\isacommand{fun}\isamarkupfalse%
\ wt{\isacharunderscore}observable\ {\isacharcolon}{\isacharcolon}\ {\isachardoublequoteopen}observable\ {\isasymRightarrow}\ ty\ {\isasymRightarrow}\ bool{\isachardoublequoteclose}\ \isakeyword{where}\isanewline
\ \ wto{\isacharunderscore}p{\isacharcolon}\ {\isachardoublequoteopen}wt{\isacharunderscore}observable\ {\isacharparenleft}OPair\ o{\isacharprime}\ o{\isacharprime}{\isacharprime}{\isacharparenright}\ {\isacharparenleft}A\ {\isasymtimes}\ B{\isacharparenright}\ {\isacharequal}\ \isanewline
\ \ \ \ \ \ {\isacharparenleft}wt{\isacharunderscore}observable\ o{\isacharprime}\ A\ {\isasymand}\ wt{\isacharunderscore}observable\ o{\isacharprime}{\isacharprime}\ B{\isacharparenright}{\isachardoublequoteclose}\ {\isacharbar}\isanewline
\ \ wto{\isacharunderscore}f{\isacharcolon}\ {\isachardoublequoteopen}wt{\isacharunderscore}observable\ Fun\ {\isacharparenleft}A\ {\isasymrightarrow}\ B{\isacharparenright}\ {\isacharequal}\ True{\isachardoublequoteclose}\ {\isacharbar}\isanewline
\ \ wto{\isacharunderscore}c{\isacharcolon}\ {\isachardoublequoteopen}wt{\isacharunderscore}observable\ {\isacharparenleft}Con\ c{\isacharparenright}\ T\ {\isacharequal}\ {\isacharparenleft}typeof\ c\ {\isacharequal}\ T{\isacharparenright}{\isachardoublequoteclose}\ {\isacharbar}\isanewline
\ \ wto{\isacharunderscore}s{\isacharcolon}\ {\isachardoublequoteopen}wt{\isacharunderscore}observable\ OStuck\ T\ {\isacharequal}\ False{\isachardoublequoteclose}\ {\isacharbar}\isanewline
\ \ wto{\isacharunderscore}t{\isacharcolon}\ {\isachardoublequoteopen}wt{\isacharunderscore}observable\ OTimeOut\ T\ {\isacharequal}\ True{\isachardoublequoteclose}\ {\isacharbar}\isanewline
\ \ wto{\isacharunderscore}cast{\isacharcolon}\ {\isachardoublequoteopen}wt{\isacharunderscore}observable\ OCastError\ T\ {\isacharequal}\ True{\isachardoublequoteclose}\ {\isacharbar}\isanewline
\ \ wto{\isacharunderscore}a{\isacharcolon}\ {\isachardoublequoteopen}wt{\isacharunderscore}observable\ Addr\ {\isacharparenleft}RefT\ A{\isacharparenright}\ {\isacharequal}\ True{\isachardoublequoteclose}\ {\isacharbar}\isanewline
\ \ wto{\isacharunderscore}inj{\isacharcolon}\ {\isachardoublequoteopen}wt{\isacharunderscore}observable\ Inj\ DynT\ {\isacharequal}\ True{\isachardoublequoteclose}\ {\isacharbar}\isanewline
\ \ {\isachardoublequoteopen}wt{\isacharunderscore}observable\ obs\ T\ {\isacharequal}\ False{\isachardoublequoteclose}%
\isamarkupsubsection{Inversion Principles%
}
\isamarkuptrue%
\isacommand{inductive{\isacharunderscore}cases}\isamarkupfalse%
\ wtv{\isacharbrackleft}elim{\isacharbang}{\isacharbrackright}{\isacharcolon}\ {\isachardoublequoteopen}{\isasymGamma}\ {\isasymturnstile}\isactrlisub e\ Var\ x\ {\isacharcolon}\ A{\isachardoublequoteclose}\ \isakeyword{and}\isanewline
\ \ wti{\isacharbrackleft}elim{\isacharbang}{\isacharbrackright}{\isacharcolon}\ {\isachardoublequoteopen}{\isasymGamma}\ {\isasymturnstile}\isactrlisub e\ Const\ {\isacharparenleft}IntC\ n{\isacharparenright}\ {\isacharcolon}\ A{\isachardoublequoteclose}\ \isakeyword{and}\isanewline
\ \ wtb{\isacharbrackleft}elim{\isacharbang}{\isacharbrackright}{\isacharcolon}\ {\isachardoublequoteopen}{\isasymGamma}\ {\isasymturnstile}\isactrlisub e\ Const\ {\isacharparenleft}BoolC\ b{\isacharparenright}\ {\isacharcolon}\ A{\isachardoublequoteclose}\ \isakeyword{and}\isanewline
\ \ wtp{\isacharbrackleft}elim{\isacharbang}{\isacharbrackright}{\isacharcolon}\ {\isachardoublequoteopen}{\isasymGamma}\ {\isasymturnstile}\isactrlisub e\ PrimApp\ f\ e\ {\isacharcolon}\ A{\isachardoublequoteclose}\ \isakeyword{and}\isanewline
\ \ wtpair{\isacharbrackleft}elim{\isacharbang}{\isacharbrackright}{\isacharcolon}\ {\isachardoublequoteopen}{\isasymGamma}\ {\isasymturnstile}\isactrlisub e\ MkPair\ e{\isadigit{1}}\ e{\isadigit{2}}\ {\isacharcolon}\ A{\isachardoublequoteclose}\ \isakeyword{and}\isanewline
\ \ wtlam{\isacharbrackleft}elim{\isacharbang}{\isacharbrackright}{\isacharcolon}\ {\isachardoublequoteopen}{\isasymGamma}\ {\isasymturnstile}\isactrlisub e\ Lam\ x\ T\ s\ {\isacharcolon}\ A{\isachardoublequoteclose}\ \isakeyword{and}\isanewline
\ \ wtderef{\isacharbrackleft}elim{\isacharbang}{\isacharbrackright}{\isacharcolon}\ {\isachardoublequoteopen}{\isasymGamma}\ {\isasymturnstile}\isactrlisub e\ {\isacharparenleft}Deref\ e{\isacharparenright}\ {\isacharcolon}\ A{\isachardoublequoteclose}\ \isanewline
\isanewline
\isacommand{inductive{\isacharunderscore}cases}\isamarkupfalse%
\isanewline
\ \ wtlet{\isacharbrackleft}elim{\isacharbang}{\isacharbrackright}{\isacharcolon}\ {\isachardoublequoteopen}{\isasymGamma}\ {\isasymturnstile}\isactrlisub s\ SLet\ x\ e\ s\ {\isacharcolon}\ A{\isachardoublequoteclose}\ \isakeyword{and}\isanewline
\ \ wtret{\isacharbrackleft}elim{\isacharbang}{\isacharbrackright}{\isacharcolon}\ {\isachardoublequoteopen}{\isasymGamma}\ {\isasymturnstile}\isactrlisub s\ SRet\ e\ {\isacharcolon}\ A{\isachardoublequoteclose}\ \isakeyword{and}\isanewline
\ \ wtcall{\isacharbrackleft}elim{\isacharbang}{\isacharbrackright}{\isacharcolon}\ {\isachardoublequoteopen}{\isasymGamma}\ {\isasymturnstile}\isactrlisub s\ SCall\ x\ e{\isadigit{1}}\ e{\isadigit{2}}\ s\ {\isacharcolon}\ A{\isachardoublequoteclose}\ \isakeyword{and}\isanewline
\ \ wttailcall{\isacharbrackleft}elim{\isacharbang}{\isacharbrackright}{\isacharcolon}\ {\isachardoublequoteopen}{\isasymGamma}\ {\isasymturnstile}\isactrlisub s\ STailCall\ e{\isadigit{1}}\ e{\isadigit{2}}\ {\isacharcolon}\ A{\isachardoublequoteclose}\ \isakeyword{and}\isanewline
\ \ wtalloc{\isacharbrackleft}elim{\isacharbang}{\isacharbrackright}{\isacharcolon}\ {\isachardoublequoteopen}{\isasymGamma}\ {\isasymturnstile}\isactrlisub s\ SAlloc\ x\ B\ e\ s\ {\isacharcolon}\ A{\isachardoublequoteclose}\ \isakeyword{and}\isanewline
\ \ wtupdate{\isacharbrackleft}elim{\isacharbang}{\isacharbrackright}{\isacharcolon}\ {\isachardoublequoteopen}{\isasymGamma}\ {\isasymturnstile}\isactrlisub s\ SUpdate\ e{\isadigit{1}}\ e{\isadigit{2}}\ s\ {\isacharcolon}\ A{\isachardoublequoteclose}\ \isakeyword{and}\isanewline
\ \ wtdynupdate{\isacharbrackleft}elim{\isacharbang}{\isacharbrackright}{\isacharcolon}\ {\isachardoublequoteopen}{\isasymGamma}\ {\isasymturnstile}\isactrlisub s\ SDynUpdate\ e{\isadigit{1}}\ e{\isadigit{2}}\ B\ s\ {\isacharcolon}\ A{\isachardoublequoteclose}\ \ \isakeyword{and}\isanewline
\ \ wtcast{\isacharbrackleft}elim{\isacharbang}{\isacharbrackright}{\isacharcolon}\ {\isachardoublequoteopen}{\isasymGamma}\ {\isasymturnstile}\isactrlisub s\ SCast\ x\ e\ A\ B\ s\ {\isacharcolon}\ C{\isachardoublequoteclose}\ \isakeyword{and}\isanewline
\ \ wtdynderef{\isacharbrackleft}elim{\isacharbang}{\isacharbrackright}{\isacharcolon}\ {\isachardoublequoteopen}{\isasymGamma}\ {\isasymturnstile}\isactrlisub s\ SDynDeref\ x\ e\ A\ s\ {\isacharcolon}\ B{\isachardoublequoteclose}\isanewline
\isanewline
\isacommand{inductive{\isacharunderscore}cases}\isamarkupfalse%
\isanewline
\ \ bool{\isacharunderscore}int{\isacharbrackleft}elim{\isacharbang}{\isacharbrackright}{\isacharcolon}\ {\isachardoublequoteopen}{\isasymSigma}\ {\isasymturnstile}v\ {\isacharparenleft}VConst\ {\isacharparenleft}BoolC\ b{\isacharparenright}{\isacharparenright}\ {\isacharcolon}\ IntT{\isachardoublequoteclose}\ \isakeyword{and}\isanewline
\ \ inject{\isacharunderscore}int{\isacharbrackleft}elim{\isacharbang}{\isacharbrackright}{\isacharcolon}\ {\isachardoublequoteopen}{\isasymSigma}\ {\isasymturnstile}v\ {\isacharparenleft}Inject\ v\ A{\isacharparenright}\ {\isacharcolon}\ IntT{\isachardoublequoteclose}\ \isakeyword{and}\isanewline
\ \ pair{\isacharunderscore}int{\isacharbrackleft}elim{\isacharbang}{\isacharbrackright}{\isacharcolon}\ {\isachardoublequoteopen}{\isasymSigma}\ {\isasymturnstile}v\ {\isacharparenleft}VPair\ v{\isadigit{1}}\ v{\isadigit{2}}{\isacharparenright}\ {\isacharcolon}\ IntT{\isachardoublequoteclose}\ \isakeyword{and}\isanewline
\ \ int{\isacharunderscore}any{\isacharbrackleft}elim{\isacharbang}{\isacharbrackright}{\isacharcolon}\ {\isachardoublequoteopen}{\isasymSigma}\ {\isasymturnstile}v\ {\isacharparenleft}VConst\ {\isacharparenleft}IntC\ n{\isacharparenright}{\isacharparenright}\ {\isacharcolon}\ A{\isachardoublequoteclose}\ \isakeyword{and}\isanewline
\ \ bool{\isacharunderscore}any{\isacharbrackleft}elim{\isacharbang}{\isacharbrackright}{\isacharcolon}\ {\isachardoublequoteopen}{\isasymSigma}\ {\isasymturnstile}v\ {\isacharparenleft}VConst\ {\isacharparenleft}BoolC\ b{\isacharparenright}{\isacharparenright}\ {\isacharcolon}\ A{\isachardoublequoteclose}\ \isakeyword{and}\isanewline
\ \ clos{\isacharunderscore}any{\isacharbrackleft}elim{\isacharbang}{\isacharbrackright}{\isacharcolon}\ {\isachardoublequoteopen}{\isasymSigma}\ {\isasymturnstile}v\ {\isacharparenleft}Closure\ x\ T\ s\ {\isasymrho}{\isacharparenright}\ {\isacharcolon}\ A{\isachardoublequoteclose}\ \isakeyword{and}\isanewline
\ \ pair{\isacharunderscore}any{\isacharbrackleft}elim{\isacharbang}{\isacharbrackright}{\isacharcolon}\ {\isachardoublequoteopen}{\isasymSigma}\ {\isasymturnstile}v\ {\isacharparenleft}VPair\ v{\isadigit{1}}\ v{\isadigit{2}}{\isacharparenright}\ {\isacharcolon}\ A{\isachardoublequoteclose}\ \isakeyword{and}\isanewline
\ \ ref{\isacharunderscore}any{\isacharbrackleft}elim{\isacharbang}{\isacharbrackright}{\isacharcolon}\ {\isachardoublequoteopen}{\isasymSigma}\ {\isasymturnstile}v\ {\isacharparenleft}VRef\ a{\isacharparenright}\ {\isacharcolon}\ A{\isachardoublequoteclose}\ \isakeyword{and}\isanewline
\ \ inject{\isacharunderscore}any{\isacharbrackleft}elim{\isacharbang}{\isacharbrackright}{\isacharcolon}\ {\isachardoublequoteopen}{\isasymSigma}\ {\isasymturnstile}v\ {\isacharparenleft}Inject\ v\ T{\isacharparenright}\ {\isacharcolon}\ A{\isachardoublequoteclose}\isanewline
\isanewline
\isacommand{inductive{\isacharunderscore}cases}\isamarkupfalse%
\isanewline
\ \ clos{\isacharunderscore}int{\isacharbrackleft}elim{\isacharbang}{\isacharbrackright}{\isacharcolon}\ {\isachardoublequoteopen}{\isasymSigma}\ {\isasymturnstile}v\ {\isacharparenleft}Closure\ x\ T\ s\ {\isasymrho}{\isacharparenright}\ {\isacharcolon}\ IntT\ {\isachardoublequoteclose}\ \isakeyword{and}\isanewline
\ \ inj{\isacharunderscore}int{\isacharbrackleft}elim{\isacharbang}{\isacharbrackright}{\isacharcolon}\ {\isachardoublequoteopen}{\isasymSigma}\ {\isasymturnstile}v\ {\isacharparenleft}Inject\ v\ T{\isacharparenright}\ {\isacharcolon}\ IntT\ {\isachardoublequoteclose}\ \isakeyword{and}\isanewline
\ \ const{\isacharunderscore}fun{\isacharbrackleft}elim{\isacharbang}{\isacharbrackright}{\isacharcolon}\ {\isachardoublequoteopen}{\isasymSigma}\ {\isasymturnstile}v\ {\isacharparenleft}VConst\ c{\isacharparenright}\ {\isacharcolon}\ {\isacharparenleft}A\ {\isasymrightarrow}\ B{\isacharparenright}{\isachardoublequoteclose}\ \isakeyword{and}\isanewline
\ \ const{\isacharunderscore}ref{\isacharbrackleft}elim{\isacharbang}{\isacharbrackright}{\isacharcolon}\ {\isachardoublequoteopen}{\isasymSigma}\ {\isasymturnstile}v\ {\isacharparenleft}VConst\ c{\isacharparenright}\ {\isacharcolon}\ RefT\ A{\isachardoublequoteclose}\ \isakeyword{and}\isanewline
\ \ const{\isacharunderscore}any{\isacharbrackleft}elim{\isacharbang}{\isacharbrackright}{\isacharcolon}\ {\isachardoublequoteopen}{\isasymSigma}\ {\isasymturnstile}v\ {\isacharparenleft}VConst\ c{\isacharparenright}\ {\isacharcolon}\ A{\isachardoublequoteclose}\ \isakeyword{and}\isanewline
\ \ clos{\isacharunderscore}fun{\isacharbrackleft}elim{\isacharbang}{\isacharbrackright}{\isacharcolon}\ {\isachardoublequoteopen}{\isasymSigma}\ {\isasymturnstile}v\ {\isacharparenleft}Closure\ x\ T\ s\ {\isasymrho}{\isacharparenright}\ {\isacharcolon}\ {\isacharparenleft}A\ {\isasymrightarrow}\ B{\isacharparenright}{\isachardoublequoteclose}\ \isakeyword{and}\isanewline
\ \ pair{\isacharunderscore}times{\isacharbrackleft}elim{\isacharbang}{\isacharbrackright}{\isacharcolon}\ {\isachardoublequoteopen}{\isasymSigma}\ {\isasymturnstile}v\ {\isacharparenleft}VPair\ v{\isadigit{1}}\ v{\isadigit{2}}{\isacharparenright}\ {\isacharcolon}\ {\isacharparenleft}A\ {\isasymtimes}\ B{\isacharparenright}{\isachardoublequoteclose}\ \isakeyword{and}\isanewline
\ \ pairval{\isacharunderscore}inv{\isacharbrackleft}elim{\isacharbang}{\isacharbrackright}{\isacharcolon}\ {\isachardoublequoteopen}{\isasymSigma}\ {\isasymturnstile}v\ v\ {\isacharcolon}\ {\isacharparenleft}A\ {\isasymtimes}\ B{\isacharparenright}{\isachardoublequoteclose}\ \isakeyword{and}\isanewline
\ \ funval{\isacharunderscore}inv{\isacharbrackleft}elim{\isacharbang}{\isacharbrackright}{\isacharcolon}\ {\isachardoublequoteopen}{\isasymSigma}\ {\isasymturnstile}v\ v\ {\isacharcolon}\ {\isacharparenleft}A\ {\isasymrightarrow}\ B{\isacharparenright}{\isachardoublequoteclose}\ \isakeyword{and}\isanewline
\ \ refval{\isacharunderscore}inv{\isacharbrackleft}elim{\isacharbang}{\isacharbrackright}{\isacharcolon}\ {\isachardoublequoteopen}{\isasymSigma}\ {\isasymturnstile}v\ v\ {\isacharcolon}\ {\isacharparenleft}RefT\ A{\isacharparenright}{\isachardoublequoteclose}\isanewline
\isanewline
\isacommand{inductive{\isacharunderscore}cases}\isamarkupfalse%
\isanewline
\ \ cv{\isacharunderscore}val{\isacharbrackleft}elim{\isacharbang}{\isacharbrackright}{\isacharcolon}\ {\isachardoublequoteopen}{\isasymSigma}\ {\isasymturnstile}cv\ Val\ v\ {\isacharcolon}\ A{\isachardoublequoteclose}\ \isakeyword{and}\isanewline
\ \ cv{\isacharunderscore}cv{\isacharbrackleft}elim{\isacharbang}{\isacharbrackright}{\isacharcolon}\ {\isachardoublequoteopen}{\isasymSigma}\ {\isasymturnstile}cv\ VCast\ v\ A\ B\ {\isacharcolon}\ C{\isachardoublequoteclose}\ \isanewline
\isanewline
\isacommand{inductive{\isacharunderscore}cases}\isamarkupfalse%
\ \isanewline
\ \ wtek{\isacharbrackleft}elim{\isacharbang}{\isacharbrackright}{\isacharcolon}\ {\isachardoublequoteopen}{\isasymSigma}\ {\isasymturnstile}\ {\isacharbrackleft}{\isacharbrackright}\ {\isacharcolon}\ A\ {\isasymRightarrow}\ B{\isachardoublequoteclose}\ \isakeyword{and}\isanewline
\ \ wtk{\isacharbrackleft}elim{\isacharbang}{\isacharbrackright}{\isacharcolon}\ {\isachardoublequoteopen}{\isasymSigma}\ {\isasymturnstile}\ f{\isacharhash}k\ {\isacharcolon}\ A\ {\isasymRightarrow}\ B{\isachardoublequoteclose}\ \isanewline
\isanewline
\isacommand{inductive{\isacharunderscore}cases}\isamarkupfalse%
\ wts{\isacharbrackleft}elim{\isacharbang}{\isacharbrackright}{\isacharcolon}\ {\isachardoublequoteopen}wt{\isacharunderscore}state\ s\ A{\isachardoublequoteclose}%
\isamarkupsubsection{Proof of Type Safety%
}
\isamarkuptrue%
\isacommand{lemma}\isamarkupfalse%
\ static{\isacharunderscore}is{\isacharunderscore}most{\isacharunderscore}precise{\isacharcolon}\isanewline
\ \ \isakeyword{fixes}\ A{\isacharcolon}{\isacharcolon}ty\ \isakeyword{and}\ B{\isacharcolon}{\isacharcolon}ty\ \isakeyword{assumes}\ sa{\isacharcolon}\ {\isachardoublequoteopen}static\ A{\isachardoublequoteclose}\ \isakeyword{and}\ ba{\isacharcolon}\ {\isachardoublequoteopen}B\ {\isasymsqsubseteq}\ A{\isachardoublequoteclose}\ \isakeyword{shows}\ {\isachardoublequoteopen}A\ {\isacharequal}\ B{\isachardoublequoteclose}\isanewline
\isadelimproof
\ \ %
\endisadelimproof
\isatagproof
\isacommand{using}\isamarkupfalse%
\ sa\ ba\ \isacommand{apply}\isamarkupfalse%
\ {\isacharparenleft}induct\ rule{\isacharcolon}\ lesseq{\isacharunderscore}dyn{\isachardot}induct{\isacharparenright}\isanewline
\ \ \isacommand{apply}\isamarkupfalse%
\ {\isacharparenleft}case{\isacharunderscore}tac\ A{\isacharparenright}\ \isacommand{apply}\isamarkupfalse%
\ simp{\isacharplus}\ \isacommand{done}\isamarkupfalse%
\endisatagproof
{\isafoldproof}%
\isadelimproof
\isanewline
\endisadelimproof
\isanewline
\isacommand{lemma}\isamarkupfalse%
\ lookup{\isacharunderscore}dom{\isacharcolon}\isanewline
\ \ {\isachardoublequoteopen}lookup\ a\ {\isasymSigma}\ {\isacharequal}\ Result\ v\ {\isasymLongrightarrow}\ a\ {\isasymin}\ dom\ {\isasymSigma}{\isachardoublequoteclose}\isanewline
\isadelimproof
\ \ %
\endisadelimproof
\isatagproof
\isacommand{apply}\isamarkupfalse%
\ {\isacharparenleft}induct\ {\isasymSigma}{\isacharparenright}\isanewline
\ \ \isacommand{apply}\isamarkupfalse%
\ simp\ \isacommand{apply}\isamarkupfalse%
\ clarify\ \isacommand{apply}\isamarkupfalse%
\ {\isacharparenleft}case{\isacharunderscore}tac\ {\isachardoublequoteopen}a\ {\isacharequal}\ aa{\isachardoublequoteclose}{\isacharparenright}\ \isacommand{apply}\isamarkupfalse%
\ {\isacharparenleft}auto\ simp{\isacharcolon}\ dom{\isacharunderscore}def{\isacharparenright}\isanewline
\ \ \isacommand{done}\isamarkupfalse%
\endisatagproof
{\isafoldproof}%
\isadelimproof
\isanewline
\endisadelimproof
\isanewline
\isacommand{lemma}\isamarkupfalse%
\ dom{\isacharunderscore}lookup{\isacharcolon}\isanewline
\ \ {\isachardoublequoteopen}a\ {\isasymin}\ dom\ {\isasymSigma}\ {\isasymLongrightarrow}\ {\isacharparenleft}{\isasymexists}\ A{\isachardot}\ lookup\ a\ {\isasymSigma}\ {\isacharequal}\ Result\ A{\isacharparenright}{\isachardoublequoteclose}\isanewline
\isadelimproof
\ \ %
\endisadelimproof
\isatagproof
\isacommand{apply}\isamarkupfalse%
\ {\isacharparenleft}induct\ {\isasymSigma}\ arbitrary{\isacharcolon}\ a{\isacharparenright}\isanewline
\ \ \isacommand{apply}\isamarkupfalse%
\ {\isacharparenleft}simp\ add{\isacharcolon}\ dom{\isacharunderscore}def{\isacharparenright}\isanewline
\ \ \isacommand{apply}\isamarkupfalse%
\ clarify\ \isacommand{apply}\isamarkupfalse%
\ {\isacharparenleft}case{\isacharunderscore}tac\ {\isachardoublequoteopen}a\ {\isacharequal}\ aa{\isachardoublequoteclose}{\isacharparenright}\ \isacommand{apply}\isamarkupfalse%
\ simp\ \isacommand{apply}\isamarkupfalse%
\ {\isacharparenleft}simp\ add{\isacharcolon}\ dom{\isacharunderscore}def{\isacharparenright}\isanewline
\ \ \isacommand{done}\isamarkupfalse%
\endisatagproof
{\isafoldproof}%
\isadelimproof
\isanewline
\endisadelimproof
\isanewline
\isacommand{lemma}\isamarkupfalse%
\ weaken{\isacharunderscore}value{\isacharunderscore}env{\isacharcolon}\isanewline
\ \ {\isachardoublequoteopen}{\isacharparenleft}{\isasymSigma}\ {\isasymturnstile}v\ v\ {\isacharcolon}\ A\ {\isasymlongrightarrow}\ {\isacharparenleft}{\isasymforall}\ a\ B{\isachardot}\ a\ {\isasymnotin}\ dom\ {\isasymSigma}\ {\isasymlongrightarrow}\ {\isacharparenleft}a{\isacharcomma}B{\isacharparenright}{\isacharhash}{\isasymSigma}\ {\isasymturnstile}v\ v\ {\isacharcolon}\ A{\isacharparenright}{\isacharparenright}\isanewline
\ \ \ {\isasymand}\ {\isacharparenleft}{\isasymGamma}{\isacharsemicolon}{\isasymSigma}\ {\isasymturnstile}\ {\isasymrho}\ {\isasymlongrightarrow}\ {\isacharparenleft}{\isasymforall}\ a\ B{\isachardot}\ a\ {\isasymnotin}\ dom\ {\isasymSigma}\ {\isasymlongrightarrow}\ {\isasymGamma}{\isacharsemicolon}{\isacharparenleft}a{\isacharcomma}B{\isacharparenright}{\isacharhash}{\isasymSigma}\ {\isasymturnstile}\ {\isasymrho}{\isacharparenright}{\isacharparenright}{\isachardoublequoteclose}\isanewline
\isadelimproof
\ \ %
\endisadelimproof
\isatagproof
\isacommand{apply}\isamarkupfalse%
\ {\isacharparenleft}induct\ rule{\isacharcolon}\ wt{\isacharunderscore}val{\isacharunderscore}wt{\isacharunderscore}env{\isachardot}induct{\isacharparenright}\isanewline
\ \ \isacommand{using}\isamarkupfalse%
\ lookup{\isacharunderscore}dom\ \isacommand{apply}\isamarkupfalse%
\ force{\isacharplus}\ \isacommand{done}\isamarkupfalse%
\endisatagproof
{\isafoldproof}%
\isadelimproof
\isanewline
\endisadelimproof
\isanewline
\isacommand{lemma}\isamarkupfalse%
\ weaken{\isacharunderscore}stack{\isacharcolon}\isanewline
\ \ {\isachardoublequoteopen}{\isasymlbrakk}\ {\isasymSigma}\ {\isasymturnstile}\ k\ {\isacharcolon}\ A\ {\isasymRightarrow}\ B{\isacharsemicolon}\ a\ {\isasymnotin}\ dom\ {\isasymSigma}\ {\isasymrbrakk}\ {\isasymLongrightarrow}\ {\isacharparenleft}a{\isacharcomma}T{\isacharparenright}{\isacharhash}{\isasymSigma}\ {\isasymturnstile}\ k\ {\isacharcolon}\ A\ {\isasymRightarrow}\ B{\isachardoublequoteclose}\isanewline
\isadelimproof
\ \ %
\endisadelimproof
\isatagproof
\isacommand{apply}\isamarkupfalse%
\ {\isacharparenleft}induct\ k\ arbitrary{\isacharcolon}\ A\ B\ a\ {\isasymSigma}{\isacharparenright}\isanewline
\ \ \isacommand{apply}\isamarkupfalse%
\ force\isanewline
\ \ \isacommand{using}\isamarkupfalse%
\ weaken{\isacharunderscore}value{\isacharunderscore}env\ \isacommand{apply}\isamarkupfalse%
\ auto\ \isacommand{done}\isamarkupfalse%
\endisatagproof
{\isafoldproof}%
\isadelimproof
\isanewline
\endisadelimproof
\isanewline
\isacommand{lemma}\isamarkupfalse%
\ delta{\isacharunderscore}safe{\isacharcolon}\ \isanewline
\ \ \isakeyword{assumes}\ wtop{\isacharcolon}\ {\isachardoublequoteopen}typeof{\isacharunderscore}opr\ f\ {\isacharequal}\ A\ {\isasymrightarrow}\ B{\isachardoublequoteclose}\isanewline
\ \ \isakeyword{and}\ wtv{\isacharcolon}\ {\isachardoublequoteopen}{\isasymSigma}\ {\isasymturnstile}v\ v\ {\isacharcolon}\ A{\isachardoublequoteclose}\isanewline
\ \ \isakeyword{shows}\ {\isachardoublequoteopen}{\isasymexists}\ v{\isacharprime}{\isachardot}\ delta\ f\ v\ {\isacharequal}\ Result\ v{\isacharprime}\ {\isasymand}\ {\isasymSigma}\ {\isasymturnstile}v\ v{\isacharprime}\ {\isacharcolon}\ B{\isachardoublequoteclose}\isanewline
\isadelimproof
\ \ %
\endisadelimproof
\isatagproof
\isacommand{using}\isamarkupfalse%
\ wtop\ wtv\ \isacommand{apply}\isamarkupfalse%
\ {\isacharparenleft}case{\isacharunderscore}tac\ f{\isacharparenright}\ \isanewline
\ \ \isacommand{apply}\isamarkupfalse%
\ {\isacharparenleft}case{\isacharunderscore}tac\ v{\isacharcomma}\ auto{\isacharcomma}\ case{\isacharunderscore}tac\ const{\isacharcomma}\ auto{\isacharparenright}\isanewline
\ \ \isacommand{apply}\isamarkupfalse%
\ {\isacharparenleft}case{\isacharunderscore}tac\ v{\isacharcomma}\ auto{\isacharcomma}\ case{\isacharunderscore}tac\ const{\isacharcomma}\ auto{\isacharparenright}\isanewline
\ \ \isacommand{apply}\isamarkupfalse%
\ {\isacharparenleft}case{\isacharunderscore}tac\ v{\isacharcomma}\ auto{\isacharcomma}\ case{\isacharunderscore}tac\ const{\isacharcomma}\ auto{\isacharparenright}\isanewline
\ \ \isacommand{apply}\isamarkupfalse%
\ {\isacharparenleft}case{\isacharunderscore}tac\ c{\isacharcomma}\ auto{\isacharparenright}\isanewline
\ \ \isacommand{apply}\isamarkupfalse%
\ {\isacharparenleft}case{\isacharunderscore}tac\ c{\isacharcomma}\ auto{\isacharparenright}\isanewline
\ \ \isacommand{done}\isamarkupfalse%
\endisatagproof
{\isafoldproof}%
\isadelimproof
\isanewline
\endisadelimproof
\isanewline
\isacommand{lemma}\isamarkupfalse%
\ lookup{\isacharunderscore}safe{\isacharcolon}\ \isanewline
\ \ \isakeyword{assumes}\ wtg{\isacharcolon}\ {\isachardoublequoteopen}{\isasymGamma}{\isacharsemicolon}{\isasymSigma}\ {\isasymturnstile}\ {\isasymrho}{\isachardoublequoteclose}\ \isakeyword{and}\ l{\isacharcolon}\ {\isachardoublequoteopen}lookup\ x\ {\isasymGamma}\ {\isacharequal}\ Result\ A{\isachardoublequoteclose}\isanewline
\ \ \isakeyword{shows}\ {\isachardoublequoteopen}{\isasymexists}\ v{\isachardot}\ lookup\ x\ {\isasymrho}\ {\isacharequal}\ Result\ v\ {\isasymand}\ {\isasymSigma}\ {\isasymturnstile}v\ v\ {\isacharcolon}\ A{\isachardoublequoteclose}\isanewline
\isadelimproof
\ \ %
\endisadelimproof
\isatagproof
\isacommand{using}\isamarkupfalse%
\ wtg\ l\ \isacommand{by}\isamarkupfalse%
\ {\isacharparenleft}induct\ {\isasymrho}{\isacharparenright}\ force{\isacharplus}%
\endisatagproof
{\isafoldproof}%
\isadelimproof
\isanewline
\endisadelimproof
\isanewline
\isacommand{lemma}\isamarkupfalse%
\ eval{\isacharunderscore}safe{\isacharcolon}\isanewline
\ \ \isakeyword{assumes}\ wte{\isacharcolon}\ {\isachardoublequoteopen}{\isasymGamma}\ {\isasymturnstile}\isactrlisub e\ e\ {\isacharcolon}\ A{\isachardoublequoteclose}\isanewline
\ \ \isakeyword{and}\ wtg{\isacharcolon}\ {\isachardoublequoteopen}{\isasymGamma}{\isacharsemicolon}{\isasymSigma}\ {\isasymturnstile}\ {\isasymrho}{\isachardoublequoteclose}\isanewline
\ \ \isakeyword{and}\ wth{\isacharcolon}\ {\isachardoublequoteopen}wt{\isacharunderscore}heap\ {\isasymSigma}\ {\isasymmu}\ {\isacharbraceleft}{\isacharbraceright}{\isachardoublequoteclose}\isanewline
\ \ \isakeyword{shows}\ {\isachardoublequoteopen}{\isasymexists}\ v{\isachardot}\ eval\ e\ {\isasymrho}\ {\isasymmu}\ {\isacharequal}\ Result\ v\ {\isasymand}\ {\isasymSigma}\ {\isasymturnstile}v\ v\ {\isacharcolon}\ A{\isachardoublequoteclose}\isanewline
\isadelimproof
\ \ %
\endisadelimproof
\isatagproof
\isacommand{using}\isamarkupfalse%
\ wte\ wtg\ wth\isanewline
\ \ \isacommand{apply}\isamarkupfalse%
\ {\isacharparenleft}induct\ e\ {\isasymrho}\ {\isasymmu}\ arbitrary{\isacharcolon}\ A\ rule{\isacharcolon}\ eval{\isachardot}induct{\isacharparenright}\isanewline
\ \ \isacommand{using}\isamarkupfalse%
\ lookup{\isacharunderscore}safe\ \isacommand{apply}\isamarkupfalse%
\ force\isanewline
\ \ \isacommand{apply}\isamarkupfalse%
\ {\isacharparenleft}case{\isacharunderscore}tac\ c{\isacharparenright}\ \isacommand{apply}\isamarkupfalse%
\ force\ \isacommand{apply}\isamarkupfalse%
\ force\isanewline
\ \ \isacommand{using}\isamarkupfalse%
\ delta{\isacharunderscore}safe\ \isacommand{apply}\isamarkupfalse%
\ force\isanewline
\ \ \isacommand{apply}\isamarkupfalse%
\ force\isanewline
\ \ \isacommand{apply}\isamarkupfalse%
\ force\isanewline
\isacommand{proof}\isamarkupfalse%
\ {\isacharminus}\isanewline
\ \ \isacommand{fix}\isamarkupfalse%
\ e\ {\isasymrho}\ {\isasymmu}\ A\isanewline
\ \ \isacommand{assume}\isamarkupfalse%
\ IH{\isacharcolon}\ {\isachardoublequoteopen}{\isasymAnd}A{\isachardot}\ {\isasymlbrakk}{\isasymGamma}\ {\isasymturnstile}\isactrlisub e\ e\ {\isacharcolon}\ A{\isacharsemicolon}\ {\isasymGamma}{\isacharsemicolon}{\isasymSigma}\ {\isasymturnstile}\ {\isasymrho}{\isacharsemicolon}\ wt{\isacharunderscore}heap\ {\isasymSigma}\ {\isasymmu}\ {\isacharbraceleft}{\isacharbraceright}{\isasymrbrakk}\isanewline
\ \ \ \ \ \ \ \ \ \ \ \ {\isasymLongrightarrow}\ {\isasymexists}v{\isachardot}\ eval\ e\ {\isasymrho}\ {\isasymmu}\ {\isacharequal}\ Result\ v\ {\isasymand}\ {\isasymSigma}\ {\isasymturnstile}v\ v\ {\isacharcolon}\ A{\isachardoublequoteclose}\isanewline
\ \ \ \ \isakeyword{and}\ de{\isacharcolon}\ {\isachardoublequoteopen}{\isasymGamma}\ {\isasymturnstile}\isactrlisub e\ Deref\ e\ {\isacharcolon}\ A{\isachardoublequoteclose}\ \isakeyword{and}\ wtr{\isacharcolon}\ {\isachardoublequoteopen}{\isasymGamma}{\isacharsemicolon}{\isasymSigma}\ {\isasymturnstile}\ {\isasymrho}{\isachardoublequoteclose}\ \isakeyword{and}\ wth{\isacharcolon}\ {\isachardoublequoteopen}wt{\isacharunderscore}heap\ {\isasymSigma}\ {\isasymmu}\ {\isacharbraceleft}{\isacharbraceright}{\isachardoublequoteclose}\isanewline
\ \ \isacommand{from}\isamarkupfalse%
\ de\ \isacommand{have}\isamarkupfalse%
\ wte{\isacharcolon}\ {\isachardoublequoteopen}{\isasymGamma}\ {\isasymturnstile}\isactrlisub e\ e\ {\isacharcolon}\ {\isacharparenleft}RefT\ A{\isacharparenright}{\isachardoublequoteclose}\ \isacommand{by}\isamarkupfalse%
\ auto\isanewline
\ \ \isacommand{from}\isamarkupfalse%
\ de\ \isacommand{have}\isamarkupfalse%
\ sa{\isacharcolon}\ {\isachardoublequoteopen}static\ A{\isachardoublequoteclose}\ \isacommand{by}\isamarkupfalse%
\ auto\isanewline
\ \ \isacommand{from}\isamarkupfalse%
\ IH{\isacharbrackleft}of\ {\isachardoublequoteopen}RefT\ A{\isachardoublequoteclose}{\isacharbrackright}\ wte\ wtr\ wth\isanewline
\ \ \isacommand{have}\isamarkupfalse%
\ {\isachardoublequoteopen}{\isacharparenleft}{\isasymexists}v{\isachardot}\ eval\ e\ {\isasymrho}\ {\isasymmu}\ {\isacharequal}\ Result\ v\ {\isasymand}\ {\isasymSigma}\ {\isasymturnstile}v\ v\ {\isacharcolon}\ RefT\ A{\isacharparenright}{\isachardoublequoteclose}\ \isacommand{by}\isamarkupfalse%
\ simp\isanewline
\ \ \isacommand{from}\isamarkupfalse%
\ this\ \isacommand{obtain}\isamarkupfalse%
\ v\ \isakeyword{where}\ ev{\isacharcolon}\ {\isachardoublequoteopen}eval\ e\ {\isasymrho}\ {\isasymmu}\ {\isacharequal}\ Result\ v{\isachardoublequoteclose}\isanewline
\ \ \ \ \isakeyword{and}\ wtv{\isacharcolon}\ {\isachardoublequoteopen}{\isasymSigma}\ {\isasymturnstile}v\ v\ {\isacharcolon}\ RefT\ A{\isachardoublequoteclose}\ \isacommand{apply}\isamarkupfalse%
\ clarify\ \isacommand{apply}\isamarkupfalse%
\ auto\ \isacommand{done}\isamarkupfalse%
\isanewline
\ \ \isacommand{from}\isamarkupfalse%
\ wtv\ \isacommand{obtain}\isamarkupfalse%
\ a\ A{\isacharprime}\ \isakeyword{where}\ v{\isacharcolon}\ {\isachardoublequoteopen}v\ {\isacharequal}\ VRef\ a{\isachardoublequoteclose}\ \isakeyword{and}\ \isanewline
\ \ \ \ las{\isacharcolon}\ {\isachardoublequoteopen}lookup\ a\ {\isasymSigma}\ {\isacharequal}\ Result\ A{\isacharprime}{\isachardoublequoteclose}\ \isakeyword{and}\ aa{\isacharcolon}\ {\isachardoublequoteopen}A{\isacharprime}\ {\isasymsqsubseteq}\ A{\isachardoublequoteclose}\ \isacommand{apply}\isamarkupfalse%
\ auto\isanewline
\ \ \ \ \isacommand{apply}\isamarkupfalse%
\ {\isacharparenleft}case{\isacharunderscore}tac\ c{\isacharparenright}\ \isacommand{apply}\isamarkupfalse%
\ auto\ \isacommand{done}\isamarkupfalse%
\isanewline
\ \ \isacommand{from}\isamarkupfalse%
\ sa\ aa\ \isacommand{have}\isamarkupfalse%
\ \ aaeq{\isacharcolon}\ {\isachardoublequoteopen}A\ {\isacharequal}\ A{\isacharprime}{\isachardoublequoteclose}\ \isacommand{using}\isamarkupfalse%
\ static{\isacharunderscore}is{\isacharunderscore}most{\isacharunderscore}precise\ \isacommand{by}\isamarkupfalse%
\ blast\isanewline
\ \ \isacommand{from}\isamarkupfalse%
\ las\ wth\ aaeq\ \isacommand{obtain}\isamarkupfalse%
\ cv\ v{\isacharprime}\ \isakeyword{where}\ lam{\isacharcolon}\ {\isachardoublequoteopen}lookup\ a\ {\isasymmu}\ {\isacharequal}\ Result\ {\isacharparenleft}cv{\isacharcomma}A{\isacharparenright}{\isachardoublequoteclose}\isanewline
\ \ \ \ \isakeyword{and}\ wtcv{\isacharcolon}\ {\isachardoublequoteopen}{\isasymSigma}\ {\isasymturnstile}cv\ cv\ {\isacharcolon}\ A{\isachardoublequoteclose}\ \isakeyword{and}\ cv{\isacharcolon}\ {\isachardoublequoteopen}cv\ {\isacharequal}\ Val\ v{\isacharprime}{\isachardoublequoteclose}\isanewline
\ \ \ \ \isacommand{apply}\isamarkupfalse%
\ {\isacharparenleft}simp\ only{\isacharcolon}\ wt{\isacharunderscore}heap{\isacharunderscore}def{\isacharparenright}\ \isacommand{apply}\isamarkupfalse%
\ blast\ \isacommand{done}\isamarkupfalse%
\isanewline
\ \ \isacommand{from}\isamarkupfalse%
\ wtcv\ cv\ \isacommand{have}\isamarkupfalse%
\ wtvp{\isacharcolon}\ {\isachardoublequoteopen}{\isasymSigma}\ {\isasymturnstile}v\ v{\isacharprime}\ {\isacharcolon}\ A{\isachardoublequoteclose}\ \isacommand{apply}\isamarkupfalse%
\ auto\ \isacommand{done}\isamarkupfalse%
\isanewline
\ \ \isacommand{from}\isamarkupfalse%
\ ev\ lam\ cv\ wtvp\ v\ \isanewline
\ \ \isacommand{show}\isamarkupfalse%
\ {\isachardoublequoteopen}{\isacharparenleft}{\isasymexists}v{\isachardot}\ eval\ {\isacharparenleft}Deref\ e{\isacharparenright}\ {\isasymrho}\ {\isasymmu}\ {\isacharequal}\ Result\ v\ {\isasymand}\ {\isasymSigma}\ {\isasymturnstile}v\ v\ {\isacharcolon}\ A{\isacharparenright}{\isachardoublequoteclose}\ \isacommand{by}\isamarkupfalse%
\ auto\isanewline
\isacommand{qed}\isamarkupfalse%
\endisatagproof
{\isafoldproof}%
\isadelimproof
\isanewline
\endisadelimproof
\isanewline
\isacommand{lemma}\isamarkupfalse%
\ lesseq{\isacharunderscore}refl{\isacharbrackleft}simp{\isacharbrackright}{\isacharcolon}\ \isakeyword{fixes}\ A{\isacharcolon}{\isacharcolon}ty\ \isakeyword{shows}\ {\isachardoublequoteopen}A\ {\isasymsqsubseteq}\ A{\isachardoublequoteclose}\isanewline
\isadelimproof
\ \ %
\endisadelimproof
\isatagproof
\isacommand{by}\isamarkupfalse%
\ {\isacharparenleft}induct\ A{\isacharparenright}\ auto%
\endisatagproof
{\isafoldproof}%
\isadelimproof
\isanewline
\endisadelimproof
\isanewline
\isacommand{lemma}\isamarkupfalse%
\ lesseq{\isacharunderscore}env{\isacharunderscore}refl{\isacharbrackleft}simp{\isacharbrackright}{\isacharcolon}\ \isakeyword{fixes}\ {\isasymSigma}{\isacharcolon}{\isacharcolon}ty{\isacharunderscore}env\ \isakeyword{shows}\ {\isachardoublequoteopen}{\isasymSigma}\ {\isasymsqsubseteq}\ {\isasymSigma}{\isachardoublequoteclose}\isanewline
\isadelimproof
\ \ %
\endisadelimproof
\isatagproof
\isacommand{using}\isamarkupfalse%
\ lesseq{\isacharunderscore}tyenv{\isacharunderscore}def\ \isacommand{by}\isamarkupfalse%
\ simp%
\endisatagproof
{\isafoldproof}%
\isadelimproof
\isanewline
\endisadelimproof
\isanewline
\isacommand{lemma}\isamarkupfalse%
\ lesseq{\isacharunderscore}int{\isacharunderscore}pair{\isacharbrackleft}elim{\isacharbang}{\isacharbrackright}{\isacharcolon}\ {\isachardoublequoteopen}A\ {\isasymtimes}\ B\ {\isasymsqsubseteq}\ IntT\ {\isasymLongrightarrow}\ P{\isachardoublequoteclose}%
\isadelimproof
\ %
\endisadelimproof
\isatagproof
\isacommand{by}\isamarkupfalse%
\ auto%
\endisatagproof
{\isafoldproof}%
\isadelimproof
\endisadelimproof
\isanewline
\isacommand{lemma}\isamarkupfalse%
\ lesseq{\isacharunderscore}int{\isacharunderscore}fun{\isacharbrackleft}elim{\isacharbang}{\isacharbrackright}{\isacharcolon}\ {\isachardoublequoteopen}A\ {\isasymrightarrow}\ B\ {\isasymsqsubseteq}\ IntT\ {\isasymLongrightarrow}\ P{\isachardoublequoteclose}%
\isadelimproof
\ %
\endisadelimproof
\isatagproof
\isacommand{by}\isamarkupfalse%
\ auto%
\endisatagproof
{\isafoldproof}%
\isadelimproof
\endisadelimproof
\isanewline
\isacommand{lemma}\isamarkupfalse%
\ lesseq{\isacharunderscore}pair{\isacharunderscore}int{\isacharbrackleft}elim{\isacharbang}{\isacharbrackright}{\isacharcolon}\ {\isachardoublequoteopen}IntT\ {\isasymsqsubseteq}\ A\ {\isasymtimes}\ B\ {\isasymLongrightarrow}\ P{\isachardoublequoteclose}%
\isadelimproof
\ %
\endisadelimproof
\isatagproof
\isacommand{by}\isamarkupfalse%
\ auto%
\endisatagproof
{\isafoldproof}%
\isadelimproof
\endisadelimproof
\isanewline
\isacommand{lemma}\isamarkupfalse%
\ lesseq{\isacharunderscore}bool{\isacharunderscore}pair{\isacharbrackleft}elim{\isacharbang}{\isacharbrackright}{\isacharcolon}\ {\isachardoublequoteopen}A\ {\isasymtimes}\ B\ {\isasymsqsubseteq}\ BoolT\ {\isasymLongrightarrow}\ P{\isachardoublequoteclose}%
\isadelimproof
\ %
\endisadelimproof
\isatagproof
\isacommand{by}\isamarkupfalse%
\ auto%
\endisatagproof
{\isafoldproof}%
\isadelimproof
\endisadelimproof
\isanewline
\isacommand{lemma}\isamarkupfalse%
\ lesseq{\isacharunderscore}bool{\isacharunderscore}fun{\isacharbrackleft}elim{\isacharbang}{\isacharbrackright}{\isacharcolon}\ {\isachardoublequoteopen}A\ {\isasymrightarrow}\ B\ {\isasymsqsubseteq}\ BoolT\ {\isasymLongrightarrow}\ P{\isachardoublequoteclose}%
\isadelimproof
\ %
\endisadelimproof
\isatagproof
\isacommand{by}\isamarkupfalse%
\ auto%
\endisatagproof
{\isafoldproof}%
\isadelimproof
\endisadelimproof
\isanewline
\isacommand{lemma}\isamarkupfalse%
\ lesseq{\isacharunderscore}pair{\isacharunderscore}bool{\isacharbrackleft}elim{\isacharbang}{\isacharbrackright}{\isacharcolon}\ {\isachardoublequoteopen}BoolT\ {\isasymsqsubseteq}\ A\ {\isasymtimes}\ B\ {\isasymLongrightarrow}\ P{\isachardoublequoteclose}%
\isadelimproof
\ %
\endisadelimproof
\isatagproof
\isacommand{by}\isamarkupfalse%
\ auto%
\endisatagproof
{\isafoldproof}%
\isadelimproof
\endisadelimproof
\isanewline
\isacommand{lemma}\isamarkupfalse%
\ lesseq{\isacharunderscore}pair{\isacharunderscore}fun{\isacharbrackleft}elim{\isacharbang}{\isacharbrackright}{\isacharcolon}\ {\isachardoublequoteopen}C\ {\isasymrightarrow}\ D\ {\isasymsqsubseteq}\ A\ {\isasymtimes}\ B\ {\isasymLongrightarrow}\ P{\isachardoublequoteclose}%
\isadelimproof
\ %
\endisadelimproof
\isatagproof
\isacommand{by}\isamarkupfalse%
\ auto%
\endisatagproof
{\isafoldproof}%
\isadelimproof
\endisadelimproof
\isanewline
\isacommand{lemma}\isamarkupfalse%
\ lesseq{\isacharunderscore}pair{\isacharunderscore}ref{\isacharbrackleft}elim{\isacharbang}{\isacharbrackright}{\isacharcolon}\ {\isachardoublequoteopen}RefT\ C\ {\isasymsqsubseteq}\ A\ {\isasymtimes}\ B\ {\isasymLongrightarrow}\ P{\isachardoublequoteclose}%
\isadelimproof
\ %
\endisadelimproof
\isatagproof
\isacommand{by}\isamarkupfalse%
\ auto%
\endisatagproof
{\isafoldproof}%
\isadelimproof
\endisadelimproof
\isanewline
\isacommand{lemma}\isamarkupfalse%
\ lesseq{\isacharunderscore}ref{\isacharunderscore}pair{\isacharbrackleft}elim{\isacharbang}{\isacharbrackright}{\isacharcolon}\ {\isachardoublequoteopen}A\ {\isasymtimes}\ B\ {\isasymsqsubseteq}\ RefT\ C\ {\isasymLongrightarrow}\ P{\isachardoublequoteclose}%
\isadelimproof
\ %
\endisadelimproof
\isatagproof
\isacommand{by}\isamarkupfalse%
\ auto%
\endisatagproof
{\isafoldproof}%
\isadelimproof
\endisadelimproof
\isanewline
\isacommand{lemma}\isamarkupfalse%
\ lesseq{\isacharunderscore}ref{\isacharunderscore}fun{\isacharbrackleft}elim{\isacharbang}{\isacharbrackright}{\isacharcolon}\ {\isachardoublequoteopen}A\ {\isasymrightarrow}\ B\ {\isasymsqsubseteq}\ RefT\ C\ {\isasymLongrightarrow}\ P{\isachardoublequoteclose}%
\isadelimproof
\ %
\endisadelimproof
\isatagproof
\isacommand{by}\isamarkupfalse%
\ auto%
\endisatagproof
{\isafoldproof}%
\isadelimproof
\endisadelimproof
\isanewline
\isacommand{lemma}\isamarkupfalse%
\ lesseq{\isacharunderscore}pair{\isacharunderscore}dyn{\isacharbrackleft}elim{\isacharbang}{\isacharbrackright}{\isacharcolon}\ {\isachardoublequoteopen}DynT\ {\isasymsqsubseteq}\ A\ {\isasymtimes}\ B\ {\isasymLongrightarrow}\ P{\isachardoublequoteclose}%
\isadelimproof
\ %
\endisadelimproof
\isatagproof
\isacommand{by}\isamarkupfalse%
\ auto%
\endisatagproof
{\isafoldproof}%
\isadelimproof
\endisadelimproof
\isanewline
\isacommand{lemma}\isamarkupfalse%
\ lesseq{\isacharunderscore}fun{\isacharunderscore}pair{\isacharbrackleft}elim{\isacharbang}{\isacharbrackright}{\isacharcolon}\ {\isachardoublequoteopen}C\ {\isasymtimes}\ D\ {\isasymsqsubseteq}\ A\ {\isasymrightarrow}\ B\ {\isasymLongrightarrow}\ P{\isachardoublequoteclose}%
\isadelimproof
\ %
\endisadelimproof
\isatagproof
\isacommand{by}\isamarkupfalse%
\ auto%
\endisatagproof
{\isafoldproof}%
\isadelimproof
\endisadelimproof
\isanewline
\isacommand{lemma}\isamarkupfalse%
\ lesseq{\isacharunderscore}fun{\isacharunderscore}int{\isacharbrackleft}elim{\isacharbang}{\isacharbrackright}{\isacharcolon}\ {\isachardoublequoteopen}IntT\ {\isasymsqsubseteq}\ A\ {\isasymrightarrow}\ B\ {\isasymLongrightarrow}\ P{\isachardoublequoteclose}%
\isadelimproof
\ %
\endisadelimproof
\isatagproof
\isacommand{by}\isamarkupfalse%
\ auto%
\endisatagproof
{\isafoldproof}%
\isadelimproof
\endisadelimproof
\isanewline
\isacommand{lemma}\isamarkupfalse%
\ lesseq{\isacharunderscore}fun{\isacharunderscore}bool{\isacharbrackleft}elim{\isacharbang}{\isacharbrackright}{\isacharcolon}\ {\isachardoublequoteopen}BoolT\ {\isasymsqsubseteq}\ A\ {\isasymrightarrow}\ B\ {\isasymLongrightarrow}\ P{\isachardoublequoteclose}%
\isadelimproof
\ %
\endisadelimproof
\isatagproof
\isacommand{by}\isamarkupfalse%
\ auto%
\endisatagproof
{\isafoldproof}%
\isadelimproof
\endisadelimproof
\isanewline
\isacommand{lemma}\isamarkupfalse%
\ lesseq{\isacharunderscore}fun{\isacharunderscore}ref{\isacharbrackleft}elim{\isacharbang}{\isacharbrackright}{\isacharcolon}\ {\isachardoublequoteopen}RefT\ C\ {\isasymsqsubseteq}\ A\ {\isasymrightarrow}\ B\ {\isasymLongrightarrow}\ P{\isachardoublequoteclose}%
\isadelimproof
\ %
\endisadelimproof
\isatagproof
\isacommand{by}\isamarkupfalse%
\ auto%
\endisatagproof
{\isafoldproof}%
\isadelimproof
\endisadelimproof
\isanewline
\isacommand{lemma}\isamarkupfalse%
\ lesseq{\isacharunderscore}fun{\isacharunderscore}dyn{\isacharbrackleft}elim{\isacharbang}{\isacharbrackright}{\isacharcolon}\ {\isachardoublequoteopen}DynT\ {\isasymsqsubseteq}\ A\ {\isasymrightarrow}\ B\ {\isasymLongrightarrow}\ P{\isachardoublequoteclose}%
\isadelimproof
\ %
\endisadelimproof
\isatagproof
\isacommand{by}\isamarkupfalse%
\ auto%
\endisatagproof
{\isafoldproof}%
\isadelimproof
\endisadelimproof
\isanewline
\isanewline
\isacommand{lemma}\isamarkupfalse%
\ lesseq{\isacharunderscore}ref{\isacharunderscore}inv{\isacharbrackleft}elim{\isacharbang}{\isacharbrackright}{\isacharcolon}\ {\isachardoublequoteopen}{\isasymlbrakk}\ RefT\ A\ {\isasymsqsubseteq}\ RefT\ B{\isacharsemicolon}\ A\ {\isasymsqsubseteq}\ B\ {\isasymLongrightarrow}\ P\ {\isasymrbrakk}\ {\isasymLongrightarrow}\ P{\isachardoublequoteclose}%
\isadelimproof
\ %
\endisadelimproof
\isatagproof
\isacommand{by}\isamarkupfalse%
\ auto%
\endisatagproof
{\isafoldproof}%
\isadelimproof
\endisadelimproof
\isanewline
\isacommand{lemma}\isamarkupfalse%
\ lesseq{\isacharunderscore}pair{\isacharunderscore}inv{\isacharbrackleft}elim{\isacharbang}{\isacharbrackright}{\isacharcolon}\ {\isachardoublequoteopen}{\isasymlbrakk}\ {\isacharparenleft}A{\isadigit{1}}\ {\isasymtimes}\ A{\isadigit{2}}{\isacharparenright}\ {\isasymsqsubseteq}\ {\isacharparenleft}B{\isadigit{1}}\ {\isasymtimes}\ B{\isadigit{2}}{\isacharparenright}{\isacharsemicolon}\isanewline
\ \ \ \ {\isasymlbrakk}\ A{\isadigit{1}}\ {\isasymsqsubseteq}\ B{\isadigit{1}}{\isacharsemicolon}\ A{\isadigit{2}}\ {\isasymsqsubseteq}\ B{\isadigit{2}}\ {\isasymrbrakk}\ {\isasymLongrightarrow}\ P{\isasymrbrakk}\ {\isasymLongrightarrow}\ P{\isachardoublequoteclose}%
\isadelimproof
\ %
\endisadelimproof
\isatagproof
\isacommand{by}\isamarkupfalse%
\ auto%
\endisatagproof
{\isafoldproof}%
\isadelimproof
\endisadelimproof
\isanewline
\isacommand{lemma}\isamarkupfalse%
\ lesseq{\isacharunderscore}fun{\isacharunderscore}inv{\isacharbrackleft}elim{\isacharbang}{\isacharbrackright}{\isacharcolon}\ {\isachardoublequoteopen}{\isasymlbrakk}\ {\isacharparenleft}A{\isadigit{1}}\ {\isasymrightarrow}\ A{\isadigit{2}}{\isacharparenright}\ {\isasymsqsubseteq}\ {\isacharparenleft}B{\isadigit{1}}\ {\isasymrightarrow}\ B{\isadigit{2}}{\isacharparenright}{\isacharsemicolon}\ \isanewline
\ \ \ \ {\isasymlbrakk}\ A{\isadigit{1}}\ {\isasymsqsubseteq}\ B{\isadigit{1}}{\isacharsemicolon}\ A{\isadigit{2}}\ {\isasymsqsubseteq}\ B{\isadigit{2}}\ {\isasymrbrakk}\ {\isasymLongrightarrow}\ P{\isasymrbrakk}\ {\isasymLongrightarrow}\ P{\isachardoublequoteclose}%
\isadelimproof
\ %
\endisadelimproof
\isatagproof
\isacommand{by}\isamarkupfalse%
\ auto%
\endisatagproof
{\isafoldproof}%
\isadelimproof
\endisadelimproof
\isanewline
\isanewline
\isacommand{lemma}\isamarkupfalse%
\ lesseq{\isacharunderscore}dyn{\isacharunderscore}any{\isacharbrackleft}intro{\isacharbang}{\isacharbrackright}{\isacharcolon}\ {\isachardoublequoteopen}A\ {\isasymsqsubseteq}\ DynT{\isachardoublequoteclose}%
\isadelimproof
\ %
\endisadelimproof
\isatagproof
\isacommand{by}\isamarkupfalse%
\ auto%
\endisatagproof
{\isafoldproof}%
\isadelimproof
\endisadelimproof
\isanewline
\isanewline
\isacommand{lemma}\isamarkupfalse%
\ less{\isacharunderscore}eq{\isacharunderscore}fun{\isacharbrackleft}intro{\isacharbang}{\isacharbrackright}{\isacharcolon}\ {\isachardoublequoteopen}{\isasymlbrakk}\ A{\isadigit{1}}\ {\isasymsqsubseteq}\ B{\isadigit{1}}{\isacharsemicolon}\ A{\isadigit{2}}\ {\isasymsqsubseteq}\ B{\isadigit{2}}\ {\isasymrbrakk}\ {\isasymLongrightarrow}\ A{\isadigit{1}}\ {\isasymrightarrow}\ A{\isadigit{2}}\ {\isasymsqsubseteq}\ B{\isadigit{1}}\ {\isasymrightarrow}\ B{\isadigit{2}}{\isachardoublequoteclose}%
\isadelimproof
\ %
\endisadelimproof
\isatagproof
\isacommand{by}\isamarkupfalse%
\ simp%
\endisatagproof
{\isafoldproof}%
\isadelimproof
\endisadelimproof
\isanewline
\isacommand{lemma}\isamarkupfalse%
\ less{\isacharunderscore}eq{\isacharunderscore}pair{\isacharbrackleft}intro{\isacharbang}{\isacharbrackright}{\isacharcolon}\ {\isachardoublequoteopen}{\isasymlbrakk}\ A{\isadigit{1}}\ {\isasymsqsubseteq}\ B{\isadigit{1}}{\isacharsemicolon}\ A{\isadigit{2}}\ {\isasymsqsubseteq}\ B{\isadigit{2}}\ {\isasymrbrakk}\ {\isasymLongrightarrow}\ A{\isadigit{1}}\ {\isasymtimes}\ A{\isadigit{2}}\ {\isasymsqsubseteq}\ B{\isadigit{1}}\ {\isasymtimes}\ B{\isadigit{2}}{\isachardoublequoteclose}%
\isadelimproof
\ %
\endisadelimproof
\isatagproof
\isacommand{by}\isamarkupfalse%
\ simp%
\endisatagproof
{\isafoldproof}%
\isadelimproof
\endisadelimproof
\isanewline
\isanewline
\isacommand{lemma}\isamarkupfalse%
\ lesseq{\isacharunderscore}prec{\isacharunderscore}trans{\isacharbrackleft}rule{\isacharunderscore}format{\isacharcomma}trans{\isacharbrackright}{\isacharcolon}\isanewline
\ \ \isakeyword{fixes}\ B{\isacharcolon}{\isacharcolon}ty\isanewline
\ \ \isakeyword{shows}\ {\isachardoublequoteopen}{\isacharparenleft}{\isasymforall}\ A\ C{\isachardot}\ A\ {\isasymsqsubseteq}\ B\ {\isasymlongrightarrow}\ B\ {\isasymsqsubseteq}\ C\ {\isasymlongrightarrow}\ A\ {\isasymsqsubseteq}\ C{\isacharparenright}{\isachardoublequoteclose}\isanewline
\isadelimproof
\ \ %
\endisadelimproof
\isatagproof
\isacommand{apply}\isamarkupfalse%
\ {\isacharparenleft}induct\ B{\isacharparenright}\isanewline
\ \ \isacommand{apply}\isamarkupfalse%
\ clarify\ \isacommand{apply}\isamarkupfalse%
\ {\isacharparenleft}case{\isacharunderscore}tac\ C{\isacharparenright}\ \isacommand{apply}\isamarkupfalse%
\ force\ \isacommand{apply}\isamarkupfalse%
\ force\ \isacommand{apply}\isamarkupfalse%
\ force\isanewline
\ \ \ \ \isacommand{apply}\isamarkupfalse%
\ force\ \isacommand{apply}\isamarkupfalse%
\ force\ \isacommand{apply}\isamarkupfalse%
\ force\isanewline
\ \ \isacommand{apply}\isamarkupfalse%
\ clarify\ \isacommand{apply}\isamarkupfalse%
\ {\isacharparenleft}case{\isacharunderscore}tac\ C{\isacharparenright}\ \isacommand{apply}\isamarkupfalse%
\ force\ \isacommand{apply}\isamarkupfalse%
\ force\ \isacommand{apply}\isamarkupfalse%
\ force\isanewline
\ \ \ \ \isacommand{apply}\isamarkupfalse%
\ force\ \isacommand{apply}\isamarkupfalse%
\ force\ \isacommand{apply}\isamarkupfalse%
\ force\isanewline
\ \ \isacommand{defer}\isamarkupfalse%
\isanewline
\ \ \isacommand{defer}\isamarkupfalse%
\isanewline
\ \ \isacommand{apply}\isamarkupfalse%
\ clarify\ \isacommand{apply}\isamarkupfalse%
\ {\isacharparenleft}case{\isacharunderscore}tac\ C{\isacharparenright}\ \isacommand{apply}\isamarkupfalse%
\ force\ \isacommand{apply}\isamarkupfalse%
\ force\ \isacommand{apply}\isamarkupfalse%
\ force\isanewline
\ \ \ \ \isacommand{apply}\isamarkupfalse%
\ force\ \isacommand{apply}\isamarkupfalse%
\ clarify\ \isacommand{defer}\isamarkupfalse%
\ \isacommand{apply}\isamarkupfalse%
\ force\isanewline
\ \ \isacommand{apply}\isamarkupfalse%
\ clarify\ \isacommand{apply}\isamarkupfalse%
\ {\isacharparenleft}case{\isacharunderscore}tac\ C{\isacharparenright}\ \isacommand{apply}\isamarkupfalse%
\ force\ \isacommand{apply}\isamarkupfalse%
\ force\ \isacommand{apply}\isamarkupfalse%
\ force\isanewline
\ \ \ \ \isacommand{apply}\isamarkupfalse%
\ force\ \isacommand{apply}\isamarkupfalse%
\ force\ \isacommand{apply}\isamarkupfalse%
\ force\isanewline
\ \ \isacommand{prefer}\isamarkupfalse%
\ {\isadigit{3}}\isanewline
\ \ \isacommand{apply}\isamarkupfalse%
\ {\isacharparenleft}case{\isacharunderscore}tac\ A{\isacharparenright}\ \isacommand{apply}\isamarkupfalse%
\ simp\ \isacommand{apply}\isamarkupfalse%
\ simp\ \isacommand{apply}\isamarkupfalse%
\ simp\ \isacommand{apply}\isamarkupfalse%
\ simp\ \isanewline
\ \ \ \ \isacommand{apply}\isamarkupfalse%
\ clarify\ \isacommand{apply}\isamarkupfalse%
\ simp\ \isacommand{apply}\isamarkupfalse%
\ simp\isanewline
\ \ \isacommand{apply}\isamarkupfalse%
\ clarify\ \isacommand{apply}\isamarkupfalse%
\ {\isacharparenleft}case{\isacharunderscore}tac\ C{\isacharparenright}\ \isacommand{apply}\isamarkupfalse%
\ force\ \isacommand{apply}\isamarkupfalse%
\ force\isanewline
\ \ \ \ \isacommand{apply}\isamarkupfalse%
\ clarify\ \isacommand{apply}\isamarkupfalse%
\ {\isacharparenleft}case{\isacharunderscore}tac\ A{\isacharparenright}\ \isacommand{apply}\isamarkupfalse%
\ clarify\ \isacommand{apply}\isamarkupfalse%
\ force\isanewline
\ \ \ \ \isacommand{apply}\isamarkupfalse%
\ clarify\ \isacommand{defer}\isamarkupfalse%
\ \isacommand{apply}\isamarkupfalse%
\ clarify\ \isacommand{apply}\isamarkupfalse%
\ clarify\ \isacommand{apply}\isamarkupfalse%
\ clarify\isanewline
\ \ \ \ \isacommand{apply}\isamarkupfalse%
\ clarify\ \isacommand{apply}\isamarkupfalse%
\ clarify\ \isacommand{apply}\isamarkupfalse%
\ clarify\ \isanewline
\ \ \isacommand{apply}\isamarkupfalse%
\ clarify\ \isacommand{apply}\isamarkupfalse%
\ {\isacharparenleft}case{\isacharunderscore}tac\ C{\isacharparenright}\ \isacommand{apply}\isamarkupfalse%
\ clarify\ \isacommand{apply}\isamarkupfalse%
\ clarify\ \isacommand{apply}\isamarkupfalse%
\ clarify\isanewline
\ \ \ \ \isacommand{apply}\isamarkupfalse%
\ clarify\ \isacommand{apply}\isamarkupfalse%
\ {\isacharparenleft}case{\isacharunderscore}tac\ A{\isacharparenright}\ \isacommand{apply}\isamarkupfalse%
\ clarify\ \isacommand{apply}\isamarkupfalse%
\ clarify\isanewline
\ \ \ \ \isacommand{apply}\isamarkupfalse%
\ clarify\ \isacommand{apply}\isamarkupfalse%
\ clarify\ \isacommand{apply}\isamarkupfalse%
\ blast\ \isacommand{apply}\isamarkupfalse%
\ clarify\isanewline
\ \ \ \ \isacommand{apply}\isamarkupfalse%
\ clarify\ \isacommand{apply}\isamarkupfalse%
\ clarify\ \isacommand{apply}\isamarkupfalse%
\ clarify\isanewline
\ \ \isacommand{apply}\isamarkupfalse%
\ {\isacharparenleft}rule\ less{\isacharunderscore}eq{\isacharunderscore}pair{\isacharparenright}\ \isacommand{apply}\isamarkupfalse%
\ blast\ \isacommand{apply}\isamarkupfalse%
\ blast\isanewline
\ \ \isacommand{done}\isamarkupfalse%
\endisatagproof
{\isafoldproof}%
\isadelimproof
\isanewline
\endisadelimproof
\isanewline
\isacommand{lemma}\isamarkupfalse%
\ lesseq{\isacharunderscore}tyenv{\isacharunderscore}trans{\isacharcolon}\isanewline
\ \ \isakeyword{fixes}\ {\isasymSigma}{\isacharcolon}{\isacharcolon}ty{\isacharunderscore}env\ \isakeyword{and}\ {\isasymSigma}{\isacharprime}{\isacharcolon}{\isacharcolon}ty{\isacharunderscore}env\isanewline
\ \ \isakeyword{assumes}\ s{\isadigit{2}}{\isadigit{3}}{\isacharcolon}\ {\isachardoublequoteopen}{\isasymSigma}{\isacharprime}\ {\isasymsqsubseteq}\ {\isasymSigma}{\isacharprime}{\isacharprime}{\isachardoublequoteclose}\ \isakeyword{and}\ s{\isadigit{1}}{\isadigit{2}}{\isacharcolon}\ {\isachardoublequoteopen}{\isasymSigma}\ {\isasymsqsubseteq}\ {\isasymSigma}{\isacharprime}{\isachardoublequoteclose}\ \isanewline
\ \ \isakeyword{shows}\ {\isachardoublequoteopen}{\isasymSigma}\ {\isasymsqsubseteq}\ {\isasymSigma}{\isacharprime}{\isacharprime}{\isachardoublequoteclose}\isanewline
\isadelimproof
\ \ %
\endisadelimproof
\isatagproof
\isacommand{using}\isamarkupfalse%
\ s{\isadigit{1}}{\isadigit{2}}\ s{\isadigit{2}}{\isadigit{3}}\ \isacommand{apply}\isamarkupfalse%
\ {\isacharparenleft}simp\ add{\isacharcolon}\ lesseq{\isacharunderscore}tyenv{\isacharunderscore}def{\isacharparenright}\isanewline
\ \ \isacommand{apply}\isamarkupfalse%
\ clarify\isanewline
\ \ \isacommand{apply}\isamarkupfalse%
\ {\isacharparenleft}erule{\isacharunderscore}tac\ x{\isacharequal}a\ \isakeyword{in}\ allE{\isacharparenright}\isanewline
\ \ \isacommand{apply}\isamarkupfalse%
\ {\isacharparenleft}erule{\isacharunderscore}tac\ x{\isacharequal}a\ \isakeyword{in}\ allE{\isacharparenright}\isanewline
\ \ \isacommand{apply}\isamarkupfalse%
\ simp\isanewline
\ \ \isacommand{apply}\isamarkupfalse%
\ {\isacharparenleft}erule\ exE{\isacharparenright}\isanewline
\ \ \isacommand{apply}\isamarkupfalse%
\ {\isacharparenleft}erule\ conjE{\isacharparenright}\isanewline
\ \ \isacommand{apply}\isamarkupfalse%
\ {\isacharparenleft}erule{\isacharunderscore}tac\ x{\isacharequal}B\ \isakeyword{in}\ allE{\isacharparenright}\ \ \isanewline
\ \ \isacommand{using}\isamarkupfalse%
\ lesseq{\isacharunderscore}prec{\isacharunderscore}trans\ \isacommand{apply}\isamarkupfalse%
\ blast\ \isacommand{done}\isamarkupfalse%
\endisatagproof
{\isafoldproof}%
\isadelimproof
\isanewline
\endisadelimproof
\isanewline
\isacommand{lemma}\isamarkupfalse%
\ strengthen{\isacharunderscore}value{\isacharunderscore}env{\isacharcolon}\isanewline
\ \ {\isachardoublequoteopen}{\isacharparenleft}{\isasymSigma}\ {\isasymturnstile}v\ v\ {\isacharcolon}\ A\ {\isasymlongrightarrow}\ {\isacharparenleft}{\isasymforall}\ {\isasymSigma}{\isacharprime}{\isachardot}\ {\isasymSigma}{\isacharprime}\ {\isasymsqsubseteq}\ {\isasymSigma}\ {\isasymlongrightarrow}\ {\isasymSigma}{\isacharprime}\ {\isasymturnstile}v\ v\ {\isacharcolon}\ A{\isacharparenright}{\isacharparenright}\isanewline
\ \ \ {\isasymand}\ {\isacharparenleft}{\isasymGamma}{\isacharsemicolon}{\isasymSigma}\ {\isasymturnstile}\ {\isasymrho}\ {\isasymlongrightarrow}\ {\isacharparenleft}{\isasymforall}\ {\isasymSigma}{\isacharprime}{\isachardot}\ {\isasymSigma}{\isacharprime}\ {\isasymsqsubseteq}\ {\isasymSigma}\ {\isasymlongrightarrow}\ {\isasymGamma}{\isacharsemicolon}{\isasymSigma}{\isacharprime}\ {\isasymturnstile}\ {\isasymrho}{\isacharparenright}{\isacharparenright}{\isachardoublequoteclose}\isanewline
\isadelimproof
\ \ %
\endisadelimproof
\isatagproof
\isacommand{apply}\isamarkupfalse%
\ {\isacharparenleft}induct\ rule{\isacharcolon}\ wt{\isacharunderscore}val{\isacharunderscore}wt{\isacharunderscore}env{\isachardot}induct{\isacharparenright}\isanewline
\ \ \isacommand{apply}\isamarkupfalse%
\ force{\isacharplus}\ \isacommand{defer}\isamarkupfalse%
\ \isacommand{apply}\isamarkupfalse%
\ force{\isacharplus}\isanewline
\ \ \isacommand{apply}\isamarkupfalse%
\ {\isacharparenleft}simp\ only{\isacharcolon}\ lesseq{\isacharunderscore}tyenv{\isacharunderscore}def{\isacharparenright}\isanewline
\ \ \isacommand{apply}\isamarkupfalse%
\ clarify\ \isacommand{apply}\isamarkupfalse%
\ {\isacharparenleft}erule{\isacharunderscore}tac\ x{\isacharequal}a\ \isakeyword{in}\ allE{\isacharparenright}\ \isacommand{apply}\isamarkupfalse%
\ {\isacharparenleft}erule{\isacharunderscore}tac\ x{\isacharequal}A\ \isakeyword{in}\ allE{\isacharparenright}\isanewline
\ \ \isacommand{apply}\isamarkupfalse%
\ clarify\ \isacommand{apply}\isamarkupfalse%
\ {\isacharparenleft}rule\ wt{\isacharunderscore}ref{\isacharparenright}\ \isacommand{apply}\isamarkupfalse%
\ simp\isanewline
\ \ \isacommand{apply}\isamarkupfalse%
\ {\isacharparenleft}rule\ lesseq{\isacharunderscore}prec{\isacharunderscore}trans{\isacharparenright}\ \isacommand{apply}\isamarkupfalse%
\ blast\ \isacommand{apply}\isamarkupfalse%
\ blast\isanewline
\ \ \isacommand{done}\isamarkupfalse%
\endisatagproof
{\isafoldproof}%
\isadelimproof
\isanewline
\endisadelimproof
\isanewline
\isacommand{lemma}\isamarkupfalse%
\ strengthen{\isacharunderscore}casted{\isacharunderscore}value{\isacharcolon}\isanewline
\ \ \isakeyword{fixes}\ {\isasymSigma}{\isacharcolon}{\isacharcolon}ty{\isacharunderscore}env\isanewline
\ \ \isakeyword{assumes}\ wtcv{\isacharcolon}\ {\isachardoublequoteopen}{\isasymSigma}\ {\isasymturnstile}cv\ cv\ {\isacharcolon}\ A{\isachardoublequoteclose}\ \isakeyword{and}\ ss{\isacharcolon}\ {\isachardoublequoteopen}{\isasymSigma}{\isacharprime}\ {\isasymsqsubseteq}\ {\isasymSigma}{\isachardoublequoteclose}\isanewline
\ \ \isakeyword{shows}\ {\isachardoublequoteopen}{\isasymSigma}{\isacharprime}\ {\isasymturnstile}cv\ cv\ {\isacharcolon}\ A{\isachardoublequoteclose}\isanewline
\isadelimproof
\ \ %
\endisadelimproof
\isatagproof
\isacommand{using}\isamarkupfalse%
\ wtcv\ ss\ \isacommand{apply}\isamarkupfalse%
\ {\isacharparenleft}induct\ arbitrary{\isacharcolon}\ {\isasymSigma}{\isacharprime}\ rule{\isacharcolon}\ wt{\isacharunderscore}casted{\isacharunderscore}val{\isachardot}induct{\isacharparenright}\isanewline
\ \ \isacommand{using}\isamarkupfalse%
\ strengthen{\isacharunderscore}value{\isacharunderscore}env\ \isacommand{apply}\isamarkupfalse%
\ blast\isanewline
\ \ \isacommand{using}\isamarkupfalse%
\ strengthen{\isacharunderscore}value{\isacharunderscore}env\ \isacommand{apply}\isamarkupfalse%
\ blast\isanewline
\ \ \isacommand{done}\isamarkupfalse%
\endisatagproof
{\isafoldproof}%
\isadelimproof
\isanewline
\endisadelimproof
\isanewline
\isacommand{lemma}\isamarkupfalse%
\ strengthen{\isacharunderscore}stack{\isacharcolon}\isanewline
\ \ \isakeyword{fixes}\ {\isasymSigma}{\isacharcolon}{\isacharcolon}ty{\isacharunderscore}env\isanewline
\ \ \isakeyword{assumes}\ wtk{\isacharcolon}\ {\isachardoublequoteopen}{\isasymSigma}\ {\isasymturnstile}\ k\ {\isacharcolon}\ A\ {\isasymRightarrow}\ B{\isachardoublequoteclose}\ \isakeyword{and}\ ss{\isacharcolon}\ {\isachardoublequoteopen}{\isasymSigma}{\isacharprime}\ {\isasymsqsubseteq}\ {\isasymSigma}{\isachardoublequoteclose}\ \isakeyword{shows}\ {\isachardoublequoteopen}{\isasymSigma}{\isacharprime}\ {\isasymturnstile}\ k\ {\isacharcolon}\ A\ {\isasymRightarrow}\ B{\isachardoublequoteclose}\isanewline
\isadelimproof
\ \ %
\endisadelimproof
\isatagproof
\isacommand{using}\isamarkupfalse%
\ wtk\ ss\ \isacommand{apply}\isamarkupfalse%
\ {\isacharparenleft}induct\ arbitrary{\isacharcolon}\ {\isasymSigma}{\isacharprime}\ rule{\isacharcolon}\ wt{\isacharunderscore}stack{\isachardot}induct{\isacharparenright}\isanewline
\ \ \isacommand{apply}\isamarkupfalse%
\ blast\isanewline
\ \ \isacommand{using}\isamarkupfalse%
\ strengthen{\isacharunderscore}value{\isacharunderscore}env\ \isacommand{apply}\isamarkupfalse%
\ blast\ \isacommand{done}\isamarkupfalse%
\endisatagproof
{\isafoldproof}%
\isadelimproof
\isanewline
\endisadelimproof
\isanewline
\isacommand{lemma}\isamarkupfalse%
\ meet{\isacharunderscore}safe{\isacharcolon}\ {\isachardoublequoteopen}{\isacharparenleft}{\isacharparenleft}{\isasymexists}\ C{\isachardot}\ meet\ A\ B\ {\isacharequal}\ Result\ C{\isacharparenright}\ {\isasymor}\ meet\ A\ B\ {\isacharequal}\ CastError{\isacharparenright}{\isachardoublequoteclose}\isanewline
\isadelimproof
\ \ %
\endisadelimproof
\isatagproof
\isacommand{apply}\isamarkupfalse%
\ {\isacharparenleft}induct\ rule{\isacharcolon}\ meet{\isachardot}induct\ {\isacharparenright}\ \isacommand{by}\isamarkupfalse%
\ auto%
\endisatagproof
{\isafoldproof}%
\isadelimproof
\isanewline
\endisadelimproof
\isanewline
\isacommand{lemma}\isamarkupfalse%
\ meet{\isacharunderscore}is{\isacharunderscore}meet{\isacharunderscore}aux{\isacharcolon}\isanewline
\ \ {\isachardoublequoteopen}{\isacharparenleft}{\isasymexists}\ C{\isachardot}\ meet\ A\ B\ {\isacharequal}\ Result\ C\ {\isasymand}\ C\ {\isasymsqsubseteq}\ A\ {\isasymand}\ C\ {\isasymsqsubseteq}\ B{\isacharparenright}\ {\isasymor}\ meet\ A\ B\ {\isacharequal}\ CastError{\isachardoublequoteclose}\isanewline
\isadelimproof
\ \ %
\endisadelimproof
\isatagproof
\isacommand{apply}\isamarkupfalse%
\ {\isacharparenleft}induct\ rule{\isacharcolon}\ meet{\isachardot}induct{\isacharparenright}\ \isacommand{apply}\isamarkupfalse%
\ auto\ \isacommand{done}\isamarkupfalse%
\endisatagproof
{\isafoldproof}%
\isadelimproof
\isanewline
\endisadelimproof
\isanewline
\isacommand{lemma}\isamarkupfalse%
\ meet{\isacharunderscore}is{\isacharunderscore}meet{\isacharcolon}\isanewline
\ \ {\isachardoublequoteopen}meet\ A\ B\ {\isacharequal}\ Result\ C\ {\isasymLongrightarrow}\ C\ {\isasymsqsubseteq}\ A\ {\isasymand}\ C\ {\isasymsqsubseteq}\ B{\isachardoublequoteclose}\isanewline
\isadelimproof
\ \ %
\endisadelimproof
\isatagproof
\isacommand{using}\isamarkupfalse%
\ meet{\isacharunderscore}is{\isacharunderscore}meet{\isacharunderscore}aux{\isacharbrackleft}of\ A\ B{\isacharbrackright}\ \isacommand{apply}\isamarkupfalse%
\ auto\ \isacommand{done}\isamarkupfalse%
\endisatagproof
{\isafoldproof}%
\isadelimproof
\isanewline
\endisadelimproof
\isanewline
\isacommand{lemma}\isamarkupfalse%
\ dom{\isacharunderscore}heap{\isacharcolon}\ \isanewline
\ \ \isakeyword{assumes}\ as{\isacharcolon}\ {\isachardoublequoteopen}a\ {\isasymin}\ dom\ {\isasymSigma}\ {\isachardoublequoteclose}\ \isakeyword{and}\ wth{\isacharcolon}\ {\isachardoublequoteopen}wt{\isacharunderscore}heap\ {\isasymSigma}\ {\isasymmu}\ ads{\isachardoublequoteclose}\ \isakeyword{shows}\ {\isachardoublequoteopen}a\ {\isasymin}\ dom\ {\isasymmu}{\isachardoublequoteclose}\isanewline
\isadelimproof
\endisadelimproof
\isatagproof
\isacommand{proof}\isamarkupfalse%
\ {\isacharminus}\isanewline
\ \ \isacommand{from}\isamarkupfalse%
\ as\ \isacommand{obtain}\isamarkupfalse%
\ A\ \isakeyword{where}\ las{\isacharcolon}\ {\isachardoublequoteopen}lookup\ a\ {\isasymSigma}\ {\isacharequal}\ Result\ A{\isachardoublequoteclose}\isanewline
\ \ \ \ \isacommand{using}\isamarkupfalse%
\ dom{\isacharunderscore}lookup{\isacharbrackleft}of\ a\ {\isasymSigma}{\isacharbrackright}\ \isacommand{by}\isamarkupfalse%
\ blast\isanewline
\ \ \isacommand{from}\isamarkupfalse%
\ las\ wth\ \isacommand{obtain}\isamarkupfalse%
\ cv\ \isakeyword{where}\isanewline
\ \ \ \ lam{\isadigit{2}}{\isacharcolon}\ {\isachardoublequoteopen}lookup\ a\ {\isasymmu}\ {\isacharequal}\ Result\ {\isacharparenleft}cv{\isacharcomma}A{\isacharparenright}{\isachardoublequoteclose}\ \isanewline
\ \ \ \ \isacommand{using}\isamarkupfalse%
\ wt{\isacharunderscore}heap{\isacharunderscore}def{\isacharbrackleft}of\ {\isasymSigma}\ {\isasymmu}\ ads{\isacharbrackright}\ \isacommand{apply}\isamarkupfalse%
\ blast\ \isacommand{done}\isamarkupfalse%
\isanewline
\ \ \isacommand{from}\isamarkupfalse%
\ lam{\isadigit{2}}\ \isacommand{show}\isamarkupfalse%
\ {\isacharquery}thesis\ \isacommand{using}\isamarkupfalse%
\ lookup{\isacharunderscore}dom{\isacharbrackleft}of\ a\ {\isasymmu}{\isacharbrackright}\ \isacommand{apply}\isamarkupfalse%
\ simp\ \isacommand{done}\isamarkupfalse%
\isanewline
\isacommand{qed}\isamarkupfalse%
\endisatagproof
{\isafoldproof}%
\isadelimproof
\isanewline
\endisadelimproof
\isanewline
\isacommand{fun}\isamarkupfalse%
\ cval{\isacharunderscore}ads\ {\isacharcolon}{\isacharcolon}\ {\isachardoublequoteopen}casted{\isacharunderscore}val\ {\isasymRightarrow}\ nat\ {\isasymRightarrow}\ nat\ set\ {\isasymRightarrow}\ nat\ set{\isachardoublequoteclose}\ \isakeyword{where}\isanewline
\ \ cvads{\isadigit{1}}{\isacharcolon}\ {\isachardoublequoteopen}cval{\isacharunderscore}ads\ {\isacharparenleft}Val\ v{\isacharparenright}\ a\ ads\ {\isacharequal}\ ads\ {\isacharminus}\ {\isacharbraceleft}a{\isacharbraceright}{\isachardoublequoteclose}\ {\isacharbar}\isanewline
\ \ cvads{\isadigit{2}}{\isacharcolon}\ {\isachardoublequoteopen}cval{\isacharunderscore}ads\ {\isacharparenleft}VCast\ v\ A\ B{\isacharparenright}\ a\ ads\ {\isacharequal}\ ads\ {\isasymunion}\ {\isacharbraceleft}a{\isacharbraceright}{\isachardoublequoteclose}\isanewline
\isanewline
\isacommand{lemma}\isamarkupfalse%
\ update{\isacharunderscore}heap{\isacharunderscore}val{\isacharcolon}\isanewline
\ \ \isakeyword{fixes}\ A{\isacharcolon}{\isacharcolon}ty\ \isakeyword{and}\ B{\isacharcolon}{\isacharcolon}ty\isanewline
\ \ \isakeyword{assumes}\ wth{\isacharcolon}\ {\isachardoublequoteopen}wt{\isacharunderscore}heap\ {\isasymSigma}\ {\isasymmu}\ ads{\isachardoublequoteclose}\ \isakeyword{and}\ las{\isacharcolon}\ {\isachardoublequoteopen}lookup\ a\ {\isasymSigma}\ {\isacharequal}\ Result\ A{\isachardoublequoteclose}\ \isakeyword{and}\ ab{\isacharcolon}\ {\isachardoublequoteopen}B\ {\isasymsqsubseteq}\ A{\isachardoublequoteclose}\isanewline
\ \ \isakeyword{and}\ wtv{\isacharcolon}\ {\isachardoublequoteopen}{\isasymSigma}\ {\isasymturnstile}cv\ cv\ {\isacharcolon}\ B{\isachardoublequoteclose}\isanewline
\ \ \isakeyword{shows}\ {\isachardoublequoteopen}wt{\isacharunderscore}heap\ {\isacharparenleft}{\isacharparenleft}a{\isacharcomma}B{\isacharparenright}{\isacharhash}{\isasymSigma}{\isacharparenright}\ {\isacharparenleft}{\isacharparenleft}a{\isacharcomma}{\isacharparenleft}cv{\isacharcomma}B{\isacharparenright}{\isacharparenright}{\isacharhash}{\isasymmu}{\isacharparenright}\ {\isacharparenleft}cval{\isacharunderscore}ads\ cv\ a\ ads{\isacharparenright}{\isachardoublequoteclose}\isanewline
\isadelimproof
\ \ %
\endisadelimproof
\isatagproof
\isacommand{apply}\isamarkupfalse%
\ {\isacharparenleft}simp\ only{\isacharcolon}\ wt{\isacharunderscore}heap{\isacharunderscore}def{\isacharparenright}\ \isacommand{apply}\isamarkupfalse%
\ {\isacharparenleft}rule\ conjI{\isacharparenright}\isanewline
\ \ \isacommand{apply}\isamarkupfalse%
\ clarify\ \isacommand{defer}\isamarkupfalse%
\ \isacommand{apply}\isamarkupfalse%
\ {\isacharparenleft}rule\ conjI{\isacharparenright}\ \isacommand{apply}\isamarkupfalse%
\ clarify\ \isacommand{defer}\isamarkupfalse%
\isanewline
\ \ \isacommand{using}\isamarkupfalse%
\ wth\ \isacommand{apply}\isamarkupfalse%
\ {\isacharparenleft}cases\ cv{\isacharparenright}\ \isacommand{apply}\isamarkupfalse%
\ {\isacharparenleft}simp\ add{\isacharcolon}\ dom{\isacharunderscore}def\ wt{\isacharunderscore}heap{\isacharunderscore}def{\isacharparenright}\ \isacommand{apply}\isamarkupfalse%
\ blast\isanewline
\ \ \ \ \isacommand{apply}\isamarkupfalse%
\ {\isacharparenleft}simp\ add{\isacharcolon}\ dom{\isacharunderscore}def\ wt{\isacharunderscore}heap{\isacharunderscore}def{\isacharparenright}\ \isacommand{apply}\isamarkupfalse%
\ blast\isanewline
\isacommand{proof}\isamarkupfalse%
\ {\isacharminus}\isanewline
\ \ \isacommand{fix}\isamarkupfalse%
\ a{\isacharprime}\ A{\isacharprime}\ \isacommand{assume}\isamarkupfalse%
\ las{\isadigit{2}}{\isacharcolon}\ {\isachardoublequoteopen}lookup\ a{\isacharprime}\ {\isacharparenleft}{\isacharparenleft}a{\isacharcomma}B{\isacharparenright}{\isacharhash}{\isasymSigma}{\isacharparenright}\ {\isacharequal}\ Result\ A{\isacharprime}{\isachardoublequoteclose}\isanewline
\ \ \isacommand{let}\isamarkupfalse%
\ {\isacharquery}S{\isadigit{2}}\ {\isacharequal}\ {\isachardoublequoteopen}{\isacharparenleft}a{\isacharcomma}B{\isacharparenright}\ {\isacharhash}\ {\isasymSigma}{\isachardoublequoteclose}\isanewline
\ \ \isacommand{from}\isamarkupfalse%
\ ab\ las\ \isacommand{have}\isamarkupfalse%
\ ss{\isacharcolon}\ {\isachardoublequoteopen}{\isacharquery}S{\isadigit{2}}\ {\isasymsqsubseteq}\ {\isasymSigma}{\isachardoublequoteclose}\isanewline
\ \ \ \ \isacommand{using}\isamarkupfalse%
\ lookup{\isacharunderscore}dom{\isacharbrackleft}of\ a\ {\isasymSigma}\ A{\isacharbrackright}\isanewline
\ \ \ \ \isacommand{apply}\isamarkupfalse%
\ {\isacharparenleft}simp\ add{\isacharcolon}\ lesseq{\isacharunderscore}tyenv{\isacharunderscore}def\ dom{\isacharunderscore}def{\isacharparenright}\ \isacommand{apply}\isamarkupfalse%
\ auto\ \isacommand{done}\isamarkupfalse%
\isanewline
\ \ \isacommand{show}\isamarkupfalse%
\ {\isachardoublequoteopen}{\isasymexists}cv{\isacharprime}{\isachardot}\ lookup\ a{\isacharprime}\ {\isacharparenleft}{\isacharparenleft}a{\isacharcomma}\ cv{\isacharcomma}\ B{\isacharparenright}\ {\isacharhash}\ {\isasymmu}{\isacharparenright}\ {\isacharequal}\ Result\ {\isacharparenleft}cv{\isacharprime}{\isacharcomma}\ A{\isacharprime}{\isacharparenright}\ {\isasymand}\isanewline
\ \ \ \ \ \ {\isacharparenleft}a{\isacharcomma}\ B{\isacharparenright}\ {\isacharhash}\ {\isasymSigma}\ {\isasymturnstile}cv\ cv{\isacharprime}\ {\isacharcolon}\ A{\isacharprime}\ {\isasymand}\ {\isacharparenleft}a{\isacharprime}\ {\isasymnotin}\ cval{\isacharunderscore}ads\ cv\ a\ ads\ {\isasymlongrightarrow}\ {\isacharparenleft}{\isasymexists}v{\isachardot}\ cv{\isacharprime}\ {\isacharequal}\ Val\ v{\isacharparenright}{\isacharparenright}{\isachardoublequoteclose}\isanewline
\ \ \isacommand{proof}\isamarkupfalse%
\ {\isacharparenleft}cases\ {\isachardoublequoteopen}a{\isacharprime}\ {\isacharequal}\ a{\isachardoublequoteclose}{\isacharparenright}\isanewline
\ \ \ \ \isacommand{assume}\isamarkupfalse%
\ aa{\isacharcolon}\ {\isachardoublequoteopen}a{\isacharprime}\ {\isacharequal}\ a{\isachardoublequoteclose}\isanewline
\ \ \ \ \isacommand{from}\isamarkupfalse%
\ aa\ las{\isadigit{2}}\ \isacommand{have}\isamarkupfalse%
\ abc{\isacharcolon}\ {\isachardoublequoteopen}A{\isacharprime}\ {\isacharequal}\ B{\isachardoublequoteclose}\ \isacommand{apply}\isamarkupfalse%
\ simp\ \isacommand{done}\isamarkupfalse%
\isanewline
\ \ \ \ \isacommand{from}\isamarkupfalse%
\ wtv\ ss\ \isacommand{have}\isamarkupfalse%
\ wtv{\isadigit{2}}{\isacharcolon}\ {\isachardoublequoteopen}{\isacharquery}S{\isadigit{2}}\ {\isasymturnstile}cv\ cv\ {\isacharcolon}\ B{\isachardoublequoteclose}\ \isanewline
\ \ \ \ \ \ \isacommand{using}\isamarkupfalse%
\ strengthen{\isacharunderscore}casted{\isacharunderscore}value\ \isacommand{apply}\isamarkupfalse%
\ blast\ \isacommand{done}\isamarkupfalse%
\isanewline
\ \ \ \ \isacommand{from}\isamarkupfalse%
\ aa\ abc\ wtv{\isadigit{2}}\ \isacommand{show}\isamarkupfalse%
\ {\isacharquery}thesis\ \isacommand{apply}\isamarkupfalse%
\ auto\ \isanewline
\ \ \ \ \ \ \isacommand{apply}\isamarkupfalse%
\ {\isacharparenleft}case{\isacharunderscore}tac\ cv{\isacharparenright}\ \isacommand{apply}\isamarkupfalse%
\ auto\ \isacommand{done}\isamarkupfalse%
\ \isanewline
\ \ \isacommand{next}\isamarkupfalse%
\isanewline
\ \ \ \ \isacommand{assume}\isamarkupfalse%
\ aa{\isacharcolon}\ {\isachardoublequoteopen}a{\isacharprime}\ {\isasymnoteq}\ a{\isachardoublequoteclose}\isanewline
\ \ \ \ \isacommand{from}\isamarkupfalse%
\ las{\isadigit{2}}\ aa\ \isacommand{have}\isamarkupfalse%
\ las{\isadigit{3}}{\isacharcolon}\ {\isachardoublequoteopen}lookup\ a{\isacharprime}\ {\isasymSigma}\ {\isacharequal}\ Result\ A{\isacharprime}{\isachardoublequoteclose}\ \isacommand{by}\isamarkupfalse%
\ simp\isanewline
\ \ \ \ \isacommand{from}\isamarkupfalse%
\ las{\isadigit{3}}\ wth\ \isacommand{obtain}\isamarkupfalse%
\ cv{\isacharprime}\ v{\isacharprime}\ \isakeyword{where}\isanewline
\ \ \ \ \ \ lam{\isadigit{2}}{\isacharcolon}\ {\isachardoublequoteopen}lookup\ a{\isacharprime}\ {\isasymmu}\ {\isacharequal}\ Result\ {\isacharparenleft}cv{\isacharprime}{\isacharcomma}A{\isacharprime}{\isacharparenright}{\isachardoublequoteclose}\ \isakeyword{and}\ wtcv{\isadigit{3}}{\isacharcolon}\ {\isachardoublequoteopen}{\isasymSigma}\ {\isasymturnstile}cv\ cv{\isacharprime}\ {\isacharcolon}\ A{\isacharprime}{\isachardoublequoteclose}\ \isanewline
\ \ \ \ \ \ \isakeyword{and}\ cval{\isacharcolon}\ {\isachardoublequoteopen}a{\isacharprime}\ {\isasymnotin}\ ads\ {\isasymlongrightarrow}\ {\isacharparenleft}{\isasymexists}\ v{\isachardot}\ cv{\isacharprime}\ {\isacharequal}\ Val\ v{\isacharparenright}{\isachardoublequoteclose}\isanewline
\ \ \ \ \ \ \isacommand{using}\isamarkupfalse%
\ wt{\isacharunderscore}heap{\isacharunderscore}def{\isacharbrackleft}of\ {\isasymSigma}\ {\isasymmu}\ ads{\isacharbrackright}\ \isacommand{apply}\isamarkupfalse%
\ blast\ \isacommand{done}\isamarkupfalse%
\isanewline
\ \ \ \ \isacommand{from}\isamarkupfalse%
\ wtcv{\isadigit{3}}\ ss\ \isacommand{have}\isamarkupfalse%
\ wtcv{\isadigit{4}}{\isacharcolon}\ {\isachardoublequoteopen}{\isacharquery}S{\isadigit{2}}\ {\isasymturnstile}cv\ cv{\isacharprime}\ {\isacharcolon}\ A{\isacharprime}{\isachardoublequoteclose}\isanewline
\ \ \ \ \ \ \isacommand{using}\isamarkupfalse%
\ strengthen{\isacharunderscore}casted{\isacharunderscore}value\ \isacommand{by}\isamarkupfalse%
\ blast\isanewline
\ \ \ \ \isacommand{from}\isamarkupfalse%
\ aa\ wtcv{\isadigit{4}}\ lam{\isadigit{2}}\ cval\ \isacommand{show}\isamarkupfalse%
\ {\isacharquery}thesis\ \isanewline
\ \ \ \ \ \ \isacommand{apply}\isamarkupfalse%
\ {\isacharparenleft}case{\isacharunderscore}tac\ cv{\isacharparenright}\ \isacommand{apply}\isamarkupfalse%
\ auto\ \isacommand{done}\isamarkupfalse%
\isanewline
\ \ \isacommand{qed}\isamarkupfalse%
\isanewline
\isacommand{next}\isamarkupfalse%
\isanewline
\ \ \isacommand{fix}\isamarkupfalse%
\ a{\isacharprime}\ \isacommand{assume}\isamarkupfalse%
\ ad{\isacharcolon}\ {\isachardoublequoteopen}a{\isacharprime}\ {\isasymin}\ dom\ {\isacharparenleft}{\isacharparenleft}a{\isacharcomma}{\isacharparenleft}cv{\isacharcomma}B{\isacharparenright}{\isacharparenright}{\isacharhash}{\isasymmu}{\isacharparenright}{\isachardoublequoteclose}\isanewline
\ \ \isacommand{let}\isamarkupfalse%
\ {\isacharquery}M{\isadigit{2}}\ {\isacharequal}\ {\isachardoublequoteopen}{\isacharparenleft}{\isacharparenleft}a{\isacharcomma}{\isacharparenleft}cv{\isacharcomma}B{\isacharparenright}{\isacharparenright}{\isacharhash}{\isasymmu}{\isacharparenright}{\isachardoublequoteclose}\isanewline
\ \ \isacommand{show}\isamarkupfalse%
\ {\isachardoublequoteopen}a{\isacharprime}\ {\isacharless}\ length\ {\isacharquery}M{\isadigit{2}}{\isachardoublequoteclose}\isanewline
\ \ \isacommand{proof}\isamarkupfalse%
\ {\isacharparenleft}cases\ {\isachardoublequoteopen}a{\isacharprime}\ {\isacharequal}\ a{\isachardoublequoteclose}{\isacharparenright}\isanewline
\ \ \ \ \isacommand{assume}\isamarkupfalse%
\ aa{\isacharcolon}\ {\isachardoublequoteopen}a{\isacharprime}\ {\isacharequal}\ a{\isachardoublequoteclose}\isanewline
\ \ \ \ \isacommand{from}\isamarkupfalse%
\ las\ \isacommand{have}\isamarkupfalse%
\ adom{\isacharcolon}\ {\isachardoublequoteopen}a\ {\isasymin}\ dom\ {\isasymSigma}{\isachardoublequoteclose}\ \isacommand{using}\isamarkupfalse%
\ lookup{\isacharunderscore}dom{\isacharbrackleft}of\ a\ {\isasymSigma}{\isacharbrackright}\ \isacommand{by}\isamarkupfalse%
\ blast\isanewline
\ \ \ \ \isacommand{from}\isamarkupfalse%
\ wth\ adom\ \isacommand{have}\isamarkupfalse%
\ {\isachardoublequoteopen}a\ {\isasymin}\ dom\ {\isasymmu}{\isachardoublequoteclose}\ \isacommand{using}\isamarkupfalse%
\ dom{\isacharunderscore}heap\ \isacommand{by}\isamarkupfalse%
\ blast\isanewline
\ \ \ \ \isacommand{with}\isamarkupfalse%
\ aa\ wth\ \isacommand{show}\isamarkupfalse%
\ {\isacharquery}thesis\ \isanewline
\ \ \ \ \ \ \isacommand{apply}\isamarkupfalse%
\ {\isacharparenleft}simp\ add{\isacharcolon}\ wt{\isacharunderscore}heap{\isacharunderscore}def{\isacharparenright}\ \isacommand{apply}\isamarkupfalse%
\ auto\ \isacommand{done}\isamarkupfalse%
\isanewline
\ \ \isacommand{next}\isamarkupfalse%
\isanewline
\ \ \ \ \isacommand{assume}\isamarkupfalse%
\ aa{\isacharcolon}\ {\isachardoublequoteopen}a{\isacharprime}\ {\isasymnoteq}\ a{\isachardoublequoteclose}\isanewline
\ \ \ \ \isacommand{from}\isamarkupfalse%
\ this\ ad\ wth\ \isacommand{show}\isamarkupfalse%
\ {\isacharquery}thesis\ \isanewline
\ \ \ \ \ \ \isacommand{apply}\isamarkupfalse%
\ {\isacharparenleft}simp\ add{\isacharcolon}\ dom{\isacharunderscore}def\ wt{\isacharunderscore}heap{\isacharunderscore}def{\isacharparenright}\isanewline
\ \ \ \ \ \ \isacommand{apply}\isamarkupfalse%
\ auto\ \isacommand{done}\isamarkupfalse%
\isanewline
\ \ \isacommand{qed}\isamarkupfalse%
\isanewline
\isacommand{qed}\isamarkupfalse%
\endisatagproof
{\isafoldproof}%
\isadelimproof
\isanewline
\endisadelimproof
\isanewline
\isacommand{lemma}\isamarkupfalse%
\ cast{\isacharunderscore}safe{\isacharcolon}\isanewline
\ \ \isakeyword{fixes}\ v{\isacharcolon}{\isacharcolon}val\isanewline
\ \ \isakeyword{assumes}\ wtv{\isacharcolon}\ {\isachardoublequoteopen}{\isasymSigma}\ {\isasymturnstile}v\ v\ {\isacharcolon}\ A{\isachardoublequoteclose}\ \ \isakeyword{and}\ wth{\isacharcolon}\ {\isachardoublequoteopen}wt{\isacharunderscore}heap\ {\isasymSigma}\ {\isasymmu}\ {\isacharparenleft}set\ ads{\isadigit{1}}{\isacharparenright}{\isachardoublequoteclose}\isanewline
\ \ \isakeyword{shows}\ {\isachardoublequoteopen}{\isacharparenleft}{\isasymexists}\ v{\isacharprime}\ {\isasymSigma}{\isacharprime}\ {\isasymmu}{\isacharprime}\ ads{\isadigit{2}}{\isachardot}\ cast\ v\ A\ B\ {\isasymmu}\ ads{\isadigit{1}}\ {\isacharequal}\ Result\ {\isacharparenleft}v{\isacharprime}{\isacharcomma}{\isasymmu}{\isacharprime}{\isacharcomma}ads{\isadigit{2}}{\isacharparenright}\ {\isasymand}\ {\isasymSigma}{\isacharprime}\ {\isasymturnstile}v\ v{\isacharprime}\ {\isacharcolon}\ B\isanewline
\ \ \ \ \ \ \ \ \ \ \ \ {\isasymand}\ wt{\isacharunderscore}heap\ {\isasymSigma}{\isacharprime}\ {\isasymmu}{\isacharprime}\ {\isacharparenleft}set\ ads{\isadigit{2}}{\isacharparenright}\ {\isasymand}\ {\isasymSigma}{\isacharprime}\ {\isasymsqsubseteq}\ {\isasymSigma}{\isacharparenright}\isanewline
\ \ \ \ \ \ \ \ \ {\isasymor}\ {\isacharparenleft}cast\ v\ A\ B\ {\isasymmu}\ ads{\isadigit{1}}\ {\isacharequal}\ CastError{\isacharparenright}{\isachardoublequoteclose}\ \isanewline
\isadelimproof
\ \ %
\endisadelimproof
\isatagproof
\isacommand{using}\isamarkupfalse%
\ wtv\ wth\ \isacommand{apply}\isamarkupfalse%
\ {\isacharparenleft}induct\ arbitrary{\isacharcolon}\ {\isasymSigma}\ rule{\isacharcolon}\ cast{\isachardot}induct{\isacharparenright}\isanewline
\ \ \isacommand{apply}\isamarkupfalse%
\ force\isanewline
\ \ \isacommand{apply}\isamarkupfalse%
\ force\isanewline
\ \ \isacommand{apply}\isamarkupfalse%
\ force\isanewline
\ \ \isacommand{apply}\isamarkupfalse%
\ force\isanewline
\ \ \isacommand{defer}\isamarkupfalse%
\isanewline
\ \ \isacommand{defer}\isamarkupfalse%
\isanewline
\ \ \isacommand{apply}\isamarkupfalse%
\ simp\ \isacommand{apply}\isamarkupfalse%
\ {\isacharparenleft}case{\isacharunderscore}tac\ T{\isadigit{1}}{\isacharparenright}\ \isacommand{apply}\isamarkupfalse%
\ force\ \isacommand{apply}\isamarkupfalse%
\ force\ \isacommand{apply}\isamarkupfalse%
\ force\isanewline
\ \ \ \ \isacommand{apply}\isamarkupfalse%
\ force\ \isacommand{apply}\isamarkupfalse%
\ force\ \isacommand{apply}\isamarkupfalse%
\ force\isanewline
\ \ \isacommand{apply}\isamarkupfalse%
\ simp\ \isacommand{apply}\isamarkupfalse%
\ {\isacharparenleft}case{\isacharunderscore}tac\ T{\isadigit{1}}{\isacharparenright}\ \isacommand{apply}\isamarkupfalse%
\ force\ \isacommand{apply}\isamarkupfalse%
\ force\ \isacommand{apply}\isamarkupfalse%
\ force\isanewline
\ \ \ \ \isacommand{apply}\isamarkupfalse%
\ force\ \isacommand{apply}\isamarkupfalse%
\ force\ \isacommand{apply}\isamarkupfalse%
\ force\isanewline
\ \ \isacommand{apply}\isamarkupfalse%
\ simp\ \isacommand{apply}\isamarkupfalse%
\ {\isacharparenleft}case{\isacharunderscore}tac\ T{\isadigit{1}}{\isacharparenright}\ \isacommand{apply}\isamarkupfalse%
\ force\ \isacommand{apply}\isamarkupfalse%
\ force\ \isacommand{apply}\isamarkupfalse%
\ force\isanewline
\ \ \ \ \isacommand{apply}\isamarkupfalse%
\ force\ \isacommand{apply}\isamarkupfalse%
\ force\ \isacommand{apply}\isamarkupfalse%
\ force\isanewline
\ \ \isacommand{apply}\isamarkupfalse%
\ simp\ \isacommand{apply}\isamarkupfalse%
\ {\isacharparenleft}case{\isacharunderscore}tac\ T{\isadigit{1}}{\isacharparenright}\ \isacommand{apply}\isamarkupfalse%
\ force\ \isacommand{apply}\isamarkupfalse%
\ force\ \isacommand{apply}\isamarkupfalse%
\ force\isanewline
\ \ \ \ \isacommand{apply}\isamarkupfalse%
\ force\ \isacommand{apply}\isamarkupfalse%
\ force\ \isacommand{apply}\isamarkupfalse%
\ force\isanewline
\ \ \isacommand{apply}\isamarkupfalse%
\ simp\ \isacommand{apply}\isamarkupfalse%
\ {\isacharparenleft}case{\isacharunderscore}tac\ T{\isadigit{1}}{\isacharparenright}\ \isacommand{apply}\isamarkupfalse%
\ force\ \isacommand{apply}\isamarkupfalse%
\ force\ \isacommand{apply}\isamarkupfalse%
\ force\isanewline
\ \ \ \ \isacommand{apply}\isamarkupfalse%
\ force\ \isacommand{apply}\isamarkupfalse%
\ force\ \isacommand{apply}\isamarkupfalse%
\ force\isanewline
\ \ \isacommand{apply}\isamarkupfalse%
\ force\ \isacommand{apply}\isamarkupfalse%
\ force\ \isacommand{apply}\isamarkupfalse%
\ force\ \isacommand{apply}\isamarkupfalse%
\ force\ \isacommand{apply}\isamarkupfalse%
\ force\ \isacommand{apply}\isamarkupfalse%
\ force\isanewline
\ \ \isacommand{apply}\isamarkupfalse%
\ force\ \isacommand{apply}\isamarkupfalse%
\ force\ \isacommand{apply}\isamarkupfalse%
\ force\ \isacommand{apply}\isamarkupfalse%
\ force\ \isacommand{apply}\isamarkupfalse%
\ force\ \isacommand{apply}\isamarkupfalse%
\ force\isanewline
\ \ \isacommand{apply}\isamarkupfalse%
\ force\ \isacommand{apply}\isamarkupfalse%
\ force\ \isacommand{apply}\isamarkupfalse%
\ force\ \isacommand{apply}\isamarkupfalse%
\ force\ \isacommand{apply}\isamarkupfalse%
\ force\ \isacommand{apply}\isamarkupfalse%
\ force\isanewline
\ \ \isacommand{apply}\isamarkupfalse%
\ force\ \isacommand{apply}\isamarkupfalse%
\ force\ \isacommand{apply}\isamarkupfalse%
\ force\ \isacommand{apply}\isamarkupfalse%
\ force\ \isacommand{apply}\isamarkupfalse%
\ force\ \isacommand{apply}\isamarkupfalse%
\ force\isanewline
\ \ \isacommand{apply}\isamarkupfalse%
\ force\ \isacommand{apply}\isamarkupfalse%
\ force\ \isacommand{apply}\isamarkupfalse%
\ force\ \isacommand{apply}\isamarkupfalse%
\ force\ \isacommand{apply}\isamarkupfalse%
\ force\ \isacommand{apply}\isamarkupfalse%
\ force\isanewline
\ \ \isacommand{apply}\isamarkupfalse%
\ force\ \isacommand{apply}\isamarkupfalse%
\ force\ \isacommand{apply}\isamarkupfalse%
\ force\ \isacommand{apply}\isamarkupfalse%
\ force\ \isacommand{apply}\isamarkupfalse%
\ force\ \isacommand{apply}\isamarkupfalse%
\ force\isanewline
\ \ \isacommand{apply}\isamarkupfalse%
\ force\ \isacommand{apply}\isamarkupfalse%
\ force\ \isacommand{apply}\isamarkupfalse%
\ force\ \isacommand{apply}\isamarkupfalse%
\ force\ \isacommand{apply}\isamarkupfalse%
\ force\ \isacommand{apply}\isamarkupfalse%
\ force\isanewline
\ \ \isacommand{apply}\isamarkupfalse%
\ force\ \isacommand{apply}\isamarkupfalse%
\ force\ \isacommand{apply}\isamarkupfalse%
\ force\ \isacommand{apply}\isamarkupfalse%
\ force\ \isacommand{apply}\isamarkupfalse%
\ force\ \isacommand{apply}\isamarkupfalse%
\ force\isanewline
\ \ \isacommand{apply}\isamarkupfalse%
\ force\ \isacommand{apply}\isamarkupfalse%
\ force\ \isacommand{apply}\isamarkupfalse%
\ force\ \isacommand{apply}\isamarkupfalse%
\ force\ \isacommand{apply}\isamarkupfalse%
\ force\ \isacommand{apply}\isamarkupfalse%
\ force\isanewline
\ \ \isacommand{apply}\isamarkupfalse%
\ force\ \isacommand{apply}\isamarkupfalse%
\ force\ \isacommand{apply}\isamarkupfalse%
\ force\ \isacommand{apply}\isamarkupfalse%
\ force\ \isacommand{apply}\isamarkupfalse%
\ force\ \isacommand{apply}\isamarkupfalse%
\ force\isanewline
\ \ \isacommand{apply}\isamarkupfalse%
\ force\ \isacommand{apply}\isamarkupfalse%
\ force\ \isacommand{apply}\isamarkupfalse%
\ force\ \isacommand{apply}\isamarkupfalse%
\ force\ \isacommand{apply}\isamarkupfalse%
\ force\ \isacommand{apply}\isamarkupfalse%
\ force\isanewline
\ \ \isacommand{apply}\isamarkupfalse%
\ force\ \isacommand{apply}\isamarkupfalse%
\ force\ \isacommand{apply}\isamarkupfalse%
\ force\ \isacommand{apply}\isamarkupfalse%
\ force\ \isacommand{apply}\isamarkupfalse%
\ force\ \isacommand{apply}\isamarkupfalse%
\ force\isanewline
\ \ \isacommand{apply}\isamarkupfalse%
\ force\ \isacommand{apply}\isamarkupfalse%
\ force\ \isacommand{apply}\isamarkupfalse%
\ force\ \isacommand{apply}\isamarkupfalse%
\ force\ \isacommand{apply}\isamarkupfalse%
\ force\ \isacommand{apply}\isamarkupfalse%
\ force\isanewline
\ \ \isacommand{apply}\isamarkupfalse%
\ force\ \isacommand{apply}\isamarkupfalse%
\ force\ \isacommand{apply}\isamarkupfalse%
\ force\ \isacommand{apply}\isamarkupfalse%
\ force\ \isacommand{apply}\isamarkupfalse%
\ force\ \isacommand{apply}\isamarkupfalse%
\ force\isanewline
\ \ \isacommand{apply}\isamarkupfalse%
\ force\ \isacommand{apply}\isamarkupfalse%
\ force\ \isacommand{apply}\isamarkupfalse%
\ force\ \isacommand{apply}\isamarkupfalse%
\ force\ \isacommand{apply}\isamarkupfalse%
\ force\ \isacommand{apply}\isamarkupfalse%
\ force\isanewline
\ \ \isacommand{apply}\isamarkupfalse%
\ force\ \isacommand{apply}\isamarkupfalse%
\ force\ \isacommand{apply}\isamarkupfalse%
\ force\ \isacommand{apply}\isamarkupfalse%
\ force\ \isacommand{apply}\isamarkupfalse%
\ force\ \isacommand{apply}\isamarkupfalse%
\ force\isanewline
\ \ \isacommand{apply}\isamarkupfalse%
\ force\ \isacommand{apply}\isamarkupfalse%
\ force\ \isacommand{apply}\isamarkupfalse%
\ force\ \isacommand{apply}\isamarkupfalse%
\ force\ \isacommand{apply}\isamarkupfalse%
\ force\ \isacommand{apply}\isamarkupfalse%
\ force\isanewline
\ \ \isacommand{apply}\isamarkupfalse%
\ force\ \isacommand{apply}\isamarkupfalse%
\ force\ \isacommand{apply}\isamarkupfalse%
\ force\ \isacommand{apply}\isamarkupfalse%
\ force\ \isacommand{apply}\isamarkupfalse%
\ force\ \isacommand{apply}\isamarkupfalse%
\ force\isanewline
\ \ \isacommand{apply}\isamarkupfalse%
\ force\ \isacommand{apply}\isamarkupfalse%
\ force\ \isacommand{apply}\isamarkupfalse%
\ force\ \isacommand{apply}\isamarkupfalse%
\ force\ \isacommand{apply}\isamarkupfalse%
\ force\ \isacommand{apply}\isamarkupfalse%
\ force\isanewline
\ \ \isacommand{apply}\isamarkupfalse%
\ force\ \isacommand{apply}\isamarkupfalse%
\ force\ \isacommand{apply}\isamarkupfalse%
\ force\ \isacommand{apply}\isamarkupfalse%
\ force\ \isacommand{apply}\isamarkupfalse%
\ force\ \isacommand{apply}\isamarkupfalse%
\ force\isanewline
\ \ \isacommand{apply}\isamarkupfalse%
\ force\ \isacommand{apply}\isamarkupfalse%
\ force\ \isacommand{apply}\isamarkupfalse%
\ force\ \isacommand{apply}\isamarkupfalse%
\ force\ \isacommand{apply}\isamarkupfalse%
\ force\ \isacommand{apply}\isamarkupfalse%
\ force\isanewline
\ \ \isacommand{apply}\isamarkupfalse%
\ force\ \isacommand{apply}\isamarkupfalse%
\ force\ \isacommand{apply}\isamarkupfalse%
\ force\ \isanewline
\ \ \isacommand{defer}\isamarkupfalse%
\isanewline
\isacommand{proof}\isamarkupfalse%
\ {\isacharminus}\isanewline
\ \ \isacommand{fix}\isamarkupfalse%
\ a\ A\ B\ {\isasymmu}\ ads{\isadigit{1}}\ {\isasymSigma}\ \isanewline
\ \ \isacommand{assume}\isamarkupfalse%
\ wta{\isacharcolon}\ {\isachardoublequoteopen}{\isasymSigma}\ {\isasymturnstile}v\ VRef\ a\ {\isacharcolon}\ RefT\ A{\isachardoublequoteclose}\ \isakeyword{and}\ wth{\isacharcolon}\ {\isachardoublequoteopen}wt{\isacharunderscore}heap\ {\isasymSigma}\ {\isasymmu}\ {\isacharparenleft}set\ ads{\isadigit{1}}{\isacharparenright}{\isachardoublequoteclose}\isanewline
\ \ \isacommand{from}\isamarkupfalse%
\ wta\ \isacommand{obtain}\isamarkupfalse%
\ C\ \isakeyword{where}\ las{\isacharcolon}\ {\isachardoublequoteopen}lookup\ a\ {\isasymSigma}\ {\isacharequal}\ Result\ C{\isachardoublequoteclose}\ \isakeyword{and}\ aa{\isacharcolon}\ {\isachardoublequoteopen}C\ {\isasymsqsubseteq}\ A{\isachardoublequoteclose}\ \isacommand{by}\isamarkupfalse%
\ auto\isanewline
\ \ \isacommand{from}\isamarkupfalse%
\ wth\ las\ \isacommand{obtain}\isamarkupfalse%
\ cv\ \isakeyword{where}\ lam{\isacharcolon}\ {\isachardoublequoteopen}lookup\ a\ {\isasymmu}\ {\isacharequal}\ Result\ {\isacharparenleft}cv{\isacharcomma}C{\isacharparenright}{\isachardoublequoteclose}\isanewline
\ \ \ \ \isakeyword{and}\ wtcv{\isacharcolon}\ {\isachardoublequoteopen}{\isasymSigma}\ {\isasymturnstile}cv\ cv\ {\isacharcolon}\ C{\isachardoublequoteclose}\ \isacommand{using}\isamarkupfalse%
\ wt{\isacharunderscore}heap{\isacharunderscore}def\ \isacommand{apply}\isamarkupfalse%
\ force\ \isacommand{done}\isamarkupfalse%
\isanewline
\ \ \isacommand{from}\isamarkupfalse%
\ meet{\isacharunderscore}safe{\isacharbrackleft}of\ B\ C{\isacharbrackright}\ \isanewline
\ \ \isacommand{show}\isamarkupfalse%
\ {\isachardoublequoteopen}{\isacharparenleft}{\isasymexists}v{\isacharprime}\ {\isasymSigma}{\isacharprime}\ {\isasymmu}{\isacharprime}\ ads{\isadigit{2}}{\isachardot}\isanewline
\ \ \ \ \ \ \ \ \ \ \ \ \ \ cast\ {\isacharparenleft}VRef\ a{\isacharparenright}\ {\isacharparenleft}RefT\ A{\isacharparenright}\ {\isacharparenleft}RefT\ B{\isacharparenright}\ {\isasymmu}\ ads{\isadigit{1}}\ {\isacharequal}\ Result\ {\isacharparenleft}v{\isacharprime}{\isacharcomma}\ {\isasymmu}{\isacharprime}{\isacharcomma}\ ads{\isadigit{2}}{\isacharparenright}\ {\isasymand}\isanewline
\ \ \ \ \ \ \ \ \ \ \ \ \ \ {\isasymSigma}{\isacharprime}\ {\isasymturnstile}v\ v{\isacharprime}\ {\isacharcolon}\ RefT\ B\ {\isasymand}\isanewline
\ \ \ \ \ \ \ \ \ \ \ \ \ \ wt{\isacharunderscore}heap\ {\isasymSigma}{\isacharprime}\ {\isasymmu}{\isacharprime}\ {\isacharparenleft}set\ ads{\isadigit{2}}{\isacharparenright}\ {\isasymand}\ {\isasymSigma}{\isacharprime}\ {\isasymsqsubseteq}\ {\isasymSigma}{\isacharparenright}\isanewline
\ \ \ \ \ \ \ \ {\isasymor}\ cast\ {\isacharparenleft}VRef\ a{\isacharparenright}\ {\isacharparenleft}RefT\ A{\isacharparenright}\ {\isacharparenleft}RefT\ B{\isacharparenright}\ {\isasymmu}\ ads{\isadigit{1}}\ {\isacharequal}\ CastError{\isachardoublequoteclose}\isanewline
\ \ \isacommand{proof}\isamarkupfalse%
\isanewline
\ \ \ \ \isacommand{assume}\isamarkupfalse%
\ {\isachardoublequoteopen}{\isasymexists}\ BC{\isachardot}\ meet\ B\ C\ {\isacharequal}\ Result\ BC{\isachardoublequoteclose}\isanewline
\ \ \ \ \isacommand{from}\isamarkupfalse%
\ this\ \isacommand{obtain}\isamarkupfalse%
\ BC\ \isakeyword{where}\ bc{\isacharcolon}\ {\isachardoublequoteopen}meet\ B\ C\ {\isacharequal}\ Result\ BC{\isachardoublequoteclose}\ \isacommand{by}\isamarkupfalse%
\ blast\isanewline
\ \ \ \ \isacommand{from}\isamarkupfalse%
\ bc\ \isacommand{have}\isamarkupfalse%
\ leq{\isacharunderscore}bbc{\isacharcolon}\ {\isachardoublequoteopen}BC\ {\isasymsqsubseteq}\ B{\isachardoublequoteclose}\ \isacommand{using}\isamarkupfalse%
\ meet{\isacharunderscore}is{\isacharunderscore}meet\ \isacommand{by}\isamarkupfalse%
\ blast\isanewline
\ \ \ \ \isacommand{show}\isamarkupfalse%
\ {\isacharquery}thesis\isanewline
\ \ \ \ \isacommand{proof}\isamarkupfalse%
\ {\isacharparenleft}cases\ {\isachardoublequoteopen}C\ {\isasymsqsubseteq}\ BC{\isachardoublequoteclose}{\isacharparenright}\isanewline
\ \ \ \ \ \ \isacommand{assume}\isamarkupfalse%
\ bcc{\isacharcolon}\ {\isachardoublequoteopen}C\ {\isasymsqsubseteq}\ BC{\isachardoublequoteclose}\isanewline
\ \ \ \ \ \ \isacommand{from}\isamarkupfalse%
\ bc\ bcc\ lam\isanewline
\ \ \ \ \ \ \isacommand{have}\isamarkupfalse%
\ ca{\isacharcolon}\ {\isachardoublequoteopen}cast\ {\isacharparenleft}VRef\ a{\isacharparenright}\ {\isacharparenleft}RefT\ A{\isacharparenright}\ {\isacharparenleft}RefT\ B{\isacharparenright}\ {\isasymmu}\ ads{\isadigit{1}}\ {\isacharequal}\ Result\ {\isacharparenleft}VRef\ a{\isacharcomma}{\isasymmu}{\isacharcomma}ads{\isadigit{1}}{\isacharparenright}{\isachardoublequoteclose}\ \isacommand{by}\isamarkupfalse%
\ simp\isanewline
\ \ \ \ \ \ \isacommand{from}\isamarkupfalse%
\ leq{\isacharunderscore}bbc\ bcc\ \isacommand{have}\isamarkupfalse%
\ leq{\isacharunderscore}bc{\isacharcolon}\ {\isachardoublequoteopen}C\ {\isasymsqsubseteq}\ B{\isachardoublequoteclose}\ \isacommand{using}\isamarkupfalse%
\ lesseq{\isacharunderscore}prec{\isacharunderscore}trans\ \isacommand{by}\isamarkupfalse%
\ blast\isanewline
\ \ \ \ \ \ \isacommand{from}\isamarkupfalse%
\ las\ leq{\isacharunderscore}bc\ \isacommand{have}\isamarkupfalse%
\ wta{\isadigit{2}}{\isacharcolon}\ {\isachardoublequoteopen}{\isasymSigma}\ {\isasymturnstile}v\ VRef\ a\ {\isacharcolon}\ RefT\ B{\isachardoublequoteclose}\ \isacommand{by}\isamarkupfalse%
\ {\isacharparenleft}rule\ wt{\isacharunderscore}ref{\isacharparenright}\isanewline
\ \ \ \ \ \ \isacommand{from}\isamarkupfalse%
\ ca\ wta{\isadigit{2}}\ wth\ \isacommand{show}\isamarkupfalse%
\ {\isacharquery}thesis\ \isacommand{by}\isamarkupfalse%
\ auto\isanewline
\ \ \ \ \isacommand{next}\isamarkupfalse%
\isanewline
\ \ \ \ \ \ \isacommand{assume}\isamarkupfalse%
\ bcc{\isacharcolon}\ {\isachardoublequoteopen}{\isasymnot}\ {\isacharparenleft}C\ {\isasymsqsubseteq}\ BC{\isacharparenright}{\isachardoublequoteclose}\isanewline
\ \ \ \ \ \ \isacommand{let}\isamarkupfalse%
\ {\isacharquery}VC\ {\isacharequal}\ {\isachardoublequoteopen}mk{\isacharunderscore}vcast\ cv\ C\ BC{\isachardoublequoteclose}\isanewline
\ \ \ \ \ \ \isacommand{let}\isamarkupfalse%
\ {\isacharquery}M{\isadigit{2}}\ {\isacharequal}\ {\isachardoublequoteopen}{\isacharparenleft}a{\isacharcomma}\ {\isacharquery}VC{\isacharcomma}\ BC{\isacharparenright}{\isacharhash}{\isasymmu}{\isachardoublequoteclose}\isanewline
\ \ \ \ \ \ \isacommand{from}\isamarkupfalse%
\ bc\ bcc\ lam\isanewline
\ \ \ \ \ \ \isacommand{have}\isamarkupfalse%
\ ca{\isacharcolon}\ {\isachardoublequoteopen}cast\ {\isacharparenleft}VRef\ a{\isacharparenright}\ {\isacharparenleft}RefT\ A{\isacharparenright}\ {\isacharparenleft}RefT\ B{\isacharparenright}\ {\isasymmu}\ ads{\isadigit{1}}\ {\isacharequal}\ Result\ {\isacharparenleft}VRef\ a{\isacharcomma}{\isacharquery}M{\isadigit{2}}{\isacharcomma}a{\isacharhash}ads{\isadigit{1}}{\isacharparenright}{\isachardoublequoteclose}\ \isacommand{by}\isamarkupfalse%
\ simp\isanewline
\ \ \ \ \ \ \isacommand{let}\isamarkupfalse%
\ {\isacharquery}S{\isadigit{2}}\ {\isacharequal}\ {\isachardoublequoteopen}{\isacharparenleft}a{\isacharcomma}BC{\isacharparenright}{\isacharhash}{\isasymSigma}{\isachardoublequoteclose}\isanewline
\ \ \ \ \ \ \isacommand{from}\isamarkupfalse%
\ bc\ \isacommand{have}\isamarkupfalse%
\ cbc{\isacharcolon}\ {\isachardoublequoteopen}BC\ {\isasymsqsubseteq}\ C{\isachardoublequoteclose}\ \isacommand{using}\isamarkupfalse%
\ meet{\isacharunderscore}is{\isacharunderscore}meet\ \isacommand{by}\isamarkupfalse%
\ blast\isanewline
\ \ \ \ \ \ \isacommand{from}\isamarkupfalse%
\ las\ \isacommand{have}\isamarkupfalse%
\ adom{\isacharcolon}\ {\isachardoublequoteopen}a\ {\isasymin}\ dom\ {\isasymSigma}{\isachardoublequoteclose}\ \isacommand{using}\isamarkupfalse%
\ lookup{\isacharunderscore}dom{\isacharbrackleft}of\ a\ {\isasymSigma}{\isacharbrackright}\ \isacommand{by}\isamarkupfalse%
\ blast\isanewline
\ \ \ \ \ \ \isacommand{from}\isamarkupfalse%
\ las\ adom\ cbc\ \isacommand{have}\isamarkupfalse%
\ ss{\isadigit{2}}{\isacharcolon}\ {\isachardoublequoteopen}{\isacharquery}S{\isadigit{2}}\ {\isasymsqsubseteq}\ {\isasymSigma}{\isachardoublequoteclose}\ \isacommand{using}\isamarkupfalse%
\ lesseq{\isacharunderscore}tyenv{\isacharunderscore}def{\isacharbrackleft}of\ {\isacharquery}S{\isadigit{2}}\ {\isasymSigma}{\isacharbrackright}\isanewline
\ \ \ \ \ \ \ \ \isacommand{apply}\isamarkupfalse%
\ {\isacharparenleft}simp\ add{\isacharcolon}\ dom{\isacharunderscore}def{\isacharparenright}\ \isacommand{apply}\isamarkupfalse%
\ blast\ \isacommand{done}\isamarkupfalse%
\isanewline
\ \ \ \ \ \ \isacommand{from}\isamarkupfalse%
\ wtcv\ ss{\isadigit{2}}\ \isacommand{have}\isamarkupfalse%
\ wtcv{\isadigit{2}}{\isacharcolon}\ {\isachardoublequoteopen}{\isacharquery}S{\isadigit{2}}\ {\isasymturnstile}cv\ cv\ {\isacharcolon}\ C{\isachardoublequoteclose}\isanewline
\ \ \ \ \ \ \ \ \isacommand{using}\isamarkupfalse%
\ strengthen{\isacharunderscore}casted{\isacharunderscore}value\ \isacommand{apply}\isamarkupfalse%
\ blast\ \isacommand{done}\isamarkupfalse%
\isanewline
\ \ \ \ \ \ \isacommand{from}\isamarkupfalse%
\ wtcv\ cbc\ \isacommand{have}\isamarkupfalse%
\ wtcvbc{\isacharcolon}\ {\isachardoublequoteopen}{\isasymSigma}\ {\isasymturnstile}cv\ {\isacharquery}VC\ {\isacharcolon}\ BC{\isachardoublequoteclose}\ \isanewline
\ \ \ \ \ \ \ \ \isacommand{apply}\isamarkupfalse%
\ {\isacharparenleft}case{\isacharunderscore}tac\ cv{\isacharparenright}\isanewline
\ \ \ \ \ \ \ \ \isacommand{apply}\isamarkupfalse%
\ simp\ \isacommand{apply}\isamarkupfalse%
\ auto\isanewline
\ \ \ \ \ \ \ \ \isacommand{using}\isamarkupfalse%
\ lesseq{\isacharunderscore}prec{\isacharunderscore}trans\ \isacommand{apply}\isamarkupfalse%
\ blast\ \isacommand{done}\isamarkupfalse%
\ \ \ \ \ \ \ \ \isanewline
\ \ \ \ \ \ \isacommand{from}\isamarkupfalse%
\ wth\ las\ wtcvbc\ cbc\isanewline
\ \ \ \ \ \ \isacommand{have}\isamarkupfalse%
\ wth{\isadigit{2}}{\isacharcolon}\ {\isachardoublequoteopen}wt{\isacharunderscore}heap\ {\isacharquery}S{\isadigit{2}}\ {\isacharquery}M{\isadigit{2}}\ {\isacharparenleft}set\ {\isacharparenleft}a{\isacharhash}ads{\isadigit{1}}{\isacharparenright}{\isacharparenright}{\isachardoublequoteclose}\isanewline
\ \ \ \ \ \ \ \ \isacommand{using}\isamarkupfalse%
\ update{\isacharunderscore}heap{\isacharunderscore}val{\isacharbrackleft}of\ {\isasymSigma}\ {\isasymmu}\ {\isachardoublequoteopen}set\ ads{\isadigit{1}}{\isachardoublequoteclose}\ a\ C\ BC\ {\isacharquery}VC{\isacharbrackright}\ \isanewline
\ \ \ \ \ \ \ \ \isacommand{apply}\isamarkupfalse%
\ {\isacharparenleft}case{\isacharunderscore}tac\ cv{\isacharparenright}\ \isacommand{apply}\isamarkupfalse%
\ auto\ \isacommand{done}\isamarkupfalse%
\isanewline
\ \ \ \ \ \ \isacommand{have}\isamarkupfalse%
\ las{\isadigit{2}}{\isacharcolon}\ {\isachardoublequoteopen}lookup\ a\ {\isacharquery}S{\isadigit{2}}\ {\isacharequal}\ Result\ BC{\isachardoublequoteclose}\ \isacommand{by}\isamarkupfalse%
\ simp\isanewline
\ \ \ \ \ \ \isacommand{from}\isamarkupfalse%
\ las{\isadigit{2}}\ leq{\isacharunderscore}bbc\ \isacommand{have}\isamarkupfalse%
\ wta{\isadigit{2}}{\isacharcolon}\ {\isachardoublequoteopen}{\isacharquery}S{\isadigit{2}}\ {\isasymturnstile}v\ VRef\ a\ {\isacharcolon}\ RefT\ B{\isachardoublequoteclose}\ \isanewline
\ \ \ \ \ \ \ \ \isacommand{using}\isamarkupfalse%
\ wt{\isacharunderscore}ref{\isacharbrackleft}of\ a\ {\isacharquery}S{\isadigit{2}}\ BC\ B{\isacharbrackright}\ \isacommand{apply}\isamarkupfalse%
\ simp\ \isacommand{done}\isamarkupfalse%
\isanewline
\ \ \ \ \ \ \isacommand{from}\isamarkupfalse%
\ wta{\isadigit{2}}\ ca\ wth{\isadigit{2}}\ ss{\isadigit{2}}\ \isacommand{show}\isamarkupfalse%
\ {\isacharquery}thesis\ \isacommand{by}\isamarkupfalse%
\ {\isacharparenleft}auto\ simp{\isacharcolon}\ dom{\isacharunderscore}def{\isacharparenright}\isanewline
\ \ \ \ \isacommand{qed}\isamarkupfalse%
\isanewline
\ \ \isacommand{next}\isamarkupfalse%
\isanewline
\ \ \ \ \isacommand{assume}\isamarkupfalse%
\ {\isachardoublequoteopen}meet\ B\ C\ {\isacharequal}\ CastError{\isachardoublequoteclose}\isanewline
\ \ \ \ \isacommand{from}\isamarkupfalse%
\ this\ lam\ \isacommand{show}\isamarkupfalse%
\ {\isacharquery}thesis\ \isacommand{by}\isamarkupfalse%
\ simp\isanewline
\ \ \isacommand{qed}\isamarkupfalse%
\isanewline
\isacommand{next}\isamarkupfalse%
\isanewline
\ \ \isacommand{fix}\isamarkupfalse%
\ \ v{\isadigit{1}}\ v{\isadigit{2}}\ A\ B\ C\ D\ {\isasymmu}\ {\isasymSigma}\ ads{\isadigit{1}}\isanewline
\ \ \isacommand{let}\isamarkupfalse%
\ {\isacharquery}P\ {\isacharequal}\ {\isachardoublequoteopen}{\isasymlambda}\ v\ A\ B\ {\isasymmu}\ ads{\isadigit{1}}\ {\isasymSigma}{\isachardot}\isanewline
\ \ \ \ \ \ \ \ \ {\isacharparenleft}{\isasymexists}\ v{\isacharprime}\ {\isasymSigma}{\isacharprime}\ {\isasymmu}{\isacharprime}\ ads{\isadigit{2}}{\isachardot}\ cast\ v\ A\ B\ {\isasymmu}\ ads{\isadigit{1}}\ {\isacharequal}\ Result\ {\isacharparenleft}v{\isacharprime}{\isacharcomma}{\isasymmu}{\isacharprime}{\isacharcomma}ads{\isadigit{2}}{\isacharparenright}\ {\isasymand}\ {\isasymSigma}{\isacharprime}\ {\isasymturnstile}v\ v{\isacharprime}\ {\isacharcolon}\ B\isanewline
\ \ \ \ \ \ \ \ \ \ \ \ {\isasymand}\ wt{\isacharunderscore}heap\ {\isasymSigma}{\isacharprime}\ {\isasymmu}{\isacharprime}\ {\isacharparenleft}set\ ads{\isadigit{2}}{\isacharparenright}\ {\isasymand}\ {\isasymSigma}{\isacharprime}\ {\isasymsqsubseteq}\ {\isasymSigma}{\isacharparenright}{\isachardoublequoteclose}\isanewline
\ \ \isacommand{assume}\isamarkupfalse%
\ IH{\isadigit{1}}{\isacharcolon}\ {\isachardoublequoteopen}{\isasymAnd}{\isasymSigma}{\isachardot}\ {\isasymlbrakk}{\isasymSigma}\ {\isasymturnstile}v\ v{\isadigit{1}}\ {\isacharcolon}\ A{\isacharsemicolon}\ wt{\isacharunderscore}heap\ {\isasymSigma}\ {\isasymmu}\ {\isacharparenleft}set\ ads{\isadigit{1}}{\isacharparenright}{\isasymrbrakk}\ {\isasymLongrightarrow}\isanewline
\ \ \ \ \ \ \ \ {\isacharparenleft}{\isacharquery}P\ v{\isadigit{1}}\ A\ C\ {\isasymmu}\ ads{\isadigit{1}}\ {\isasymSigma}{\isacharparenright}\ {\isasymor}\ {\isacharparenleft}cast\ v{\isadigit{1}}\ A\ C\ {\isasymmu}\ ads{\isadigit{1}}\ {\isacharequal}\ CastError{\isacharparenright}{\isachardoublequoteclose}\isanewline
\ \ \ \ \isakeyword{and}\ IH{\isadigit{2}}{\isacharcolon}\ {\isachardoublequoteopen}{\isasymAnd}x\ xa\ xb\ {\isasymSigma}{\isachardot}\ {\isasymlbrakk}\ {\isasymSigma}\ {\isasymturnstile}v\ v{\isadigit{2}}\ {\isacharcolon}\ B{\isacharsemicolon}\ wt{\isacharunderscore}heap\ {\isasymSigma}\ xa\ {\isacharparenleft}set\ xb{\isacharparenright}{\isasymrbrakk}\ {\isasymLongrightarrow}\isanewline
\ \ \ \ \ \ \ \ {\isacharparenleft}{\isacharquery}P\ v{\isadigit{2}}\ B\ D\ xa\ xb\ {\isasymSigma}{\isacharparenright}\ {\isasymor}\ {\isacharparenleft}cast\ v{\isadigit{2}}\ B\ D\ xa\ xb\ {\isacharequal}\ CastError{\isacharparenright}{\isachardoublequoteclose}\isanewline
\ \ \ \ \isakeyword{and}\ wtp{\isacharcolon}\ {\isachardoublequoteopen}{\isasymSigma}\ {\isasymturnstile}v\ VPair\ v{\isadigit{1}}\ v{\isadigit{2}}\ {\isacharcolon}\ A\ {\isasymtimes}\ B{\isachardoublequoteclose}\ \isakeyword{and}\ wth{\isacharcolon}\ {\isachardoublequoteopen}wt{\isacharunderscore}heap\ {\isasymSigma}\ {\isasymmu}\ {\isacharparenleft}set\ ads{\isadigit{1}}{\isacharparenright}{\isachardoublequoteclose}\isanewline
\ \ \isacommand{from}\isamarkupfalse%
\ wtp\ \isacommand{have}\isamarkupfalse%
\ wtv{\isadigit{1}}{\isacharcolon}\ {\isachardoublequoteopen}{\isasymSigma}\ {\isasymturnstile}v\ v{\isadigit{1}}\ {\isacharcolon}\ A{\isachardoublequoteclose}\ \isacommand{by}\isamarkupfalse%
\ auto\isanewline
\ \ \isacommand{from}\isamarkupfalse%
\ wtp\ \isacommand{have}\isamarkupfalse%
\ wtv{\isadigit{2}}{\isacharcolon}\ {\isachardoublequoteopen}{\isasymSigma}\ {\isasymturnstile}v\ v{\isadigit{2}}\ {\isacharcolon}\ B{\isachardoublequoteclose}\ \isacommand{by}\isamarkupfalse%
\ auto\isanewline
\ \ \isacommand{from}\isamarkupfalse%
\ wtv{\isadigit{1}}\ wth\ IH{\isadigit{1}}\ \isacommand{have}\isamarkupfalse%
\ IH{\isadigit{1}}conc{\isacharcolon}\isanewline
\ \ \ \ {\isachardoublequoteopen}{\isacharquery}P\ v{\isadigit{1}}\ A\ C\ {\isasymmu}\ ads{\isadigit{1}}\ {\isasymSigma}\ \ {\isasymor}\ {\isacharparenleft}cast\ v{\isadigit{1}}\ A\ C\ {\isasymmu}\ ads{\isadigit{1}}\ {\isacharequal}\ CastError{\isacharparenright}{\isachardoublequoteclose}\ \isacommand{by}\isamarkupfalse%
\ blast\isanewline
\ \ \isacommand{from}\isamarkupfalse%
\ IH{\isadigit{1}}conc\isanewline
\ \ \isacommand{show}\isamarkupfalse%
\ {\isachardoublequoteopen}{\isacharparenleft}{\isacharquery}P\ {\isacharparenleft}VPair\ v{\isadigit{1}}\ v{\isadigit{2}}{\isacharparenright}\ {\isacharparenleft}A\ {\isasymtimes}\ B{\isacharparenright}\ {\isacharparenleft}C\ {\isasymtimes}\ D{\isacharparenright}\ {\isasymmu}\ ads{\isadigit{1}}\ {\isasymSigma}{\isacharparenright}\ \isanewline
\ \ \ \ \ \ \ \ {\isasymor}\ cast\ {\isacharparenleft}VPair\ v{\isadigit{1}}\ v{\isadigit{2}}{\isacharparenright}\ {\isacharparenleft}A\ {\isasymtimes}\ B{\isacharparenright}\ {\isacharparenleft}C\ {\isasymtimes}\ D{\isacharparenright}\ {\isasymmu}\ ads{\isadigit{1}}\ {\isacharequal}\ CastError{\isachardoublequoteclose}\isanewline
\ \ \isacommand{proof}\isamarkupfalse%
\isanewline
\ \ \ \ \isacommand{assume}\isamarkupfalse%
\ {\isachardoublequoteopen}{\isacharquery}P\ v{\isadigit{1}}\ A\ C\ {\isasymmu}\ ads{\isadigit{1}}\ {\isasymSigma}{\isachardoublequoteclose}\isanewline
\ \ \ \ \isacommand{from}\isamarkupfalse%
\ this\ \isacommand{obtain}\isamarkupfalse%
\ v{\isadigit{1}}{\isacharprime}\ {\isasymSigma}{\isadigit{1}}\ {\isasymmu}{\isadigit{1}}\ ads{\isadigit{2}}\ \isakeyword{where}\isanewline
\ \ \ \ \ \ cv{\isadigit{1}}{\isacharcolon}\ {\isachardoublequoteopen}cast\ v{\isadigit{1}}\ A\ C\ {\isasymmu}\ ads{\isadigit{1}}\ {\isacharequal}\ Result\ {\isacharparenleft}v{\isadigit{1}}{\isacharprime}{\isacharcomma}\ {\isasymmu}{\isadigit{1}}{\isacharcomma}\ ads{\isadigit{2}}{\isacharparenright}{\isachardoublequoteclose}\isanewline
\ \ \ \ \ \ \isakeyword{and}\ wtv{\isadigit{1}}{\isacharcolon}\ {\isachardoublequoteopen}{\isasymSigma}{\isadigit{1}}\ {\isasymturnstile}v\ v{\isadigit{1}}{\isacharprime}\ {\isacharcolon}\ C{\isachardoublequoteclose}\ \isakeyword{and}\ wth{\isadigit{1}}{\isacharcolon}\ {\isachardoublequoteopen}wt{\isacharunderscore}heap\ {\isasymSigma}{\isadigit{1}}\ {\isasymmu}{\isadigit{1}}\ {\isacharparenleft}set\ ads{\isadigit{2}}{\isacharparenright}{\isachardoublequoteclose}\ \isanewline
\ \ \ \ \ \ \isakeyword{and}\ s{\isadigit{1}}s{\isacharcolon}\ {\isachardoublequoteopen}{\isasymSigma}{\isadigit{1}}\ {\isasymsqsubseteq}\ {\isasymSigma}{\isachardoublequoteclose}\ \isacommand{by}\isamarkupfalse%
\ blast\isanewline
\ \ \ \ \isacommand{from}\isamarkupfalse%
\ wtv{\isadigit{2}}\ s{\isadigit{1}}s\ \isacommand{have}\isamarkupfalse%
\ wtv{\isadigit{2}}p{\isacharcolon}\ {\isachardoublequoteopen}{\isasymSigma}{\isadigit{1}}\ {\isasymturnstile}v\ v{\isadigit{2}}\ {\isacharcolon}\ B{\isachardoublequoteclose}\ \isacommand{using}\isamarkupfalse%
\ strengthen{\isacharunderscore}value{\isacharunderscore}env\ \isacommand{apply}\isamarkupfalse%
\ blast\ \isacommand{done}\isamarkupfalse%
\isanewline
\ \ \ \ \isacommand{from}\isamarkupfalse%
\ wtv{\isadigit{2}}p\ wth{\isadigit{1}}\ IH{\isadigit{2}}\ \isacommand{have}\isamarkupfalse%
\ IH{\isadigit{2}}conc{\isacharcolon}\isanewline
\ \ \ \ \ \ {\isachardoublequoteopen}{\isacharquery}P\ v{\isadigit{2}}\ B\ D\ {\isasymmu}{\isadigit{1}}\ ads{\isadigit{2}}\ {\isasymSigma}{\isadigit{1}}\ {\isasymor}\ {\isacharparenleft}cast\ v{\isadigit{2}}\ B\ D\ {\isasymmu}{\isadigit{1}}\ ads{\isadigit{2}}\ {\isacharequal}\ CastError{\isacharparenright}{\isachardoublequoteclose}\ \isacommand{apply}\isamarkupfalse%
\ blast\ \isacommand{done}\isamarkupfalse%
\isanewline
\ \ \ \ \isacommand{from}\isamarkupfalse%
\ IH{\isadigit{2}}conc\isanewline
\ \ \ \ \isacommand{show}\isamarkupfalse%
\ {\isacharquery}thesis\isanewline
\ \ \ \ \isacommand{proof}\isamarkupfalse%
\isanewline
\ \ \ \ \ \ \isacommand{assume}\isamarkupfalse%
\ {\isachardoublequoteopen}{\isacharquery}P\ v{\isadigit{2}}\ B\ D\ {\isasymmu}{\isadigit{1}}\ ads{\isadigit{2}}\ {\isasymSigma}{\isadigit{1}}{\isachardoublequoteclose}\isanewline
\ \ \ \ \ \ \isacommand{from}\isamarkupfalse%
\ this\ \isacommand{obtain}\isamarkupfalse%
\ v{\isadigit{2}}{\isacharprime}\ {\isasymSigma}{\isadigit{2}}\ {\isasymmu}{\isadigit{2}}\ ads{\isadigit{3}}\ \isakeyword{where}\isanewline
\ \ \ \ \ \ \ \ cv{\isadigit{2}}{\isacharcolon}\ {\isachardoublequoteopen}cast\ v{\isadigit{2}}\ B\ D\ {\isasymmu}{\isadigit{1}}\ ads{\isadigit{2}}\ {\isacharequal}\ Result\ {\isacharparenleft}v{\isadigit{2}}{\isacharprime}{\isacharcomma}\ {\isasymmu}{\isadigit{2}}{\isacharcomma}\ ads{\isadigit{3}}{\isacharparenright}{\isachardoublequoteclose}\isanewline
\ \ \ \ \ \ \ \ \isakeyword{and}\ wtv{\isadigit{2}}{\isacharcolon}\ {\isachardoublequoteopen}{\isasymSigma}{\isadigit{2}}\ {\isasymturnstile}v\ v{\isadigit{2}}{\isacharprime}\ {\isacharcolon}\ D{\isachardoublequoteclose}\ \isakeyword{and}\ wth{\isadigit{2}}{\isacharcolon}\ {\isachardoublequoteopen}wt{\isacharunderscore}heap\ {\isasymSigma}{\isadigit{2}}\ {\isasymmu}{\isadigit{2}}\ {\isacharparenleft}set\ ads{\isadigit{3}}{\isacharparenright}{\isachardoublequoteclose}\isanewline
\ \ \ \ \ \ \ \ \isakeyword{and}\ s{\isadigit{2}}s{\isadigit{1}}{\isacharcolon}\ {\isachardoublequoteopen}{\isasymSigma}{\isadigit{2}}\ {\isasymsqsubseteq}\ {\isasymSigma}{\isadigit{1}}{\isachardoublequoteclose}\ \isacommand{apply}\isamarkupfalse%
\ fast\ \isacommand{done}\isamarkupfalse%
\isanewline
\ \ \ \ \ \ \isacommand{let}\isamarkupfalse%
\ {\isacharquery}V\ {\isacharequal}\ {\isachardoublequoteopen}VPair\ v{\isadigit{1}}{\isacharprime}\ v{\isadigit{2}}{\isacharprime}{\isachardoublequoteclose}\isanewline
\ \ \ \ \ \ \isacommand{from}\isamarkupfalse%
\ cv{\isadigit{1}}\ cv{\isadigit{2}}\isanewline
\ \ \ \ \ \ \isacommand{have}\isamarkupfalse%
\ cvp{\isacharcolon}\ {\isachardoublequoteopen}cast\ {\isacharparenleft}VPair\ v{\isadigit{1}}\ v{\isadigit{2}}{\isacharparenright}\ {\isacharparenleft}A\ {\isasymtimes}\ B{\isacharparenright}\ {\isacharparenleft}C\ {\isasymtimes}\ D{\isacharparenright}\ {\isasymmu}\ ads{\isadigit{1}}\ {\isacharequal}\ Result\ {\isacharparenleft}{\isacharquery}V{\isacharcomma}\ {\isasymmu}{\isadigit{2}}{\isacharcomma}\ ads{\isadigit{3}}{\isacharparenright}{\isachardoublequoteclose}\ \isacommand{by}\isamarkupfalse%
\ simp\isanewline
\ \ \ \ \ \ \isacommand{from}\isamarkupfalse%
\ wtv{\isadigit{1}}\ s{\isadigit{2}}s{\isadigit{1}}\ \isacommand{have}\isamarkupfalse%
\ wtv{\isadigit{1}}p{\isacharcolon}\ {\isachardoublequoteopen}{\isasymSigma}{\isadigit{2}}\ {\isasymturnstile}v\ v{\isadigit{1}}{\isacharprime}\ {\isacharcolon}\ C{\isachardoublequoteclose}\ \isacommand{using}\isamarkupfalse%
\ strengthen{\isacharunderscore}value{\isacharunderscore}env\ \isacommand{by}\isamarkupfalse%
\ blast\isanewline
\ \ \ \ \ \ \isacommand{from}\isamarkupfalse%
\ wtv{\isadigit{1}}p\ wtv{\isadigit{2}}\ \isacommand{have}\isamarkupfalse%
\ wtp{\isacharcolon}\ {\isachardoublequoteopen}{\isasymSigma}{\isadigit{2}}\ {\isasymturnstile}v\ {\isacharquery}V\ {\isacharcolon}\ {\isacharparenleft}C\ {\isasymtimes}\ D{\isacharparenright}{\isachardoublequoteclose}\ \isacommand{by}\isamarkupfalse%
\ blast\isanewline
\ \ \ \ \ \ \isacommand{from}\isamarkupfalse%
\ s{\isadigit{1}}s\ s{\isadigit{2}}s{\isadigit{1}}\ \isacommand{have}\isamarkupfalse%
\ s{\isadigit{2}}s{\isacharcolon}\ {\isachardoublequoteopen}{\isasymSigma}{\isadigit{2}}\ {\isasymsqsubseteq}\ {\isasymSigma}{\isachardoublequoteclose}\ \isacommand{using}\isamarkupfalse%
\ lesseq{\isacharunderscore}tyenv{\isacharunderscore}trans\ \isacommand{by}\isamarkupfalse%
\ blast\isanewline
\ \ \ \ \ \ \isacommand{from}\isamarkupfalse%
\ wtp\ wth{\isadigit{2}}\ s{\isadigit{2}}s\ cvp\ \isacommand{show}\isamarkupfalse%
\ {\isacharquery}thesis\ \isacommand{by}\isamarkupfalse%
\ blast\isanewline
\ \ \ \ \isacommand{next}\isamarkupfalse%
\isanewline
\ \ \ \ \ \ \isacommand{assume}\isamarkupfalse%
\ {\isachardoublequoteopen}cast\ v{\isadigit{2}}\ B\ D\ {\isasymmu}{\isadigit{1}}\ ads{\isadigit{2}}\ {\isacharequal}\ CastError{\isachardoublequoteclose}\isanewline
\ \ \ \ \ \ \isacommand{with}\isamarkupfalse%
\ cv{\isadigit{1}}\ \isacommand{show}\isamarkupfalse%
\ {\isacharquery}thesis\ \isacommand{by}\isamarkupfalse%
\ simp\isanewline
\ \ \ \ \isacommand{qed}\isamarkupfalse%
\isanewline
\ \ \isacommand{next}\isamarkupfalse%
\isanewline
\ \ \ \ \isacommand{assume}\isamarkupfalse%
\ {\isachardoublequoteopen}cast\ v{\isadigit{1}}\ A\ C\ {\isasymmu}\ ads{\isadigit{1}}\ {\isacharequal}\ CastError{\isachardoublequoteclose}\isanewline
\ \ \ \ \isacommand{from}\isamarkupfalse%
\ this\ \isacommand{show}\isamarkupfalse%
\ {\isacharquery}thesis\ \isacommand{apply}\isamarkupfalse%
\ auto\ \isacommand{done}\isamarkupfalse%
\isanewline
\ \ \isacommand{qed}\isamarkupfalse%
\isanewline
\isacommand{qed}\isamarkupfalse%
\endisatagproof
{\isafoldproof}%
\isadelimproof
\isanewline
\endisadelimproof
\isanewline
\isanewline
\isacommand{lemma}\isamarkupfalse%
\ step{\isacharunderscore}safe{\isacharcolon}\ \isanewline
\ \ \isakeyword{assumes}\ wtsA{\isacharcolon}\ {\isachardoublequoteopen}wt{\isacharunderscore}state\ s\ A{\isachardoublequoteclose}\isanewline
\ \ \isakeyword{shows}\ {\isachardoublequoteopen}final\ s\ {\isasymor}\ {\isacharparenleft}{\isasymexists}\ s{\isacharprime}{\isachardot}\ step\ s\ {\isacharequal}\ Result\ s{\isacharprime}\ {\isasymand}\ wt{\isacharunderscore}state\ s{\isacharprime}\ A{\isacharparenright}\isanewline
\ \ \ \ {\isasymor}\ step\ s\ {\isacharequal}\ CastError{\isachardoublequoteclose}\isanewline
\isadelimproof
\ \ %
\endisadelimproof
\isatagproof
\isacommand{using}\isamarkupfalse%
\ wtsA\ \isanewline
\isacommand{proof}\isamarkupfalse%
\ {\isacharparenleft}rule\ wts{\isacharparenright}\isanewline
\ \ \isacommand{fix}\isamarkupfalse%
\ {\isasymSigma}\ {\isasymmu}\ ads\ {\isasymGamma}\ {\isasymrho}\ st\ A{\isacharprime}\ k\isanewline
\ \ \isacommand{assume}\isamarkupfalse%
\ st{\isacharcolon}\ {\isachardoublequoteopen}s\ {\isacharequal}\ {\isacharparenleft}st{\isacharcomma}\ {\isasymrho}{\isacharcomma}\ k{\isacharcomma}\ {\isasymmu}{\isacharcomma}\ ads{\isacharparenright}{\isachardoublequoteclose}\ \isakeyword{and}\ gr{\isacharcolon}\ {\isachardoublequoteopen}{\isasymGamma}{\isacharsemicolon}{\isasymSigma}\ {\isasymturnstile}\ {\isasymrho}{\isachardoublequoteclose}\ \isakeyword{and}\ wts{\isacharcolon}\ {\isachardoublequoteopen}{\isasymGamma}\ {\isasymturnstile}\isactrlisub s\ st\ {\isacharcolon}\ A{\isacharprime}{\isachardoublequoteclose}\isanewline
\ \ \ \ \isakeyword{and}\ wt{\isacharunderscore}k{\isacharcolon}\ {\isachardoublequoteopen}{\isasymSigma}\ {\isasymturnstile}\ k\ {\isacharcolon}\ A{\isacharprime}\ {\isasymRightarrow}\ A{\isachardoublequoteclose}\ \isakeyword{and}\ wt{\isacharunderscore}h{\isacharcolon}\ {\isachardoublequoteopen}wt{\isacharunderscore}heap\ {\isasymSigma}\ {\isasymmu}\ {\isacharparenleft}set\ ads{\isacharparenright}{\isachardoublequoteclose}\isanewline
\ \ \ \ \isanewline
\ \ \isacommand{show}\isamarkupfalse%
\ {\isacharquery}thesis\isanewline
\ \ \isacommand{proof}\isamarkupfalse%
\ {\isacharparenleft}cases\ ads{\isacharparenright}\isanewline
\ \ \ \ \isacommand{case}\isamarkupfalse%
\ {\isacharparenleft}Cons\ a\ ads{\isacharprime}{\isacharparenright}\isanewline
\ \ \ \ \isacommand{from}\isamarkupfalse%
\ wt{\isacharunderscore}h\ Cons\ \isacommand{have}\isamarkupfalse%
\ adoms{\isacharcolon}\ {\isachardoublequoteopen}a\ {\isasymin}\ dom\ {\isasymSigma}{\isachardoublequoteclose}\ \isacommand{apply}\isamarkupfalse%
\ {\isacharparenleft}simp\ add{\isacharcolon}\ wt{\isacharunderscore}heap{\isacharunderscore}def{\isacharparenright}\ \isacommand{done}\isamarkupfalse%
\isanewline
\ \ \ \ \isacommand{from}\isamarkupfalse%
\ adoms\ \isacommand{obtain}\isamarkupfalse%
\ B\ \isakeyword{where}\ las{\isacharcolon}\ {\isachardoublequoteopen}lookup\ a\ {\isasymSigma}\ {\isacharequal}\ Result\ B{\isachardoublequoteclose}\isanewline
\ \ \ \ \ \ \isacommand{using}\isamarkupfalse%
\ dom{\isacharunderscore}lookup{\isacharbrackleft}of\ a\ {\isasymSigma}{\isacharbrackright}\ \isacommand{by}\isamarkupfalse%
\ auto\isanewline
\ \ \ \ \isacommand{from}\isamarkupfalse%
\ las\ wt{\isacharunderscore}h\ \isacommand{obtain}\isamarkupfalse%
\ cv\ \isakeyword{where}\isanewline
\ \ \ \ \ \ lam{\isacharcolon}\ {\isachardoublequoteopen}lookup\ a\ {\isasymmu}\ {\isacharequal}\ Result\ {\isacharparenleft}cv{\isacharcomma}B{\isacharparenright}{\isachardoublequoteclose}\ \isakeyword{and}\ wtcv{\isacharcolon}\ {\isachardoublequoteopen}{\isasymSigma}\ {\isasymturnstile}cv\ cv\ {\isacharcolon}\ B{\isachardoublequoteclose}\ \isanewline
\ \ \ \ \ \ \isakeyword{and}\ cval{\isacharcolon}\ {\isachardoublequoteopen}a\ {\isasymnotin}\ {\isacharparenleft}set\ ads{\isacharparenright}\ {\isasymlongrightarrow}\ {\isacharparenleft}{\isasymexists}\ v{\isachardot}\ cv\ {\isacharequal}\ Val\ v{\isacharparenright}{\isachardoublequoteclose}\ \isanewline
\ \ \ \ \ \ \isacommand{using}\isamarkupfalse%
\ wt{\isacharunderscore}heap{\isacharunderscore}def{\isacharbrackleft}of\ {\isasymSigma}\ {\isasymmu}\ {\isachardoublequoteopen}set\ ads{\isachardoublequoteclose}{\isacharbrackright}\ \isacommand{apply}\isamarkupfalse%
\ blast\ \isacommand{done}\isamarkupfalse%
\isanewline
\ \ \ \ \isacommand{show}\isamarkupfalse%
\ {\isacharquery}thesis\isanewline
\ \ \ \ \isacommand{proof}\isamarkupfalse%
\ {\isacharparenleft}cases\ cv{\isacharparenright}\isanewline
\ \ \ \ \ \ \isacommand{case}\isamarkupfalse%
\ {\isacharparenleft}Val\ v{\isacharparenright}\isanewline
\ \ \ \ \ \ \isacommand{have}\isamarkupfalse%
\ wth{\isadigit{2}}{\isacharcolon}\ {\isachardoublequoteopen}wt{\isacharunderscore}heap\ {\isasymSigma}\ {\isasymmu}\ {\isacharparenleft}set\ ads{\isacharprime}{\isacharparenright}{\isachardoublequoteclose}\isanewline
\ \ \ \ \ \ \ \ \isacommand{apply}\isamarkupfalse%
\ {\isacharparenleft}simp\ only{\isacharcolon}\ wt{\isacharunderscore}heap{\isacharunderscore}def{\isacharparenright}\isanewline
\ \ \ \ \ \ \ \ \isacommand{apply}\isamarkupfalse%
\ {\isacharparenleft}rule\ conjI{\isacharparenright}\isanewline
\ \ \ \ \ \ \ \ \isacommand{apply}\isamarkupfalse%
\ clarify\ \isacommand{defer}\isamarkupfalse%
\ \isanewline
\ \ \ \ \ \ \ \ \isacommand{apply}\isamarkupfalse%
\ {\isacharparenleft}rule\ conjI{\isacharparenright}\isanewline
\ \ \ \ \ \ \ \ \isacommand{apply}\isamarkupfalse%
\ clarify\ \isacommand{defer}\isamarkupfalse%
\isanewline
\ \ \ \ \ \ \ \ \isacommand{using}\isamarkupfalse%
\ wt{\isacharunderscore}h\ Cons\ \isacommand{apply}\isamarkupfalse%
\ {\isacharparenleft}simp\ add{\isacharcolon}\ wt{\isacharunderscore}heap{\isacharunderscore}def{\isacharparenright}\ \isanewline
\ \ \ \ \ \ \isacommand{proof}\isamarkupfalse%
\ {\isacharminus}\isanewline
\ \ \ \ \ \ \ \ \isacommand{fix}\isamarkupfalse%
\ a{\isacharprime}\ A{\isacharprime}\ \isacommand{assume}\isamarkupfalse%
\ las{\isadigit{2}}{\isacharcolon}\ {\isachardoublequoteopen}lookup\ a{\isacharprime}\ {\isasymSigma}\ {\isacharequal}\ Result\ A{\isacharprime}{\isachardoublequoteclose}\isanewline
\ \ \ \ \ \ \ \ \isacommand{from}\isamarkupfalse%
\ las{\isadigit{2}}\ wt{\isacharunderscore}h\ \isacommand{obtain}\isamarkupfalse%
\ cv{\isacharprime}\ \isakeyword{where}\isanewline
\ \ \ \ \ \ \ \ \ \ lam{\isadigit{2}}{\isacharcolon}\ {\isachardoublequoteopen}lookup\ a{\isacharprime}\ {\isasymmu}\ {\isacharequal}\ Result\ {\isacharparenleft}cv{\isacharprime}{\isacharcomma}A{\isacharprime}{\isacharparenright}{\isachardoublequoteclose}\ \isakeyword{and}\ wtcv{\isadigit{2}}{\isacharcolon}\ {\isachardoublequoteopen}{\isasymSigma}\ {\isasymturnstile}cv\ cv{\isacharprime}\ {\isacharcolon}\ A{\isacharprime}{\isachardoublequoteclose}\ \isanewline
\ \ \ \ \ \ \ \ \ \ \isakeyword{and}\ cval{\isadigit{2}}{\isacharcolon}\ {\isachardoublequoteopen}a{\isacharprime}\ {\isasymnotin}\ {\isacharparenleft}set\ ads{\isacharparenright}\ {\isasymlongrightarrow}\ {\isacharparenleft}{\isasymexists}\ v{\isachardot}\ cv{\isacharprime}\ {\isacharequal}\ Val\ v{\isacharparenright}{\isachardoublequoteclose}\ \isanewline
\ \ \ \ \ \ \ \ \ \ \isacommand{using}\isamarkupfalse%
\ wt{\isacharunderscore}heap{\isacharunderscore}def{\isacharbrackleft}of\ {\isasymSigma}\ {\isasymmu}\ {\isachardoublequoteopen}set\ ads{\isachardoublequoteclose}{\isacharbrackright}\ \isacommand{apply}\isamarkupfalse%
\ blast\ \isacommand{done}\isamarkupfalse%
\isanewline
\ \ \ \ \ \ \ \ \isacommand{have}\isamarkupfalse%
\ ap{\isacharunderscore}ads{\isacharcolon}\ {\isachardoublequoteopen}{\isacharparenleft}a{\isacharprime}\ {\isasymnotin}\ set\ ads{\isacharprime}\ {\isasymlongrightarrow}\ {\isacharparenleft}{\isasymexists}\ v{\isachardot}\ cv{\isacharprime}\ {\isacharequal}\ Val\ v{\isacharparenright}{\isacharparenright}{\isachardoublequoteclose}\ \isanewline
\ \ \ \ \ \ \ \ \isacommand{proof}\isamarkupfalse%
\isanewline
\ \ \ \ \ \ \ \ \ \ \isacommand{assume}\isamarkupfalse%
\ ap{\isacharunderscore}adsp{\isacharcolon}\ {\isachardoublequoteopen}a{\isacharprime}\ {\isasymnotin}\ set\ ads{\isacharprime}{\isachardoublequoteclose}\isanewline
\ \ \ \ \ \ \ \ \ \ \isacommand{show}\isamarkupfalse%
\ {\isachardoublequoteopen}{\isasymexists}\ v{\isachardot}\ cv{\isacharprime}\ {\isacharequal}\ Val\ v{\isachardoublequoteclose}\isanewline
\ \ \ \ \ \ \ \ \ \ \isacommand{proof}\isamarkupfalse%
\ {\isacharparenleft}cases\ {\isachardoublequoteopen}a{\isacharprime}\ {\isacharequal}\ a{\isachardoublequoteclose}{\isacharparenright}\isanewline
\ \ \ \ \ \ \ \ \ \ \ \ \isacommand{assume}\isamarkupfalse%
\ aa{\isacharcolon}\ {\isachardoublequoteopen}a{\isacharprime}\ {\isacharequal}\ a{\isachardoublequoteclose}\isanewline
\ \ \ \ \ \ \ \ \ \ \ \ \isacommand{from}\isamarkupfalse%
\ lam{\isadigit{2}}\ Val\ lam\ aa\ \isacommand{show}\isamarkupfalse%
\ {\isacharquery}thesis\ \isacommand{apply}\isamarkupfalse%
\ auto\ \isacommand{done}\isamarkupfalse%
\isanewline
\ \ \ \ \ \ \ \ \ \ \isacommand{next}\isamarkupfalse%
\isanewline
\ \ \ \ \ \ \ \ \ \ \ \ \isacommand{assume}\isamarkupfalse%
\ aa{\isacharcolon}\ {\isachardoublequoteopen}a{\isacharprime}\ {\isasymnoteq}\ a{\isachardoublequoteclose}\isanewline
\ \ \ \ \ \ \ \ \ \ \ \ \isacommand{from}\isamarkupfalse%
\ Cons\ ap{\isacharunderscore}adsp\ aa\ \isacommand{have}\isamarkupfalse%
\ {\isachardoublequoteopen}a{\isacharprime}\ {\isasymnotin}\ set\ ads{\isachardoublequoteclose}\ \isacommand{apply}\isamarkupfalse%
\ auto\ \isacommand{done}\isamarkupfalse%
\isanewline
\ \ \ \ \ \ \ \ \ \ \ \ \isacommand{with}\isamarkupfalse%
\ cval{\isadigit{2}}\ \isacommand{show}\isamarkupfalse%
\ {\isacharquery}thesis\ \isacommand{by}\isamarkupfalse%
\ simp\isanewline
\ \ \ \ \ \ \ \ \ \ \isacommand{qed}\isamarkupfalse%
\isanewline
\ \ \ \ \ \ \ \ \isacommand{qed}\isamarkupfalse%
\isanewline
\ \ \ \ \ \ \ \ \isacommand{from}\isamarkupfalse%
\ lam{\isadigit{2}}\ wtcv{\isadigit{2}}\ ap{\isacharunderscore}ads\isanewline
\ \ \ \ \ \ \ \ \isacommand{show}\isamarkupfalse%
\ {\isachardoublequoteopen}{\isasymexists}\ cv{\isachardot}\ lookup\ a{\isacharprime}\ {\isasymmu}\ {\isacharequal}\ Result\ {\isacharparenleft}cv{\isacharcomma}A{\isacharprime}{\isacharparenright}\ {\isasymand}\ {\isasymSigma}\ {\isasymturnstile}cv\ cv\ {\isacharcolon}\ A{\isacharprime}\isanewline
\ \ \ \ \ \ \ \ \ \ {\isasymand}\ {\isacharparenleft}a{\isacharprime}\ {\isasymnotin}\ set\ ads{\isacharprime}\ {\isasymlongrightarrow}\ {\isacharparenleft}{\isasymexists}\ v{\isachardot}\ cv\ {\isacharequal}\ Val\ v{\isacharparenright}{\isacharparenright}{\isachardoublequoteclose}\ \isacommand{by}\isamarkupfalse%
\ auto\isanewline
\ \ \ \ \ \ \isacommand{next}\isamarkupfalse%
\isanewline
\ \ \ \ \ \ \ \ \isacommand{fix}\isamarkupfalse%
\ a\ \isacommand{assume}\isamarkupfalse%
\ {\isachardoublequoteopen}a\ {\isasymin}\ dom\ {\isasymmu}{\isachardoublequoteclose}\isanewline
\ \ \ \ \ \ \ \ \isacommand{with}\isamarkupfalse%
\ wt{\isacharunderscore}h\ \isacommand{show}\isamarkupfalse%
\ {\isachardoublequoteopen}a\ {\isacharless}\ length\ {\isasymmu}{\isachardoublequoteclose}\ \isacommand{by}\isamarkupfalse%
\ {\isacharparenleft}simp\ add{\isacharcolon}\ wt{\isacharunderscore}heap{\isacharunderscore}def{\isacharparenright}\ \isanewline
\ \ \ \ \ \ \isacommand{qed}\isamarkupfalse%
\isanewline
\ \ \ \ \ \ \isacommand{from}\isamarkupfalse%
\ lam\ st\ Cons\ Val\ wth{\isadigit{2}}\ wt{\isacharunderscore}k\ gr\ wts\ \isacommand{show}\isamarkupfalse%
\ {\isacharquery}thesis\ \isacommand{by}\isamarkupfalse%
\ auto\isanewline
\ \ \ \ \isacommand{next}\isamarkupfalse%
\isanewline
\ \ \ \ \ \ \isacommand{case}\isamarkupfalse%
\ {\isacharparenleft}VCast\ v\ C\ B{\isacharprime}{\isacharparenright}\isanewline
\ \ \ \ \ \ \isacommand{from}\isamarkupfalse%
\ wtcv\ VCast\ \isacommand{have}\isamarkupfalse%
\ wtv{\isacharcolon}\ {\isachardoublequoteopen}{\isasymSigma}\ {\isasymturnstile}v\ v\ {\isacharcolon}\ C{\isachardoublequoteclose}\ \isacommand{apply}\isamarkupfalse%
\ blast\ \isacommand{done}\isamarkupfalse%
\isanewline
\ \ \ \ \ \ \isacommand{from}\isamarkupfalse%
\ wtcv\ VCast\ \isacommand{have}\isamarkupfalse%
\ bb{\isacharcolon}\ {\isachardoublequoteopen}B{\isacharprime}\ {\isacharequal}\ B{\isachardoublequoteclose}\ \isacommand{by}\isamarkupfalse%
\ blast\isanewline
\ \ \ \ \ \ \isacommand{from}\isamarkupfalse%
\ wtcv\ VCast\ \isacommand{have}\isamarkupfalse%
\ cb{\isacharcolon}\ {\isachardoublequoteopen}B\ {\isasymsqsubseteq}\ C{\isachardoublequoteclose}\ \isacommand{by}\isamarkupfalse%
\ blast\isanewline
\ \ \ \ \ \ \isacommand{from}\isamarkupfalse%
\ wtv\ wt{\isacharunderscore}h\ Cons\ \isacommand{have}\isamarkupfalse%
\ {\isachardoublequoteopen}{\isacharparenleft}{\isasymexists}v{\isacharprime}\ {\isasymSigma}{\isacharprime}\ {\isasymmu}{\isacharprime}\ ads{\isadigit{2}}{\isachardot}\isanewline
\ \ \ \ \ \ \ \ \ cast\ v\ C\ B{\isacharprime}\ {\isasymmu}\ {\isacharparenleft}a{\isacharhash}ads{\isacharprime}{\isacharparenright}\ {\isacharequal}\ Result\ {\isacharparenleft}v{\isacharprime}{\isacharcomma}\ {\isasymmu}{\isacharprime}{\isacharcomma}\ ads{\isadigit{2}}{\isacharparenright}\ {\isasymand}\isanewline
\ \ \ \ \ \ \ \ \ {\isasymSigma}{\isacharprime}\ {\isasymturnstile}v\ v{\isacharprime}\ {\isacharcolon}\ B{\isacharprime}\ {\isasymand}\ wt{\isacharunderscore}heap\ {\isasymSigma}{\isacharprime}\ {\isasymmu}{\isacharprime}\ {\isacharparenleft}set\ ads{\isadigit{2}}{\isacharparenright}\ {\isasymand}\ {\isasymSigma}{\isacharprime}\ {\isasymsqsubseteq}\ {\isasymSigma}{\isacharparenright}\ {\isasymor}\isanewline
\ \ \ \ \ \ \ \ \ cast\ v\ C\ B{\isacharprime}\ {\isasymmu}\ {\isacharparenleft}a{\isacharhash}ads{\isacharprime}{\isacharparenright}\ {\isacharequal}\ CastError{\isachardoublequoteclose}\isanewline
\ \ \ \ \ \ \ \ \isacommand{using}\isamarkupfalse%
\ cast{\isacharunderscore}safe{\isacharbrackleft}of\ {\isasymSigma}\ v\ C\ {\isasymmu}\ {\isachardoublequoteopen}a{\isacharhash}ads{\isacharprime}{\isachardoublequoteclose}\ B{\isacharprime}{\isacharbrackright}\ \isacommand{by}\isamarkupfalse%
\ simp\isanewline
\ \ \ \ \ \ \isacommand{thus}\isamarkupfalse%
\ {\isacharquery}thesis\isanewline
\ \ \ \ \ \ \isacommand{proof}\isamarkupfalse%
\isanewline
\ \ \ \ \ \ \ \ \isacommand{assume}\isamarkupfalse%
\ {\isachardoublequoteopen}{\isacharparenleft}{\isasymexists}v{\isacharprime}\ {\isasymSigma}{\isacharprime}\ {\isasymmu}{\isacharprime}\ ads{\isadigit{2}}{\isachardot}\isanewline
\ \ \ \ \ \ \ \ \ cast\ v\ C\ B{\isacharprime}\ {\isasymmu}\ {\isacharparenleft}a{\isacharhash}ads{\isacharprime}{\isacharparenright}\ {\isacharequal}\ Result\ {\isacharparenleft}v{\isacharprime}{\isacharcomma}\ {\isasymmu}{\isacharprime}{\isacharcomma}\ ads{\isadigit{2}}{\isacharparenright}\ {\isasymand}\isanewline
\ \ \ \ \ \ \ \ \ {\isasymSigma}{\isacharprime}\ {\isasymturnstile}v\ v{\isacharprime}\ {\isacharcolon}\ B{\isacharprime}\ {\isasymand}\ wt{\isacharunderscore}heap\ {\isasymSigma}{\isacharprime}\ {\isasymmu}{\isacharprime}\ {\isacharparenleft}set\ ads{\isadigit{2}}{\isacharparenright}\ {\isasymand}\ {\isasymSigma}{\isacharprime}\ {\isasymsqsubseteq}\ {\isasymSigma}{\isacharparenright}{\isachardoublequoteclose}\isanewline
\ \ \ \ \ \ \ \ \isacommand{from}\isamarkupfalse%
\ this\ \isacommand{obtain}\isamarkupfalse%
\ v{\isacharprime}\ {\isasymSigma}{\isacharprime}\ {\isasymmu}{\isacharprime}\ ads{\isadigit{2}}\ \isakeyword{where}\isanewline
\ \ \ \ \ \ \ \ \ \ castv{\isacharcolon}\ {\isachardoublequoteopen}cast\ v\ C\ B{\isacharprime}\ {\isasymmu}\ {\isacharparenleft}a{\isacharhash}ads{\isacharprime}{\isacharparenright}\ {\isacharequal}\ Result\ {\isacharparenleft}v{\isacharprime}{\isacharcomma}{\isasymmu}{\isacharprime}{\isacharcomma}ads{\isadigit{2}}{\isacharparenright}{\isachardoublequoteclose}\ \isakeyword{and}\isanewline
\ \ \ \ \ \ \ \ \ \ wtvp{\isacharcolon}\ {\isachardoublequoteopen}{\isasymSigma}{\isacharprime}\ {\isasymturnstile}v\ v{\isacharprime}\ {\isacharcolon}\ B{\isacharprime}{\isachardoublequoteclose}\ \isakeyword{and}\ wth{\isadigit{2}}{\isacharcolon}\ {\isachardoublequoteopen}wt{\isacharunderscore}heap\ {\isasymSigma}{\isacharprime}\ {\isasymmu}{\isacharprime}\ {\isacharparenleft}set\ ads{\isadigit{2}}{\isacharparenright}{\isachardoublequoteclose}\ \isakeyword{and}\isanewline
\ \ \ \ \ \ \ \ \ \ ss{\isacharcolon}\ {\isachardoublequoteopen}{\isasymSigma}{\isacharprime}\ {\isasymsqsubseteq}\ {\isasymSigma}{\isachardoublequoteclose}\ \isacommand{apply}\isamarkupfalse%
\ blast\ \isacommand{done}\isamarkupfalse%
\isanewline
\ \ \ \ \ \ \ \ \isacommand{from}\isamarkupfalse%
\ las\ ss\ \isacommand{obtain}\isamarkupfalse%
\ B{\isadigit{2}}\ \isakeyword{where}\ las{\isadigit{2}}{\isacharcolon}\ {\isachardoublequoteopen}lookup\ a\ {\isasymSigma}{\isacharprime}\ {\isacharequal}\ Result\ B{\isadigit{2}}{\isachardoublequoteclose}\isanewline
\ \ \ \ \ \ \ \ \ \ \isakeyword{and}\ bb{\isadigit{2}}{\isacharcolon}\ {\isachardoublequoteopen}B{\isadigit{2}}\ {\isasymsqsubseteq}\ B{\isachardoublequoteclose}\ \isacommand{apply}\isamarkupfalse%
\ {\isacharparenleft}simp\ add{\isacharcolon}\ lesseq{\isacharunderscore}tyenv{\isacharunderscore}def\ dom{\isacharunderscore}def{\isacharparenright}\isanewline
\ \ \ \ \ \ \ \ \ \ \isacommand{apply}\isamarkupfalse%
\ auto\ \isacommand{apply}\isamarkupfalse%
\ blast\ \isacommand{done}\isamarkupfalse%
\isanewline
\ \ \ \ \ \ \ \ \isacommand{from}\isamarkupfalse%
\ las{\isadigit{2}}\ wth{\isadigit{2}}\ \isacommand{obtain}\isamarkupfalse%
\ cv{\isadigit{2}}\ \isakeyword{where}\ lam{\isadigit{2}}{\isacharcolon}\ {\isachardoublequoteopen}lookup\ a\ {\isasymmu}{\isacharprime}\ {\isacharequal}\ Result\ {\isacharparenleft}cv{\isadigit{2}}{\isacharcomma}B{\isadigit{2}}{\isacharparenright}{\isachardoublequoteclose}\ \isanewline
\ \ \ \ \ \ \ \ \ \ \isacommand{apply}\isamarkupfalse%
\ {\isacharparenleft}simp\ add{\isacharcolon}\ wt{\isacharunderscore}heap{\isacharunderscore}def{\isacharparenright}\ \isacommand{apply}\isamarkupfalse%
\ fast\ \isacommand{done}\isamarkupfalse%
\isanewline
\ \ \ \ \ \ \ \ \isacommand{show}\isamarkupfalse%
\ {\isacharquery}thesis\isanewline
\ \ \ \ \ \ \ \ \isacommand{proof}\isamarkupfalse%
\ {\isacharparenleft}cases\ {\isachardoublequoteopen}B\ {\isasymsqsubseteq}\ B{\isadigit{2}}{\isachardoublequoteclose}{\isacharparenright}\isanewline
\ \ \ \ \ \ \ \ \ \ \isacommand{assume}\isamarkupfalse%
\ b{\isadigit{2}}b{\isacharcolon}\ {\isachardoublequoteopen}B\ {\isasymsqsubseteq}\ B{\isadigit{2}}{\isachardoublequoteclose}\isanewline
\ \ \ \ \ \ \ \ \ \ \isacommand{let}\isamarkupfalse%
\ {\isacharquery}M{\isadigit{2}}\ {\isacharequal}\ {\isachardoublequoteopen}{\isacharparenleft}a{\isacharcomma}{\isacharparenleft}Val\ v{\isacharprime}{\isacharcomma}B{\isacharparenright}{\isacharparenright}{\isacharhash}{\isasymmu}{\isacharprime}{\isachardoublequoteclose}\isanewline
\ \ \ \ \ \ \ \ \ \ \isacommand{let}\isamarkupfalse%
\ {\isacharquery}S{\isadigit{2}}\ {\isacharequal}\ {\isachardoublequoteopen}{\isacharparenleft}a{\isacharcomma}B{\isacharparenright}{\isacharhash}{\isasymSigma}{\isacharprime}{\isachardoublequoteclose}\isanewline
\ \ \ \ \ \ \ \ \ \ \isacommand{let}\isamarkupfalse%
\ {\isacharquery}ads\ {\isacharequal}\ {\isachardoublequoteopen}removeAll\ a\ ads{\isadigit{2}}{\isachardoublequoteclose}\isanewline
\ \ \ \ \ \ \ \ \ \ \isacommand{let}\isamarkupfalse%
\ {\isacharquery}S\ {\isacharequal}\ {\isachardoublequoteopen}{\isacharparenleft}st{\isacharcomma}\ {\isasymrho}{\isacharcomma}\ k{\isacharcomma}\ {\isacharquery}M{\isadigit{2}}{\isacharcomma}\ {\isacharquery}ads{\isacharparenright}{\isachardoublequoteclose}\isanewline
\ \ \ \ \ \ \ \ \ \ \isacommand{from}\isamarkupfalse%
\ Cons\ st\ VCast\ lam\ lam{\isadigit{2}}\ castv\ b{\isadigit{2}}b\isanewline
\ \ \ \ \ \ \ \ \ \ \isacommand{have}\isamarkupfalse%
\ steps{\isacharcolon}\ {\isachardoublequoteopen}step\ s\ {\isacharequal}\ Result\ {\isacharquery}S{\isachardoublequoteclose}\ \isacommand{by}\isamarkupfalse%
\ simp\isanewline
\ \ \ \ \ \ \ \ \ \ \isacommand{from}\isamarkupfalse%
\ wtvp\ bb\ \isacommand{have}\isamarkupfalse%
\ wtvp{\isadigit{2}}{\isacharcolon}\ {\isachardoublequoteopen}{\isasymSigma}{\isacharprime}\ {\isasymturnstile}cv\ Val\ v{\isacharprime}\ {\isacharcolon}\ B{\isachardoublequoteclose}\ \isacommand{by}\isamarkupfalse%
\ blast\isanewline
\isanewline
\ \ \ \ \ \ \ \ \ \ \isacommand{from}\isamarkupfalse%
\ wth{\isadigit{2}}\ las{\isadigit{2}}\ b{\isadigit{2}}b\isanewline
\ \ \ \ \ \ \ \ \ \ \isacommand{have}\isamarkupfalse%
\ {\isachardoublequoteopen}wt{\isacharunderscore}heap\ {\isacharquery}S{\isadigit{2}}\ {\isacharquery}M{\isadigit{2}}\ {\isacharparenleft}cval{\isacharunderscore}ads\ {\isacharparenleft}Val\ v{\isacharprime}{\isacharparenright}\ a\ {\isacharparenleft}set\ ads{\isadigit{2}}{\isacharparenright}{\isacharparenright}{\isachardoublequoteclose}\ \isanewline
\ \ \ \ \ \ \ \ \ \ \ \ \isacommand{apply}\isamarkupfalse%
\ {\isacharparenleft}rule\ update{\isacharunderscore}heap{\isacharunderscore}val{\isacharparenright}\ \isacommand{using}\isamarkupfalse%
\ wtvp\ bb\ \isacommand{apply}\isamarkupfalse%
\ auto\ \isacommand{done}\isamarkupfalse%
\isanewline
\ \ \ \ \ \ \ \ \ \ \isacommand{hence}\isamarkupfalse%
\ wth{\isadigit{3}}{\isacharcolon}\ {\isachardoublequoteopen}wt{\isacharunderscore}heap\ {\isacharquery}S{\isadigit{2}}\ {\isacharquery}M{\isadigit{2}}\ {\isacharparenleft}set\ {\isacharquery}ads{\isacharparenright}{\isachardoublequoteclose}\ \isacommand{apply}\isamarkupfalse%
\ simp\ \isacommand{done}\isamarkupfalse%
\isanewline
\ \ \ \ \ \ \ \ \ \ \isacommand{from}\isamarkupfalse%
\ ss\ las{\isadigit{2}}\ b{\isadigit{2}}b\ \isacommand{have}\isamarkupfalse%
\ ss{\isadigit{2}}{\isacharcolon}\ {\isachardoublequoteopen}{\isacharquery}S{\isadigit{2}}\ {\isasymsqsubseteq}\ {\isasymSigma}{\isachardoublequoteclose}\isanewline
\ \ \ \ \ \ \ \ \ \ \ \ \isacommand{apply}\isamarkupfalse%
\ {\isacharparenleft}simp\ only{\isacharcolon}\ lesseq{\isacharunderscore}tyenv{\isacharunderscore}def{\isacharparenright}\isanewline
\ \ \ \ \ \ \ \ \ \ \ \ \isacommand{apply}\isamarkupfalse%
\ {\isacharparenleft}frule\ lookup{\isacharunderscore}dom{\isacharparenright}\isanewline
\ \ \ \ \ \ \ \ \ \ \ \ \isacommand{apply}\isamarkupfalse%
\ {\isacharparenleft}rule\ conjI{\isacharparenright}\ \isacommand{apply}\isamarkupfalse%
\ {\isacharparenleft}simp\ add{\isacharcolon}\ dom{\isacharunderscore}def{\isacharparenright}\ \isacommand{apply}\isamarkupfalse%
\ force\isanewline
\ \ \ \ \ \ \ \ \ \ \ \ \isacommand{apply}\isamarkupfalse%
\ auto\isanewline
\ \ \ \ \ \ \ \ \ \ \ \ \isacommand{apply}\isamarkupfalse%
\ {\isacharparenleft}erule{\isacharunderscore}tac\ x{\isacharequal}a\ \isakeyword{in}\ allE{\isacharparenright}\isanewline
\ \ \ \ \ \ \ \ \ \ \ \ \isacommand{apply}\isamarkupfalse%
\ {\isacharparenleft}erule{\isacharunderscore}tac\ x{\isacharequal}A\ \isakeyword{in}\ allE{\isacharparenright}\isanewline
\ \ \ \ \ \ \ \ \ \ \ \ \isacommand{apply}\isamarkupfalse%
\ {\isacharparenleft}erule\ impE{\isacharparenright}\ \isacommand{apply}\isamarkupfalse%
\ simp\isanewline
\ \ \ \ \ \ \ \ \ \ \ \ \isacommand{apply}\isamarkupfalse%
\ {\isacharparenleft}erule\ exE{\isacharparenright}\ \isacommand{apply}\isamarkupfalse%
\ {\isacharparenleft}erule\ conjE{\isacharparenright}\ \isacommand{apply}\isamarkupfalse%
\ simp\ \isanewline
\ \ \ \ \ \ \ \ \ \ \ \ \isacommand{using}\isamarkupfalse%
\ lesseq{\isacharunderscore}prec{\isacharunderscore}trans\ \isacommand{apply}\isamarkupfalse%
\ blast\ \isacommand{done}\isamarkupfalse%
\isanewline
\ \ \ \ \ \ \ \ \ \ \isacommand{from}\isamarkupfalse%
\ gr\ ss{\isadigit{2}}\ \isacommand{have}\isamarkupfalse%
\ gr{\isadigit{2}}{\isacharcolon}\ {\isachardoublequoteopen}{\isasymGamma}{\isacharsemicolon}\ {\isacharquery}S{\isadigit{2}}\ {\isasymturnstile}\ {\isasymrho}{\isachardoublequoteclose}\ \isanewline
\ \ \ \ \ \ \ \ \ \ \ \ \isacommand{using}\isamarkupfalse%
\ strengthen{\isacharunderscore}value{\isacharunderscore}env\ \isacommand{apply}\isamarkupfalse%
\ blast\ \isacommand{done}\isamarkupfalse%
\isanewline
\ \ \ \ \ \ \ \ \ \ \isacommand{from}\isamarkupfalse%
\ wt{\isacharunderscore}k\ ss{\isadigit{2}}\ \isacommand{have}\isamarkupfalse%
\ wtk{\isadigit{2}}{\isacharcolon}\ {\isachardoublequoteopen}{\isacharquery}S{\isadigit{2}}\ {\isasymturnstile}\ k\ {\isacharcolon}\ A{\isacharprime}\ {\isasymRightarrow}\ A{\isachardoublequoteclose}\ \isanewline
\ \ \ \ \ \ \ \ \ \ \ \ \isacommand{using}\isamarkupfalse%
\ strengthen{\isacharunderscore}stack\ \isacommand{apply}\isamarkupfalse%
\ blast\ \isacommand{done}\isamarkupfalse%
\isanewline
\ \ \ \ \ \ \ \ \ \ \isacommand{from}\isamarkupfalse%
\ wth{\isadigit{3}}\ gr{\isadigit{2}}\ wts\ wtk{\isadigit{2}}\isanewline
\ \ \ \ \ \ \ \ \ \ \isacommand{have}\isamarkupfalse%
\ wtS{\isacharcolon}\ {\isachardoublequoteopen}wt{\isacharunderscore}state\ {\isacharquery}S\ A{\isachardoublequoteclose}\ \isacommand{by}\isamarkupfalse%
\ {\isacharparenleft}rule\ wts{\isacharunderscore}intro{\isacharparenright}\isanewline
\ \ \ \ \ \ \ \ \ \ \isacommand{from}\isamarkupfalse%
\ steps\ wtS\ \isacommand{show}\isamarkupfalse%
\ {\isacharquery}thesis\ \isacommand{by}\isamarkupfalse%
\ blast\isanewline
\ \ \ \ \ \ \ \ \isacommand{next}\isamarkupfalse%
\isanewline
\ \ \ \ \ \ \ \ \ \ \isacommand{assume}\isamarkupfalse%
\ b{\isadigit{2}}b{\isacharcolon}\ {\isachardoublequoteopen}{\isasymnot}\ B\ {\isasymsqsubseteq}\ B{\isadigit{2}}{\isachardoublequoteclose}\isanewline
\ \ \ \ \ \ \ \ \ \ \isacommand{let}\isamarkupfalse%
\ {\isacharquery}S\ {\isacharequal}\ {\isachardoublequoteopen}{\isacharparenleft}st{\isacharcomma}\ {\isasymrho}{\isacharcomma}\ k{\isacharcomma}\ {\isasymmu}{\isacharprime}{\isacharcomma}\ ads{\isadigit{2}}{\isacharparenright}{\isachardoublequoteclose}\isanewline
\ \ \ \ \ \ \ \ \ \ \isacommand{let}\isamarkupfalse%
\ {\isacharquery}ads\ {\isacharequal}\ {\isachardoublequoteopen}ads{\isadigit{2}}{\isachardoublequoteclose}\isanewline
\ \ \ \ \ \ \ \ \ \ \isacommand{from}\isamarkupfalse%
\ Cons\ st\ VCast\ lam\ lam{\isadigit{2}}\ castv\ b{\isadigit{2}}b\isanewline
\ \ \ \ \ \ \ \ \ \ \isacommand{have}\isamarkupfalse%
\ steps{\isacharcolon}\ {\isachardoublequoteopen}step\ s\ {\isacharequal}\ Result\ {\isacharquery}S{\isachardoublequoteclose}\ \isacommand{by}\isamarkupfalse%
\ simp\isanewline
\ \ \ \ \ \ \ \ \ \ \isanewline
\ \ \ \ \ \ \ \ \ \ \isacommand{have}\isamarkupfalse%
\ wtS{\isacharcolon}\ {\isachardoublequoteopen}wt{\isacharunderscore}state\ {\isacharquery}S\ A{\isachardoublequoteclose}\ \isanewline
\ \ \ \ \ \ \ \ \ \ \ \ \isacommand{apply}\isamarkupfalse%
\ {\isacharparenleft}rule\ wts{\isacharunderscore}intro{\isacharparenright}\isanewline
\ \ \ \ \ \ \ \ \ \ \ \ \isacommand{using}\isamarkupfalse%
\ wth{\isadigit{2}}\ \isacommand{apply}\isamarkupfalse%
\ assumption\isanewline
\ \ \ \ \ \ \ \ \ \ \ \ \isacommand{using}\isamarkupfalse%
\ gr\ ss\ strengthen{\isacharunderscore}value{\isacharunderscore}env\ \isacommand{apply}\isamarkupfalse%
\ force\isanewline
\ \ \ \ \ \ \ \ \ \ \ \ \isacommand{using}\isamarkupfalse%
\ wts\ \isacommand{apply}\isamarkupfalse%
\ assumption\isanewline
\ \ \ \ \ \ \ \ \ \ \ \ \isacommand{using}\isamarkupfalse%
\ wt{\isacharunderscore}k\ ss\ strengthen{\isacharunderscore}stack\ \isacommand{apply}\isamarkupfalse%
\ force\ \isacommand{done}\isamarkupfalse%
\isanewline
\ \ \ \ \ \ \ \ \ \ \isacommand{from}\isamarkupfalse%
\ steps\ wtS\ \isacommand{show}\isamarkupfalse%
\ {\isacharquery}thesis\ \isacommand{by}\isamarkupfalse%
\ blast\isanewline
\ \ \ \ \ \ \ \ \isacommand{qed}\isamarkupfalse%
\isanewline
\ \ \ \ \ \ \isacommand{next}\isamarkupfalse%
\isanewline
\ \ \ \ \ \ \ \ \isacommand{assume}\isamarkupfalse%
\ {\isachardoublequoteopen}cast\ v\ C\ B{\isacharprime}\ {\isasymmu}\ {\isacharparenleft}a{\isacharhash}ads{\isacharprime}{\isacharparenright}\ {\isacharequal}\ CastError{\isachardoublequoteclose}\isanewline
\ \ \ \ \ \ \ \ \isacommand{with}\isamarkupfalse%
\ st\ Cons\ lam\ VCast\ \isacommand{have}\isamarkupfalse%
\ {\isachardoublequoteopen}step\ s\ {\isacharequal}\ CastError{\isachardoublequoteclose}\ \isacommand{by}\isamarkupfalse%
\ simp\isanewline
\ \ \ \ \ \ \ \ \isacommand{thus}\isamarkupfalse%
\ {\isacharquery}thesis\ \isacommand{by}\isamarkupfalse%
\ simp\isanewline
\ \ \ \ \ \ \isacommand{qed}\isamarkupfalse%
\isanewline
\ \ \ \ \isacommand{qed}\isamarkupfalse%
\isanewline
\ \ \isacommand{next}\isamarkupfalse%
\isanewline
\ \ \ \ \isacommand{case}\isamarkupfalse%
\ Nil\isanewline
\ \ \ \ \isacommand{show}\isamarkupfalse%
\ {\isacharquery}thesis\isanewline
\ \ \ \ \isacommand{proof}\isamarkupfalse%
\ {\isacharparenleft}cases\ st{\isacharparenright}\isanewline
\ \ \ \ \ \ \isacommand{case}\isamarkupfalse%
\ {\isacharparenleft}SLet\ x\ e\ s{\isadigit{2}}{\isacharparenright}\isanewline
\ \ \ \ \ \ \isacommand{from}\isamarkupfalse%
\ wts\ SLet\ \isacommand{obtain}\isamarkupfalse%
\ B\ \isakeyword{where}\ wte{\isacharcolon}\ {\isachardoublequoteopen}{\isasymGamma}\ {\isasymturnstile}\isactrlisub e\ e\ {\isacharcolon}\ B{\isachardoublequoteclose}\ \isanewline
\ \ \ \ \ \ \ \ \isakeyword{and}\ wts{\isadigit{2}}{\isacharcolon}\ {\isachardoublequoteopen}{\isacharparenleft}x{\isacharcomma}B{\isacharparenright}{\isacharhash}{\isasymGamma}\ {\isasymturnstile}\isactrlisub s\ s{\isadigit{2}}\ {\isacharcolon}\ A{\isacharprime}{\isachardoublequoteclose}\ \isacommand{by}\isamarkupfalse%
\ blast\isanewline
\ \ \ \ \ \ \isacommand{from}\isamarkupfalse%
\ wte\ gr\ wt{\isacharunderscore}h\ Nil\ \isacommand{have}\isamarkupfalse%
\ {\isachardoublequoteopen}{\isacharparenleft}{\isasymexists}v{\isachardot}\ eval\ e\ {\isasymrho}\ {\isasymmu}\ {\isacharequal}\ Result\ v\ {\isasymand}\ {\isasymSigma}\ {\isasymturnstile}v\ v\ {\isacharcolon}\ B{\isacharparenright}{\isachardoublequoteclose}\isanewline
\ \ \ \ \ \ \ \ \isacommand{using}\isamarkupfalse%
\ eval{\isacharunderscore}safe{\isacharbrackleft}of\ {\isasymGamma}\ e\ B\ {\isasymSigma}\ {\isasymrho}\ {\isasymmu}{\isacharbrackright}\ \isacommand{by}\isamarkupfalse%
\ simp\isanewline
\ \ \ \ \ \ \isacommand{from}\isamarkupfalse%
\ this\ \isacommand{obtain}\isamarkupfalse%
\ v\ \isakeyword{where}\ v{\isacharcolon}\ {\isachardoublequoteopen}eval\ e\ {\isasymrho}\ {\isasymmu}\ {\isacharequal}\ Result\ v{\isachardoublequoteclose}\isanewline
\ \ \ \ \ \ \ \ \isakeyword{and}\ wtv{\isacharcolon}\ {\isachardoublequoteopen}{\isasymSigma}\ {\isasymturnstile}v\ v\ {\isacharcolon}\ B{\isachardoublequoteclose}\ \isacommand{by}\isamarkupfalse%
\ blast\isanewline
\ \ \ \ \ \ \isacommand{from}\isamarkupfalse%
\ v\ SLet\ st\ gr\ wts{\isadigit{2}}\ wt{\isacharunderscore}k\ wtv\ wt{\isacharunderscore}h\ Nil\ \isacommand{show}\isamarkupfalse%
\ {\isacharquery}thesis\ \isacommand{by}\isamarkupfalse%
\ auto\isanewline
\ \ \ \ \isacommand{next}\isamarkupfalse%
\isanewline
\ \ \ \ \ \ \isacommand{case}\isamarkupfalse%
\ {\isacharparenleft}SRet\ e{\isacharparenright}\isanewline
\ \ \ \ \ \ \isacommand{show}\isamarkupfalse%
\ {\isacharquery}thesis\isanewline
\ \ \ \ \ \ \isacommand{proof}\isamarkupfalse%
\ {\isacharparenleft}cases\ k{\isacharparenright}\isanewline
\ \ \ \ \ \ \ \ \isacommand{assume}\isamarkupfalse%
\ k{\isacharcolon}\ {\isachardoublequoteopen}k\ {\isacharequal}\ {\isacharbrackleft}{\isacharbrackright}{\isachardoublequoteclose}\ \isacommand{from}\isamarkupfalse%
\ Nil\ SRet\ st\ k\ \isacommand{have}\isamarkupfalse%
\ {\isachardoublequoteopen}final\ s{\isachardoublequoteclose}\ \isacommand{by}\isamarkupfalse%
\ simp\isanewline
\ \ \ \ \ \ \ \ \isacommand{thus}\isamarkupfalse%
\ {\isacharquery}thesis\ \isacommand{by}\isamarkupfalse%
\ blast\isanewline
\ \ \ \ \ \ \isacommand{next}\isamarkupfalse%
\isanewline
\ \ \ \ \ \ \ \ \isacommand{fix}\isamarkupfalse%
\ f\ k{\isacharprime}\ \isacommand{assume}\isamarkupfalse%
\ k{\isacharcolon}\ {\isachardoublequoteopen}k\ {\isacharequal}\ f{\isacharhash}k{\isacharprime}{\isachardoublequoteclose}\isanewline
\ \ \ \ \ \ \ \ \isacommand{from}\isamarkupfalse%
\ wts\ SRet\ \isacommand{have}\isamarkupfalse%
\ wte{\isacharcolon}\ {\isachardoublequoteopen}{\isasymGamma}\ {\isasymturnstile}\isactrlisub e\ e\ {\isacharcolon}\ A{\isacharprime}{\isachardoublequoteclose}\ \isacommand{by}\isamarkupfalse%
\ blast\isanewline
\ \ \ \ \ \ \ \ \isacommand{from}\isamarkupfalse%
\ wte\ gr\ wt{\isacharunderscore}h\ Nil\ \isacommand{have}\isamarkupfalse%
\ {\isachardoublequoteopen}{\isacharparenleft}{\isasymexists}v{\isachardot}\ eval\ e\ {\isasymrho}\ {\isasymmu}\ {\isacharequal}\ Result\ v\ {\isasymand}\ {\isasymSigma}\ {\isasymturnstile}v\ v\ {\isacharcolon}\ A{\isacharprime}{\isacharparenright}{\isachardoublequoteclose}\isanewline
\ \ \ \ \ \ \ \ \ \ \isacommand{using}\isamarkupfalse%
\ eval{\isacharunderscore}safe{\isacharbrackleft}of\ {\isasymGamma}\ e\ A{\isacharprime}\ {\isasymSigma}\ {\isasymrho}\ {\isasymmu}{\isacharbrackright}\ \isacommand{by}\isamarkupfalse%
\ simp\isanewline
\ \ \ \ \ \ \ \ \isacommand{from}\isamarkupfalse%
\ this\ \isacommand{obtain}\isamarkupfalse%
\ v\ \isakeyword{where}\ v{\isacharcolon}\ {\isachardoublequoteopen}eval\ e\ {\isasymrho}\ {\isasymmu}\ {\isacharequal}\ Result\ v{\isachardoublequoteclose}\isanewline
\ \ \ \ \ \ \ \ \ \ \isakeyword{and}\ wtv{\isacharcolon}\ {\isachardoublequoteopen}{\isasymSigma}\ {\isasymturnstile}v\ v\ {\isacharcolon}\ A{\isacharprime}{\isachardoublequoteclose}\ \isacommand{by}\isamarkupfalse%
\ blast\isanewline
\ \ \ \ \ \ \ \ \isacommand{from}\isamarkupfalse%
\ Nil\ SRet\ st\ v\ wtv\ k\ wt{\isacharunderscore}k\ wt{\isacharunderscore}h\ \isacommand{show}\isamarkupfalse%
\ {\isacharquery}thesis\ \isacommand{by}\isamarkupfalse%
\ auto\isanewline
\ \ \ \ \ \ \isacommand{qed}\isamarkupfalse%
\isanewline
\ \ \ \ \isacommand{next}\isamarkupfalse%
\isanewline
\ \ \ \ \ \ \isacommand{case}\isamarkupfalse%
\ {\isacharparenleft}SCall\ x\ e{\isadigit{1}}\ e{\isadigit{2}}\ s{\isadigit{2}}{\isacharparenright}\isanewline
\ \ \ \ \ \ \isacommand{from}\isamarkupfalse%
\ wts\ SCall\ \isacommand{obtain}\isamarkupfalse%
\ B\ C\ \isakeyword{where}\ wte{\isadigit{1}}{\isacharcolon}\ {\isachardoublequoteopen}{\isasymGamma}\ {\isasymturnstile}\isactrlisub e\ e{\isadigit{1}}\ {\isacharcolon}\ B\ {\isasymrightarrow}\ C{\isachardoublequoteclose}\ \isanewline
\ \ \ \ \ \ \ \ \isakeyword{and}\ wte{\isadigit{2}}{\isacharcolon}\ {\isachardoublequoteopen}{\isasymGamma}\ {\isasymturnstile}\isactrlisub e\ e{\isadigit{2}}\ {\isacharcolon}\ B{\isachardoublequoteclose}\ \isakeyword{and}\ wts{\isadigit{2}}{\isacharcolon}\ {\isachardoublequoteopen}{\isacharparenleft}x{\isacharcomma}C{\isacharparenright}{\isacharhash}{\isasymGamma}\ {\isasymturnstile}\isactrlisub s\ s{\isadigit{2}}\ {\isacharcolon}\ A{\isacharprime}{\isachardoublequoteclose}\ \isacommand{by}\isamarkupfalse%
\ blast\isanewline
\ \ \ \ \ \ \isacommand{from}\isamarkupfalse%
\ wte{\isadigit{1}}\ gr\ wt{\isacharunderscore}h\ Nil\ \isacommand{have}\isamarkupfalse%
\ {\isachardoublequoteopen}{\isacharparenleft}{\isasymexists}v{\isadigit{1}}{\isachardot}\ eval\ e{\isadigit{1}}\ {\isasymrho}\ {\isasymmu}\ {\isacharequal}\ Result\ v{\isadigit{1}}\ {\isasymand}\ {\isasymSigma}\ {\isasymturnstile}v\ v{\isadigit{1}}\ {\isacharcolon}\ B{\isasymrightarrow}C{\isacharparenright}{\isachardoublequoteclose}\isanewline
\ \ \ \ \ \ \ \ \isacommand{using}\isamarkupfalse%
\ eval{\isacharunderscore}safe{\isacharbrackleft}of\ {\isasymGamma}\ e{\isadigit{1}}\ {\isachardoublequoteopen}B{\isasymrightarrow}C{\isachardoublequoteclose}\ {\isasymSigma}\ {\isasymrho}\ {\isasymmu}{\isacharbrackright}\ \isacommand{by}\isamarkupfalse%
\ simp\isanewline
\ \ \ \ \ \ \isacommand{from}\isamarkupfalse%
\ this\ \isacommand{obtain}\isamarkupfalse%
\ v{\isadigit{1}}\ \isakeyword{where}\ v{\isadigit{1}}{\isacharcolon}\ {\isachardoublequoteopen}eval\ e{\isadigit{1}}\ {\isasymrho}\ {\isasymmu}\ {\isacharequal}\ Result\ v{\isadigit{1}}{\isachardoublequoteclose}\isanewline
\ \ \ \ \ \ \ \ \isakeyword{and}\ wtv{\isadigit{1}}{\isacharcolon}\ {\isachardoublequoteopen}{\isasymSigma}\ {\isasymturnstile}v\ v{\isadigit{1}}\ {\isacharcolon}\ {\isacharparenleft}B\ {\isasymrightarrow}\ C{\isacharparenright}{\isachardoublequoteclose}\ \isacommand{by}\isamarkupfalse%
\ blast\isanewline
\ \ \ \ \ \ \isacommand{from}\isamarkupfalse%
\ wte{\isadigit{2}}\ gr\ wt{\isacharunderscore}h\ Nil\ \isacommand{have}\isamarkupfalse%
\ {\isachardoublequoteopen}{\isacharparenleft}{\isasymexists}v{\isadigit{2}}{\isachardot}\ eval\ e{\isadigit{2}}\ {\isasymrho}\ {\isasymmu}\ {\isacharequal}\ Result\ v{\isadigit{2}}\ {\isasymand}\ {\isasymSigma}\ {\isasymturnstile}v\ v{\isadigit{2}}\ {\isacharcolon}\ B{\isacharparenright}{\isachardoublequoteclose}\isanewline
\ \ \ \ \ \ \ \ \isacommand{using}\isamarkupfalse%
\ eval{\isacharunderscore}safe{\isacharbrackleft}of\ {\isasymGamma}\ e{\isadigit{2}}\ {\isachardoublequoteopen}B{\isachardoublequoteclose}\ {\isasymSigma}\ {\isasymrho}\ {\isasymmu}{\isacharbrackright}\ \isacommand{by}\isamarkupfalse%
\ simp\isanewline
\ \ \ \ \ \ \isacommand{from}\isamarkupfalse%
\ this\ \isacommand{obtain}\isamarkupfalse%
\ v{\isadigit{2}}\ \isakeyword{where}\ v{\isadigit{2}}{\isacharcolon}\ {\isachardoublequoteopen}eval\ e{\isadigit{2}}\ {\isasymrho}\ {\isasymmu}\ {\isacharequal}\ Result\ v{\isadigit{2}}{\isachardoublequoteclose}\isanewline
\ \ \ \ \ \ \ \ \isakeyword{and}\ wtv{\isadigit{2}}{\isacharcolon}\ {\isachardoublequoteopen}{\isasymSigma}\ {\isasymturnstile}v\ v{\isadigit{2}}\ {\isacharcolon}\ B{\isachardoublequoteclose}\ \isacommand{by}\isamarkupfalse%
\ blast\isanewline
\ \ \ \ \ \ \isacommand{from}\isamarkupfalse%
\ SCall\ st\ v{\isadigit{1}}\ v{\isadigit{2}}\ wtv{\isadigit{1}}\ gr\ wtv{\isadigit{2}}\ wts{\isadigit{2}}\ wt{\isacharunderscore}k\ wt{\isacharunderscore}h\ Nil\ \isacommand{show}\isamarkupfalse%
\ {\isacharquery}thesis\isanewline
\ \ \ \ \ \ \ \ \isacommand{by}\isamarkupfalse%
\ {\isacharparenleft}case{\isacharunderscore}tac\ v{\isadigit{1}}{\isacharcomma}\ auto{\isacharcomma}\ case{\isacharunderscore}tac\ {\isachardoublequoteopen}const{\isachardoublequoteclose}{\isacharcomma}\ auto{\isacharparenright}\isanewline
\ \ \ \ \isacommand{next}\isamarkupfalse%
\isanewline
\ \ \ \ \ \ \isacommand{case}\isamarkupfalse%
\ {\isacharparenleft}STailCall\ e{\isadigit{1}}\ e{\isadigit{2}}{\isacharparenright}\isanewline
\ \ \ \ \ \ \isacommand{from}\isamarkupfalse%
\ wts\ STailCall\ \isacommand{obtain}\isamarkupfalse%
\ B\ \isakeyword{where}\ wte{\isadigit{1}}{\isacharcolon}\ {\isachardoublequoteopen}{\isasymGamma}\ {\isasymturnstile}\isactrlisub e\ e{\isadigit{1}}\ {\isacharcolon}\ B\ {\isasymrightarrow}\ A{\isacharprime}{\isachardoublequoteclose}\ \isanewline
\ \ \ \ \ \ \ \ \isakeyword{and}\ wte{\isadigit{2}}{\isacharcolon}\ {\isachardoublequoteopen}{\isasymGamma}\ {\isasymturnstile}\isactrlisub e\ e{\isadigit{2}}\ {\isacharcolon}\ B{\isachardoublequoteclose}\ \isacommand{by}\isamarkupfalse%
\ blast\isanewline
\ \ \ \ \ \ \isacommand{from}\isamarkupfalse%
\ wte{\isadigit{1}}\ gr\ wt{\isacharunderscore}h\ Nil\ \isacommand{have}\isamarkupfalse%
\ {\isachardoublequoteopen}{\isacharparenleft}{\isasymexists}v{\isadigit{1}}{\isachardot}\ eval\ e{\isadigit{1}}\ {\isasymrho}\ {\isasymmu}\ {\isacharequal}\ Result\ v{\isadigit{1}}\ {\isasymand}\ {\isasymSigma}\ {\isasymturnstile}v\ v{\isadigit{1}}\ {\isacharcolon}\ B{\isasymrightarrow}A{\isacharprime}{\isacharparenright}{\isachardoublequoteclose}\isanewline
\ \ \ \ \ \ \ \ \isacommand{using}\isamarkupfalse%
\ eval{\isacharunderscore}safe{\isacharbrackleft}of\ {\isasymGamma}\ e{\isadigit{1}}\ {\isachardoublequoteopen}B{\isasymrightarrow}A{\isacharprime}{\isachardoublequoteclose}\ {\isasymSigma}\ {\isasymrho}\ {\isasymmu}{\isacharbrackright}\ \isacommand{by}\isamarkupfalse%
\ simp\isanewline
\ \ \ \ \ \ \isacommand{from}\isamarkupfalse%
\ this\ \isacommand{obtain}\isamarkupfalse%
\ v{\isadigit{1}}\ \isakeyword{where}\ v{\isadigit{1}}{\isacharcolon}\ {\isachardoublequoteopen}eval\ e{\isadigit{1}}\ {\isasymrho}\ {\isasymmu}\ {\isacharequal}\ Result\ v{\isadigit{1}}{\isachardoublequoteclose}\isanewline
\ \ \ \ \ \ \ \ \isakeyword{and}\ wtv{\isadigit{1}}{\isacharcolon}\ {\isachardoublequoteopen}{\isasymSigma}\ {\isasymturnstile}v\ v{\isadigit{1}}\ {\isacharcolon}\ {\isacharparenleft}B\ {\isasymrightarrow}\ A{\isacharprime}{\isacharparenright}{\isachardoublequoteclose}\ \isacommand{by}\isamarkupfalse%
\ blast\isanewline
\ \ \ \ \ \ \isacommand{from}\isamarkupfalse%
\ wte{\isadigit{2}}\ gr\ wt{\isacharunderscore}h\ Nil\ \isacommand{have}\isamarkupfalse%
\ {\isachardoublequoteopen}{\isacharparenleft}{\isasymexists}v{\isadigit{2}}{\isachardot}\ eval\ e{\isadigit{2}}\ {\isasymrho}\ {\isasymmu}\ {\isacharequal}\ Result\ v{\isadigit{2}}\ {\isasymand}\ {\isasymSigma}\ {\isasymturnstile}v\ v{\isadigit{2}}\ {\isacharcolon}\ B{\isacharparenright}{\isachardoublequoteclose}\isanewline
\ \ \ \ \ \ \ \ \isacommand{using}\isamarkupfalse%
\ eval{\isacharunderscore}safe{\isacharbrackleft}of\ {\isasymGamma}\ e{\isadigit{2}}\ {\isachardoublequoteopen}B{\isachardoublequoteclose}\ {\isasymSigma}\ {\isasymrho}\ {\isasymmu}{\isacharbrackright}\ \isacommand{by}\isamarkupfalse%
\ simp\isanewline
\ \ \ \ \ \ \isacommand{from}\isamarkupfalse%
\ this\ \isacommand{obtain}\isamarkupfalse%
\ v{\isadigit{2}}\ \isakeyword{where}\ v{\isadigit{2}}{\isacharcolon}\ {\isachardoublequoteopen}eval\ e{\isadigit{2}}\ {\isasymrho}\ {\isasymmu}\ {\isacharequal}\ Result\ v{\isadigit{2}}{\isachardoublequoteclose}\isanewline
\ \ \ \ \ \ \ \ \isakeyword{and}\ wtv{\isadigit{2}}{\isacharcolon}\ {\isachardoublequoteopen}{\isasymSigma}\ {\isasymturnstile}v\ v{\isadigit{2}}\ {\isacharcolon}\ B{\isachardoublequoteclose}\ \isacommand{by}\isamarkupfalse%
\ blast\isanewline
\ \ \ \ \ \ \isacommand{from}\isamarkupfalse%
\ wtv{\isadigit{1}}\ \isacommand{show}\isamarkupfalse%
\ {\isacharquery}thesis\isanewline
\ \ \ \ \ \ \isacommand{proof}\isamarkupfalse%
\ {\isacharparenleft}rule\ funval{\isacharunderscore}inv{\isacharparenright}\isanewline
\ \ \ \ \ \ \ \ \isacommand{fix}\isamarkupfalse%
\ c\ \isacommand{assume}\isamarkupfalse%
\ {\isachardoublequoteopen}typeof\ c\ {\isacharequal}\ B\ {\isasymrightarrow}\ A{\isacharprime}{\isachardoublequoteclose}\ \isacommand{hence}\isamarkupfalse%
\ {\isachardoublequoteopen}False{\isachardoublequoteclose}\isanewline
\ \ \ \ \ \ \ \ \ \ \isacommand{apply}\isamarkupfalse%
\ {\isacharparenleft}case{\isacharunderscore}tac\ c{\isacharparenright}\ \isacommand{apply}\isamarkupfalse%
\ auto\ \isacommand{done}\isamarkupfalse%
\isanewline
\ \ \ \ \ \ \ \ \isacommand{thus}\isamarkupfalse%
\ {\isacharquery}thesis\ \isacommand{by}\isamarkupfalse%
\ simp\isanewline
\ \ \ \ \ \ \isacommand{next}\isamarkupfalse%
\isanewline
\ \ \ \ \ \ \ \ \isacommand{fix}\isamarkupfalse%
\ {\isasymGamma}{\isacharprime}\ {\isasymrho}{\isacharprime}\ x\ s{\isacharprime}\ \isacommand{assume}\isamarkupfalse%
\ clos{\isacharcolon}\ {\isachardoublequoteopen}v{\isadigit{1}}\ {\isacharequal}\ Closure\ x\ B\ s{\isacharprime}\ {\isasymrho}{\isacharprime}{\isachardoublequoteclose}\ \isakeyword{and}\ ge{\isacharcolon}\ {\isachardoublequoteopen}{\isasymGamma}{\isacharprime}{\isacharsemicolon}{\isasymSigma}\ {\isasymturnstile}\ {\isasymrho}{\isacharprime}{\isachardoublequoteclose}\isanewline
\ \ \ \ \ \ \ \ \ \ \isakeyword{and}\ wtsp{\isacharcolon}\ {\isachardoublequoteopen}{\isacharparenleft}x{\isacharcomma}B{\isacharparenright}{\isacharhash}{\isasymGamma}{\isacharprime}\ {\isasymturnstile}\isactrlisub s\ s{\isacharprime}\ {\isacharcolon}\ A{\isacharprime}{\isachardoublequoteclose}\isanewline
\ \ \ \ \ \ \ \ \isacommand{show}\isamarkupfalse%
\ {\isacharquery}thesis\ \isacommand{apply}\isamarkupfalse%
\ {\isacharparenleft}rule\ disjI{\isadigit{2}}{\isacharparenright}\ \isacommand{apply}\isamarkupfalse%
\ {\isacharparenleft}rule\ disjI{\isadigit{1}}{\isacharparenright}\isanewline
\ \ \ \ \ \ \ \ \isacommand{proof}\isamarkupfalse%
\ {\isacharminus}\isanewline
\ \ \ \ \ \ \ \ \ \ \isacommand{from}\isamarkupfalse%
\ clos\ v{\isadigit{1}}\ v{\isadigit{2}}\ Nil\ \isacommand{have}\isamarkupfalse%
\ s{\isacharcolon}\isanewline
\ \ \ \ \ \ \ \ \ \ \ \ {\isachardoublequoteopen}step\ {\isacharparenleft}STailCall\ e{\isadigit{1}}\ e{\isadigit{2}}{\isacharcomma}\ {\isasymrho}{\isacharcomma}\ k{\isacharcomma}\ {\isasymmu}{\isacharcomma}{\isacharbrackleft}{\isacharbrackright}{\isacharparenright}\ {\isacharequal}\ Result\ {\isacharparenleft}s{\isacharprime}{\isacharcomma}\ {\isacharparenleft}x{\isacharcomma}v{\isadigit{2}}{\isacharparenright}{\isacharhash}{\isasymrho}{\isacharprime}{\isacharcomma}\ k{\isacharcomma}\ {\isasymmu}{\isacharcomma}{\isacharbrackleft}{\isacharbrackright}{\isacharparenright}{\isachardoublequoteclose}\isanewline
\ \ \ \ \ \ \ \ \ \ \ \ \isacommand{by}\isamarkupfalse%
\ simp\isanewline
\ \ \ \ \ \ \ \ \ \ \isacommand{from}\isamarkupfalse%
\ ge\ wtsp\ wt{\isacharunderscore}k\ wtv{\isadigit{2}}\ wt{\isacharunderscore}h\ Nil\isanewline
\ \ \ \ \ \ \ \ \ \ \isacommand{have}\isamarkupfalse%
\ wtns{\isacharcolon}\ {\isachardoublequoteopen}wt{\isacharunderscore}state\ {\isacharparenleft}s{\isacharprime}{\isacharcomma}\ {\isacharparenleft}x{\isacharcomma}v{\isadigit{2}}{\isacharparenright}{\isacharhash}{\isasymrho}{\isacharprime}{\isacharcomma}\ k{\isacharcomma}\ {\isasymmu}{\isacharcomma}{\isacharbrackleft}{\isacharbrackright}{\isacharparenright}\ A{\isachardoublequoteclose}\ \isacommand{by}\isamarkupfalse%
\ auto\isanewline
\ \ \ \ \ \ \ \ \ \ \isacommand{from}\isamarkupfalse%
\ s\ wtns\ st\ STailCall\ Nil\isanewline
\ \ \ \ \ \ \ \ \ \ \isacommand{show}\isamarkupfalse%
\ {\isachardoublequoteopen}{\isasymexists}s{\isacharprime}{\isachardot}\ step\ s\ {\isacharequal}\ Result\ s{\isacharprime}\ {\isasymand}\ wt{\isacharunderscore}state\ s{\isacharprime}\ A{\isachardoublequoteclose}\ \isanewline
\ \ \ \ \ \ \ \ \ \ \ \ \isacommand{by}\isamarkupfalse%
\ auto\isanewline
\ \ \ \ \ \ \ \ \isacommand{qed}\isamarkupfalse%
\isanewline
\ \ \ \ \ \ \isacommand{qed}\isamarkupfalse%
\isanewline
\ \ \ \ \isacommand{next}\isamarkupfalse%
\isanewline
\ \ \ \ \ \ \isacommand{case}\isamarkupfalse%
\ {\isacharparenleft}SAlloc\ x\ B\ e\ s{\isacharprime}{\isacharparenright}\isanewline
\ \ \ \ \ \ \isacommand{from}\isamarkupfalse%
\ wts\ SAlloc\ \isacommand{have}\isamarkupfalse%
\ wte{\isacharcolon}\ {\isachardoublequoteopen}{\isasymGamma}\ {\isasymturnstile}\isactrlisub e\ e\ {\isacharcolon}\ B{\isachardoublequoteclose}\ \isacommand{by}\isamarkupfalse%
\ fast\isanewline
\ \ \ \ \ \ \isacommand{from}\isamarkupfalse%
\ wts\ SAlloc\ Nil\ \isacommand{have}\isamarkupfalse%
\ wts{\isadigit{2}}{\isacharcolon}\ {\isachardoublequoteopen}{\isacharparenleft}x{\isacharcomma}RefT\ B{\isacharparenright}{\isacharhash}{\isasymGamma}\ {\isasymturnstile}\isactrlisub s\ s{\isacharprime}\ {\isacharcolon}\ A{\isacharprime}{\isachardoublequoteclose}\ \isacommand{by}\isamarkupfalse%
\ fast\ \isanewline
\ \ \ \ \ \ \isacommand{from}\isamarkupfalse%
\ wte\ gr\ wt{\isacharunderscore}h\ Nil\ \isacommand{have}\isamarkupfalse%
\ {\isachardoublequoteopen}{\isacharparenleft}{\isasymexists}v{\isachardot}\ eval\ e\ {\isasymrho}\ {\isasymmu}\ {\isacharequal}\ Result\ v\ {\isasymand}\ {\isasymSigma}\ {\isasymturnstile}v\ v\ {\isacharcolon}\ B{\isacharparenright}{\isachardoublequoteclose}\isanewline
\ \ \ \ \ \ \ \ \isacommand{using}\isamarkupfalse%
\ eval{\isacharunderscore}safe{\isacharbrackleft}of\ {\isasymGamma}\ e\ B\ {\isasymSigma}\ {\isasymrho}\ {\isasymmu}{\isacharbrackright}\ \isacommand{by}\isamarkupfalse%
\ simp\isanewline
\ \ \ \ \ \ \isacommand{from}\isamarkupfalse%
\ this\ \isacommand{obtain}\isamarkupfalse%
\ v\ \isakeyword{where}\ v{\isacharcolon}\ {\isachardoublequoteopen}eval\ e\ {\isasymrho}\ {\isasymmu}\ {\isacharequal}\ Result\ v{\isachardoublequoteclose}\isanewline
\ \ \ \ \ \ \ \ \isakeyword{and}\ wtv{\isacharcolon}\ {\isachardoublequoteopen}{\isasymSigma}\ {\isasymturnstile}v\ v\ {\isacharcolon}\ B{\isachardoublequoteclose}\ \isacommand{using}\isamarkupfalse%
\ eval{\isacharunderscore}safe\ \isacommand{by}\isamarkupfalse%
\ blast\isanewline
\ \ \ \ \ \ \isacommand{let}\isamarkupfalse%
\ {\isacharquery}a\ {\isacharequal}\ {\isachardoublequoteopen}length\ {\isasymmu}{\isachardoublequoteclose}\isanewline
\ \ \ \ \ \ \isacommand{let}\isamarkupfalse%
\ {\isacharquery}S{\isadigit{2}}\ {\isacharequal}\ {\isachardoublequoteopen}{\isacharparenleft}{\isacharquery}a{\isacharcomma}B{\isacharparenright}{\isacharhash}{\isasymSigma}{\isachardoublequoteclose}\ \isakeyword{and}\ {\isacharquery}M{\isadigit{2}}\ {\isacharequal}\ {\isachardoublequoteopen}{\isacharparenleft}{\isacharquery}a{\isacharcomma}{\isacharparenleft}Val\ v{\isacharcomma}\ B{\isacharparenright}{\isacharparenright}{\isacharhash}{\isasymmu}{\isachardoublequoteclose}\isanewline
\ \ \ \ \ \ \isacommand{from}\isamarkupfalse%
\ wt{\isacharunderscore}h\ dom{\isacharunderscore}lookup{\isacharbrackleft}of\ {\isacharquery}a\ {\isasymSigma}{\isacharbrackright}\ wt{\isacharunderscore}heap{\isacharunderscore}def{\isacharbrackleft}of\ {\isasymSigma}\ {\isasymmu}{\isacharbrackright}\ lookup{\isacharunderscore}dom{\isacharbrackleft}of\ {\isacharquery}a\ {\isasymmu}{\isacharbrackright}\isanewline
\ \ \ \ \ \ \isacommand{have}\isamarkupfalse%
\ ads{\isacharcolon}\ {\isachardoublequoteopen}{\isacharquery}a\ {\isasymnotin}\ dom\ {\isasymSigma}{\isachardoublequoteclose}\ \isacommand{apply}\isamarkupfalse%
\ blast\ \isacommand{done}\isamarkupfalse%
\isanewline
\ \ \ \ \ \ \isacommand{from}\isamarkupfalse%
\ wtv\ ads\ \isacommand{have}\isamarkupfalse%
\ wtv{\isacharunderscore}{\isadigit{2}}{\isacharcolon}\ {\isachardoublequoteopen}{\isacharquery}S{\isadigit{2}}\ {\isasymturnstile}v\ v\ {\isacharcolon}\ B{\isachardoublequoteclose}\ \isanewline
\ \ \ \ \ \ \ \ \isacommand{using}\isamarkupfalse%
\ weaken{\isacharunderscore}value{\isacharunderscore}env{\isacharbrackleft}of\ {\isasymSigma}\ v\ B{\isacharbrackright}\ \isacommand{apply}\isamarkupfalse%
\ fast\ \isacommand{done}\isamarkupfalse%
\isanewline
\ \ \ \ \ \ \isacommand{from}\isamarkupfalse%
\ ads\ wtv{\isacharunderscore}{\isadigit{2}}\ wt{\isacharunderscore}h\ Nil\ \isacommand{have}\isamarkupfalse%
\ wt{\isacharunderscore}h{\isadigit{2}}{\isacharcolon}\ {\isachardoublequoteopen}wt{\isacharunderscore}heap\ {\isacharquery}S{\isadigit{2}}\ {\isacharquery}M{\isadigit{2}}\ {\isacharbraceleft}{\isacharbraceright}{\isachardoublequoteclose}\isanewline
\ \ \ \ \ \ \ \ \isacommand{apply}\isamarkupfalse%
\ {\isacharparenleft}simp\ only{\isacharcolon}\ wt{\isacharunderscore}heap{\isacharunderscore}def{\isacharparenright}\ \isacommand{apply}\isamarkupfalse%
\ {\isacharparenleft}rule\ conjI{\isacharparenright}\isanewline
\ \ \ \ \ \ \ \ \isacommand{apply}\isamarkupfalse%
\ clarify\ \isacommand{apply}\isamarkupfalse%
\ {\isacharparenleft}case{\isacharunderscore}tac\ {\isachardoublequoteopen}a\ {\isacharequal}\ length\ {\isasymmu}{\isachardoublequoteclose}{\isacharparenright}\isanewline
\ \ \ \ \ \ \ \ \isacommand{apply}\isamarkupfalse%
\ simp\ \isacommand{using}\isamarkupfalse%
\ weaken{\isacharunderscore}value{\isacharunderscore}env\ \isacommand{apply}\isamarkupfalse%
\ force\isanewline
\ \ \ \ \ \ \ \ \isacommand{apply}\isamarkupfalse%
\ {\isacharparenleft}erule{\isacharunderscore}tac\ x{\isacharequal}a\ \isakeyword{in}\ allE{\isacharparenright}\isanewline
\ \ \ \ \ \ \ \ \isacommand{apply}\isamarkupfalse%
\ {\isacharparenleft}erule{\isacharunderscore}tac\ x{\isacharequal}a\ \isakeyword{in}\ allE{\isacharparenright}\isanewline
\ \ \ \ \ \ \ \ \isacommand{apply}\isamarkupfalse%
\ {\isacharparenleft}erule{\isacharunderscore}tac\ x{\isacharequal}A\ \isakeyword{in}\ allE{\isacharparenright}\isanewline
\ \ \ \ \ \ \ \ \isacommand{apply}\isamarkupfalse%
\ {\isacharparenleft}erule\ impE{\isacharparenright}\ \isacommand{apply}\isamarkupfalse%
\ {\isacharparenleft}erule\ impE{\isacharparenright}\isanewline
\ \ \ \ \ \ \ \ \isacommand{apply}\isamarkupfalse%
\ simp\ \isacommand{apply}\isamarkupfalse%
\ {\isacharparenleft}erule\ exE{\isacharparenright}\ \isacommand{apply}\isamarkupfalse%
\ clarify\isanewline
\ \ \ \ \ \ \ \ \isacommand{using}\isamarkupfalse%
\ lookup{\isacharunderscore}dom\ \isacommand{apply}\isamarkupfalse%
\ force\isanewline
\ \ \ \ \ \ \ \ \isacommand{apply}\isamarkupfalse%
\ auto\isanewline
\ \ \ \ \ \ \ \ \isacommand{using}\isamarkupfalse%
\ weaken{\isacharunderscore}value{\isacharunderscore}env\ \isacommand{apply}\isamarkupfalse%
\ force\isanewline
\ \ \ \ \ \ \ \ \isacommand{apply}\isamarkupfalse%
\ {\isacharparenleft}simp\ add{\isacharcolon}\ dom{\isacharunderscore}def{\isacharparenright}\ \isacommand{apply}\isamarkupfalse%
\ auto\ \isacommand{done}\isamarkupfalse%
\isanewline
\ \ \ \ \ \ \isacommand{from}\isamarkupfalse%
\ ads\ wt{\isacharunderscore}k\ \isacommand{have}\isamarkupfalse%
\ wt{\isacharunderscore}k{\isadigit{2}}{\isacharcolon}\ {\isachardoublequoteopen}{\isacharquery}S{\isadigit{2}}\ {\isasymturnstile}\ k\ {\isacharcolon}\ A{\isacharprime}\ {\isasymRightarrow}\ A{\isachardoublequoteclose}\ \isacommand{using}\isamarkupfalse%
\ weaken{\isacharunderscore}stack\ \isacommand{by}\isamarkupfalse%
\ force\isanewline
\ \ \ \ \ \ \isacommand{from}\isamarkupfalse%
\ ads\ gr\ \isacommand{have}\isamarkupfalse%
\ gr{\isacharunderscore}{\isadigit{2}}{\isacharcolon}\ {\isachardoublequoteopen}{\isasymGamma}{\isacharsemicolon}\ {\isacharquery}S{\isadigit{2}}\ {\isasymturnstile}\ {\isasymrho}{\isachardoublequoteclose}\isanewline
\ \ \ \ \ \ \ \ \isacommand{using}\isamarkupfalse%
\ weaken{\isacharunderscore}value{\isacharunderscore}env\ \isacommand{apply}\isamarkupfalse%
\ auto\ \isacommand{done}\isamarkupfalse%
\isanewline
\ \ \ \ \ \ \isacommand{let}\isamarkupfalse%
\ {\isacharquery}s{\isadigit{2}}\ {\isacharequal}\ {\isachardoublequoteopen}{\isacharparenleft}s{\isacharprime}{\isacharcomma}\ {\isacharparenleft}x{\isacharcomma}VRef\ {\isacharquery}a{\isacharparenright}{\isacharhash}{\isasymrho}{\isacharcomma}\ k{\isacharcomma}\ {\isacharparenleft}{\isacharquery}a{\isacharcomma}{\isacharparenleft}Val\ v{\isacharcomma}B{\isacharparenright}{\isacharparenright}{\isacharhash}{\isasymmu}{\isacharcomma}\ {\isacharbrackleft}{\isacharbrackright}{\isacharparenright}{\isachardoublequoteclose}\isanewline
\ \ \ \ \ \ \isacommand{from}\isamarkupfalse%
\ Nil\ v\ SAlloc\ st\ \isacommand{have}\isamarkupfalse%
\ step{\isacharunderscore}s{\isacharcolon}\ {\isachardoublequoteopen}step\ s\ {\isacharequal}\ Result\ {\isacharquery}s{\isadigit{2}}{\isachardoublequoteclose}\isanewline
\ \ \ \ \ \ \ \ \isacommand{apply}\isamarkupfalse%
\ {\isacharparenleft}simp\ add{\isacharcolon}\ Let{\isacharunderscore}def{\isacharparenright}\ \isacommand{done}\isamarkupfalse%
\isanewline
\ \ \ \ \ \ \isacommand{from}\isamarkupfalse%
\ gr{\isacharunderscore}{\isadigit{2}}\ wts{\isadigit{2}}\ wt{\isacharunderscore}k{\isadigit{2}}\ wtv{\isacharunderscore}{\isadigit{2}}\ wt{\isacharunderscore}h{\isadigit{2}}\ \isanewline
\ \ \ \ \ \ \isacommand{have}\isamarkupfalse%
\ wts{\isadigit{2}}{\isacharcolon}\ {\isachardoublequoteopen}wt{\isacharunderscore}state\ {\isacharquery}s{\isadigit{2}}\ A{\isachardoublequoteclose}\ \isacommand{by}\isamarkupfalse%
\ auto\isanewline
\ \ \ \ \ \ \isacommand{from}\isamarkupfalse%
\ step{\isacharunderscore}s\ wts{\isadigit{2}}\ \isacommand{show}\isamarkupfalse%
\ {\isacharquery}thesis\ \isacommand{by}\isamarkupfalse%
\ fast\isanewline
\ \ \ \ \isacommand{next}\isamarkupfalse%
\isanewline
\ \ \ \ \ \ \isacommand{case}\isamarkupfalse%
\ {\isacharparenleft}SUpdate\ e{\isadigit{1}}\ e{\isadigit{2}}\ s{\isacharprime}{\isacharparenright}\isanewline
\ \ \ \ \ \ \isacommand{from}\isamarkupfalse%
\ wts\ SUpdate\ \isacommand{obtain}\isamarkupfalse%
\ B\ \isakeyword{where}\ wte{\isadigit{1}}{\isacharcolon}\ {\isachardoublequoteopen}{\isasymGamma}\ {\isasymturnstile}\isactrlisub e\ e{\isadigit{1}}\ {\isacharcolon}\ RefT\ B{\isachardoublequoteclose}\isanewline
\ \ \ \ \ \ \ \ \isakeyword{and}\ wte{\isadigit{2}}{\isacharcolon}\ {\isachardoublequoteopen}{\isasymGamma}\ {\isasymturnstile}\isactrlisub e\ e{\isadigit{2}}\ {\isacharcolon}\ B{\isachardoublequoteclose}\ \isakeyword{and}\ wts{\isadigit{2}}{\isacharcolon}\ {\isachardoublequoteopen}{\isasymGamma}\ {\isasymturnstile}\isactrlisub s\ s{\isacharprime}\ {\isacharcolon}\ A{\isacharprime}{\isachardoublequoteclose}\ \isanewline
\ \ \ \ \ \ \ \ \isakeyword{and}\ sb{\isacharcolon}\ {\isachardoublequoteopen}static\ B{\isachardoublequoteclose}\ \isacommand{by}\isamarkupfalse%
\ fast\isanewline
\ \ \ \ \ \ \isacommand{from}\isamarkupfalse%
\ wte{\isadigit{1}}\ gr\ wt{\isacharunderscore}h\ Nil\ \isacommand{have}\isamarkupfalse%
\ {\isachardoublequoteopen}{\isacharparenleft}{\isasymexists}v{\isadigit{1}}{\isachardot}\ eval\ e{\isadigit{1}}\ {\isasymrho}\ {\isasymmu}\ {\isacharequal}\ Result\ v{\isadigit{1}}\ {\isasymand}\ {\isasymSigma}\ {\isasymturnstile}v\ v{\isadigit{1}}\ {\isacharcolon}\ RefT\ B{\isacharparenright}{\isachardoublequoteclose}\isanewline
\ \ \ \ \ \ \ \ \isacommand{using}\isamarkupfalse%
\ eval{\isacharunderscore}safe{\isacharbrackleft}of\ {\isasymGamma}\ e{\isadigit{1}}\ {\isachardoublequoteopen}RefT\ B{\isachardoublequoteclose}\ {\isasymSigma}\ {\isasymrho}\ {\isasymmu}{\isacharbrackright}\ \isacommand{by}\isamarkupfalse%
\ simp\isanewline
\ \ \ \ \ \ \isacommand{from}\isamarkupfalse%
\ this\ \isacommand{obtain}\isamarkupfalse%
\ v{\isadigit{1}}\ \isakeyword{where}\ v{\isadigit{1}}{\isacharcolon}\ {\isachardoublequoteopen}eval\ e{\isadigit{1}}\ {\isasymrho}\ {\isasymmu}\ {\isacharequal}\ Result\ v{\isadigit{1}}{\isachardoublequoteclose}\isanewline
\ \ \ \ \ \ \ \ \isakeyword{and}\ wtv{\isadigit{1}}{\isacharcolon}\ {\isachardoublequoteopen}{\isasymSigma}\ {\isasymturnstile}v\ v{\isadigit{1}}\ {\isacharcolon}\ RefT\ B{\isachardoublequoteclose}\ \isacommand{apply}\isamarkupfalse%
\ clarify\ \isacommand{apply}\isamarkupfalse%
\ assumption\ \isacommand{done}\isamarkupfalse%
\isanewline
\ \ \ \ \ \ \isacommand{from}\isamarkupfalse%
\ wte{\isadigit{2}}\ gr\ wt{\isacharunderscore}h\ Nil\ \isacommand{have}\isamarkupfalse%
\ {\isachardoublequoteopen}{\isacharparenleft}{\isasymexists}v{\isadigit{2}}{\isachardot}\ eval\ e{\isadigit{2}}\ {\isasymrho}\ {\isasymmu}\ {\isacharequal}\ Result\ v{\isadigit{2}}\ {\isasymand}\ {\isasymSigma}\ {\isasymturnstile}v\ v{\isadigit{2}}\ {\isacharcolon}\ B{\isacharparenright}{\isachardoublequoteclose}\isanewline
\ \ \ \ \ \ \ \ \isacommand{using}\isamarkupfalse%
\ eval{\isacharunderscore}safe{\isacharbrackleft}of\ {\isasymGamma}\ e{\isadigit{2}}\ {\isachardoublequoteopen}B{\isachardoublequoteclose}\ {\isasymSigma}\ {\isasymrho}\ {\isasymmu}{\isacharbrackright}\ \isacommand{by}\isamarkupfalse%
\ simp\isanewline
\ \ \ \ \ \ \isacommand{from}\isamarkupfalse%
\ this\ \isacommand{obtain}\isamarkupfalse%
\ v{\isadigit{2}}\ \isakeyword{where}\ v{\isadigit{2}}{\isacharcolon}\ {\isachardoublequoteopen}eval\ e{\isadigit{2}}\ {\isasymrho}\ {\isasymmu}\ {\isacharequal}\ Result\ v{\isadigit{2}}{\isachardoublequoteclose}\isanewline
\ \ \ \ \ \ \ \ \isakeyword{and}\ wtv{\isadigit{2}}{\isacharcolon}\ {\isachardoublequoteopen}{\isasymSigma}\ {\isasymturnstile}v\ v{\isadigit{2}}\ {\isacharcolon}\ B{\isachardoublequoteclose}\ \isacommand{by}\isamarkupfalse%
\ fast\isanewline
\ \ \ \ \ \ \isacommand{from}\isamarkupfalse%
\ wtv{\isadigit{1}}\ \isacommand{obtain}\isamarkupfalse%
\ a\ B{\isacharprime}\ \isakeyword{where}\ v{\isadigit{1}}a{\isacharcolon}\ {\isachardoublequoteopen}v{\isadigit{1}}\ {\isacharequal}\ VRef\ a{\isachardoublequoteclose}\isanewline
\ \ \ \ \ \ \ \ \isakeyword{and}\ las{\isacharcolon}\ {\isachardoublequoteopen}lookup\ a\ {\isasymSigma}\ {\isacharequal}\ Result\ B{\isacharprime}{\isachardoublequoteclose}\ \isakeyword{and}\ bb{\isacharcolon}\ {\isachardoublequoteopen}B{\isacharprime}\ {\isasymsqsubseteq}\ B{\isachardoublequoteclose}\isanewline
\ \ \ \ \ \ \ \ \isacommand{apply}\isamarkupfalse%
\ auto\ \isacommand{apply}\isamarkupfalse%
\ {\isacharparenleft}case{\isacharunderscore}tac\ c{\isacharparenright}\ \isacommand{apply}\isamarkupfalse%
\ auto\ \isacommand{done}\isamarkupfalse%
\isanewline
\ \ \ \ \ \ \isacommand{from}\isamarkupfalse%
\ sb\ bb\ \isacommand{have}\isamarkupfalse%
\ bbeq{\isacharcolon}\ {\isachardoublequoteopen}B\ {\isacharequal}\ B{\isacharprime}{\isachardoublequoteclose}\isanewline
\ \ \ \ \ \ \ \ \isacommand{using}\isamarkupfalse%
\ static{\isacharunderscore}is{\isacharunderscore}most{\isacharunderscore}precise\ \isacommand{apply}\isamarkupfalse%
\ blast\ \isacommand{done}\isamarkupfalse%
\isanewline
\ \ \ \ \ \ \isacommand{from}\isamarkupfalse%
\ las\ wt{\isacharunderscore}h\ Nil\ \isacommand{obtain}\isamarkupfalse%
\ v\ \isakeyword{where}\ lam{\isacharcolon}\ {\isachardoublequoteopen}lookup\ a\ {\isasymmu}\ {\isacharequal}\ Result\ {\isacharparenleft}Val\ v{\isacharcomma}B{\isacharprime}{\isacharparenright}{\isachardoublequoteclose}\isanewline
\ \ \ \ \ \ \ \ \isakeyword{and}\ wtv{\isacharcolon}\ {\isachardoublequoteopen}{\isasymSigma}\ {\isasymturnstile}v\ v\ {\isacharcolon}\ B{\isacharprime}{\isachardoublequoteclose}\isanewline
\ \ \ \ \ \ \ \ \isacommand{apply}\isamarkupfalse%
\ {\isacharparenleft}simp\ only{\isacharcolon}\ wt{\isacharunderscore}heap{\isacharunderscore}def{\isacharparenright}\ \isacommand{apply}\isamarkupfalse%
\ auto\isanewline
\ \ \ \ \ \ \ \ \isacommand{apply}\isamarkupfalse%
\ {\isacharparenleft}erule{\isacharunderscore}tac\ x{\isacharequal}a\ \isakeyword{in}\ allE{\isacharparenright}\isanewline
\ \ \ \ \ \ \ \ \isacommand{apply}\isamarkupfalse%
\ {\isacharparenleft}erule{\isacharunderscore}tac\ x{\isacharequal}a\ \isakeyword{in}\ allE{\isacharparenright}\isanewline
\ \ \ \ \ \ \ \ \isacommand{apply}\isamarkupfalse%
\ {\isacharparenleft}erule{\isacharunderscore}tac\ x{\isacharequal}B{\isacharprime}\ \isakeyword{in}\ allE{\isacharparenright}\isanewline
\ \ \ \ \ \ \ \ \isacommand{apply}\isamarkupfalse%
\ auto\ \isacommand{done}\isamarkupfalse%
\isanewline
\ \ \ \ \ \ \isacommand{let}\isamarkupfalse%
\ {\isacharquery}M{\isadigit{2}}\ {\isacharequal}\ {\isachardoublequoteopen}{\isacharparenleft}a{\isacharcomma}\ {\isacharparenleft}Val\ v{\isadigit{2}}{\isacharcomma}\ B{\isacharprime}{\isacharparenright}{\isacharparenright}\ {\isacharhash}\ {\isasymmu}{\isachardoublequoteclose}\isanewline
\ \ \ \ \ \ \isacommand{let}\isamarkupfalse%
\ {\isacharquery}S{\isadigit{2}}\ {\isacharequal}\ {\isachardoublequoteopen}{\isacharparenleft}a{\isacharcomma}B{\isacharprime}{\isacharparenright}{\isacharhash}{\isasymSigma}{\isachardoublequoteclose}\isanewline
\ \ \ \ \ \ \isacommand{from}\isamarkupfalse%
\ las\ \isacommand{have}\isamarkupfalse%
\ ss{\isacharcolon}\ {\isachardoublequoteopen}{\isacharquery}S{\isadigit{2}}\ {\isasymsqsubseteq}\ {\isasymSigma}{\isachardoublequoteclose}\ \isanewline
\ \ \ \ \ \ \ \ \isacommand{apply}\isamarkupfalse%
\ {\isacharparenleft}simp\ add{\isacharcolon}\ lesseq{\isacharunderscore}tyenv{\isacharunderscore}def\ dom{\isacharunderscore}def{\isacharparenright}\ \isacommand{apply}\isamarkupfalse%
\ auto\isanewline
\ \ \ \ \ \ \ \ \isacommand{using}\isamarkupfalse%
\ lookup{\isacharunderscore}dom{\isacharbrackleft}of\ a\ {\isasymSigma}{\isacharbrackright}\ dom{\isacharunderscore}def\ \isacommand{apply}\isamarkupfalse%
\ force\ \isacommand{done}\isamarkupfalse%
\isanewline
\ \ \ \ \ \ \isacommand{from}\isamarkupfalse%
\ Nil\ wt{\isacharunderscore}h\ wtv{\isadigit{2}}\ las\ bbeq\ \isacommand{have}\isamarkupfalse%
\ wth{\isadigit{2}}{\isacharcolon}\ {\isachardoublequoteopen}wt{\isacharunderscore}heap\ {\isacharquery}S{\isadigit{2}}\ {\isacharquery}M{\isadigit{2}}\ {\isacharbraceleft}{\isacharbraceright}{\isachardoublequoteclose}\ \isanewline
\ \ \ \ \ \ \ \ \isacommand{using}\isamarkupfalse%
\ update{\isacharunderscore}heap{\isacharunderscore}val{\isacharbrackleft}of\ {\isasymSigma}\ {\isasymmu}\ {\isachardoublequoteopen}set\ ads{\isachardoublequoteclose}\ a\ B{\isacharprime}\ B{\isacharprime}\ {\isachardoublequoteopen}Val\ v{\isadigit{2}}{\isachardoublequoteclose}{\isacharbrackright}\isanewline
\ \ \ \ \ \ \ \ \isacommand{apply}\isamarkupfalse%
\ simp\ \isacommand{apply}\isamarkupfalse%
\ auto\ \isacommand{done}\isamarkupfalse%
\isanewline
\ \ \ \ \ \ \isacommand{have}\isamarkupfalse%
\ wts{\isadigit{2}}{\isacharcolon}\ {\isachardoublequoteopen}wt{\isacharunderscore}state\ {\isacharparenleft}s{\isacharprime}{\isacharcomma}\ {\isasymrho}{\isacharcomma}\ k{\isacharcomma}\ {\isacharquery}M{\isadigit{2}}{\isacharcomma}\ {\isacharbrackleft}{\isacharbrackright}{\isacharparenright}\ A{\isachardoublequoteclose}\ \isanewline
\ \ \ \ \ \ \ \ \isacommand{apply}\isamarkupfalse%
\ {\isacharparenleft}rule\ wts{\isacharunderscore}intro{\isacharparenright}\isanewline
\ \ \ \ \ \ \ \ \isacommand{apply}\isamarkupfalse%
\ simp\isanewline
\ \ \ \ \ \ \ \ \isacommand{using}\isamarkupfalse%
\ wth{\isadigit{2}}\ \isacommand{apply}\isamarkupfalse%
\ simp\isanewline
\ \ \ \ \ \ \ \ \isacommand{using}\isamarkupfalse%
\ gr\ ss\ strengthen{\isacharunderscore}value{\isacharunderscore}env\ \isacommand{apply}\isamarkupfalse%
\ blast\isanewline
\ \ \ \ \ \ \ \ \isacommand{using}\isamarkupfalse%
\ wts{\isadigit{2}}\ \isacommand{apply}\isamarkupfalse%
\ blast\isanewline
\ \ \ \ \ \ \ \ \isacommand{using}\isamarkupfalse%
\ wt{\isacharunderscore}k\ ss\ strengthen{\isacharunderscore}stack\ \isacommand{apply}\isamarkupfalse%
\ blast\isanewline
\ \ \ \ \ \ \ \ \isacommand{done}\isamarkupfalse%
\isanewline
\ \ \ \ \ \ \isacommand{from}\isamarkupfalse%
\ st\ Nil\ v{\isadigit{1}}\ v{\isadigit{2}}\ SUpdate\ lam\ v{\isadigit{1}}a\ wts{\isadigit{2}}\ \isacommand{show}\isamarkupfalse%
\ {\isacharquery}thesis\ \isacommand{by}\isamarkupfalse%
\ simp\isanewline
\ \ \ \ \isacommand{next}\isamarkupfalse%
\isanewline
\ \ \ \ \ \ \isacommand{case}\isamarkupfalse%
\ {\isacharparenleft}SDynUpdate\ e{\isadigit{1}}\ e{\isadigit{2}}\ B\ s{\isacharprime}{\isacharparenright}\isanewline
\ \ \ \ \ \ \isacommand{from}\isamarkupfalse%
\ wts\ SDynUpdate\ \isacommand{have}\isamarkupfalse%
\ wte{\isadigit{1}}{\isacharcolon}\ {\isachardoublequoteopen}{\isasymGamma}\ {\isasymturnstile}\isactrlisub e\ e{\isadigit{1}}\ {\isacharcolon}\ RefT\ B{\isachardoublequoteclose}\ \isacommand{by}\isamarkupfalse%
\ fast\isanewline
\ \ \ \ \ \ \isacommand{from}\isamarkupfalse%
\ \ wts\ SDynUpdate\ \isacommand{have}\isamarkupfalse%
\ wte{\isadigit{2}}{\isacharcolon}\ {\isachardoublequoteopen}{\isasymGamma}\ {\isasymturnstile}\isactrlisub e\ e{\isadigit{2}}\ {\isacharcolon}\ B{\isachardoublequoteclose}\ \isacommand{by}\isamarkupfalse%
\ fast\isanewline
\ \ \ \ \ \ \isacommand{from}\isamarkupfalse%
\ wts\ SDynUpdate\ \isacommand{have}\isamarkupfalse%
\ wts{\isadigit{2}}{\isacharcolon}\ {\isachardoublequoteopen}{\isasymGamma}\ {\isasymturnstile}\isactrlisub s\ s{\isacharprime}\ {\isacharcolon}\ A{\isacharprime}{\isachardoublequoteclose}\ \ \isacommand{by}\isamarkupfalse%
\ fast\ \isanewline
\ \ \ \ \ \ \isacommand{from}\isamarkupfalse%
\ wte{\isadigit{1}}\ gr\ wt{\isacharunderscore}h\ Nil\ \isacommand{have}\isamarkupfalse%
\ {\isachardoublequoteopen}{\isacharparenleft}{\isasymexists}v{\isadigit{1}}{\isachardot}\ eval\ e{\isadigit{1}}\ {\isasymrho}\ {\isasymmu}\ {\isacharequal}\ Result\ v{\isadigit{1}}\ {\isasymand}\ {\isasymSigma}\ {\isasymturnstile}v\ v{\isadigit{1}}\ {\isacharcolon}\ RefT\ B{\isacharparenright}{\isachardoublequoteclose}\isanewline
\ \ \ \ \ \ \ \ \isacommand{using}\isamarkupfalse%
\ eval{\isacharunderscore}safe{\isacharbrackleft}of\ {\isasymGamma}\ e{\isadigit{1}}\ {\isachardoublequoteopen}RefT\ B{\isachardoublequoteclose}\ {\isasymSigma}\ {\isasymrho}\ {\isasymmu}{\isacharbrackright}\ \isacommand{by}\isamarkupfalse%
\ simp\isanewline
\ \ \ \ \ \ \isacommand{from}\isamarkupfalse%
\ this\ \isacommand{obtain}\isamarkupfalse%
\ v{\isadigit{1}}\ \isakeyword{where}\ v{\isadigit{1}}{\isacharcolon}\ {\isachardoublequoteopen}eval\ e{\isadigit{1}}\ {\isasymrho}\ {\isasymmu}\ {\isacharequal}\ Result\ v{\isadigit{1}}{\isachardoublequoteclose}\isanewline
\ \ \ \ \ \ \ \ \isakeyword{and}\ wtv{\isadigit{1}}{\isacharcolon}\ {\isachardoublequoteopen}{\isasymSigma}\ {\isasymturnstile}v\ v{\isadigit{1}}\ {\isacharcolon}\ RefT\ B{\isachardoublequoteclose}\ \isacommand{apply}\isamarkupfalse%
\ clarify\ \isacommand{apply}\isamarkupfalse%
\ assumption\ \isacommand{done}\isamarkupfalse%
\isanewline
\ \ \ \ \ \ \isacommand{from}\isamarkupfalse%
\ wte{\isadigit{2}}\ gr\ wt{\isacharunderscore}h\ Nil\ \isacommand{have}\isamarkupfalse%
\ {\isachardoublequoteopen}{\isacharparenleft}{\isasymexists}v{\isadigit{2}}{\isachardot}\ eval\ e{\isadigit{2}}\ {\isasymrho}\ {\isasymmu}\ {\isacharequal}\ Result\ v{\isadigit{2}}\ {\isasymand}\ {\isasymSigma}\ {\isasymturnstile}v\ v{\isadigit{2}}\ {\isacharcolon}\ B{\isacharparenright}{\isachardoublequoteclose}\isanewline
\ \ \ \ \ \ \ \ \isacommand{using}\isamarkupfalse%
\ eval{\isacharunderscore}safe{\isacharbrackleft}of\ {\isasymGamma}\ e{\isadigit{2}}\ {\isachardoublequoteopen}B{\isachardoublequoteclose}\ {\isasymSigma}\ {\isasymrho}\ {\isasymmu}{\isacharbrackright}\ \isacommand{by}\isamarkupfalse%
\ simp\isanewline
\ \ \ \ \ \ \isacommand{from}\isamarkupfalse%
\ this\ \isacommand{obtain}\isamarkupfalse%
\ v{\isadigit{2}}\ \isakeyword{where}\ v{\isadigit{2}}{\isacharcolon}\ {\isachardoublequoteopen}eval\ e{\isadigit{2}}\ {\isasymrho}\ {\isasymmu}\ {\isacharequal}\ Result\ v{\isadigit{2}}{\isachardoublequoteclose}\isanewline
\ \ \ \ \ \ \ \ \isakeyword{and}\ wtv{\isadigit{2}}{\isacharcolon}\ {\isachardoublequoteopen}{\isasymSigma}\ {\isasymturnstile}v\ v{\isadigit{2}}\ {\isacharcolon}\ B{\isachardoublequoteclose}\ \isacommand{by}\isamarkupfalse%
\ fast\isanewline
\ \ \ \ \ \ \isacommand{from}\isamarkupfalse%
\ wtv{\isadigit{1}}\ \isacommand{obtain}\isamarkupfalse%
\ a\ B{\isacharprime}\ \isakeyword{where}\ v{\isadigit{1}}a{\isacharcolon}\ {\isachardoublequoteopen}v{\isadigit{1}}\ {\isacharequal}\ VRef\ a{\isachardoublequoteclose}\isanewline
\ \ \ \ \ \ \ \ \isakeyword{and}\ las{\isacharcolon}\ {\isachardoublequoteopen}lookup\ a\ {\isasymSigma}\ {\isacharequal}\ Result\ B{\isacharprime}{\isachardoublequoteclose}\ \isakeyword{and}\ bb{\isacharcolon}\ {\isachardoublequoteopen}B{\isacharprime}\ {\isasymsqsubseteq}\ B{\isachardoublequoteclose}\isanewline
\ \ \ \ \ \ \ \ \isacommand{apply}\isamarkupfalse%
\ auto\ \isacommand{apply}\isamarkupfalse%
\ {\isacharparenleft}case{\isacharunderscore}tac\ c{\isacharparenright}\ \isacommand{apply}\isamarkupfalse%
\ auto\ \isacommand{done}\isamarkupfalse%
\isanewline
\ \ \ \ \ \ \isacommand{from}\isamarkupfalse%
\ las\ wt{\isacharunderscore}h\ Nil\ \isacommand{obtain}\isamarkupfalse%
\ v\ \isakeyword{where}\ lam{\isacharcolon}\ {\isachardoublequoteopen}lookup\ a\ {\isasymmu}\ {\isacharequal}\ Result\ {\isacharparenleft}Val\ v{\isacharcomma}B{\isacharprime}{\isacharparenright}{\isachardoublequoteclose}\isanewline
\ \ \ \ \ \ \ \ \isakeyword{and}\ wtv{\isacharcolon}\ {\isachardoublequoteopen}{\isasymSigma}\ {\isasymturnstile}v\ v\ {\isacharcolon}\ B{\isacharprime}{\isachardoublequoteclose}\isanewline
\ \ \ \ \ \ \ \ \isacommand{apply}\isamarkupfalse%
\ {\isacharparenleft}simp\ only{\isacharcolon}\ wt{\isacharunderscore}heap{\isacharunderscore}def{\isacharparenright}\ \isacommand{apply}\isamarkupfalse%
\ auto\isanewline
\ \ \ \ \ \ \ \ \isacommand{apply}\isamarkupfalse%
\ {\isacharparenleft}erule{\isacharunderscore}tac\ x{\isacharequal}a\ \isakeyword{in}\ allE{\isacharparenright}\isanewline
\ \ \ \ \ \ \ \ \isacommand{apply}\isamarkupfalse%
\ {\isacharparenleft}erule{\isacharunderscore}tac\ x{\isacharequal}a\ \isakeyword{in}\ allE{\isacharparenright}\isanewline
\ \ \ \ \ \ \ \ \isacommand{apply}\isamarkupfalse%
\ {\isacharparenleft}erule{\isacharunderscore}tac\ x{\isacharequal}B{\isacharprime}\ \isakeyword{in}\ allE{\isacharparenright}\isanewline
\ \ \ \ \ \ \ \ \isacommand{apply}\isamarkupfalse%
\ auto\ \isacommand{done}\isamarkupfalse%
\isanewline
\ \ \ \ \ \ \isacommand{from}\isamarkupfalse%
\ lam\ Nil\ v{\isadigit{1}}\ v{\isadigit{2}}\ SDynUpdate\ st\ v{\isadigit{1}}a\isanewline
\ \ \ \ \ \ \isacommand{have}\isamarkupfalse%
\ steps{\isacharcolon}\ {\isachardoublequoteopen}step\ s\ {\isacharequal}\ Result\ {\isacharparenleft}s{\isacharprime}{\isacharcomma}\ {\isasymrho}{\isacharcomma}\ k{\isacharcomma}\ {\isacharparenleft}a{\isacharcomma}\ VCast\ v{\isadigit{2}}\ B\ B{\isacharprime}{\isacharcomma}\ B{\isacharprime}{\isacharparenright}\ {\isacharhash}\ {\isasymmu}{\isacharcomma}\ {\isacharbrackleft}a{\isacharbrackright}{\isacharparenright}{\isachardoublequoteclose}\isanewline
\ \ \ \ \ \ \ \ \isacommand{by}\isamarkupfalse%
\ simp\isanewline
\ \ \ \ \ \ \isacommand{from}\isamarkupfalse%
\ wtv{\isadigit{2}}\ bb\ \isacommand{have}\isamarkupfalse%
\ wtcv{\isadigit{2}}{\isacharcolon}\ {\isachardoublequoteopen}{\isasymSigma}\ {\isasymturnstile}cv\ VCast\ v{\isadigit{2}}\ B\ B{\isacharprime}\ {\isacharcolon}\ B{\isacharprime}{\isachardoublequoteclose}\ \isacommand{by}\isamarkupfalse%
\ blast\isanewline
\ \ \ \ \ \ \isacommand{let}\isamarkupfalse%
\ {\isacharquery}S{\isadigit{2}}\ {\isacharequal}\ {\isachardoublequoteopen}{\isacharparenleft}a{\isacharcomma}\ B{\isacharprime}{\isacharparenright}\ {\isacharhash}\ {\isasymSigma}{\isachardoublequoteclose}\isanewline
\ \ \ \ \ \ \isacommand{let}\isamarkupfalse%
\ {\isacharquery}M{\isadigit{2}}\ {\isacharequal}\ {\isachardoublequoteopen}{\isacharparenleft}a{\isacharcomma}\ VCast\ v{\isadigit{2}}\ B\ B{\isacharprime}{\isacharcomma}\ B{\isacharprime}{\isacharparenright}\ {\isacharhash}\ {\isasymmu}{\isachardoublequoteclose}\isanewline
\ \ \ \ \ \ \isacommand{from}\isamarkupfalse%
\ wt{\isacharunderscore}h\ Nil\ las\ wtcv{\isadigit{2}}\ \isacommand{have}\isamarkupfalse%
\ wt{\isacharunderscore}h{\isadigit{2}}{\isacharcolon}\ {\isachardoublequoteopen}wt{\isacharunderscore}heap\ {\isacharquery}S{\isadigit{2}}\ {\isacharquery}M{\isadigit{2}}\ {\isacharbraceleft}a{\isacharbraceright}{\isachardoublequoteclose}\isanewline
\ \ \ \ \ \ \ \ \isacommand{using}\isamarkupfalse%
\ update{\isacharunderscore}heap{\isacharunderscore}val{\isacharbrackleft}of\ {\isasymSigma}\ {\isasymmu}\ {\isachardoublequoteopen}{\isacharbraceleft}{\isacharbraceright}{\isachardoublequoteclose}\ a\ B{\isacharprime}\ B{\isacharprime}\ {\isachardoublequoteopen}VCast\ v{\isadigit{2}}\ B\ B{\isacharprime}{\isachardoublequoteclose}{\isacharbrackright}\ \isacommand{by}\isamarkupfalse%
\ auto\isanewline
\ \ \ \ \ \ \isacommand{from}\isamarkupfalse%
\ las\ \isacommand{have}\isamarkupfalse%
\ ss{\isacharcolon}\ {\isachardoublequoteopen}{\isacharquery}S{\isadigit{2}}\ {\isasymsqsubseteq}\ {\isasymSigma}{\isachardoublequoteclose}\isanewline
\ \ \ \ \ \ \ \ \isacommand{apply}\isamarkupfalse%
\ {\isacharparenleft}simp\ add{\isacharcolon}\ lesseq{\isacharunderscore}tyenv{\isacharunderscore}def{\isacharparenright}\ \isanewline
\ \ \ \ \ \ \ \ \isacommand{apply}\isamarkupfalse%
\ {\isacharparenleft}simp\ add{\isacharcolon}\ dom{\isacharunderscore}def{\isacharparenright}\isanewline
\ \ \ \ \ \ \ \ \isacommand{apply}\isamarkupfalse%
\ {\isacharparenleft}frule\ lookup{\isacharunderscore}dom{\isacharparenright}\isanewline
\ \ \ \ \ \ \ \ \isacommand{apply}\isamarkupfalse%
\ {\isacharparenleft}simp\ add{\isacharcolon}\ dom{\isacharunderscore}def{\isacharparenright}\ \isacommand{apply}\isamarkupfalse%
\ blast\ \isacommand{done}\isamarkupfalse%
\isanewline
\ \ \ \ \ \ \isacommand{have}\isamarkupfalse%
\ wt{\isacharunderscore}s{\isacharcolon}\ {\isachardoublequoteopen}wt{\isacharunderscore}state\ {\isacharparenleft}s{\isacharprime}{\isacharcomma}\ {\isasymrho}{\isacharcomma}\ k{\isacharcomma}\ {\isacharquery}M{\isadigit{2}}{\isacharcomma}\ {\isacharbrackleft}a{\isacharbrackright}{\isacharparenright}\ A{\isachardoublequoteclose}\isanewline
\ \ \ \ \ \ \ \ \isacommand{apply}\isamarkupfalse%
\ {\isacharparenleft}rule\ wts{\isacharunderscore}intro{\isacharparenright}\isanewline
\ \ \ \ \ \ \ \ \isacommand{using}\isamarkupfalse%
\ wt{\isacharunderscore}h{\isadigit{2}}\ \isacommand{apply}\isamarkupfalse%
\ simp\isanewline
\ \ \ \ \ \ \ \ \isacommand{using}\isamarkupfalse%
\ gr\ ss\ strengthen{\isacharunderscore}value{\isacharunderscore}env\ \isacommand{apply}\isamarkupfalse%
\ blast\isanewline
\ \ \ \ \ \ \ \ \isacommand{using}\isamarkupfalse%
\ wts{\isadigit{2}}\ \isacommand{apply}\isamarkupfalse%
\ simp\isanewline
\ \ \ \ \ \ \ \ \isacommand{using}\isamarkupfalse%
\ wt{\isacharunderscore}k\ ss\ strengthen{\isacharunderscore}stack\ \isacommand{apply}\isamarkupfalse%
\ blast\ \isacommand{done}\isamarkupfalse%
\isanewline
\ \ \ \ \ \ \isacommand{from}\isamarkupfalse%
\ steps\ wt{\isacharunderscore}s\ \isacommand{show}\isamarkupfalse%
\ {\isacharquery}thesis\ \isacommand{by}\isamarkupfalse%
\ simp\isanewline
\ \ \ \ \isacommand{next}\isamarkupfalse%
\isanewline
\ \ \ \ \ \ \isacommand{case}\isamarkupfalse%
\ {\isacharparenleft}SCast\ x\ e\ B\ C\ s{\isacharprime}{\isacharparenright}\isanewline
\ \ \ \ \ \ \isacommand{from}\isamarkupfalse%
\ wts\ SCast\ \isacommand{have}\isamarkupfalse%
\ wte{\isacharcolon}\ {\isachardoublequoteopen}{\isasymGamma}\ {\isasymturnstile}\isactrlisub e\ e\ {\isacharcolon}\ B{\isachardoublequoteclose}\ \isacommand{by}\isamarkupfalse%
\ auto\isanewline
\ \ \ \ \ \ \isacommand{from}\isamarkupfalse%
\ wts\ SCast\ \isacommand{have}\isamarkupfalse%
\ wts{\isadigit{2}}{\isacharcolon}\ {\isachardoublequoteopen}{\isacharparenleft}x{\isacharcomma}C{\isacharparenright}{\isacharhash}{\isasymGamma}\ {\isasymturnstile}\isactrlisub s\ s{\isacharprime}\ {\isacharcolon}\ A{\isacharprime}{\isachardoublequoteclose}\ \isacommand{by}\isamarkupfalse%
\ auto\isanewline
\ \ \ \ \ \ \isacommand{from}\isamarkupfalse%
\ wte\ gr\ wt{\isacharunderscore}h\ Nil\ \isacommand{have}\isamarkupfalse%
\ {\isachardoublequoteopen}{\isacharparenleft}{\isasymexists}v{\isachardot}\ eval\ e\ {\isasymrho}\ {\isasymmu}\ {\isacharequal}\ Result\ v\ {\isasymand}\ {\isasymSigma}\ {\isasymturnstile}v\ v\ {\isacharcolon}\ B{\isacharparenright}{\isachardoublequoteclose}\isanewline
\ \ \ \ \ \ \ \ \isacommand{using}\isamarkupfalse%
\ eval{\isacharunderscore}safe{\isacharbrackleft}of\ {\isasymGamma}\ e\ B\ {\isasymSigma}\ {\isasymrho}\ {\isasymmu}{\isacharbrackright}\ \isacommand{by}\isamarkupfalse%
\ simp\isanewline
\ \ \ \ \ \ \isacommand{from}\isamarkupfalse%
\ this\ \isacommand{obtain}\isamarkupfalse%
\ v\ \isakeyword{where}\ v{\isacharcolon}\ {\isachardoublequoteopen}eval\ e\ {\isasymrho}\ {\isasymmu}\ {\isacharequal}\ Result\ v{\isachardoublequoteclose}\isanewline
\ \ \ \ \ \ \ \ \isakeyword{and}\ wtv{\isacharcolon}\ {\isachardoublequoteopen}{\isasymSigma}\ {\isasymturnstile}v\ v\ {\isacharcolon}\ B{\isachardoublequoteclose}\ \isacommand{using}\isamarkupfalse%
\ eval{\isacharunderscore}safe\ \isacommand{by}\isamarkupfalse%
\ blast\isanewline
\ \ \ \ \ \ \isacommand{from}\isamarkupfalse%
\ wtv\ wt{\isacharunderscore}h\ \isacommand{have}\isamarkupfalse%
\ {\isachardoublequoteopen}{\isacharparenleft}{\isasymexists}v{\isacharprime}\ {\isasymSigma}{\isacharprime}\ {\isasymmu}{\isacharprime}\ ads{\isadigit{2}}{\isachardot}\isanewline
\ \ \ \ \ \ \ \ cast\ v\ B\ C\ {\isasymmu}\ ads\ {\isacharequal}\ Result\ {\isacharparenleft}v{\isacharprime}{\isacharcomma}\ {\isasymmu}{\isacharprime}{\isacharcomma}\ ads{\isadigit{2}}{\isacharparenright}\ {\isasymand}\isanewline
\ \ \ \ \ \ \ \ {\isasymSigma}{\isacharprime}\ {\isasymturnstile}v\ v{\isacharprime}\ {\isacharcolon}\ C\ {\isasymand}\ wt{\isacharunderscore}heap\ {\isasymSigma}{\isacharprime}\ {\isasymmu}{\isacharprime}\ {\isacharparenleft}set\ ads{\isadigit{2}}{\isacharparenright}\ {\isasymand}\ {\isasymSigma}{\isacharprime}\ {\isasymsqsubseteq}\ {\isasymSigma}{\isacharparenright}\ {\isasymor}\isanewline
\ \ \ \ \ \ \ \ cast\ v\ B\ C\ {\isasymmu}\ ads\ {\isacharequal}\ CastError{\isachardoublequoteclose}\ \isacommand{using}\isamarkupfalse%
\ cast{\isacharunderscore}safe\ \isacommand{by}\isamarkupfalse%
\ blast\isanewline
\ \ \ \ \ \ \isacommand{thus}\isamarkupfalse%
\ {\isacharquery}thesis\isanewline
\ \ \ \ \ \ \isacommand{proof}\isamarkupfalse%
\isanewline
\ \ \ \ \ \ \ \ \isacommand{assume}\isamarkupfalse%
\ {\isachardoublequoteopen}{\isacharparenleft}{\isasymexists}v{\isacharprime}\ {\isasymSigma}{\isacharprime}\ {\isasymmu}{\isacharprime}\ ads{\isadigit{2}}{\isachardot}\isanewline
\ \ \ \ \ \ \ \ \ \ cast\ v\ B\ C\ {\isasymmu}\ ads\ {\isacharequal}\ Result\ {\isacharparenleft}v{\isacharprime}{\isacharcomma}\ {\isasymmu}{\isacharprime}{\isacharcomma}\ ads{\isadigit{2}}{\isacharparenright}\ {\isasymand}\isanewline
\ \ \ \ \ \ \ \ \ \ {\isasymSigma}{\isacharprime}\ {\isasymturnstile}v\ v{\isacharprime}\ {\isacharcolon}\ C\ {\isasymand}\ wt{\isacharunderscore}heap\ {\isasymSigma}{\isacharprime}\ {\isasymmu}{\isacharprime}\ {\isacharparenleft}set\ ads{\isadigit{2}}{\isacharparenright}\ {\isasymand}\ {\isasymSigma}{\isacharprime}\ {\isasymsqsubseteq}\ {\isasymSigma}{\isacharparenright}{\isachardoublequoteclose}\isanewline
\ \ \ \ \ \ \ \ \isacommand{from}\isamarkupfalse%
\ this\ \isacommand{obtain}\isamarkupfalse%
\ v{\isacharprime}\ {\isasymSigma}{\isacharprime}\ {\isasymmu}{\isacharprime}\ ads{\isadigit{2}}\ \isakeyword{where}\isanewline
\ \ \ \ \ \ \ \ \ \ castv{\isacharcolon}\ {\isachardoublequoteopen}cast\ v\ B\ C\ {\isasymmu}\ ads\ {\isacharequal}\ Result\ {\isacharparenleft}v{\isacharprime}{\isacharcomma}{\isasymmu}{\isacharprime}{\isacharcomma}ads{\isadigit{2}}{\isacharparenright}{\isachardoublequoteclose}\ \isakeyword{and}\isanewline
\ \ \ \ \ \ \ \ \ \ wtvp{\isacharcolon}\ {\isachardoublequoteopen}{\isasymSigma}{\isacharprime}\ {\isasymturnstile}v\ v{\isacharprime}\ {\isacharcolon}\ C{\isachardoublequoteclose}\ \isakeyword{and}\ wth{\isadigit{2}}{\isacharcolon}\ {\isachardoublequoteopen}wt{\isacharunderscore}heap\ {\isasymSigma}{\isacharprime}\ {\isasymmu}{\isacharprime}\ {\isacharparenleft}set\ ads{\isadigit{2}}{\isacharparenright}{\isachardoublequoteclose}\ \isakeyword{and}\isanewline
\ \ \ \ \ \ \ \ \ \ ss{\isacharcolon}\ {\isachardoublequoteopen}{\isasymSigma}{\isacharprime}\ {\isasymsqsubseteq}\ {\isasymSigma}{\isachardoublequoteclose}\ \isacommand{by}\isamarkupfalse%
\ blast\isanewline
\ \ \ \ \ \ \ \ \isacommand{let}\isamarkupfalse%
\ {\isacharquery}R{\isadigit{2}}\ {\isacharequal}\ {\isachardoublequoteopen}{\isacharparenleft}x{\isacharcomma}v{\isacharprime}{\isacharparenright}{\isacharhash}{\isasymrho}{\isachardoublequoteclose}\isanewline
\ \ \ \ \ \ \ \ \isacommand{let}\isamarkupfalse%
\ {\isacharquery}G{\isadigit{2}}\ {\isacharequal}\ {\isachardoublequoteopen}{\isacharparenleft}x{\isacharcomma}C{\isacharparenright}{\isacharhash}{\isasymGamma}{\isachardoublequoteclose}\isanewline
\ \ \ \ \ \ \ \ \isacommand{from}\isamarkupfalse%
\ gr\ ss\ \isacommand{have}\isamarkupfalse%
\ gr{\isadigit{2}}{\isacharcolon}\ {\isachardoublequoteopen}{\isasymGamma}{\isacharsemicolon}{\isasymSigma}{\isacharprime}\ {\isasymturnstile}\ {\isasymrho}{\isachardoublequoteclose}\ \isacommand{using}\isamarkupfalse%
\ strengthen{\isacharunderscore}value{\isacharunderscore}env\ \isacommand{by}\isamarkupfalse%
\ blast\isanewline
\ \ \ \ \ \ \ \ \isacommand{from}\isamarkupfalse%
\ wtvp\ gr{\isadigit{2}}\ \isacommand{have}\isamarkupfalse%
\ gr{\isadigit{3}}{\isacharcolon}\ {\isachardoublequoteopen}{\isacharquery}G{\isadigit{2}}{\isacharsemicolon}{\isasymSigma}{\isacharprime}\ {\isasymturnstile}\ {\isacharquery}R{\isadigit{2}}{\isachardoublequoteclose}\ \isacommand{by}\isamarkupfalse%
\ {\isacharparenleft}rule\ wt{\isacharunderscore}cons{\isacharparenright}\isanewline
\ \ \ \ \ \ \ \ \isacommand{from}\isamarkupfalse%
\ castv\ SCast\ Nil\ v\ st\isanewline
\ \ \ \ \ \ \ \ \isacommand{have}\isamarkupfalse%
\ steps{\isacharcolon}\ {\isachardoublequoteopen}step\ s\ {\isacharequal}\ Result\ {\isacharparenleft}s{\isacharprime}{\isacharcomma}\ {\isacharparenleft}x{\isacharcomma}\ v{\isacharprime}{\isacharparenright}\ {\isacharhash}\ {\isasymrho}{\isacharcomma}\ k{\isacharcomma}\ {\isasymmu}{\isacharprime}{\isacharcomma}\ ads{\isadigit{2}}{\isacharparenright}{\isachardoublequoteclose}\ \isacommand{by}\isamarkupfalse%
\ simp\isanewline
\ \ \ \ \ \ \ \ \isacommand{have}\isamarkupfalse%
\ wt{\isacharunderscore}s{\isacharcolon}\ {\isachardoublequoteopen}wt{\isacharunderscore}state\ {\isacharparenleft}s{\isacharprime}{\isacharcomma}\ {\isacharparenleft}x{\isacharcomma}\ v{\isacharprime}{\isacharparenright}\ {\isacharhash}\ {\isasymrho}{\isacharcomma}\ k{\isacharcomma}\ {\isasymmu}{\isacharprime}{\isacharcomma}\ ads{\isadigit{2}}{\isacharparenright}\ A{\isachardoublequoteclose}\ \isanewline
\ \ \ \ \ \ \ \ \ \ \isacommand{apply}\isamarkupfalse%
\ {\isacharparenleft}rule\ wts{\isacharunderscore}intro{\isacharparenright}\isanewline
\ \ \ \ \ \ \ \ \ \ \isacommand{using}\isamarkupfalse%
\ wth{\isadigit{2}}\ Nil\ \isacommand{apply}\isamarkupfalse%
\ simp\isanewline
\ \ \ \ \ \ \ \ \ \ \isacommand{using}\isamarkupfalse%
\ gr{\isadigit{3}}\ \isacommand{apply}\isamarkupfalse%
\ simp\isanewline
\ \ \ \ \ \ \ \ \ \ \isacommand{using}\isamarkupfalse%
\ wts{\isadigit{2}}\ \isacommand{apply}\isamarkupfalse%
\ simp\isanewline
\ \ \ \ \ \ \ \ \ \ \isacommand{using}\isamarkupfalse%
\ wt{\isacharunderscore}k\ ss\ strengthen{\isacharunderscore}stack\ \isacommand{apply}\isamarkupfalse%
\ blast\ \isacommand{done}\isamarkupfalse%
\isanewline
\ \ \ \ \ \ \ \ \isacommand{from}\isamarkupfalse%
\ steps\ wt{\isacharunderscore}s\ \isacommand{show}\isamarkupfalse%
\ {\isacharquery}thesis\ \isacommand{by}\isamarkupfalse%
\ blast\isanewline
\ \ \ \ \ \ \isacommand{next}\isamarkupfalse%
\isanewline
\ \ \ \ \ \ \ \ \isacommand{assume}\isamarkupfalse%
\ {\isachardoublequoteopen}cast\ v\ B\ C\ {\isasymmu}\ ads\ {\isacharequal}\ CastError{\isachardoublequoteclose}\isanewline
\ \ \ \ \ \ \ \ \isacommand{from}\isamarkupfalse%
\ this\ SCast\ Nil\ st\ v\ \isacommand{show}\isamarkupfalse%
\ {\isacharquery}thesis\ \isacommand{by}\isamarkupfalse%
\ simp\isanewline
\ \ \ \ \ \ \isacommand{qed}\isamarkupfalse%
\isanewline
\ \ \isacommand{next}\isamarkupfalse%
\isanewline
\ \ \ \ \isacommand{case}\isamarkupfalse%
\ {\isacharparenleft}SDynDeref\ x\ e\ B\ s{\isacharprime}{\isacharparenright}\isanewline
\ \ \ \ \isacommand{from}\isamarkupfalse%
\ wts\ SDynDeref\ \isacommand{have}\isamarkupfalse%
\ wte{\isacharcolon}\ {\isachardoublequoteopen}{\isasymGamma}\ {\isasymturnstile}\isactrlisub e\ e\ {\isacharcolon}\ RefT\ B{\isachardoublequoteclose}\ \isacommand{by}\isamarkupfalse%
\ blast\isanewline
\ \ \ \ \isacommand{from}\isamarkupfalse%
\ wts\ SDynDeref\ \isacommand{have}\isamarkupfalse%
\ wtsp{\isacharcolon}\ {\isachardoublequoteopen}{\isacharparenleft}x{\isacharcomma}B{\isacharparenright}{\isacharhash}{\isasymGamma}\ {\isasymturnstile}\isactrlisub s\ s{\isacharprime}\ {\isacharcolon}\ A{\isacharprime}{\isachardoublequoteclose}\ \isacommand{by}\isamarkupfalse%
\ blast\isanewline
\ \ \ \ \isacommand{from}\isamarkupfalse%
\ wte\ gr\ wt{\isacharunderscore}h\ Nil\ \isacommand{have}\isamarkupfalse%
\ {\isachardoublequoteopen}{\isacharparenleft}{\isasymexists}v{\isachardot}\ eval\ e\ {\isasymrho}\ {\isasymmu}\ {\isacharequal}\ Result\ v\ {\isasymand}\ {\isasymSigma}\ {\isasymturnstile}v\ v\ {\isacharcolon}\ RefT\ B{\isacharparenright}{\isachardoublequoteclose}\isanewline
\ \ \ \ \ \ \isacommand{using}\isamarkupfalse%
\ eval{\isacharunderscore}safe{\isacharbrackleft}of\ {\isasymGamma}\ e\ {\isachardoublequoteopen}RefT\ B{\isachardoublequoteclose}\ {\isasymSigma}\ {\isasymrho}\ {\isasymmu}{\isacharbrackright}\ \isacommand{by}\isamarkupfalse%
\ simp\isanewline
\ \ \ \ \isacommand{from}\isamarkupfalse%
\ this\ \isacommand{obtain}\isamarkupfalse%
\ v\ \isakeyword{where}\ v{\isacharcolon}\ {\isachardoublequoteopen}eval\ e\ {\isasymrho}\ {\isasymmu}\ {\isacharequal}\ Result\ v{\isachardoublequoteclose}\isanewline
\ \ \ \ \ \ \isakeyword{and}\ wtv{\isacharcolon}\ {\isachardoublequoteopen}{\isasymSigma}\ {\isasymturnstile}v\ v\ {\isacharcolon}\ RefT\ B{\isachardoublequoteclose}\ \isacommand{using}\isamarkupfalse%
\ eval{\isacharunderscore}safe\ \isacommand{by}\isamarkupfalse%
\ blast\isanewline
\ \ \ \ \isacommand{from}\isamarkupfalse%
\ wtv\ \isacommand{obtain}\isamarkupfalse%
\ a\ B{\isacharprime}\ \isakeyword{where}\ va{\isacharcolon}\ {\isachardoublequoteopen}v\ {\isacharequal}\ VRef\ a{\isachardoublequoteclose}\isanewline
\ \ \ \ \ \ \isakeyword{and}\ las{\isacharcolon}\ {\isachardoublequoteopen}lookup\ a\ {\isasymSigma}\ {\isacharequal}\ Result\ B{\isacharprime}{\isachardoublequoteclose}\ \isakeyword{and}\ bb{\isacharcolon}\ {\isachardoublequoteopen}B{\isacharprime}\ {\isasymsqsubseteq}\ B{\isachardoublequoteclose}\ \isacommand{apply}\isamarkupfalse%
\ auto\ \isanewline
\ \ \ \ \ \ \isacommand{apply}\isamarkupfalse%
\ {\isacharparenleft}case{\isacharunderscore}tac\ c{\isacharparenright}\ \isacommand{apply}\isamarkupfalse%
\ auto\ \isacommand{done}\isamarkupfalse%
\isanewline
\ \ \ \ \isacommand{from}\isamarkupfalse%
\ las\ wt{\isacharunderscore}h\ Nil\ \isacommand{obtain}\isamarkupfalse%
\ v{\isadigit{2}}\ \isakeyword{where}\ lam{\isacharcolon}\ {\isachardoublequoteopen}lookup\ a\ {\isasymmu}\ {\isacharequal}\ Result\ {\isacharparenleft}Val\ v{\isadigit{2}}{\isacharcomma}B{\isacharprime}{\isacharparenright}{\isachardoublequoteclose}\isanewline
\ \ \ \ \ \ \isakeyword{and}\ wtv{\isadigit{2}}{\isacharcolon}\ {\isachardoublequoteopen}{\isasymSigma}\ {\isasymturnstile}v\ v{\isadigit{2}}\ {\isacharcolon}\ B{\isacharprime}{\isachardoublequoteclose}\ \isacommand{apply}\isamarkupfalse%
\ {\isacharparenleft}simp\ add{\isacharcolon}\ wt{\isacharunderscore}heap{\isacharunderscore}def{\isacharparenright}\ \isacommand{by}\isamarkupfalse%
\ blast\isanewline
\ \ \ \ \isacommand{from}\isamarkupfalse%
\ wtv{\isadigit{2}}\ wt{\isacharunderscore}h\ las\ \isacommand{have}\isamarkupfalse%
\ {\isachardoublequoteopen}{\isacharparenleft}{\isasymexists}v{\isacharprime}\ {\isasymSigma}{\isacharprime}\ {\isasymmu}{\isacharprime}\ ads{\isadigit{2}}{\isachardot}\isanewline
\ \ \ \ \ \ cast\ v{\isadigit{2}}\ B{\isacharprime}\ B\ {\isasymmu}\ ads\ {\isacharequal}\ Result\ {\isacharparenleft}v{\isacharprime}{\isacharcomma}\ {\isasymmu}{\isacharprime}{\isacharcomma}\ ads{\isadigit{2}}{\isacharparenright}\ {\isasymand}\isanewline
\ \ \ \ \ \ {\isasymSigma}{\isacharprime}\ {\isasymturnstile}v\ v{\isacharprime}\ {\isacharcolon}\ B\ {\isasymand}\ wt{\isacharunderscore}heap\ {\isasymSigma}{\isacharprime}\ {\isasymmu}{\isacharprime}\ {\isacharparenleft}set\ ads{\isadigit{2}}{\isacharparenright}\ {\isasymand}\ {\isasymSigma}{\isacharprime}\ {\isasymsqsubseteq}\ {\isasymSigma}{\isacharparenright}\ {\isasymor}\isanewline
\ \ \ \ \ \ cast\ v{\isadigit{2}}\ B{\isacharprime}\ B\ {\isasymmu}\ ads\ {\isacharequal}\ CastError{\isachardoublequoteclose}\ \isacommand{using}\isamarkupfalse%
\ cast{\isacharunderscore}safe\ \isacommand{by}\isamarkupfalse%
\ blast\isanewline
\ \ \ \ \isacommand{thus}\isamarkupfalse%
\ {\isacharquery}thesis\isanewline
\ \ \ \ \isacommand{proof}\isamarkupfalse%
\isanewline
\ \ \ \ \ \ \isacommand{assume}\isamarkupfalse%
\ {\isachardoublequoteopen}{\isacharparenleft}{\isasymexists}v{\isacharprime}\ {\isasymSigma}{\isacharprime}\ {\isasymmu}{\isacharprime}\ ads{\isadigit{2}}{\isachardot}\isanewline
\ \ \ \ \ \ \ \ cast\ v{\isadigit{2}}\ B{\isacharprime}\ B\ {\isasymmu}\ ads\ {\isacharequal}\ Result\ {\isacharparenleft}v{\isacharprime}{\isacharcomma}\ {\isasymmu}{\isacharprime}{\isacharcomma}\ ads{\isadigit{2}}{\isacharparenright}\ {\isasymand}\isanewline
\ \ \ \ \ \ \ \ {\isasymSigma}{\isacharprime}\ {\isasymturnstile}v\ v{\isacharprime}\ {\isacharcolon}\ B\ {\isasymand}\ wt{\isacharunderscore}heap\ {\isasymSigma}{\isacharprime}\ {\isasymmu}{\isacharprime}\ {\isacharparenleft}set\ ads{\isadigit{2}}{\isacharparenright}\ {\isasymand}\ {\isasymSigma}{\isacharprime}\ {\isasymsqsubseteq}\ {\isasymSigma}{\isacharparenright}{\isachardoublequoteclose}\isanewline
\ \ \ \ \ \ \isacommand{from}\isamarkupfalse%
\ this\ \isacommand{obtain}\isamarkupfalse%
\ v{\isacharprime}\ {\isasymSigma}{\isacharprime}\ {\isasymmu}{\isacharprime}\ ads{\isadigit{2}}\ \isakeyword{where}\isanewline
\ \ \ \ \ \ \ \ castv{\isadigit{2}}{\isacharcolon}\ {\isachardoublequoteopen}cast\ v{\isadigit{2}}\ B{\isacharprime}\ B\ {\isasymmu}\ ads\ {\isacharequal}\ Result\ {\isacharparenleft}v{\isacharprime}{\isacharcomma}\ {\isasymmu}{\isacharprime}{\isacharcomma}\ ads{\isadigit{2}}{\isacharparenright}{\isachardoublequoteclose}\isanewline
\ \ \ \ \ \ \ \ \isakeyword{and}\ wtvp{\isacharcolon}\ {\isachardoublequoteopen}{\isasymSigma}{\isacharprime}\ {\isasymturnstile}v\ v{\isacharprime}\ {\isacharcolon}\ B{\isachardoublequoteclose}\isanewline
\ \ \ \ \ \ \ \ \isakeyword{and}\ wth{\isadigit{2}}{\isacharcolon}\ {\isachardoublequoteopen}wt{\isacharunderscore}heap\ {\isasymSigma}{\isacharprime}\ {\isasymmu}{\isacharprime}\ {\isacharparenleft}set\ ads{\isadigit{2}}{\isacharparenright}{\isachardoublequoteclose}\isanewline
\ \ \ \ \ \ \ \ \isakeyword{and}\ ss{\isacharcolon}\ {\isachardoublequoteopen}{\isasymSigma}{\isacharprime}\ {\isasymsqsubseteq}\ {\isasymSigma}{\isachardoublequoteclose}\ \isacommand{by}\isamarkupfalse%
\ blast\isanewline
\ \ \ \ \ \ \isacommand{from}\isamarkupfalse%
\ st\ Nil\ SDynDeref\ castv{\isadigit{2}}\ v\ va\ lam\isanewline
\ \ \ \ \ \ \isacommand{have}\isamarkupfalse%
\ steps{\isacharcolon}\ {\isachardoublequoteopen}step\ s\ {\isacharequal}\ Result\ {\isacharparenleft}s{\isacharprime}{\isacharcomma}\ {\isacharparenleft}x{\isacharcomma}\ v{\isacharprime}{\isacharparenright}\ {\isacharhash}\ {\isasymrho}{\isacharcomma}\ k{\isacharcomma}\ {\isasymmu}{\isacharprime}{\isacharcomma}\ ads{\isadigit{2}}{\isacharparenright}{\isachardoublequoteclose}\ \isacommand{by}\isamarkupfalse%
\ simp\isanewline
\ \ \ \ \ \ \isacommand{from}\isamarkupfalse%
\ gr\ ss\ \isacommand{have}\isamarkupfalse%
\ gr{\isadigit{2}}{\isacharcolon}\ {\isachardoublequoteopen}{\isasymGamma}{\isacharsemicolon}\ {\isasymSigma}{\isacharprime}\ {\isasymturnstile}\ {\isasymrho}{\isachardoublequoteclose}\ \isacommand{using}\isamarkupfalse%
\ strengthen{\isacharunderscore}value{\isacharunderscore}env\ \isacommand{by}\isamarkupfalse%
\ blast\isanewline
\ \ \ \ \ \ \isacommand{from}\isamarkupfalse%
\ gr{\isadigit{2}}\ wtvp\ \isacommand{have}\isamarkupfalse%
\ gr{\isadigit{3}}{\isacharcolon}\ {\isachardoublequoteopen}{\isacharparenleft}x{\isacharcomma}B{\isacharparenright}{\isacharhash}{\isasymGamma}{\isacharsemicolon}{\isasymSigma}{\isacharprime}\ {\isasymturnstile}\ {\isacharparenleft}x{\isacharcomma}v{\isacharprime}{\isacharparenright}{\isacharhash}{\isasymrho}{\isachardoublequoteclose}\ \isacommand{by}\isamarkupfalse%
\ blast\isanewline
\ \ \ \ \ \ \isacommand{have}\isamarkupfalse%
\ wt{\isacharunderscore}s{\isacharcolon}\ {\isachardoublequoteopen}wt{\isacharunderscore}state\ {\isacharparenleft}s{\isacharprime}{\isacharcomma}\ {\isacharparenleft}x{\isacharcomma}\ v{\isacharprime}{\isacharparenright}\ {\isacharhash}\ {\isasymrho}{\isacharcomma}\ k{\isacharcomma}\ {\isasymmu}{\isacharprime}{\isacharcomma}\ ads{\isadigit{2}}{\isacharparenright}\ A{\isachardoublequoteclose}\ \isanewline
\ \ \ \ \ \ \ \ \isacommand{apply}\isamarkupfalse%
\ {\isacharparenleft}rule\ wts{\isacharunderscore}intro{\isacharparenright}\isanewline
\ \ \ \ \ \ \ \ \isacommand{using}\isamarkupfalse%
\ wth{\isadigit{2}}\ Nil\ \isacommand{apply}\isamarkupfalse%
\ simp\isanewline
\ \ \ \ \ \ \ \ \isacommand{using}\isamarkupfalse%
\ gr{\isadigit{3}}\ \isacommand{apply}\isamarkupfalse%
\ blast\isanewline
\ \ \ \ \ \ \ \ \isacommand{using}\isamarkupfalse%
\ wtsp\ \isacommand{apply}\isamarkupfalse%
\ blast\isanewline
\ \ \ \ \ \ \ \ \isacommand{using}\isamarkupfalse%
\ wt{\isacharunderscore}k\ ss\ strengthen{\isacharunderscore}stack\ \isacommand{apply}\isamarkupfalse%
\ blast\ \isacommand{done}\isamarkupfalse%
\isanewline
\ \ \ \ \ \ \isacommand{from}\isamarkupfalse%
\ steps\ wt{\isacharunderscore}s\ \isacommand{show}\isamarkupfalse%
\ {\isacharquery}thesis\ \isacommand{by}\isamarkupfalse%
\ simp\isanewline
\ \ \ \ \isacommand{next}\isamarkupfalse%
\isanewline
\ \ \ \ \ \ \isacommand{assume}\isamarkupfalse%
\ {\isachardoublequoteopen}cast\ v{\isadigit{2}}\ B{\isacharprime}\ B\ {\isasymmu}\ ads\ {\isacharequal}\ CastError{\isachardoublequoteclose}\isanewline
\ \ \ \ \ \ \isacommand{from}\isamarkupfalse%
\ st\ Nil\ SDynDeref\ this\ v\ va\ lam\ \isacommand{show}\isamarkupfalse%
\ {\isacharquery}thesis\ \isacommand{by}\isamarkupfalse%
\ simp\isanewline
\ \ \ \ \isacommand{qed}\isamarkupfalse%
\isanewline
\ \ \isacommand{qed}\isamarkupfalse%
\isanewline
\isacommand{qed}\isamarkupfalse%
\isanewline
\isacommand{qed}\isamarkupfalse%
\endisatagproof
{\isafoldproof}%
\isadelimproof
\isanewline
\endisadelimproof
\isanewline
\isacommand{lemma}\isamarkupfalse%
\ observe{\isacharunderscore}safe{\isacharcolon}\isanewline
\ \ \isakeyword{assumes}\ wtv{\isacharcolon}\ {\isachardoublequoteopen}{\isasymSigma}\ {\isasymturnstile}v\ v\ {\isacharcolon}\ A{\isachardoublequoteclose}\isanewline
\ \ \isakeyword{shows}\ {\isachardoublequoteopen}wt{\isacharunderscore}observable\ {\isacharparenleft}observe\ v{\isacharparenright}\ A{\isachardoublequoteclose}\isanewline
\isadelimproof
\ \ %
\endisadelimproof
\isatagproof
\isacommand{using}\isamarkupfalse%
\ wtv\ \isacommand{apply}\isamarkupfalse%
\ {\isacharparenleft}induct\ v\ arbitrary{\isacharcolon}\ {\isasymSigma}\ A{\isacharparenright}\isanewline
\ \ \isacommand{apply}\isamarkupfalse%
\ {\isacharparenleft}case{\isacharunderscore}tac\ const{\isacharparenright}\ \isacommand{apply}\isamarkupfalse%
\ force{\isacharplus}\ \isacommand{done}\isamarkupfalse%
\endisatagproof
{\isafoldproof}%
\isadelimproof
\endisadelimproof
\begin{isamarkuptext}%
For this lemma, we choose not to use the induction rule for steps
  because that induction rule is a bit messy, with lots of cases
  that can be dealt with in a similar fashion. In the following,
  we just do proof by induction on n.%
\end{isamarkuptext}%
\isamarkuptrue%
\isacommand{lemma}\isamarkupfalse%
\ steps{\isacharunderscore}safe{\isacharcolon}\isanewline
\ \ \isakeyword{assumes}\ wtsA{\isacharcolon}\ {\isachardoublequoteopen}wt{\isacharunderscore}state\ s\ A{\isachardoublequoteclose}\isanewline
\ \ \isakeyword{shows}\ {\isachardoublequoteopen}{\isasymexists}\ r{\isachardot}\ steps\ n\ s\ {\isacharequal}\ r\ {\isasymand}\ wt{\isacharunderscore}observable\ r\ A{\isachardoublequoteclose}\isanewline
\isadelimproof
\ \ %
\endisadelimproof
\isatagproof
\isacommand{using}\isamarkupfalse%
\ wtsA\isanewline
\isacommand{proof}\isamarkupfalse%
\ {\isacharparenleft}induct\ n\ arbitrary{\isacharcolon}\ s{\isacharparenright}\isanewline
\ \ \isacommand{fix}\isamarkupfalse%
\ s\ \isacommand{have}\isamarkupfalse%
\ {\isachardoublequoteopen}steps\ {\isadigit{0}}\ s\ {\isacharequal}\ OTimeOut{\isachardoublequoteclose}\ \isacommand{by}\isamarkupfalse%
\ simp\isanewline
\ \ \isacommand{thus}\isamarkupfalse%
\ {\isachardoublequoteopen}{\isasymexists}r{\isachardot}\ steps\ {\isadigit{0}}\ s\ {\isacharequal}\ r\ {\isasymand}\ wt{\isacharunderscore}observable\ r\ A{\isachardoublequoteclose}\ \isacommand{by}\isamarkupfalse%
\ auto\isanewline
\isacommand{next}\isamarkupfalse%
\isanewline
\ \ \isacommand{fix}\isamarkupfalse%
\ n\ s\isanewline
\ \ \isacommand{assume}\isamarkupfalse%
\ IH{\isacharcolon}\ {\isachardoublequoteopen}{\isasymAnd}s{\isachardot}\ wt{\isacharunderscore}state\ s\ A\ {\isasymLongrightarrow}\isanewline
\ \ \ \ \ \ \ \ \ \ \ \ \ \ \ {\isacharparenleft}{\isasymexists}r{\isachardot}\ steps\ n\ s\ {\isacharequal}\ r\ {\isasymand}\ wt{\isacharunderscore}observable\ r\ A{\isacharparenright}{\isachardoublequoteclose}\isanewline
\ \ \ \ \isakeyword{and}\ wts{\isacharcolon}\ {\isachardoublequoteopen}wt{\isacharunderscore}state\ s\ A{\isachardoublequoteclose}\isanewline
\ \ \isacommand{{\isacharbraceleft}}\isamarkupfalse%
\ \isacommand{assume}\isamarkupfalse%
\ {\isachardoublequoteopen}final\ s{\isachardoublequoteclose}\isanewline
\ \ \ \ \isacommand{from}\isamarkupfalse%
\ this\ \isacommand{obtain}\isamarkupfalse%
\ e\ {\isasymrho}\ {\isasymmu}\ \isakeyword{where}\ s{\isacharcolon}\ {\isachardoublequoteopen}s\ {\isacharequal}\ {\isacharparenleft}SRet\ e{\isacharcomma}\ {\isasymrho}{\isacharcomma}\ {\isacharbrackleft}{\isacharbrackright}{\isacharcomma}\ {\isasymmu}{\isacharcomma}{\isacharbrackleft}{\isacharbrackright}{\isacharparenright}{\isachardoublequoteclose}\isanewline
\ \ \ \ \ \ \isacommand{apply}\isamarkupfalse%
\ {\isacharparenleft}case{\isacharunderscore}tac\ s{\isacharparenright}\ \isacommand{apply}\isamarkupfalse%
\ {\isacharparenleft}case{\isacharunderscore}tac\ a{\isacharparenright}\ \isacommand{apply}\isamarkupfalse%
\ auto\isanewline
\ \ \ \ \ \ \isacommand{apply}\isamarkupfalse%
\ {\isacharparenleft}case{\isacharunderscore}tac\ c{\isacharparenright}\ \isacommand{apply}\isamarkupfalse%
\ auto\ \isacommand{apply}\isamarkupfalse%
\ {\isacharparenleft}case{\isacharunderscore}tac\ e{\isacharparenright}\ \isacommand{apply}\isamarkupfalse%
\ auto\ \isacommand{done}\isamarkupfalse%
\isanewline
\ \ \ \ \isacommand{from}\isamarkupfalse%
\ wts\ s\ \isacommand{obtain}\isamarkupfalse%
\ {\isasymGamma}\ {\isasymSigma}\ \isakeyword{where}\ wte{\isacharcolon}\ {\isachardoublequoteopen}{\isasymGamma}\ {\isasymturnstile}\isactrlisub e\ e\ {\isacharcolon}\ A{\isachardoublequoteclose}\ \isakeyword{and}\ wtg{\isacharcolon}\ {\isachardoublequoteopen}{\isasymGamma}{\isacharsemicolon}{\isasymSigma}\ {\isasymturnstile}\ {\isasymrho}{\isachardoublequoteclose}\ \isanewline
\ \ \ \ \ \ \isakeyword{and}\ wth{\isacharcolon}\ {\isachardoublequoteopen}wt{\isacharunderscore}heap\ {\isasymSigma}\ {\isasymmu}\ {\isacharbraceleft}{\isacharbraceright}{\isachardoublequoteclose}\ \isacommand{by}\isamarkupfalse%
\ auto\isanewline
\ \ \ \ \isacommand{from}\isamarkupfalse%
\ wte\ wtg\ wth\ \isacommand{have}\isamarkupfalse%
\ {\isachardoublequoteopen}{\isacharparenleft}{\isasymexists}v{\isachardot}\ eval\ e\ {\isasymrho}\ {\isasymmu}\ {\isacharequal}\ Result\ v\ {\isasymand}\ {\isasymSigma}\ {\isasymturnstile}v\ v\ {\isacharcolon}\ A{\isacharparenright}\isanewline
\ \ \ \ \ \ {\isasymor}\ eval\ e\ {\isasymrho}\ {\isasymmu}\ {\isacharequal}\ CastError{\isachardoublequoteclose}\isanewline
\ \ \ \ \ \ \isacommand{using}\isamarkupfalse%
\ eval{\isacharunderscore}safe{\isacharbrackleft}of\ {\isasymGamma}\ e\ A\ {\isasymSigma}\ {\isasymrho}\ {\isasymmu}{\isacharbrackright}\ \isacommand{by}\isamarkupfalse%
\ simp\isanewline
\ \ \ \ \isacommand{hence}\isamarkupfalse%
\ {\isachardoublequoteopen}{\isasymexists}r{\isachardot}\ steps\ {\isacharparenleft}Suc\ n{\isacharparenright}\ s\ {\isacharequal}\ r\ {\isasymand}\ wt{\isacharunderscore}observable\ r\ A{\isachardoublequoteclose}\isanewline
\ \ \ \ \isacommand{proof}\isamarkupfalse%
\isanewline
\ \ \ \ \ \ \isacommand{assume}\isamarkupfalse%
\ {\isachardoublequoteopen}{\isasymexists}v{\isachardot}\ eval\ e\ {\isasymrho}\ {\isasymmu}\ {\isacharequal}\ Result\ v\ {\isasymand}\ {\isasymSigma}\ {\isasymturnstile}v\ v\ {\isacharcolon}\ A{\isachardoublequoteclose}\isanewline
\ \ \ \ \ \ \isacommand{from}\isamarkupfalse%
\ this\ \isacommand{obtain}\isamarkupfalse%
\ v\ \isakeyword{where}\ ev{\isacharcolon}\ {\isachardoublequoteopen}eval\ e\ {\isasymrho}\ {\isasymmu}\ {\isacharequal}\ Result\ v{\isachardoublequoteclose}\ \isanewline
\ \ \ \ \ \ \ \ \isakeyword{and}\ wtv{\isacharcolon}\ {\isachardoublequoteopen}{\isasymSigma}\ {\isasymturnstile}v\ v\ {\isacharcolon}\ A{\isachardoublequoteclose}\ \isacommand{by}\isamarkupfalse%
\ blast\isanewline
\ \ \ \ \ \ \isacommand{from}\isamarkupfalse%
\ wtv\ \isacommand{have}\isamarkupfalse%
\ {\isachardoublequoteopen}wt{\isacharunderscore}observable\ {\isacharparenleft}observe\ v{\isacharparenright}\ A{\isachardoublequoteclose}\ \isacommand{using}\isamarkupfalse%
\ observe{\isacharunderscore}safe\ \isacommand{by}\isamarkupfalse%
\ blast\isanewline
\ \ \ \ \ \ \isacommand{with}\isamarkupfalse%
\ s\ ev\ \isacommand{show}\isamarkupfalse%
\ {\isachardoublequoteopen}{\isasymexists}r{\isachardot}\ steps\ {\isacharparenleft}Suc\ n{\isacharparenright}\ s\ {\isacharequal}\ r\ {\isasymand}\ wt{\isacharunderscore}observable\ r\ A{\isachardoublequoteclose}\ \isacommand{by}\isamarkupfalse%
\ auto\isanewline
\ \ \ \ \isacommand{next}\isamarkupfalse%
\isanewline
\ \ \ \ \ \ \isacommand{assume}\isamarkupfalse%
\ {\isachardoublequoteopen}eval\ e\ {\isasymrho}\ {\isasymmu}\ {\isacharequal}\ CastError{\isachardoublequoteclose}\isanewline
\ \ \ \ \ \ \isacommand{with}\isamarkupfalse%
\ s\ \isacommand{show}\isamarkupfalse%
\ {\isacharquery}thesis\ \isacommand{apply}\isamarkupfalse%
\ simp\ \isacommand{done}\isamarkupfalse%
\isanewline
\ \ \ \ \isacommand{qed}\isamarkupfalse%
\isanewline
\ \ \isacommand{{\isacharbraceright}}\isamarkupfalse%
\ \isacommand{moreover}\isamarkupfalse%
\ \isacommand{{\isacharbraceleft}}\isamarkupfalse%
\ \isacommand{assume}\isamarkupfalse%
\ fs{\isacharcolon}\ {\isachardoublequoteopen}{\isasymnot}\ final\ s{\isachardoublequoteclose}\isanewline
\ \ \ \ \isacommand{from}\isamarkupfalse%
\ fs\ wts\ \isacommand{have}\isamarkupfalse%
\ {\isachardoublequoteopen}{\isacharparenleft}{\isasymexists}s{\isacharprime}{\isachardot}\ step\ s\ {\isacharequal}\ Result\ s{\isacharprime}\ {\isasymand}\ wt{\isacharunderscore}state\ s{\isacharprime}\ A{\isacharparenright}\isanewline
\ \ \ \ \ \ {\isasymor}\ step\ s\ {\isacharequal}\ CastError{\isachardoublequoteclose}\ \isacommand{using}\isamarkupfalse%
\ step{\isacharunderscore}safe{\isacharbrackleft}of\ s\ A{\isacharbrackright}\ \isacommand{by}\isamarkupfalse%
\ simp\isanewline
\ \ \ \ \isacommand{hence}\isamarkupfalse%
\ {\isachardoublequoteopen}{\isasymexists}r{\isachardot}\ steps\ {\isacharparenleft}Suc\ n{\isacharparenright}\ s\ {\isacharequal}\ r\ {\isasymand}\ wt{\isacharunderscore}observable\ r\ A{\isachardoublequoteclose}\isanewline
\ \ \ \ \isacommand{proof}\isamarkupfalse%
\isanewline
\ \ \ \ \ \ \isacommand{assume}\isamarkupfalse%
\ {\isachardoublequoteopen}{\isasymexists}s{\isacharprime}{\isachardot}\ step\ s\ {\isacharequal}\ Result\ s{\isacharprime}\ {\isasymand}\ wt{\isacharunderscore}state\ s{\isacharprime}\ A{\isachardoublequoteclose}\isanewline
\ \ \ \ \ \ \isacommand{from}\isamarkupfalse%
\ this\ \isacommand{obtain}\isamarkupfalse%
\ s{\isacharprime}\ \isakeyword{where}\ st{\isacharcolon}\ {\isachardoublequoteopen}step\ s\ {\isacharequal}\ Result\ s{\isacharprime}{\isachardoublequoteclose}\ \isanewline
\ \ \ \ \ \ \ \ \isakeyword{and}\ wtsp{\isacharcolon}\ {\isachardoublequoteopen}wt{\isacharunderscore}state\ s{\isacharprime}\ A{\isachardoublequoteclose}\ \isacommand{by}\isamarkupfalse%
\ blast\isanewline
\ \ \ \ \ \ \isacommand{from}\isamarkupfalse%
\ wtsp\ IH\ \isacommand{have}\isamarkupfalse%
\ ssp{\isacharcolon}\isanewline
\ \ \ \ \ \ \ \ {\isachardoublequoteopen}{\isasymexists}r{\isachardot}\ steps\ n\ s{\isacharprime}\ {\isacharequal}\ r\ {\isasymand}\ wt{\isacharunderscore}observable\ r\ A{\isachardoublequoteclose}\ \isacommand{by}\isamarkupfalse%
\ blast\isanewline
\ \ \ \ \ \ \isacommand{from}\isamarkupfalse%
\ fs\ st\ \isacommand{have}\isamarkupfalse%
\ ss{\isacharcolon}\ {\isachardoublequoteopen}steps\ {\isacharparenleft}Suc\ n{\isacharparenright}\ s\ {\isacharequal}\ steps\ n\ s{\isacharprime}{\isachardoublequoteclose}\isanewline
\ \ \ \ \ \ \ \ \isacommand{apply}\isamarkupfalse%
\ auto\ \isacommand{apply}\isamarkupfalse%
\ {\isacharparenleft}case{\isacharunderscore}tac\ a{\isacharparenright}\ \isacommand{apply}\isamarkupfalse%
\ auto\ \isacommand{apply}\isamarkupfalse%
\ {\isacharparenleft}case{\isacharunderscore}tac\ ab{\isacharparenright}\ \isacommand{apply}\isamarkupfalse%
\ auto\isanewline
\ \ \ \ \ \ \ \ \isacommand{apply}\isamarkupfalse%
\ {\isacharparenleft}case{\isacharunderscore}tac\ b{\isacharparenright}\ \isacommand{apply}\isamarkupfalse%
\ auto\ \isacommand{done}\isamarkupfalse%
\isanewline
\ \ \ \ \ \ \isacommand{from}\isamarkupfalse%
\ ssp\ ss\ \isacommand{show}\isamarkupfalse%
\ {\isacharquery}thesis\ \isacommand{by}\isamarkupfalse%
\ simp\isanewline
\ \ \ \ \isacommand{next}\isamarkupfalse%
\isanewline
\ \ \ \ \ \ \isacommand{assume}\isamarkupfalse%
\ {\isachardoublequoteopen}step\ s\ {\isacharequal}\ CastError{\isachardoublequoteclose}\isanewline
\ \ \ \ \ \ \isacommand{from}\isamarkupfalse%
\ fs\ this\ \isacommand{show}\isamarkupfalse%
\ {\isacharquery}thesis\ \isacommand{apply}\isamarkupfalse%
\ simp\isanewline
\ \ \ \ \ \ \ \ \isacommand{apply}\isamarkupfalse%
\ {\isacharparenleft}case{\isacharunderscore}tac\ s{\isacharparenright}\ \isacommand{apply}\isamarkupfalse%
\ simp\ \isacommand{apply}\isamarkupfalse%
\ {\isacharparenleft}case{\isacharunderscore}tac\ a{\isacharparenright}\ \isacommand{apply}\isamarkupfalse%
\ auto\isanewline
\ \ \ \ \ \ \ \ \isacommand{apply}\isamarkupfalse%
\ {\isacharparenleft}case{\isacharunderscore}tac\ c{\isacharparenright}\ \isacommand{apply}\isamarkupfalse%
\ auto\ \ \isacommand{apply}\isamarkupfalse%
\ {\isacharparenleft}case{\isacharunderscore}tac\ e{\isacharparenright}\ \isacommand{apply}\isamarkupfalse%
\ auto\ \isacommand{done}\isamarkupfalse%
\isanewline
\ \ \ \ \isacommand{qed}\isamarkupfalse%
\isanewline
\ \ \isacommand{{\isacharbraceright}}\isamarkupfalse%
\ \isacommand{ultimately}\isamarkupfalse%
\ \isacommand{show}\isamarkupfalse%
\ {\isachardoublequoteopen}{\isasymexists}r{\isachardot}\ steps\ {\isacharparenleft}Suc\ n{\isacharparenright}\ s\ {\isacharequal}\ r\ {\isasymand}\ wt{\isacharunderscore}observable\ r\ A{\isachardoublequoteclose}\ \isacommand{by}\isamarkupfalse%
\ blast\isanewline
\isacommand{qed}\isamarkupfalse%
\endisatagproof
{\isafoldproof}%
\isadelimproof
\isanewline
\endisadelimproof
\isanewline
\isacommand{theorem}\isamarkupfalse%
\ type{\isacharunderscore}safety{\isacharcolon}\isanewline
\ \ \isakeyword{assumes}\ wts{\isacharcolon}\ {\isachardoublequoteopen}{\isacharbrackleft}{\isacharbrackright}\ {\isasymturnstile}\isactrlisub s\ s\ {\isacharcolon}\ A{\isachardoublequoteclose}\isanewline
\ \ \isakeyword{shows}\ {\isachardoublequoteopen}{\isasymexists}\ r{\isachardot}\ run\ s\ {\isacharequal}\ r\ {\isasymand}\ wt{\isacharunderscore}observable\ r\ A{\isachardoublequoteclose}\isanewline
\isadelimproof
\endisadelimproof
\isatagproof
\isacommand{proof}\isamarkupfalse%
\ {\isacharminus}\isanewline
\ \ \isacommand{have}\isamarkupfalse%
\ wtg{\isacharcolon}\ {\isachardoublequoteopen}{\isacharbrackleft}{\isacharbrackright}{\isacharsemicolon}{\isacharbrackleft}{\isacharbrackright}\ {\isasymturnstile}\ {\isacharbrackleft}{\isacharbrackright}{\isachardoublequoteclose}\ \isacommand{by}\isamarkupfalse%
\ fast\isanewline
\ \ \isacommand{from}\isamarkupfalse%
\ wtg\ wts\ \isacommand{have}\isamarkupfalse%
\ {\isadigit{1}}{\isacharcolon}\ {\isachardoublequoteopen}wt{\isacharunderscore}state\ {\isacharparenleft}s{\isacharcomma}{\isacharbrackleft}{\isacharbrackright}{\isacharcomma}{\isacharbrackleft}{\isacharbrackright}{\isacharcomma}{\isacharbrackleft}{\isacharbrackright}{\isacharcomma}{\isacharbrackleft}{\isacharbrackright}{\isacharparenright}\ A{\isachardoublequoteclose}\isanewline
\ \ \ \ \isacommand{by}\isamarkupfalse%
\ {\isacharparenleft}auto\ simp{\isacharcolon}\ wt{\isacharunderscore}heap{\isacharunderscore}def\ dom{\isacharunderscore}def{\isacharparenright}\isanewline
\ \ \isacommand{let}\isamarkupfalse%
\ {\isacharquery}n\ {\isacharequal}\ {\isachardoublequoteopen}{\isadigit{1}}{\isadigit{0}}{\isadigit{0}}{\isadigit{0}}{\isadigit{0}}{\isadigit{0}}{\isadigit{0}}{\isachardoublequoteclose}\ \isakeyword{and}\ {\isacharquery}s\ {\isacharequal}\ {\isachardoublequoteopen}{\isacharparenleft}s{\isacharcomma}{\isacharbrackleft}{\isacharbrackright}{\isacharcomma}{\isacharbrackleft}{\isacharbrackright}{\isacharcomma}{\isacharbrackleft}{\isacharbrackright}{\isacharcomma}{\isacharbrackleft}{\isacharbrackright}{\isacharparenright}{\isachardoublequoteclose}\isanewline
\ \ \isacommand{from}\isamarkupfalse%
\ {\isadigit{1}}\ \isacommand{have}\isamarkupfalse%
\ {\isadigit{2}}{\isacharcolon}\ {\isachardoublequoteopen}{\isasymexists}\ r{\isachardot}\ steps\ {\isacharquery}n\ {\isacharquery}s\ {\isacharequal}\ r\ {\isasymand}\ wt{\isacharunderscore}observable\ r\ A{\isachardoublequoteclose}\ \isacommand{by}\isamarkupfalse%
\ {\isacharparenleft}rule\ steps{\isacharunderscore}safe{\isacharparenright}\isanewline
\ \ \isacommand{from}\isamarkupfalse%
\ {\isadigit{2}}\ \isacommand{obtain}\isamarkupfalse%
\ r\ \isakeyword{where}\ {\isadigit{4}}{\isacharcolon}\ {\isachardoublequoteopen}steps\ {\isacharquery}n\ {\isacharquery}s\ {\isacharequal}\ r{\isachardoublequoteclose}\ \isakeyword{and}\ {\isadigit{5}}{\isacharcolon}\ {\isachardoublequoteopen}wt{\isacharunderscore}observable\ r\ A{\isachardoublequoteclose}\ \isacommand{by}\isamarkupfalse%
\ blast\isanewline
\ \ \isacommand{from}\isamarkupfalse%
\ {\isadigit{4}}\ \isacommand{have}\isamarkupfalse%
\ {\isadigit{6}}{\isacharcolon}\ {\isachardoublequoteopen}run\ s\ {\isacharequal}\ r{\isachardoublequoteclose}\ \isacommand{by}\isamarkupfalse%
\ {\isacharparenleft}simp\ add{\isacharcolon}\ run{\isacharunderscore}def{\isacharparenright}\isanewline
\ \ \isacommand{from}\isamarkupfalse%
\ {\isadigit{6}}\ {\isadigit{5}}\ \isacommand{show}\isamarkupfalse%
\ {\isachardoublequoteopen}{\isasymexists}\ r{\isachardot}\ run\ s\ {\isacharequal}\ r\ {\isasymand}\ wt{\isacharunderscore}observable\ r\ A{\isachardoublequoteclose}\ \isacommand{by}\isamarkupfalse%
\ blast\isanewline
\isacommand{qed}\isamarkupfalse%
\endisatagproof
{\isafoldproof}%
\isadelimproof
\isanewline
\endisadelimproof
\isadelimtheory
\isanewline
\endisadelimtheory
\isatagtheory
\isacommand{end}\isamarkupfalse%
\endisatagtheory
{\isafoldtheory}%
\isadelimtheory
\endisadelimtheory
\end{isabellebody}%